\documentclass[12pt]{iopart}

\usepackage{graphicx}
\input epsf
\input iopams.sty

\begin{document}

\jl{19}

\review{Extrinsic Magnetotransport Phenomena in Ferromagnetic Oxides}

\author{M Ziese \footnote{Present address: Department of
Superconductivity and Magnetism, University of Leipzig,
Linn\'estrasse 5, 04103 Leipzig, Germany}}

\address{Department of Physics and Astronomy, University of Sheffield, 
Sheffield S3 7RH, United Kingdom}

\date{\today}

\ead{ziese@physik.uni-leipzig.de}

\begin{abstract}
Magnetic oxides show a variety of extrinsic magnetotransport
phenomena: grain-boundary--, tunnelling-- and domain-wall
magnetoresistance. In view of these phenomena the role of some
magnetic oxides is outstanding: these are believed to be
half-metallic having only one spin-subband at the Fermi level. These
fully spin-polarized oxides have great potential for applications in
spin-electronic devices and have, accordingly, attracted an intense
research activity in recent years.

This review is focused on extrinsic magnetotransport effects in
ferromagnetic oxides. It consists of two parts; the second part is
devoted to an overview of experimental data and theoretical models for
extrinsic magnetotransport phenomena. Here a critical discussion of
domain-wall scattering is given. Results on surfacial and interfacial
magnetism in oxides are presented. Spin-polarized tunnelling in
ferromagnetic junctions is reviewed and grain-boundary
magnetoresistance is interpreted within a model of spin-polarized
tunnelling through natural oxide barriers. The situation in
ferromagnetic oxides is compared with data and models for
conventional ferromagnets. The first part of the review summarizes
basic material properties, especially data on the spin-polarization
and evidence for half-metallicity. Furthermore, intrinsic conduction
mechanisms are discussed. An outlook on the further development of
oxide spin-electronics concludes this review.
\end{abstract}
\submitted

\maketitle

\tableofcontents

\clearpage

\section{Introduction \label{intro}}
Although the study of magnetoresistance in ferromagnets started as
early as 1857 with the measurements of anisotropic magnetoresistance
in nickel and iron by William Thomson (Lord Kelvin), recent years have
witnessed a tremendous interest into magnetotransport phenomena in
magnetic oxides. Studies were stimulated by the discovery of
``colossal magnetoresistance'' (CMR) in ferromagnetic perovskites 
of the type $\rm La_{1-x}Sr_xMnO_3$. The so-called manganites display a rich
phase diagram as a function of temperature, magnetic field and doping
that is due to the intricate interplay of charge, spin, orbital and
lattice degrees of freedom. Colossal magnetoresistance is found on a
magnetic field scale of several Teslas being not very appealing for
applications. Accordingly, many research groups focused on the
investigation of extrinsic magnetoresistance effects found in various
magnetic oxides, since these promised a large magnetoresistance ratio
in low magnetic fields. To a large extent this research is driven by
the rapid increase of data storage density in magnetic storage
devices. Since read heads for hard disks employ magnetoresistive
read-out techniques, progressive miniaturization of sensors requires
materials or heterostructures with increasing magnetoresistive
effect. The development of hard disk storage media is currently
very rapid with a doubling of storage density about every nine
months. Therefore, the need for more efficient magnetoresistive
sensors will persist in the future. It has to be clear that room
temperature performance is the most vital criterion in judging new
magnetoresistive materials.

In this review intrinsic and extrinsic magnetoresistive effects are
distinguished. Whereas intrinsic effects are found in the bulk of the
ferromagnetic material and are determined by material parameters,
extrinsic effects are only found at defect structures, in suitable
artificial heterostructures and devices. This distinction is not
unique, since impurity scattering plays a vital role in some of the
effects and is per se an extrinsic effect. 

Extrinsic magnetotransport effects in ferromagnetic oxides fall into
three broad classes, namely grain-boundary magnetoresistance, 
spin-polarized transport in ferromagnetic tunnelling junctions 
and domain-wall magnetoresistance. It
was realized that manganite samples containing a large number of
extended defects such as grain boundaries display a huge low field
magnetoresistance much larger than the intrinsic magnetoresistance
(Hwang \etal 1996, Gupta \etal 1996). Various samples have been
studied, namely polycrystalline ceramics and films, pressed powders,
single grain boundaries on bi-crystal substrates, scratched
substrates, step-edge and laser-patterned junctions. The extrinsic
effect therefore appears under various names such as grain-boundary
magnetoresistance, junction magnetoresistance (JMR) and powder
magnetoresistance (PMR). However, the basic physical mechanism behind
these phenomena appears to be the same. This mechanism is likely to be
spin-polarized tunnelling, although a final consensus on the
interpretaion of grain boundary magnetoresistance has not yet been
reached; this conclusion links the investigations of transport
properties near extended defects to another class of extrinsic
magnetoresistance: spin-polarized tunnelling in
heterostructures. Such a heterostructure consists of two ferromagnetic
layers separated by an insulating layer. The tunnelling current is
found to depend sensitively on the relative magnetization direction in the
ferromagnetic electrodes. This is related to the spin-dependent
density of states at the Fermi level. It is useful to define a
spin-polarization at the Fermi level that quantifies the
spin-imbalance in itinerant ferromagnets; the elemental ferromagnets
Fe, Co and Ni have spin-polarizations of 45\%, 42\% and 31\%,
respectively. Spin-polarized tunnelling in a thin film heterostructure
and the basic theory was first discussed by Julliere (1975). However,
an even earlier work on magnetic tunnelling junctions using manganite
electrodes was reported by van den Brom and Volger (1968). Early
experiments yielded only small magnetoresistive effects and the
reproducibility of the junctions was poor. A breakthrough was achieved
by the work of Moodera \etal (1995) reporting magnetoresistance ratios
in excess of 20\% in permalloy/$\rm Al_2O_3$/CoFe junctions at low
temperature in agreement with Julliere's model; at room temperature
effects of about 10\% were recorded. Experimental work on manganite
tunnelling junctions started at about the same time (Sun \etal
1996). These studies were stimulated by band-structure calculations
that indicated the half-metallic nature of ferromagnetic manganite
oxides. 

The concept of a half-metal was introduced by de Groot \etal
(1983) motivated by band-structure calculations of Mn-based Heusler
alloys; a half-metal is defined as having a metallic density of states
in one of the spin channels and a gap in the density of states in the
other spin channel; within any reasonable definition of
spin-polarization this corresponds to a spin-polarization value of
100\%. The concept of spin-polarization is central to the theory of
spin-polarized tunnelling; the only parameters entering Julliere's
model for the tunnelling magnetoresistance are the spin-polarizations
of the ferromagnetic electrodes; in the case of half-metallic metals
the magnetoresistance ratio is expected to reach 100\%. This is
intuitively clear, since in the absence of spin-flips, a tunnelling
current cannot flow between half-metallic electrodes with antiparallel
magnetization directions, since there is no density of states at the
Fermi level of the other electrode into which electrons could be
tunnelling. Therefore, spin-polarized tunnelling in heterostructures
using half-metallic electrodes is promising for applications and, at
the same time, theoretically challenging, since the dependence of the
tunnelling current on the band structure and barrier properties is not
yet fully understood. The first experiments on
manganite-insulator-manganite tunnelling junctions showed large
magnetotunnelling effects at low temperatures (Lu \etal 1996). The
magnetoresistance ratio, however, decreases rapidly with
temperature. A future experimental challenge is the achievement of a
huge room temperature magnetoresistance in magnetic oxide
heterostructures. Apart from the manganites, CrO$_2$, magnetite 
($\rm Fe_3O_4$) and $\rm Sr_2MoFeO_6$ are suspected to be
half-metallic magnets and are investigated in view of device
applications.  

A long-standing problem is the question of the magnetoresistance of a
domain wall. Within a domain the magnetization vector and therefore
the majority and minority currents have spatially independent
orientations. Accordingly, when charge carriers cross a domain wall,
majority electrons may end up in the minority channel and vice versa
leading to an additional magnetoresistance. An obvious measure of the
scattering efficiency of a domain wall is the ratio between the time
of passage through the domain wall and the precession period of the
spin around the exchange field. In the adiabatic limit scattering
effects are likely to be small, whereas these are expected to become
appreciable in the ballistic regime.

This review is divided into two parts. The first part starts with
a brief overview over the materials as well as the definition and
measurement of the spin-polarization in section~\ref{materials}. 
Section~\ref{optical} deals with the temperature dependent 
resistivity and the optical conductivity, whereas the intrinsic
magnetoresistance is analyzed in section~\ref{intrinsic}. These
sections were included in order to elucidate the basic models for
magnetotransport in the relevant oxides and to facilitate comparison
with and identification of extrinsic magnetotransport phenomena. The
presentation is restricted to a discussion of resistivity as the basic
transport coefficient and does not include other valuable transport
coefficients such as thermopower, Hall effect, etc. This is due to
space limitations and the idea to use this review as a handbook in the
evaluation of resistivity behaviours encountered in studies of
magnetic oxides. In the second part extrinsic magnetotransport
phenomena are reviewed. An overview of recent developments in the
study of domain-wall scattering in both elemental ferromagnets and the
manganites is given in section~\ref{dw}. Section~\ref{LCMOmetal}
presents a discussion of various interface properties between magnetic
oxides and normal metals, superconductors and conventional
ferromagnets. Sections~\ref{tunnelling} and \ref{JMR} contain the main
experimental data and theoretical models relating to
spin-polarized tunnelling and grain-boundary magnetoresistance,
respectively. In the section on spin-polarized tunnelling appropriate
room is given to the discussion of recent results for tunnelling
junctions made from elemental ferromagnets in order to constitute
background information for the discussion of the oxide magnets. The
discussion of grain-boundary magnetoresistance concentrates on
magnetic oxides, since grain-boundaries of elemental ferromagnets do
not form junctions with tunnelling-like characteristics. In
section~\ref{summary} a summary of significant results and an outlook
into the budding field of spin-electronics is given. 

The bulk of recent investigations on the magnetotransport properties
of oxides deals with CMR materials. The number of publications on this
subject grows exponentially and it is virtually impossible to write a
fully comprehensive review. Therefore, it might be helpful to point
out some review articles on related topics. An excellent and extensive
review on the physical properties of mixed-valence manganites was written by Coey,
Viret and von Moln\'ar (1999); this review is mainly focused on
intrinsic properties. Two further extensive reviews on intrinsic conduction
mechanisms by Dagotto, Hotta and Moreo (2001) and Nagaev (2001)
focused on phase separation and the magnetoimpurity theory,
respectively. Short reviews on experimental work
and the basic theory of colossal magnetoresistance were published by
Ramirez (1997), Fontcuberta (1999) and Tokura and Tomioka
(1999). Jaime and Salamon (1999) discussed intrinsic transport
properties of the manganites and Lyanda-Geller \etal (2001) Hall
effect studies. The broad field of metal-insulator transitions was
addressed by Imada, Fujimori and Tokura (1998). An overview of
investigations on extrinsic magnetoresistance was compiled by Gupta
and Sun (1999). A review of theoretical studies of double exchange
systems was given by Furukawa (1998). Ziese (2000a) reviewed recent
developments of oxide spin-electronics. A review on spin-dependent
transport in magnetic nanostructures covering a few results on oxides
was given by Ansermet (1998). Readers interested in spin-polarized
tunnelling in conventional metallic systems are referred to Moodera
and Mathon (1999). A comprehensive survey of spintronics can be found
in the book ``Spin Electronics'' edited by M Ziese and M J Thornton.
\section*{Preliminaries: Materials and intrinsic magnetotransport properties} 
\addcontentsline{toc}{section}
{\protect{Part 1: Preliminaries: Materials and intrinsic magnetotransport  properties}}
\section{Materials, half-metallicity and spin-polarization \label{materials}}
\subsection{Overview of the materials}
Ferro- and ferrimagnetic oxides that are thought to be half-metallic
are the most interesting class of materials for the study of extrinsic
magnetotransport phenomena. In this section the definitions of
half-metallicity and spin-polarization are reviewed; some measurements
of the spin-polarization using various techniques are discussed and
compared. First, however, a brief overview of materials is given
(crystal structures are shown in figure~\ref{crystal}):

\begin{enumerate}

\item
The colossal magnetoresistance manganites of the form 
$\rm RE_{1-x}A_xMnO_3$ have attracted most of the research
efforts. RE stands for a rare earth ion
such as La, Nd, Pr or Gd and A denotes a divalent ion such as Ca, Sr or Ba.
The manganites crystallize in the perovskite structure, see
figure~\ref{crystal}a.
Depending on doping, these compounds show a complex magnetic phase
diagram, see Coey \etal (1999). Ferromagnetism is found in the doping
range $0.15 < x < 0.5$ and, in view of an understanding of extrinsic
magnetotransport, this review will be restricted to the discussion of
the magnetic and transport properties of manganites in this range. The
highest Curie temperatures are found in the archetypal compound $\rm
La_{1-x}A_xMnO_3$ at a doping level $x \sim 1/3$ with $T_C$ of 270~K
(Ca substitution), 360~K (Sr) and 330~K (Ba). The manganites show a
metal-insulator transition accompanying the ferromagnetic
transition. Ferromagnetic order in the mixed-valence manganites is
induced by the double exchange mechanism proposed by Zener (1951). The
intrinsic resistivity and magnetoresistance will be discussed in later
sections. Replacing Mn by Co, Ni, Fe, ... leads to related families of
oxides, see Goodenough and Longo. Especially the cobaltites show an
appreciable magnetoresistance, see e.g.~Yamaguchi \etal (1995) for
single crystal work. These oxides will not be discussed here, since
extrinsic magnetoresistance effects have not been studied in these. 

\begin{figure}[t]
\begin{center}
\vspace*{0.2cm}
\hspace*{2.0cm}\includegraphics[width=0.75\textwidth]{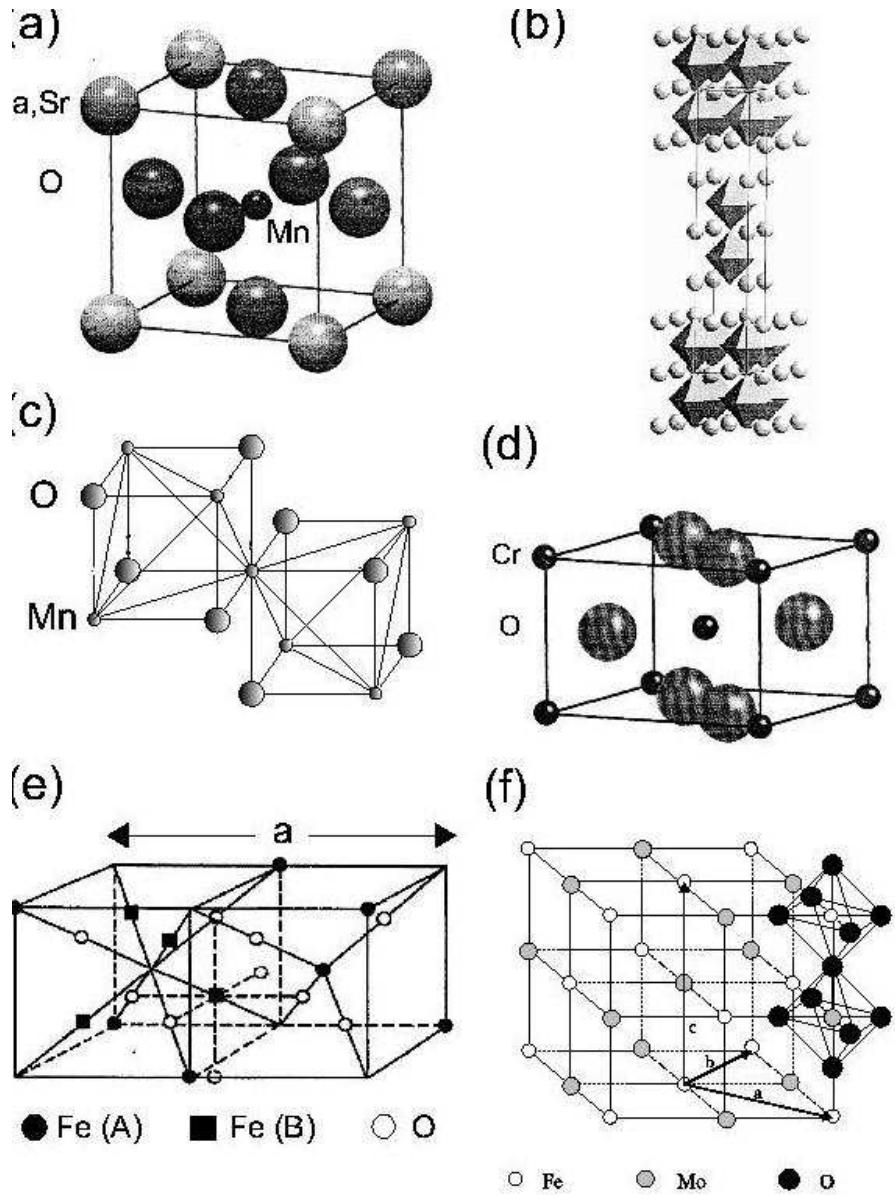}
\vspace{-1.5cm}
\end{center}
\caption{Crystal structures of the most important oxides discussed in
  this review: (a) perovskite stucture ($\rm La_{0.7}Sr_{0.3}MnO_3$),
  (b) $n = 2$ Ruddlesden-Popper phase ($\rm La_{1.2}Sr_{1.8}Mn_2O_7$,
  MnO$_6$ octahedra are shaded, La/Sr ions are drawn as spheres), (c)
  pyrochlore structure ($\rm Tl_2Mn_2O_7$); (d) rutile structure
  (CrO$_2$), (e) inverse spinel structure ($\rm Fe_3O_4$, for clarity
  only a quarter of the unit cell is shown) and (f) double
  perovskite structure ($\rm Sr_2FeMoO_6$).}
\label{crystal}
\end{figure}

\item
The Ruddlesden-Popper family of compounds $\rm (RE,
A)_{n+1}Mn_nO_{3n+1}$ (Moritomo \etal 1996, Kimura \etal 1996, Battle
\etal 1996, 1998) crystallize in a tetragonal structure
consisting of vertex sharing MnO$_6$ octahedra infinitely extending in the
ab-plane and having a thickness of $n$ octahedra along the c-axis, see
figure~\ref{crystal}b.
Neighbouring layers are separated by a rock-salt layer consisting of
$\rm (RE, A)_2O_2$. The manganites can be regarded as the $n = \infty$
member of the Ruddlesden-Popper series. The $n = 1$ member 
$\rm La_{0.5}Sr_{1.5}MnO_4$ shows charge ordering on the Mn sublattice, but
a significant magnetoresistance was not observed (Sternlieb \etal 1996).
A colossal magnetoresistance near the Curie temperature of 126~K was found, 
however, in the $n = 2$ compound $\rm La_{1.2}Sr_{1.8}Mn_2O_7$
(Moritomo \etal 1996). This compound is metallic at low temperatures 
and semiconducting at higher temperatures similar to the CMR materials.
The $n = 2$ compound $\rm Nd_{1+2x}Sr_{2-2x}Mn_2O_7$ shows a large 
magnetoresistance, but ferromagnetism was reported to be absent
(Battle \etal 1996). The Ruddlesden-Popper compounds have low
Curie temperatures not exceeding 150~K and are, for this reason,
not especially interesting for the fabrication of devices. The
$n = 2$ compound $\rm La_{2-2x}Sr_{1+2x}Mn_2O_7$, however,
shows an interesting anisotropy in the magnetoresistance that might be
related to an intrinsic tunnelling effect. This is further discussed
in section~\ref{tunnelling}.

\item
$\rm Tl_2Mn_2O_7$ crystallizes in the pyrochlore structure
(figure~\ref{crystal}c); this compound also shows a metal insulator-transition
and a large intrinsic magnetoresistance (Shimakawa \etal 1996, 1997)
near the Curie temperature of about 140~K. 
In contrast to the manganites, however, the carrier density is low and
the ferromagnetism arises from the super-exchange interaction between
Mn$^{4+}$ ions. Majumdar and Littlewood (1998a, 1998b)
developed a model for the magnetotransport properties of the pyrochlore
based on the assumption of a low density electron gas coupled to spin 
fluctuations. The transport properties of $\rm Tl_2Mn_2O_7$ will be
discussed in more detail in section~\ref{optical}.

\item
CrO$_2$ is a ferromagnet with a Curie temperature of about
390~K. It crystallizes in the rutile structure, see figure~\ref{crystal}d. 
This oxide is metallic both above and below the Curie temperature;
the resistivity and magnetoresistance are discussed in later sections.

\item
Magnetite, $\rm Fe_3O_4$, is a ferrimagnetic oxide crystallizing
in the inverse spinel structure (figure~\ref{crystal}e) and has the
highest Curie temperature, $T_C = 858$~K, of the magnetic oxides
discussed here. It is therefore often viewed as an ideal candidate
for room temperature applications. The temperature dependence of the
resistivity is quite complex, changing from semiconducting to metallic
behaviour slightly above room temperature and back to semiconducting
behaviour near the Curie temperature. The low temperature
resistivity and magnetoresistance will be discussed in later sections.

\item
$\rm Sr_2FeMoO_6$ and $\rm Sr_2FeReO_6$ are double perovskite 
ferromagnets with comparatively high Curie temperatures of about 420~K
and 400~K, respectively (Kobayashi \etal 1998, 1999, Manako \etal
1999). The structure is obtained by doubling the
perovskite unti cell, see figure~\ref{crystal}f. Cation pairs (Fe,Mo),
(Fe,Re) order in a rock-salt-like fashion. These compounds were
investigated in the sixties and seventies (Longo and Ward 1961,
Sleight \etal 1962, 1972, Abe \etal 1973); 
interest in these has been revived, since band-structure
calculations indicated a half-metallic state (Kobayashi \etal
1998, 1999). Often a metallic behaviour of the resistivity is observed
(Manako \etal 1999, Kobayashi \etal 1999, Asano \etal 1999);
the resistivity, however, depends sensitively on the preparation
conditions such as annealing and film growth parameters and semiconducting
behaviour is sometimes reported (Kobayashi \etal 1999, Asano \etal 1999).
The magnetic and transport properties will be further discussed in
section~\ref{optical}.

\item
SrRuO$_3$ is a metal that undergoes a ferromagnetic transition at 165~K
(Callaghan \etal 1966, Longo \etal 1968, Cao \etal 1997). It
crystallizes in an orthorhombic structure. SrRuO$_3$
is regarded as a strongly-correlated $d$-band metal (Klein \etal 1996, 
Cao \etal 1997, Fujioka \etal 1997, Okamoto \etal 1999) that falls
into the class of ``bad metals'' (Emery and Kivelson 1995). A
``bad metal'' is defined as having an unsaturated resistivity with
positive temperature coefficient that exceeds the Ioffe-Regel
limit. The ferromagnetism in SrRuO$_3$ is of itinerant character.
The intrinsic resistivity and magnetoresistance will be 
discussed in this review in order to facilitate comparison with the other
magnetic oxides.

\item
There are reports on the resistivity and magnetoresistance of 
$\rm CaCu_3Mn_4O_{12}$ (Zeng \etal 1999) with $T_C$ of 355~K, 
$\rm Na_{0.5}Ca_{0.5}Cu_{2.5}Mn_{4.5}O_{12}$ (Zeng \etal 1998)
with $T_C$ of 340~K as well as $\rm TbCu_3Mn_4O_{12}$ ($T_C = 430$~K)
and $\rm CaCu_{1.5}Mn_{5.5}O_{12}$ (Troyanchuk \etal 1998). Whereas 
Zeng \etal (1998, 1999) describe the compounds as ferromagnetic,
Troyanchuk \etal (1998) interpret magnetization data as consistent with
ferrimagnetic order. All compounds show a gradual decrease of the 
magnetoresistance in large applied fields from some 10\% at low temperature
to zero above $T_C$. Since these compounds are not 
important in device fabrication, these will not be further discussed 
in this review.

\item
The chalcogenides $\rm Fe_{1-x}Cu_xCr_2S_4$ show a moderate
magnetoresistance near the Curie temperature (Ramirez \etal 1997, Yang
\etal 2000). Theoretical studies of the electronic structure indicate
a half-metallic nature with a gap in the minority density of states
(Park \etal 1999). Since this review is focused on oxides, the
interested reader is referred to the original reports. 

\item
Stimulated by the intense research on both giant and colossal
magnetoresistance, there have been reports on magnetoresistive phenomena
in various compounds that do not fall in the classes described above.
Here only three reports are mentioned, namely the observation of a 
large positive magnetoresistance in Ag$_2$Se and
Ag$_2$Te (Xu \etal 1997, Chuprakov and Dahmen 1998) as well
as in the ferromagnetic multilayer $\rm LaMn_2Ge_2$ (Mallik \etal 1997).
A discussion of these compounds is beyond the scope of this review.

\item
At the end of this list of materials and compounds the case of GdI$_2$
should be mentioned. GdI$_2$ shows a ferromagnetic transition close to
room temperature; this transition is accompanied by a metal-insulator
transition and negative colossal magnetoresistance, see Ahn \etal
(2000). This observation is especially intriguing, since GdI$_2$ is
nominally isoelectronic to the superconductor NbSe$_2$. In the
colossal magnetoresistance manganites the situation is similar:
substitution of the magnetically active Mn by Cu leads from
ferromagnetism in $\rm La_{1-x}Sr_xMnO_3$ to superconductivity in 
$\rm La_{2-x}Sr_xCuO_4$. It would be interesting to search for similar
trends in other material classes that might establish an underlying
generic pattern.

\end{enumerate}
\subsection{Spin-polarization and half-metallicity}
\begin{figure}[t]
\begin{center}
\hspace*{2.0cm}\includegraphics[width=0.8\textwidth]{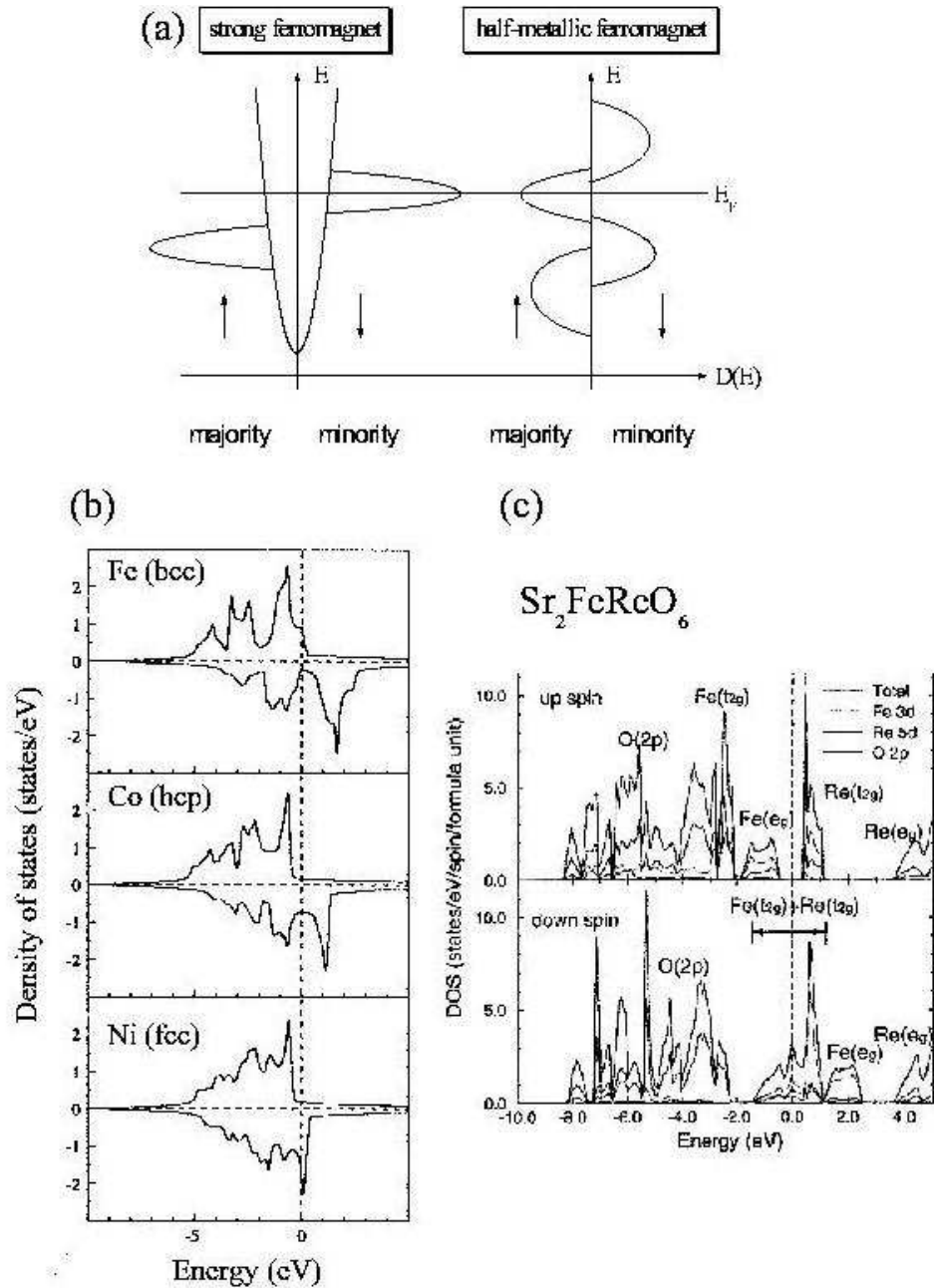}
\vspace{-1.5cm}
\end{center}
\caption{(a) Schematic density of states of a strong ferromagnet and 
a half-metallic ferromagnet. (b) Density of states of the elemental
ferromagnets Fe, Co and Ni. Co and Ni are strong ferromagnets, Fe is a
weak ferromagnet and shows a significant $d$ state contribution in the
majority spin channel. Adapted from Coey (2001). (c) Density of states
of Sr$_2$FeReO$_6$ as determined by calculations within the local
density approximation (LDA). Sr$_2$FeReO$_6$ is a half-metallic
ferromagnet with a gap in the majority density of states. Adapted from 
Kobayashi \etal (1999).} 
\label{dos}
\end{figure}

A quantity of fundamental interest for both basic physics
and device applications is the degree of spin-polarization $P$ at the
Fermi level. The band-structure of ferromagnets is spin-dependent and
two subbands are found for majority (carrier spin directed parallel to the
magnetization) and minority (spin antiparallel to the
magnetization) carriers, respectively (see Chikazumi 1997). 
The schematic density of states of a strong ferromagnet is shown in
figure~\ref{dos}(a); here the majority $d$ bands are completely filled.
In the case of an intinerant ferromagnet, within a
two-band model, the spin-polarization is often defined
as the normalized difference of the majority ($n_\uparrow$)
and minority ($n_\downarrow$) density of states at the Fermi level,
thus
\begin{equation}
P_n = \frac{n_\uparrow-n_\downarrow}{n_\uparrow+n_\downarrow}\, .
\label{Pn}
\end{equation}
This definition is somehow related to the definition of
the magnetization as the difference between the integrated majority and
minority carrier density, $M = \mu_{\rm B} \int(
n_\uparrow-n_\downarrow)dE$, and often a scaling $P(T) \propto M(T)$
is expected. The spin-polarization defined in this way might
be probed by spin-polarized photoemission.

However, as pointed out by Mazin (1999), the definition
of spin-polarization is by no means unique. 
Often transport properties are of interest, 
especially for applications. In a ferromagnet the majority and minority
carriers can be regarded as two parallel transport channels, see
Mott (1936), Campbell and Fert (1982), and a definition of the
spin-polarization in terms of the majority ($J_\uparrow$) and minority
($J_\downarrow$) current densities seems more appropriate. Within
classical Boltzmann transport theory, 
$J_{\uparrow(\downarrow)} \propto \langle nv^2\rangle_{\uparrow(\downarrow)}
\tau_{\uparrow(\downarrow)}$, where $\langle ... \rangle$ denotes a
Fermi-surface average and $\tau_{\uparrow(\downarrow)}$
the relaxation times for majority and minority carriers, respectively.
Assuming a spin-independent relaxation time, one finds
\begin{equation}
P_J = \frac{J_\uparrow-J_\downarrow}{J_\uparrow+J_\downarrow} 
= \frac{\langle nv^2\rangle_\uparrow-\langle nv^2\rangle_\downarrow}
{\langle nv^2\rangle_\uparrow+\langle nv^2\rangle_\downarrow}.
\label{PJ}
\end{equation}
This definition is fundamentally different from the definition
in terms of the carrier densities. In the case of Ni and Fe,
Mazin (1999) showed that within local spin density approximation (LSDA)
calculations, these definitions lead to significantly different
values of the spin-polarization.

Within the two-current model, the spin-polarization defined by the
majority and minority currents can be simply related to the majority
($\rho_\uparrow$) and minority ($\rho_\downarrow$) resistivities:
\begin{equation}
P_J =
\frac{\rho_\downarrow-\rho_\uparrow}{\rho_\downarrow+\rho_\uparrow}\, .
\label{PJtwo}
\end{equation}
The channel resistivities can be determined at low temperature from
deviations of Matthiesen's rule (Fert and Campbell 1968, Campbell and
Fert 1982). These depend on the impurities present in the metal and
consequently the spin-polarization can be tuned by alloying. According
to Campbell and Fert (1982) the ratio of the intrinsic resistivities 
$\mu = \rho_\downarrow/\rho_\uparrow$ characteristic for the host
material is $\mu \simeq 3.6$ in Ni and $\mu \simeq 1$ in Fe. This
yields spin-polarization values $P_J \simeq 0.56$ (Ni) and $P_J
\simeq 0$ (Fe). In contrast, the spin-polarization values at the Fermi
level as calculated by Mazin (1999) are $P_J \simeq 0$ (Ni) and $P_J
\simeq 0.2$ (Fe). Ziese (2000b) and Zhao \etal (2001) tried to develop
bulk probes for half-metallicity in $\rm La_{0.7}Ca_{0.3}MnO_3$ using the
anisotropic magnetoresistance and the temperature dependence of the
resistivity, respectively. Both experiments were consistent with a
half-metallic character of the compound.

The archetypal experiments capable of determining the transport
spin-polarization are spin-polarized
tunnelling between ferromagnetic contacts and dynamic conductance
of superconductor-ferromagnet contacts. Mazin (1999) analyzed transport
through a superconductor-ferromagnet contact. In the case of ballistic
transport without a barrier, the current through the contact is proportional
to $\langle nv\rangle$; thus, the ballistic spin-polarization is defined by
\begin{equation}
P_v = \frac{\langle nv\rangle_\uparrow-\langle nv\rangle_\downarrow}
{\langle nv\rangle_\uparrow+\langle nv\rangle_\downarrow}
\label{Pv}
\end{equation}
and does not agree with the spin-polarization Eq.~(\ref{PJ}) defined
via the currents. 
If a specular barrier is present, the tunnelling current depends in a
more complex way on both the Fermi velocity and the barrier transparency,
and the measured spin-polarization does not agree with any of the
definitions introduced so far. This analysis shows that experimental
values obtained with different techniques relate to different definitions
of the spin-polarization that need not necessarily agree.

The concept of half-metallicity was introduced by de Groot \etal (1983)
on the basis of band-structure calculations of Heusler alloys. These
calculations showed a gap in the density of states of minority
carriers. Half-metallicity is an independent particle concept and the
inclusion of many-body-effects leads to the appearance of
non-quasiparticle states extending down to the Fermi level,
see Irkhin and Katsnel'son (1994). The density of states of
a half-metallic ferromagnet as obtained within an independent particle
calculation is schematically shown in figure~\ref{dos}(b). In this case,
the spin-polarization is $P = 1$ within any reasonable definition. In
subsequent work it was found that $\rm Fe_3O_4$ ($P = -1$, Yanase and
Siratori 1984, de Groot and Buschow 1986,  P\'enicaud \etal 1992,
Yanase and Hamada 1999), CrO$_2$ ($P = 1$, Lewis \etal 1997) and
$\rm La_{0.7}Sr_{0.3}MnO_3$ ($P = 1$, Pickett and Singh 1996, de Boer
\etal 1997, Livesay \etal 1999) are half-metallic magnets.

Here, some data on the spin-polarization of elemental as well as oxide
magnets are summarized. The spin-polarization was determined by 
ferromagnet-insulator-ferromagnet (FIF) tunnelling, 
ferromagnet-insulator-superconductor (FIS)
tunnelling, Andreev reflection (AR) at a superconductor-ferromagnet interface, 
spin-polarized photoemission (SPES) and two-dimensional angular correlation
of electron-positron radiation (2D-ACAR).
Here, only spin-polarization measurements using Andreev reflection
will be discussed in greater detail; ferromagnetic tunnelling junctions
are discussed in section~\ref{tunnelling}. For a discussion of 
ferromagnet-insulator-superconductor tunnelling, spin-polarized
photoemission and 2D-ACAR, the reader is referred to the standard literature
(Meservey and Tedrow 1994, Eib and Alvarado 1976, West 1995).

Electron transport through a normal metal-superconductor interface for
energies below the superconducting gap $\Delta$ is possible
through Andreev reflection. An electron incident from the normal metal
forms a pair with another electron of opposite momentum and spin and
enters the superconductor as a Cooper pair, while a hole is
reflected. This leads to a conductivity enhancement by a factor of two
at small voltages. At large bias voltages, transport is dominated
by quasiparticle injection and the conductance approaches the normal
state conductance $G_n$. In the case of a ferromagnet-superconductor 
interface, Andreev reflection is suppressed, since not every majority
electron finds a minority electron with appropriate momentum 
to form a Cooper pair. It is evident that the zero bias conductance
should vanish in the case of a half-metallic ferromagnet. It was shown
(Soulen \etal 1998, Osofsky \etal 1999) that the conductance of
a ferromagnet-superconductor interface at zero temperature is given by
\begin{equation}
\frac{1}{G_n}\frac{dI}{dV} = 2(1-\mid P_v\mid) \qquad {\rm e}V \ll \Delta
\label{soulen}
\end{equation}
with the spin-polarization $P_v$ as defined above. Thus, from the suppression
of the conductance at zero bias the spin-polarization $P_v$ can be
determined. This method is not sensitive to the sign of the
spin-polarization. A similar method was proposed and tested on
elemental ferromagnets by Upadhyay \etal (1998).

Results of spin-polarization measurements from FIF- and FIS-measurements,
Andreev reflection and spin-polarized photoemission are summarized in 
table~\ref{spinpol} for the elemental magnets Fe, Co, Ni, Gd, Tb, Dy,
Ho, Er and Tm as well as for the oxide magnets $\rm
La_{0.7}Sr_{0.3}MnO_3$, CrO$_2$, $\rm Fe_3O_4$ and SrRuO$_3$.
Data for alloys can be found in Meservey \etal (1976), Paraskevopoulos
\etal (1977) obtained by the FIS-technique and in Nadgorny \etal
(2000) using Andreev reflection.
Although the data show some scatter, it is clear that CrO$_2$ and
$\rm La_{0.7}Sr_{0.3}MnO_3$ have a spin-polarization much larger
than that of elemental ferromagnets. The situation is not clear
for $\rm Fe_3O_4$, especially since it is not
possible to perform Andreev reflection measurements due to the
insulating nature of magnetite at low temperatures.

Note that the spin-polarization of Fe, Co and Ni as determined from
FIS and FIF measurements is positive. From the
band-structure, see figure~\ref{dos}, a negative value is expected for
Co and Ni and is indeed found in spin-polarized photoemission
studies. This discrepancy might be related to the type of bonding at the
ferromagnet-insulator interface or to interfacial scattering and will
be discussed in greater detail in section~\ref{tunnelling}. Monsma and
Parkin (2000b) reported an ageing of the spin-polarization of Ni films
as observed in tunnelling studies and related it to chemical processes
at the Ni-Al$_2$O$_3$ interface; this observation might explain the 
discrepancy between their data and the classical values of Tedrow and
Meservey (1971, 1973).

\begin{table}
\caption{Spin-polarization as determined from ferromagnet-insulator-ferromagnet 
(FIF), ferromagnet-insulator-superconductor (FIS) tunnelling,
Andreev reflection (AR) and spin resolved photoemission spectroscopy (SPES)
for elemental ferromagnets and oxide magnets. For FIF and FIS measurements an
Al$_2$O$_3$ barrier was used. The Andreev reflection
technique is not sensitive to the sign of the spin-polarization. All values
were measured at low temperatures, i.e.\ 4.2~K or below, except for
the SPES value of Ni that was measured at room temperature. The
photoemission result for CrO$_2$ by K\"amper \etal (1987) was
obtained 2~eV below the Fermi level. The abbreviations in brackets indicate
the reference with MPT80 (Meservey \etal 1980), MP00 (Monsma and
Parkin 2000a), WG00a (Worledge and Geballe 2000a), WG00b (Worledge and
Geballe 2000b), M98 (Miyazaki \etal 1998), M95 (Moodera \etal 1995),
Ji01 (Ji \etal 2001), V97 (Viret \etal 1997a), S99 (Seneor \etal
1999), U98 (Upadhyay \etal 1998), S98 (Soulen \etal 1998), S95
(Sinkovi{\'c} 1995), R95 (Rampe \etal 1995), EA76 (Eib and Alvarado
1976), P98 (Park \etal 1998a), B69 (Busch \etal 1969), K87 (K\"amper
\etal 1987) and A75 (Alvarado \etal 1975).}
\begin{indented}
\item[]\begin{tabular}{@{}lllll}
\br
Compound & FIS & FIF & AR & SPES\\
\mr
Fe & $+0.45$ (MP00) & $+0.35$ (M98) & -- & $+0.40$ (S95)\\
Co & $+0.42$ (MP00) & $+0.34$ (M95) & $0.37$ (U98) & $-0.4$ (R95)\\
Ni & $+0.31$ (MP00) &  --           & $0.32$ (U98) & $-0.3$ (EA76)\\
Gd & $+0.13$ (MPT80) & -- & -- & $+0.05$ (B69) \\
Tb & $+0.06$ (MPT80) & -- & -- & -- \\
Dy & $+0.06$ (MPT80) & -- & -- & -- \\
Ho & $+0.07$ (MPT80) & -- & -- & -- \\
Er & $+0.05$ (MPT80) & -- & -- & -- \\
Tm & $+0.03$ (MPT80) & -- & -- & -- \\
La$_{0.7}$Sr$_{0.3}$MnO$_3$ & $+0.70$ (WG00a) & $+0.83$ (V97) & $0.8$ (S98) & $+0.9$ (P98)\\
CrO$_2$ & -- & -- & $0.9$ (S98) & $+1.0$ (K87)\\
CrO$_2$ & -- & -- & $0.96$ (Ji01) & --\\
Fe$_3$O$_4$ & -- & $-0.5$ (S99) & -- & $-0.4$ (A75)\\
SrRuO$_3$ & $-0.095$ (WG00b) & -- & -- & -- \\
\br
\end{tabular}
\end{indented}
\label{spinpol}
\end{table}

\begin{figure}[t]
\vspace*{-0.5cm}
\begin{minipage}[t]{0.5\textwidth}
\centerline{\mbox{\epsfysize=7.0cm \epsffile{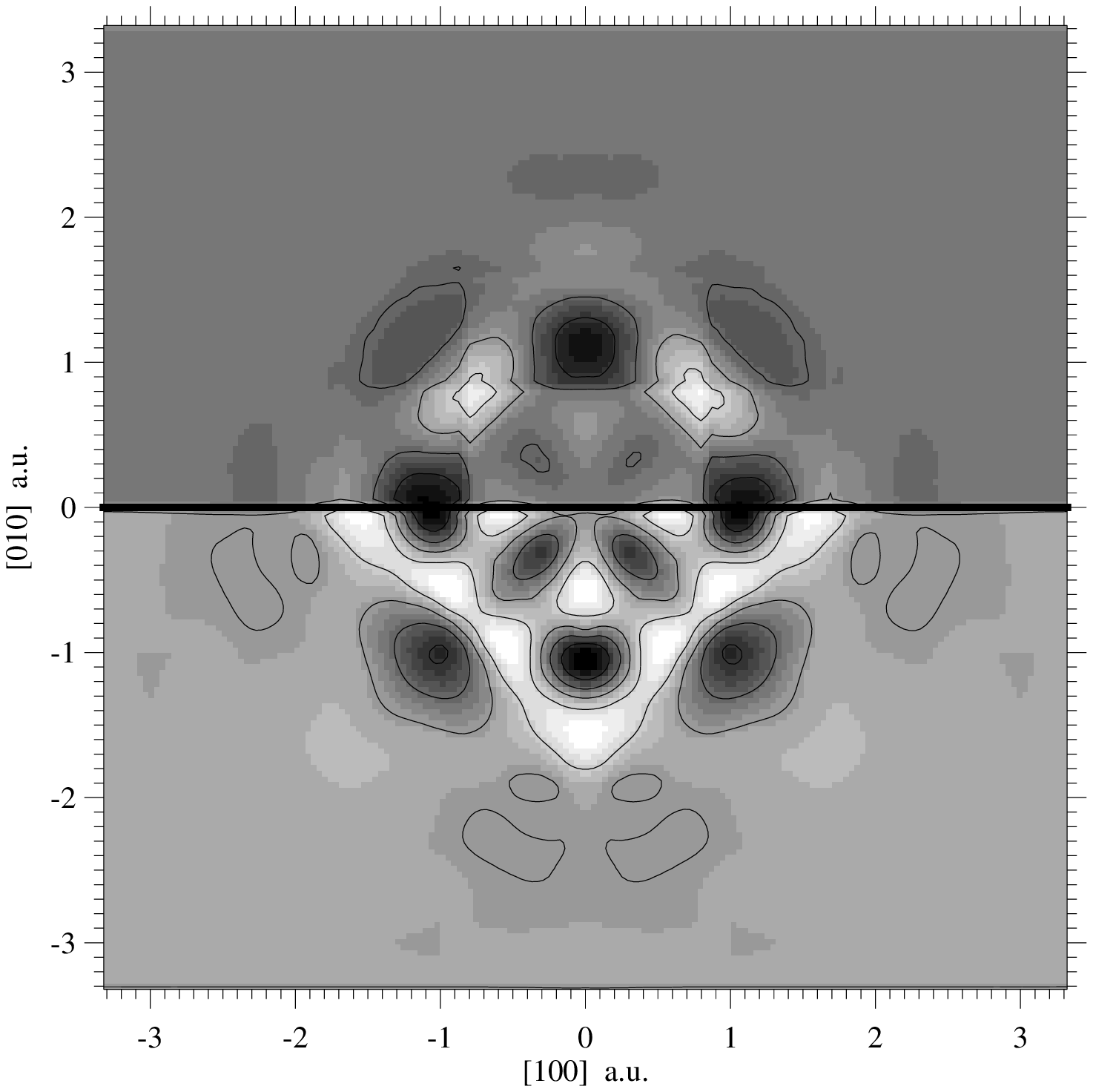}}}
\end{minipage}
\hfill
\begin{minipage}[t]{0.5\textwidth}
\centerline{\mbox{\epsfysize=7.8cm \epsffile{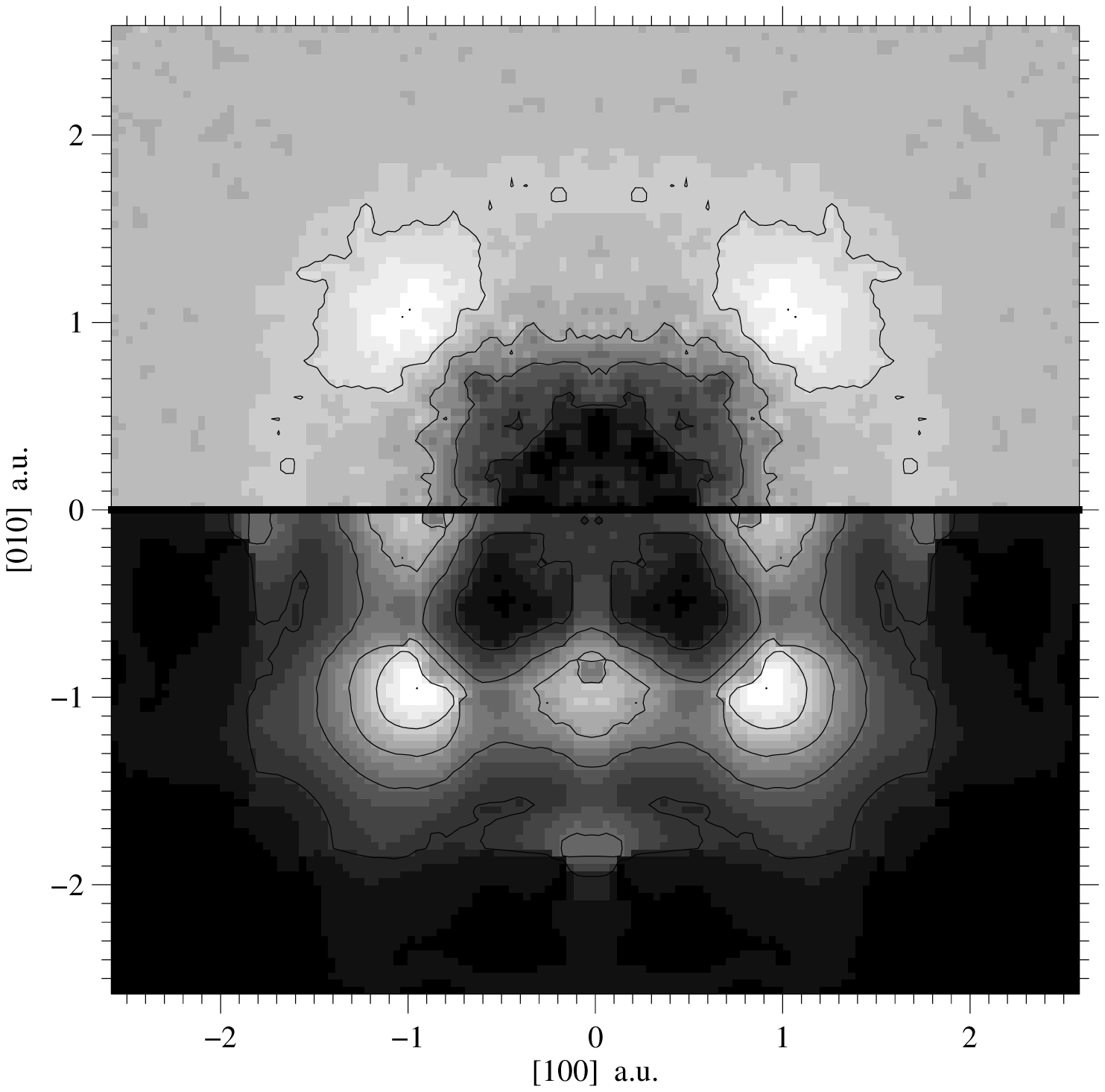}}}
\end{minipage}
\caption{(a) Radial anisotropy of the [001]-projected momentum
density of a $\rm La_{0.7}Sr_{0.3}MnO_3$ crystal, coming
from experiment (top) and band-structure calculation (bottom).
Reproduced from Livesay \etal (1999). (b) The spin density in momentum
space as seen by the positron, integrated along the [001] direction from
the experiment (top) and band-structure calculation (bottom). White
areas indicate positive spin-polarization. Unpublished, by courtesy of
S B Dugdale.}
\label{fermi}
\end{figure}

Livesay \etal (1998, 1999) used 2D-ACAR in order to determine the projection
of the momentum density and spin density of a 
$\rm La_{0.7}Sr_{0.3}MnO_3$ crystal along [001]. This yields information
on the Fermi surface topology. Experimental results are in good agreement
with band-structure calculations. Figure~\ref{fermi}(a) shows the radial
anisotropy of the [001]-projected momentum density. The measurements
are consistent with two Fermi-surface sheets, namely an electron-like sheet 
centered at the $\Gamma$ point and hole-like sheets centered at the R points.
The [001]-integrated spin density is shown in figure~\ref{fermi}(b). The
white areas indicate positive spin-polarization; these are located in the 
hole-like Fermi-surface cuboids at the R points
in agreement with their origin in the
manganese $d$-bands. These data indicate the complex nature of the
spin-polarization in the manganites.

In conclusion of this section on half-metallic magnets a remark on the
saturation magnetization is added. A stoichiometric, ordered
half-metallic ferromagnet or ferrimagnet must have a saturation moment being
an integral number of the Bohr magneton $\mu_{ B}$, since the number
of electrons per formula unit is an integer and the number of spin up
(or spin down) electrons is also integral due to the gap in either the
majority or the minority band. This argument can be extended to
non-stoichiometric ferromagnets that accordingly should have a
saturation moment equal to the spin-only moment of the
ferromagnetically coupled ions. In agreement with this rule, the
saturation moments are close to $2\mu_{\rm B}$ in CrO$_2$ (Cr$^{4+}$,
$3d^2$), $4\mu_{\rm B}$ in $\rm Fe_3O_4$ (Fe$^{2+}$, $3d^6$) and
$3.7\mu_{\rm B}$ in $\rm La_{0.7}A_{0.3}MnO_3$ (mixture of Mn$^{3+}$,
$3d^4$ and Mn$^{4+}$, $3d^3$).
\section{Intrinsic conductivity and optical properties \label{optical}}
As already discussed in section~\ref{materials}, different exchange
and transport mechanisms can be found in magnetic oxides. In this
section, the temperature dependent resistivity and the
optical conductivity are discussed and compared to recent theoretical
models. The discussion starts with the itinerant ferromagnet
SrRuO$_3$, then turns to a brief review of the super-exchange ferromagnet
$\rm Tl_2Mn_2O_7$ before presenting data and models for the double
exchange systems $\rm La_{0.7}A_{0.3}MnO_3$ and CrO$_2$. This is
followed by a brief discussion of the double perovskite $\rm
Sr_2MoFeO_6$ which seems to be an itinerant ferromagnet. At the end of
this section the transport properties of magnetite are analyzed;
magnetite is a strongly correlated ferrimagnet with significant polaronic 
effects and represents a class of its own. Since the
resistivity depends sensitively on grain boundaries, see later sections,
it is important to perform resistivity measurements on high quality
single crystals and epitaxial films. Since research activity in recent
years mainly focused on the colossal magnetoresistance perovskites,
the bulk of this section will be devoted to this system.

Apart from the specific mechanisms relating to the individual systems, the
issue of a minimal conductivity has re-emerged. This is due to the
fact that many metallic oxides such as the ferromagnetic compounds
discussed here as well as high temperature superconductors apparently
violate the Ioffe-Regel limit (Ioffe and Regel 1960, see also Mott
1978). This concept is based on the idea that the mean free path
$\ell$ cannot be smaller than the interatomic spacing $a$. In order to
fix ideas, a simple estimate that should be valid for SrRuO$_3$ in the
paramagnetic phase can be made. SrRuO$_3$ is particularly interesting,
since Shubnikov-de Haas oscillations have been observed at low
temperatures indicating Fermi-liquid behaviour (Mackenzie \etal
1998). Assuming a typical metallic carrier density $n \sim 5\times
10^{27}$~m$^{-3}$, a Fermi velocity $v_{\rm F} \sim 10^{5}$~m/s (Allen
\etal 1996) and an interatomic distance $a \sim 2.5$~{\AA}, one
finds from Drude theory a maximal resistivity 
\begin{equation}
\rho_{\rm max} \sim \frac{mv_{\rm F}}{n{\rm e}^2a} \sim
300~\mu\Omega{\rm cm}\, .
\label{rhomax}
\end{equation}
SrRuO$_3$ displays at room temperature a resistivity of about
200~$\mu\Omega$cm already close to the estimated saturation
value. Allen \etal (1996) measured the resistivity of a SrRuO$_3$
single crystal and found a linear dependence between room temperature
and 1000~K without any sign of saturation up to a resistivity value of
300~$\mu\Omega$cm. Other ferromagnetic oxides discussed here also show
resistivities of the same order of magnitude without any tendency to
saturation.

This behaviour is distinctly different from the resistivity curves
found in A-15 compounds such as Nb$_3$Sn which show an apparent
saturation towards a value $\rho_{\rm max} \simeq 150$~$\mu\Omega$cm
(Wiesmann \etal 1977). Fisk and Webb (1976) interpreted this
resistivity saturation in terms of a saturation of the mean free path
towards the interatomic distance in accordance with the Ioffe-Regel
limit. The lack of saturation seen in many oxides motivated Emery and
Kivelson (1995) to introduce the term ``bad metal'' meaning a system
with a positive resistivity coefficient, $d\rho/dT > 0$, and a mean
free path at high temperature smaller than the interatomic
distance. These authors speculated that such systems could not be
described within a Fermi-liquid picture with weak scattering.

A solution to this problem was indicated by Millis \etal (1999)
investigating electrons coupled to classical phonons with an arbitrary
strong coupling. The resistivity was calculated within dynamical mean
field theory. These calculations show a linear temperature dependence
of the resistivity at small coupling and a linear temperature
dependence with a non-zero offset at large coupling constants; these
features are reminiscent of the saturation behaviour in A-15
compounds. However, a comparison of the numerical results to second
order perturbation calculations shows that at strong electron-phonon
couplings, the scattering rate increases with temperature at a rate
much smaller than that given by perturbation theory. This indicates
that the term ``saturation'' is a misnomer and that the apparent
violation of the Ioffe-Regel limit is not per se a remarkable
feature. Actually, at high temperatures classical diffusion might be
expected leading to a linear temperature dependence of the resistivity
according to the Einstein relation
\begin{equation}
\rho = \frac{{\rm k}T}{n{\rm e}^2D}\, ,
\label{einstein}
\end{equation}
where $D$ denotes the diffusion constant.

Although the debate on a minimal metallic conductivity is certainly
not exhausted, the apparent violation of the Ioffe-Regel limit in
certain oxides with strongly coupled electron-phonon systems might not
be an experimental smoking gun in regard to exotic transport
theories.
\subsection{SrRuO$_3$}
SrRuO$_3$ is a strongly-correlated ferromagnetic $d$-band metal. 
The Curie temperature in the bulk is 165~K; for thin films reduced
Curie temperatures of 150~K were observed possibly due to strain
effects (Klein \etal 1996). Magnetization
measurements on single crystals (Cao \etal 1997) showed a magnetic
moment of about $1.5\mu_{\rm B}$ per Ru$^{4+}$ ion. Within an ionic
model, the fivefold degeneracy of the Ru ($4d^4$) states is split by the
octahedral crystalline field due to the O ions into two-thirds
occupied t$_{2g}$ and empty e$_g$ levels. Thus, Ru$^{4+}$ is in a low
spin state with an ideal value of the magnetic moment of 
$2\mu_{\rm B}$. The magnetization does not reach this ideal limit
indicating the itinerant character of the ferromagnetism.

\begin{figure}[t]
\begin{center}
\vspace*{0.3cm}
\hspace*{1.0cm} \includegraphics[width=0.8\textwidth]{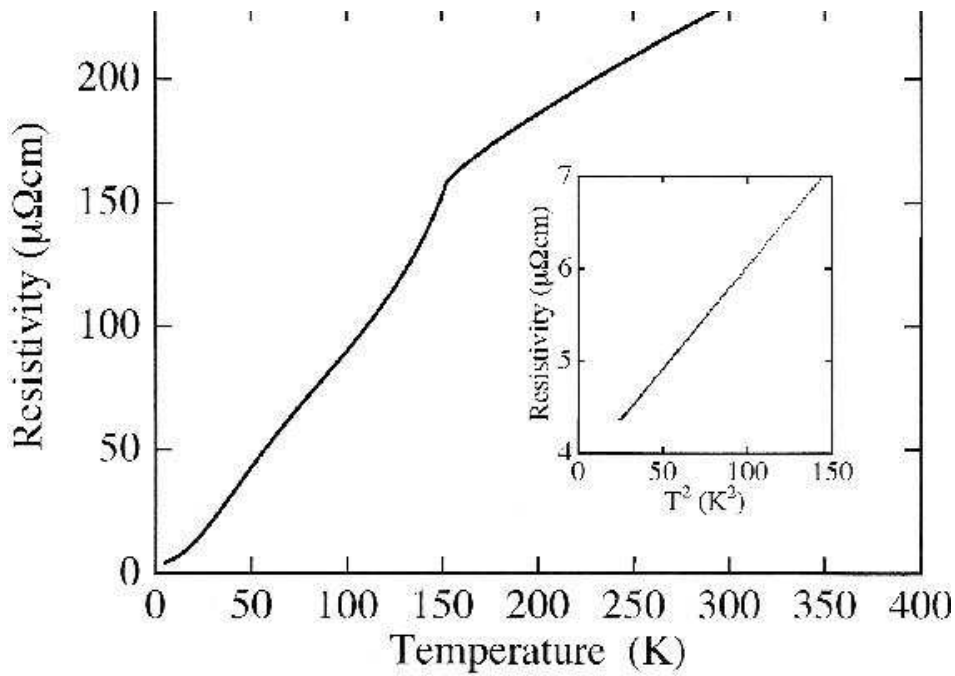}
\vspace*{-9.5cm}
\end{center}
\caption{Resistivity of a high quality SrRuO$_3$ film grown
on 2$^\circ$ miscut SrTiO$_3$. The inset shows that there is a $T^2$
scattering rate below about 10~K. Reproduced from Mackenzie \etal (1998).}
\label{mackenzie}
\end{figure}

The zero field resistivity
was found to increase proportional to $T^2$ at low temperatures below about
10~K followed by a further steep increase up to the ferromagnetic Curie
temperature that is marked by a change in slope. Above $T_C$, the resistivity
continues to increase linearly with temperature, see figure~\ref{mackenzie}.
The resistivity increases strongly from residual resistivity values
of about 4~$\mu\Omega$cm in high quality films to about 200~$\mu\Omega$cm
at room temperature. Klein \etal (1996) reported anomalously strong 
spin-scattering as evidenced by a resistivity derivative $d\rho/dT$
of SrRuO$_3$ being much larger than that of iron; these authors speculated
that this might be related to the ``bad metal'' property.

Band-structure calculations show a gap of about 0.3~eV
in the majority density of states that is only 20~mRy above the Fermi level 
(Mazin and Singh 1997, Santi and Jarlborg 1997). Thus, SrRuO$_3$
is close to being a half-metal; Mazin and Singh (1997) pointed out
that this nearly half-metallic character is important for understanding
the transport properties.

Measurements of the optical conductivity indicate deviations from
Fermi-liquid behaviour. Kostic \etal (1998) measured $\sigma(\omega)$
for temperatures between 40~K and 250~K. At low temperatures the real
part of the optical conductivity was found to decrease with frequency
as $\omega^{-1/2}$, a behaviour that is difficult to reconcile with
standard Fermi-liquid theory predicting $\Re{(\sigma)} \propto \omega^{-2}$
at comparable frequencies. At higher temperatures the low frequency
conductivity was actually found to increase with frequency indicating
the opening of an optical gap in contrast with the general
characteristics of metals. Accordingly, Kostic \etal (1998) concluded
that SrRuO$_3$ shows non-Fermi-liquid behaviour and derived frequency
dependent scattering times and effective masses. 
This work was extended by Dodge \etal (2000) who employed terahertz
time-domain and far-infrared spectroscopy to measure the conductivity
over three decades in frequency. The results can be analyzed with a
conductivity $\sigma(\omega) \propto (\tau^{-1}+{\rm i}\omega)^{-\alpha}$ 
yielding an exponent $\alpha \simeq 0.4$ in
contrast to the Drude result $\alpha = 1$. The authors conclude that
SrRuO$_3$ displays non-Fermi liquid behaviour in the 
temperature 5~K $\le T < $95~K and investigated frequency regime.
At still lower energies, however, the observation of
Shubnikov-de Haas oscillations in the resistivty of SrRuO$_3$ films
at 35~mK by Mackenzie \etal (1998) strongly suggests a ground state
that is actually a Fermi-liquid. Accordingly, the Fermi-liquid state
in SrRuO$_3$ might display some fragility leading to a breakdown of
this concept at energies much lower than in conventional metals.
The carrier mean free path was estimated by Kostic \etal (1998) using
the measured scattering times and Fermi velocities derived from band-structure
calculations. At 145~K, the mean free path for majority and minority carriers
was found to be about 0.6~nm and 1.2~nm, respectively. These values
are comparable to the size of the unit cell and clearly show that
SrRuO$_3$ is a ``bad metal''.

Mazin and Singh (1997) argued that there is a strong coupling between
electrons, phonons and magnons that probably produces significant
spin-flip scattering. Furthermore, the Fermi-surface has a complex topology
with electron- and hole-like sheets in each spin channel. Assuming
electron-phonon and electron-paramagnon scattering Mazin and Singh (1997)
argued that the resistivity is linear in temperature at high temperatures
with the coefficient proportional to the sum of the coupling constants.
The coupling constants derived from
resistivity and specific heat data are in reasonable agreement in support
of the above picture. The change in slope of the resistivity while
warming through the Curie temperature is explained by a change from
electron-magnon scattering to electron-paramagnon scattering, the
latter having a weaker coupling. The quadratic
temperature dependence of the resistivity at low temperatures is due
to electron-magnon scattering, again with an extraordinary large coupling 
constant. In summary, the transport properties of SrRuO$_3$ can be
qualitatively understood on the basis of band-structure calculations
and Boltzmann transport-theory. This is in agreement with the general
results by Millis \etal (1999) on the resistivity of a metal with
strong electron-phonon coupling, see above.
\subsection{Tl$_2$Mn$_2$O$_7$}
\begin{figure}[t]
\begin{center}
\vspace*{0.0cm}
\hspace*{2.0cm} \epsfysize=11cm \epsfbox{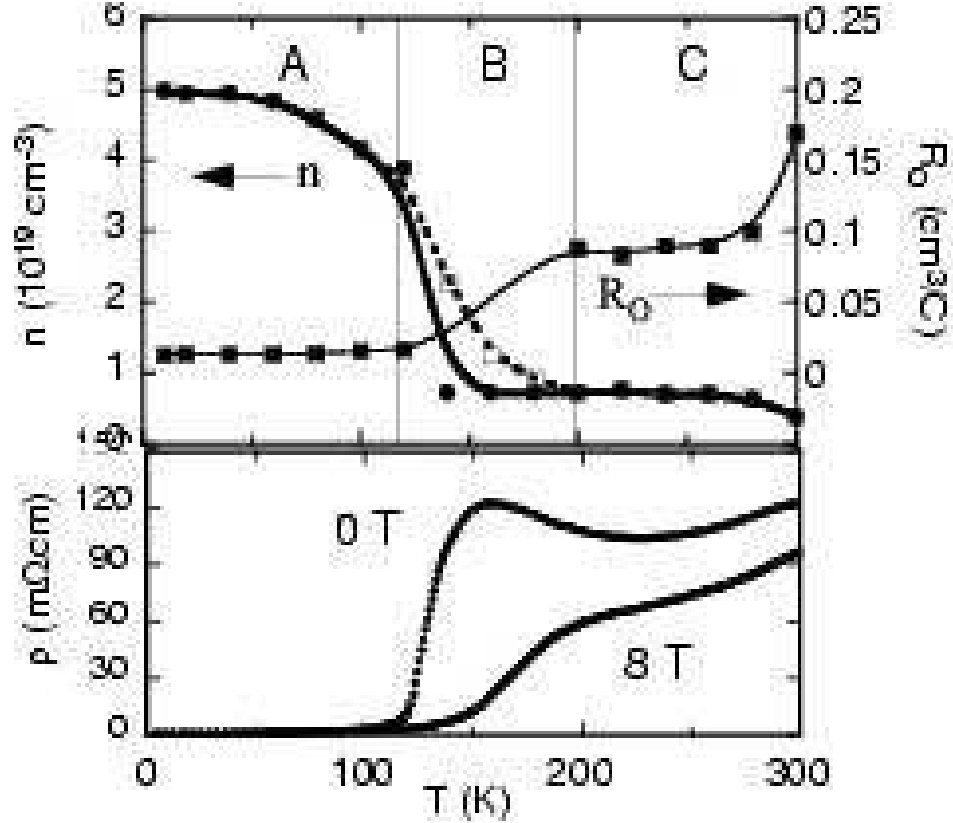}
\vspace*{0.0cm}
\end{center}
\caption{(a) Temperature dependence of the ordinary Hall coefficient,
$R_0$, (right axis) and the free-electron carrier density, $n$, (left
axis) of a polycrystalline
$\rm Tl_2Mn_2O_7$ sample. Closed circles represent the zero field
carrier density and open squares the carrier density measured at
8~T. Lines are guides for the eye. (b) The electrical resistivity of
the same sample measured as a function of temperature in zero magnetic
field and at 8~T. Reproduced from Imai \etal (2000).}
\label{imai}
\end{figure}

The pyrochlore $\rm Tl_2Mn_2O_7$ has a Curie temperature of about
140~K with a saturation magnetic moment of $3\mu_B$ per formula unit
corresponding to a ferromagnetic ordering of the Mn$^{4+}$
ions (Shimakawa \etal 1997). The transport and magnetotransport
properties are similar to the CMR manganites, namely: metallic
transport below $T_C$ crossing over to semiconducting behaviour above
$T_C$; near the Curie temperature a large magnetoresistance is
seen. Typical resistivity data are shown in figure~\ref{imai}(b).
These properties make the pyrochlore interesting in comparison to the
manganites, since they, especially the large magnetoresistance, do
not seem to arise from a double exchange mechanism (Shimakawa \etal
1997, Kwei \etal 1997, Majumdar and Littlewood 1998a, 1998b, Imai
\etal 2000). 

The carrier density of $\rm Tl_2Mn_2O_7$ is small of the order of
0.001-0.005 per formula unit (Shimakawa \etal 1996, Raju \etal 1994)
and depends on temperature and magnetic field (Imai \etal 2000), see
figure~\ref{imai}(a). This carrier density is supposed to be too small
to give rise to significant double exchange effects. X-ray and neutron
diffraction studies did not indicate any structural changes near the
Curie temperature (Shimakawa \etal 1997), thus ruling out any
spin-lattice or charge-lattice coupling. The compound does not show
any significant Jahn-Teller distortion in accordance with the idea
that manganese is in a Mn$^{4+}$ ionization state which is not
Jahn-Teller active. A comparison of the magnetic and transport
properties of $\rm A_2Mn_2O_7$ pyrochlores with A = Y, Lu, In and Tl
shows ferromagnetic order for all compounds with the Curie temperature
varying from 15~K ($\rm Y_2Mn_2O_7$ and $\rm Lu_2Mn_2O_7$) to 140~K
($\rm In_2Mn_2O_7$ and $\rm Tl_2Mn_2O_7$). Apart from $\rm
Tl_2Mn_2O_7$ all compounds are insulators. The ferromagnetic ordering
was interpreted within a super-exchange model. According to the
Goodenough-Kanamori rules (Kanamori 1958, Goodenough 1955, 1958) the
super-exchange interaction between Mn$^{4+}$ ions via oxygen ions is
antiferromagnetic for a bond angle of 180$^\circ$ and ferromagnetic
for a 90$^\circ$ bond angle. The bond angle is near 133$^\circ$ in the
pyrochlores and therefore in the crossover region. From the empirical
expression
\begin{equation}
J = J_{90^\circ}\, \sin^2\phi + J_{180^\circ}\, \cos^2\phi\, ,
\label{super}
\end{equation}
where $\phi$ denotes the bonding angle, and using values for Cr$^{3+}$
(supposed to be similar to Mn$^{4+}$),
$4S(S+1)J_{90^\circ}/{\rm k} = 380$~K, $4S(S+1)J_{180^\circ}/{\rm k} =
-420$~K, Shimakawa \etal (1999) estimated a small, but positive
ferromagnetic super-exchange coupling between the Mn$^{4+}$ ions. $S =
3/2$ denotes the Mn$^{4+}$ spin. This
is supposed to lead to the small Curie temperature in $\rm Y_2Mn_2O_7$
and $\rm Lu_2Mn_2O_7$. Electronic band-structure calculations using
the LSDA+U method showed insulating behaviour for $\rm Y_2Mn_2O_7$ and
$\rm In_2Mn_2O_7$ and metallic behaviour for $\rm
Tl_2Mn_2O_7$ (Shimakawa \etal 1999). Hybridisation between the In(5$s$), O(2$p$) and Mn(3$d$)
states in $\rm In_2Mn_2O_7$ and the Tl(6$s$), O(2$p$) and Mn(3$d$) states in
$\rm Tl_2Mn_2O_7$ was observed and interpreted as promoting the
ferromagnetism, thus leading to the high Curie temperature of these
compounds. The hybridized Tl(6$s$), O(2$p$) and Mn(3$d$) minority spin band
crosses the Fermi level and overlaps with the valence
band. Accordingly, a small free-electron like pocket is created at the
$\Gamma$ point. These electrons dominate the transport properties,
since the up-spin holes near the $\Gamma$ point have a very large
mass. Band-structure calculations by Singh (1997) using the linearized
augmented plane wave (LAPW) method yielded similar results. Highly
spin-differentiated transport with dominantly minority carriers
results. Thus, in the case of the pyrochlore one has itinerant
carriers of mainly Tl--O character interacting with the localized
Mn$^{4+}$ moments.

Majumdar and Littlewood (1998a, 1998b) proposed a model of a low density
electron gas coupled to spin fluctuations to account for the
properties of $\rm Tl_2Mn_2O_7$. The low carrier density leads to
large spin-disorder scattering within an itinerant model in agreement with
the experimental results in the metallic phase. Above $T_C$ spin
polarons are formed with a strongly field dependent binding energy;
the resistivity in this regime is semiconducting in agreement with
experiment. This model was supported by Mart{\'\i}nez \etal (1999) and
Alonso \etal (1999) through measurements of the magnetic and
magnetotransport properties as a function of carrier density in $\rm
Tl_2Mn_{2-x}Ru_xO_7$ and $\rm Tl_{2-x}Bi_xMn_2O_7$ substituion series,
respectively. Further details on specific models can be found in a
short review by Ventura and Gusm\~{a}o (2001).
\subsection{CrO$_2$}
CrO$_2$ is the simplest of the half-metallic ferromagnets. Within an
ionic model, Cr$^{4+}$ ($3d^2$) ions are ferromagnetically ordered,
yielding a magnetic moment of $2\mu_{\rm B}$ per formula unit in
agreement with magnetization measurements. The temperature
dependence of the resistivity is similar to that of SrRuO$_3$ and the basic 
transport mechanisms in both compounds might be very similar.
Band-structure calculations using the local spin-density approximation
(LSDA) (Lewis \etal 1997) and LSDA+U (Korotin \etal 1998) indicate
half-metallic behaviour with a finite density of states at the Fermi level
for the majority carriers and a gap of about 1.5~eV in the minority 
density of states. The bands crossing the Fermi energy are predominantly
of O(2$p$) character; thus, these can be viewed as charge reservoirs leading to a
non-integral occupation of the $d$-bands, a mechanism called ``self-doping''
by Korotin \etal (1998). An almost dispersionless majority spin-band
is located about 1~eV below the Fermi energy and is an almost pure $d$-band in character.
Korotin \etal (1998) therefore suggested that the charge carriers in
the extended hybridized
$p-d$ states move through localized $d$-levels near the ion cores and are
polarized by the localized moments through Hund's rule coupling.
This scenario is very similar to the double-exchange mechanism proposed
by Zener (1951); thus, CrO$_2$ might be regarded as a self-doped
double-exchange ferromagnet. According to the half-metallic band structure,
the local moment is 2$\mu_B$ in agreement with experiment.

\begin{figure}[t]
\begin{center}
\vspace*{0.0cm}
\hspace*{0.5cm} \includegraphics[width=0.8\textwidth]{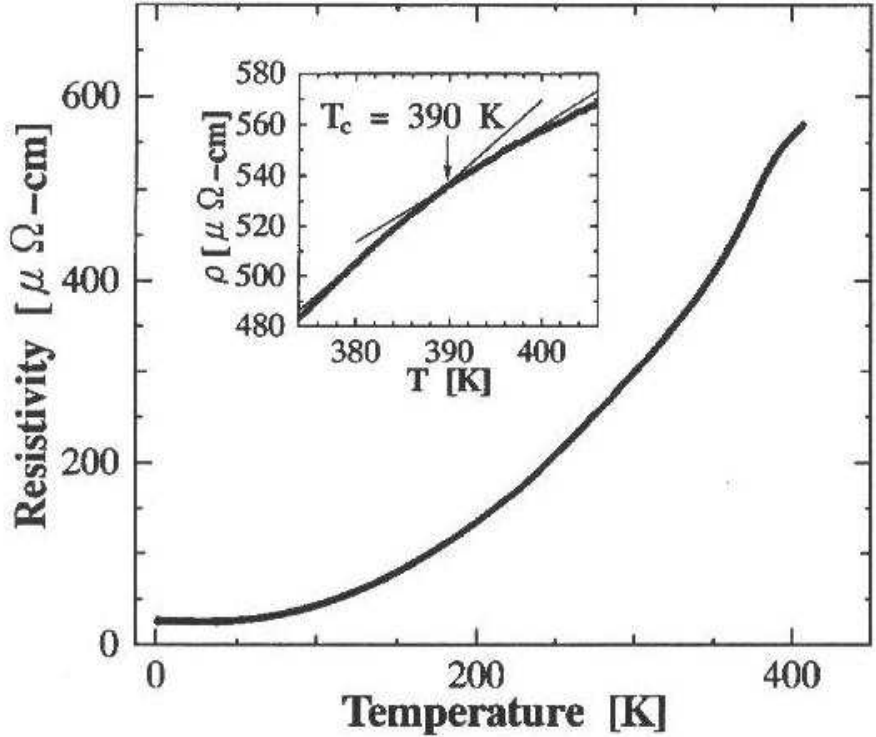}
\vspace*{-8.5cm}
\end{center}
\caption{Temperature dependence of the resistivity of a CrO$_2$ film
grown on ZrO$_2$. The inset shows the change in slope of the
resistivity near the Curie temperature of 390~K.
Reproduced from Suzuki and Tedrow (1998).}
\label{suzuki1}
\end{figure}

The measured resistivity of CrO$_2$ is indeed metallic above and below
the Curie temperature, see figure~\ref{suzuki1}, which shows the 
resistivity of a CrO$_2$ film grown on ZrO$_2$ (Suzuki and Tedrow 1998). 
A change in the slope of the resistivity is 
discernible near the Curie temperature of 390~K. 
Models for the temperature dependence of the resistivity have been 
proposed, but remain controversial.
Lewis \etal (1997) fitted Bloch-Gr\"uneisen functions to the resistivity
data of CrO$_2$ single crystals and found good agreement at temperatures
below about 200~K with a reasonable choice of parameters.
Suzuki and Tedrow (1998) reported a quadratic temperature dependence
in polycrystalline CrO$_2$ films below about 240~K. Motivated
by the band-structure calculations of Lewis \etal (1997), they fitted
their data with a model of parallel conduction of metallic and
semiconducting channels.
Barry \etal (1998) reported a dependence 
\begin{equation}
\rho(T) = \rho_0+\alpha T^2\exp\left(-\Delta/{\rm k}T\right)
\label{rcro2}
\end{equation}
indicating a gap $\Delta$ of the order of 7~meV in the excitation spectrum.
This gap was tentatively related to magnon scattering in a half-metallic
ferromagnet; since electron-magnon scattering is a spin-flip process,
it is impeded by the gap in the minority density of states.

Early spin-resolved photoemission experiments on a polycrystalline 
CrO$_2$ film by K\"amper \etal (1987) showed no spectral weight at 
the Fermi energy, in contrast to the metallic behaviour of the resistivity.
In these experiments nearly 100\% spin-polarization was found 2~eV
below the Fermi energy. More recent optical data on CrO$_2$ bulk
polycrystals by Tsujioka \etal (1997) show a small but finite density 
of states at the Fermi level. Ultraviolet photoemission and X-ray
inverse photoemission measurements show large peaks above and below
the Fermi energy that are attributed to Cr $d$-bands (Tsujioka \etal
1997); the 3$d$ band splitting is about 4.5~eV.
The position of the main Cr peaks is in good agreement with the
LSDA+U calculation, whereas the LSDA calculation yields a peak
separation that is clearly too small. This indicates the important 
influence of the strong $d-d$ Coulomb interaction on the Cr $d$-bands.
The finite density of states at the Fermi level, however, indicates
that the Cr t$_{2g}$ 3$d$-bands, being strongly hybridized with the
O(2$p$) bands, are scarcely influenced by the Coulomb repulsion; these
bands cross the Fermi level and cause the metallic transport properties.
Infrared spectroscopy shows an interband transition at 3.35~eV that
was attributed to excitations across the minority spin gap
and is in agreement with the idea of a half-metallic metal (Singley
\etal 1999). Yamamoto \etal (2000) report temperature dependent
features in the absorption spectra of CrO$_2$ films near photon
energies of 0.5, 1.0 and 1.5~eV. The spectral weight transfer was
found to scale with $(M/M_{\rm S})^2$ thus indicating a relation of
these features with the spin-polarization. The estimated minority spin
gap is 1.5~eV in agreement with band-structure calculations.

The transport and optical properties of CrO$_2$ were critically
discussed by Mazin \etal (1999). These authors suggest that correlation
effects are small and doubt the superiority of LSDA+U calculations over 
the LSDA method. Mazin \etal (1999) indeed showed that the separation
of the Cr $d$-states depends on the particular method used for
band-structure calculations. The density of states strongly
varies near the Fermi level, thus yielding the calculated carrier
density at the Fermi level somewhat uncertain. A comparison of the
specific heat data of Tsujioka \etal (1997) and the calculated
density of states yielded comparatively small effective-mass 
enhancements in the range 1.1 to 2.5. Mazin \etal (1999) suggested that
the strong temperature dependence of the resistivity is not due
to electron-magnon scattering, but caused by the strong band-structure
changes due to the coupling to the magnetism, thus leading to enhanced
electron scattering by spin fluctuations. A detailed Fermi surface
calculation was also reported by Brener \etal (2000).
\subsection{La$_{0.7}$A$_{0.3}$MnO$_3$}
\subsubsection{Resistivity and phase diagram.}
CMR manganites are oxides of the type $\rm RE_{1-x}A_xMnO_3$, where RE
denotes a rare earth and A a divalent, often alkaline earth element;
some studies with alkali element dopings, A = Na, K, ... , however, have
been made. These oxides were first investigated in the early fifties
in ceramic form (Jonker and van Santen 1950, Volger 1954, Wollan and
Koehler 1955) and at the end of the sixties in single crystal form
(Morrish \etal 1969, Leung \etal 1969, Searle and Wang 1969, Oretzki
and Gaunt 1970, Searle and Wang 1970). An account of early work can be
found in a review article by Goodenough and Longo. Research started to
focus again on these oxides in the early nineties after the discovery
of a large room temperature magnetoresistance in thin films (von Helmolt
\etal 1993). 

\begin{figure}[t]
\begin{center}
\vspace{0.0cm}
\hspace*{0.0cm} \includegraphics[width=0.75\textwidth]{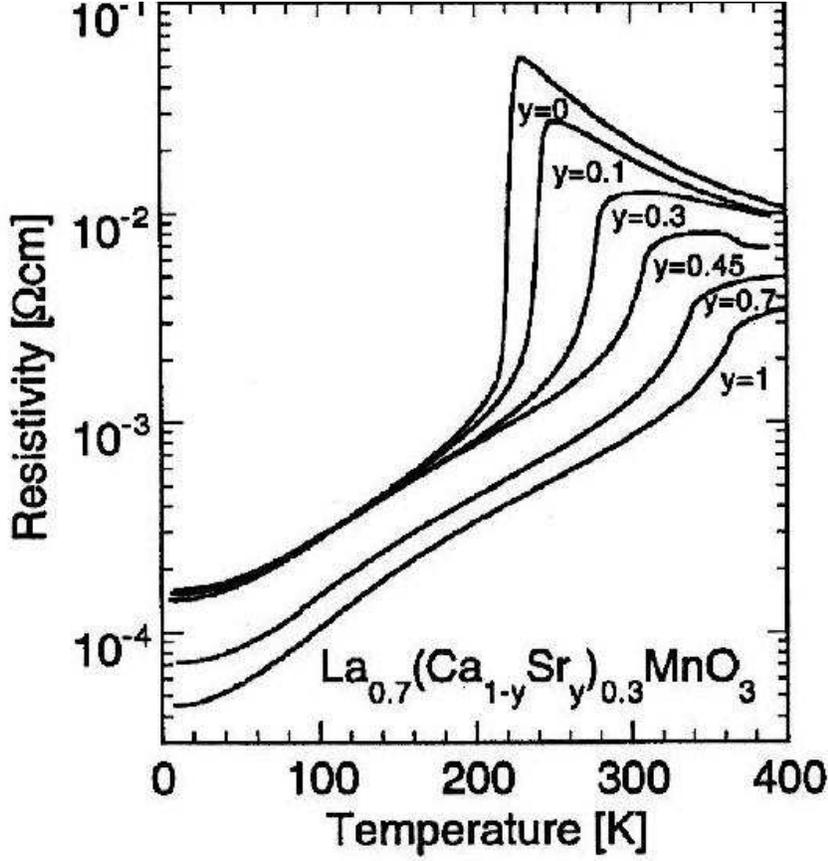}
\vspace{-5.5cm}
\end{center}
\caption{Typical resistivity versus temperature curves of 
  $\rm La_{0.7}(Ca_{1-y}Sr_y)_{0.3}MnO_3$ single crystals. The anomaly
  at a temperature of 370~K for the $y = 0.45$ doping is due to a 
  structural transition from a low temperature orthorhombic to a high
  temperature rhombohedral phase. After Tomioka \etal (2001).} 
\label{tokura1}
\end{figure}

Typical resistivity versus temperature curves for $\rm
La_{0.7}(Ca_{1-y}Sr_y)_{0.3}MnO_3$ single crystals are shown in
figure~\ref{tokura1}. At low temperatures the resistivity is metallic,
rising sharply while going through the ferromagnetic transition and showing 
semiconducting behaviour in the paramagnetic phase in the case of
Ca doping, whereas the resistivity in the case of Sr doping remains
metallic above the Curie temperature. Accordingly, the ferromagnetic
transition in this compound is accompanied by a metal-insulator
transition as evidenced by the resistivity rise and the negative
temperature coefficient of the resistivity in most compounds above $T_C$.
The residual resistivity is rather high with values between 40 and
200~$\mu\Omega$cm.

The CMR manganites crystallize in the perovskite structure as shown
in figure~\ref{crystal}(a). In the ideal perovskite structure the bond
lengths between the A and Mn cations and the O anions have the ratio
$\langle {\rm A} - {\rm O}\rangle/\langle {\rm Mn} - {\rm O}\rangle =
\sqrt{2}$, see figure~\ref{crystal}(a). In an ionic model the bond lengths
are mainly determined by the ionic radii. As a measure of deviation
from the ideal perovskite structure it is customary to define a
tolerance factor by
\begin{equation}
tol = \frac{\langle {\rm A} - {\rm O}\rangle}
{\sqrt{2}\langle {\rm Mn}-{\rm O}\rangle}\, .
\label{tol}
\end{equation}
Note that the tolerance factor depends on both temperature $T$ and
hydrostatic pressure $p$; usually $d(tol)/dT > 0$ and $d(tol)/dp < 0$
(Goodenough 1999). A tolerance factor $tol < 1$ as found in the CMR
manganites of interest in this review, places the Mn--O bonds under
compression and the A--O bonds under tension. The 
arising stresses are alleviated by a cooperative rotation of the
MnO$_6$ octahedra (Goodenough 1997). This leads to a bending of the
Mn--O--Mn bond with a decrease of the bond angle from 180$^\circ$ to
$(180^\circ-\phi)$. The cubic structure is distorted to orthorhombic
symmetry by cooperative rotations around [110] and to rhombohedral
by rotations around [111] (Goodenough and Longo).

\begin{figure}[t]
\begin{center}
\vspace{0.0cm}
\hspace*{0.0cm} \includegraphics[width=0.6\textwidth]{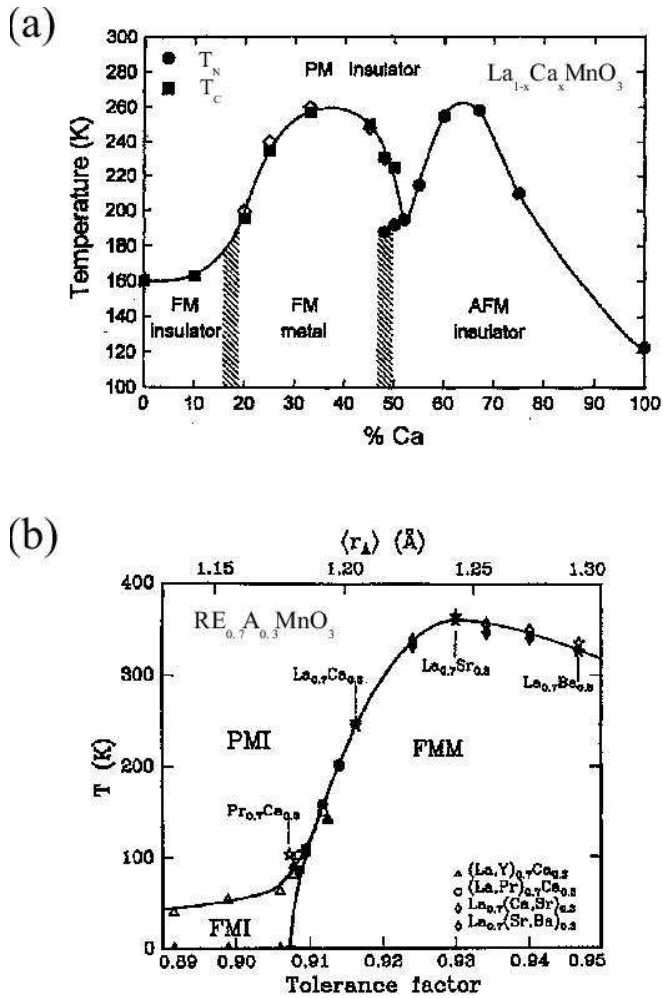}
\vspace{0.0cm}
\end{center}
\caption{(a) Phase diagram as a function of doping at constant
  tolerance factor $tol \simeq 0.918$ obtained from
  magnetization and resistivity measurements on polycrystalline $\rm
  La_{1-x}Ca_xMnO_3$ samples. This phase diagram gives an overview
  over some of the main features found in the CMR manganites. At low
  temperature, the system evolves on doping from an A-type
  antiferromagnet over insulating and metallic ferromagnetic phases
  into an insulating G-type antiferromagnet. $T_{\rm N}$ denotes the
  N\'eel, $T_{\rm C}$ the Curie temperature of the antiferromagnetic
  and ferromagnetic phases, respectively. After Schiffer \etal
  (1995). (b) Phase diagram at constant doping $x = 0.3$ als a
  function of tolerance factor. At a critical tolerance factor of
  about 0.907 the manganites become metallic at low temperature.
  After Hwang \etal (1995).} 
\label{phasedia}
\end{figure}

Much interest has been devoted to the CMR manganites, since these
display a diversified phase diagram. The phase diagram obtained from
magnetization and resistivity measurements on polycrystalline $\rm
La_{0.7}Ca_{0.3}MnO_3$ samples by Schiffer \etal (1995) is shown in
figure~\ref{phasedia}(a). This phase diagram was obtained at constant
tolerance factor. On doping with Ca the A-type antiferromagnetic,
insulating parent compound LaMnO$_3$ becomes a ferromagnetic insulator
for $x < 0.15$ and a ferromagnetic metal for $0.2 < x <0.5$, before
entering a G-type antiferromagnetic, insulating phase for $x >
0.5$. Colossal magnetoresistance appears in the doping range $0.2 < x
< 0.5$ close to the ferromagnetic transition. The Curie temperature is
maximal near $x = 0.3$. More detailed phase diagrams showing
structural transitions, charge-ordered phases etc.~can be found
e.g.~in Yamada \etal (1995) and Goodenough (1999). Here the phase
diagram will not be further discussed in detail, since the materials
of interest for extrinsic magnetoresistance effects are found near a
doping of $x = 0.3$.

In figure~\ref{phasedia}(b) the phase diagram at constant doping 
$x = 0.3$ is shown as a function of tolerance factor.
In the range of tolerance factors $0.91 \le tol
\le 0.93$ a dramatic increase of the Curie temperature was found
followed by a slow decrease in the range $0.93 \le tol \le
0.95$ (Hwang \etal 1995, Fontcuberta \etal 1996). This has been attributed
to a band narrowing on decrease of the tolerance factor. This band
narrowing weakens the ferromagnetic double exchange interaction and,
in turn, the antiferromagnetic super-exchange interaction might gain
in importance. This issue will be further discussed below. In addition
to the dependence on the tolerance factor, the Curie temperature of 
$\rm(RE_{0.7}A_{0.3})MnO_3$ perovskites at constant tolerance factor shows
a strong linear dependence upon the variance of the A site cation
radius  distribution $\sigma^2 = \langle r_{\rm A}^2\rangle - \langle
r_{\rm A}\rangle^2$ (Rodriguez-Martinez and Attfield 1996). This might
lend some support to localization models, see below.

The saturation magnetic moment at a doping $x = 0.3$ is near
$3.7\mu_{\rm B}$ (Jonker and van Santen 1950); within an
ionic picture this corresponds to the spin-only moments of 
ferromagnetically aligned 70\% Mn$^{3+}$ and 30\% Mn$^{4+}$ ions.

\subsubsection{Theoretical models.}
\paragraph{Band-structure.}
A fundamental problem arising in the study of the transport
mechanism is about the nature of the carriers. These are most often
assumed to be of Mn $3d$ character and the underlying transport
mechanism is double exchange; there are, however, some suggestions
that the CMR manganites are doped charge-transfer insulators. Although
this matter has not been fully resolved, agreement emerges that double
exchange is a basic ingredient for any model. Here I will mainly focus
on these double-exchange models and modifications due to polaronic
effects.

In this context it is informative to look at band-structure calculations.
Within the local spin density approximation (LSDA) the electronic structure of
the parent compound LaMnO$_3$ was found to be an antiferromagnetic
insulator in the orthorhombic phase (Sarma \etal 1995, Satpathy \etal
1996). The states near the top of the valence band are of Mn $3d$
character; accordingly LaMnO$_3$ appears as a Mott-Hubbard
insulator. On hole doping, band-structure calculations using LSDA
(Pickett and Singh 1996, de Boer \etal 1997, Livesay \etal 1999)
indicated a half-metallic character with a gap of about 2~eV in the
minority spin-band. Two Fermi-surface sheets were found in the
majority band, namely an electron-like sphere at the $\Gamma$ point
and hole-like cuboids at the $R$ point derived from the Mn $3d$
states. This Fermi-surface topology was corroborated by 2D-ACAR
measurements (Livesay \etal 1999), see section~\ref{materials}. On the
other hand, LSDA+U (Satpathy \etal 1996) as well as Hartree-Fock (Su
\etal 2000) calculations also yield an antiferromagnetic insulating
ground state for LaMnO$_3$, but this time the states near the top of
the valence band are of O($2p$) character indicating a charge-transfer
insulator. It seems most probable that the Mn($3d$)--O($2p$)
hybridization in the majority spin band is quite strong, but that the
$3d$ nature of the carriers is dominant such that a double-exchange
picture applies.

\paragraph{Classical double exchange.}
The transport behaviour, especially the simultaneous ferromagnetic and
metal-insulator transition, can be understood within the
double-exchange model proposed by Zener (1951) and further developed
by Anderson and Hasegawa (1955), de Gennes (1960), Searle and Wang
(1970) and Kubo and Ohata (1972).

In the perovskite structure the Mn ions are located on a simple cubic lattice,
whereas oxygen ions occupy the centers of the cube edges and the rare earth
ion or divalent dopant are located at the cube center. Thus, the Mn ions are
in an octahedral oxygen coordination and, in the ideal structure with
$tol = 1$, the Mn--O--Mn bond angle is 180$^\circ$. 
This leads to a crystal-field splitting of the Mn(3$d$) orbitals 
into low-lying $t_{2g}$ and energetically higher $e_g$ levels. 
Within the double-exchange model it is assumed that 
charge transport occurs on the Mn--O sublattice, whereas the rare earth
and earth alkaline ions act only as a charge reservoir. In the parent
compound LaMnO$_3$, the manganese ion is in a trivalent oxidation
state Mn$^{3+}$ with electronic structure 3$d^4$. According to Hund's
rules three electrons occupy the t$_{2g}$ levels and are coupled into
a core spin $S = 3/2$ by the strong intra-atomic Hund's rule
coupling. The fourth electron occupies one of the energetically
degenerate e$_g$ orbitals. Mn$^{3+}$ is known
to be a strong Jahn-Teller ion and an orthorhombic distortion
of the cubic perovskite lattice is indeed found in LaMnO$_3$. 

On doping with a divalent ion on the rare earth site, i.e.\ $\rm RE_{1-x}A_xMnO_3$,
the manganese ions become mixed valent with manganese fractions
$x$ in the tetravalent state Mn$^{4+}$ (3$d^3$) and $(1-x)$
in the trivalent state Mn$^{3+}$ (3$d^4$). Zener (1951) considered a
cluster formed from an oxygen and two Mn ions, one in the trivalent
and one in the tetravalent state. The basic idea of double
exchange is that the configurations Mn$^{3+}$--O--Mn$^{4+}$ and
Mn$^{4+}$--O--Mn$^{3+}$ are degenerate leading to a delocalization of
the hole on the Mn$^{4+}$ site:
\begin{equation}
t_{2g}^3e_g^1-O-t_{2g}^3e_g^0 \leftrightarrow
t_{2g}^3e_g^0-O-t_{2g}^3e_g^1\, .
\label{equilibrium}
\end{equation}
The transfer of a hole occurs simultaneously from Mn$^{4+}$ to O and
from O to Mn$^{3+}$; this process is a real charge transfer process
and involves overlap integrals between Mn and O orbitals. Due to the
strong Hund's rule coupling energy $J_H$, Zener (1951) suggested that the hole
transfer is only possible for parallel orientation of the core
spins. This yields the observed simultaneous occurrence of metallic
conductivity and ferromagnetism. Zener (1951) made a rough estimation
of the conductivity based on Einstein's relation and the diffusion
constant of a hole located at a Mn$^{4+}$ site.

Anderson and Hasegawa (1955) considered this three ion cluster in more
detail and calculated the transfer matrices and energy levels. They
showed that the effective hopping matrix element is given by
$t\cos(\Theta/2)$ within a classical treatment of the core spins,
where $\Theta$ denotes the angle between the core spins and $t$ is the
transfer integral. 

De Gennes (1960) was the first to formulate the double exchange
problem for a lattice and to derive a band model for the motion of
holes. An antiferromagnetic super-exchange interaction
competes with the ferromagnetic double-exchange interaction; de Gennes (1960)
predicted that this can lead to spin canting at low doping levels $x$.
Some evidence, however, suggests that phase separation into ferro- and
antiferromagnetic regions occurs at low dopings and the canted state is
not observed; this is further discussed below. 
It was shown that antiferromagnetic interactions lower the Curie
temperature and may lead to discontinuous transitions (Alonso \etal
2001) as was observed experimentally in the series $\rm
La_{0.7}A_{0.3}MnO_3$ with A = Sr, Ba, Ca by Mira \etal (1999) and
Ziese (2001a); see also the discussion on phase separation.
De Gennes (1960) further considered localized and self-trapped
carriers and showed that these give rise to a local distortion of the
spin system. This is the situation of spin-polarons to be discussed below.

The resistivity as a function of temperature was calculated within
the double-exchange model by Kubo and Ohata (1972), Calderon \etal (1999a)
and Ishizaka and Ishihara (1999). These calculations yield a metallic
state above and below the Curie temperature. This is in contrast to
the experimental results on various manganite systems, since semiconducting 
behaviour of the resistivity is found in all systems except for 
$\rm La_{0.7}Sr_{0.3}MnO_3$. Since the band-structure assumed in the
double exchange model is half-metallic, first order electron-magnon
scattering is strictly forbidden. Kubo and Ohata (1972) calculated the
low temperature resistivity due to second order electron-magnon
processes and found a contribution proportional to $T^{9/2}$.
An analysis of resistivity data in terms of power law contributions
yields an acceptable description with the resistivity given by
\begin{equation}
\rho = \rho_0+\rho_{9/2} T^{9/2}+\rho_{2} T^{2}
\label{rlcmo}
\end{equation}
up to about half the Curie temperature (Schiffer \etal 1995, Snyder
\etal 1996). In addition to the second order electron-magnon
scattering term $\rho_{9/2}T^{9/2}$, a second term $\rho_{2}T^2$
appears that might be attributed to electron-electron  
scattering. Jaime \etal (1998) argued that the coefficient
$\rho_2$ is much too large as to arise from electron-electron
scattering; it is more likely to be due to first order electron-magnon
scattering that becomes allowed at finite temperature due to a
thermally populated minority band.

Furukawa (2000) proposed an unconventional one-magnon scattering
process in half-metals that goes beyond the rigid band approximation
and takes into account spin fluctuations; this leads to a $T^3$ power
law of the low temperature resistivity. Calder\'on and Brey (2001)
considered this process in more detail and identified a $T^{3/2}$
power law dominating the $T^3$ response at low temperatures. For a
further discussion of the low temperature resistivity see the
paragraphs on localization and bi-polaron models.
\paragraph{Including electron-phonon coupling.}
Most researchers agree that double exchange is the basic mechanism
underlying the transport properties of the manganites; it seems, however,
not to be sufficient to explain the experimental results, especially
the colossal magnetoresistance discussed in the following
section. Millis \etal (1995) were among the first to promote the idea
that ``double exchange alone does not explain the resistivity of $\rm
La_{1-x}Sr_xMnO_3$''. Their argument hinges mainly on an estimate of
the Curie temperature in a pure double exchange model; this turns out
to be an order of magnitude too large. 
Moreover, Millis \etal (1995) calculated the resistivity
within the double exchange model including spin fluctuations and found
a resistivity decrease below $T_C$ and a positive magnetoresistance
above $T_C$, both features in contradiction to the experimental
results. It has to be noted, however, that this calculation does not
agree with the results of Kubo and Ohata (1972), Calder\'on \etal
(1999a) as well as Ishizaka and Ishihara (1999). Millis \etal 
(1995, 1996a, 1996b, 1996c) proposed that the electron-phonon coupling due to
the dynamic Jahn-Teller distortion has to be included. This leads to
the localization of conduction band electrons as small polarons
(Holstein polarons, see Holstein 1959) above
the Curie temperature. The basic idea is that transport in the
manganites is determined by a competition between ``self-trapping'' of
small polarons and delocalization due to the ferromagnetic
ordering. If $E_p$ denots the polaron binding energy and $t_{eff}$ the
effective hopping matrix element, then a dimensionless measure of the
electron-phonon coupling can be introduced as
\begin{equation}
\lambda = E_p/t_{eff}\, .
\label{lambda}
\end{equation}
If $\lambda$ is larger than a critical value, the
electrons are ``self-trapped'' as small polarons. On cooling through
the Curie temperature, the hopping matrix element $t_{eff}$ is
enhanced by the double exchange interaction and a polaron unbinding
transition is induced.

Strong experimental evidence for small polaron formation comes from
resistivity (Snyder \etal 1996, Ziese and Srinitiwarawong 1998), 
thermopower (Jaime \etal 1996, Palstra \etal 1997), Hall effect (Jaime
\etal 1997), optical conductivity (Quijada \etal 1998), mobility (Wang
\etal 1999), neutron scattering (Adams \etal 2000, Dai \etal 2000,
Zhang \etal 2001), volume thermal expansion (de Teresa \etal
1997), nuclear magnetic resonance (Allodi \etal 1998, Kapusta \etal
1999), a large isotope effect (Shengelaya \etal 1996, Zhao \etal
1997, Babushkina \etal 1998, Franck \etal 1998, Zhou and Goodenough
1998), X-ray absorption fine structure spectroscopy (Booth \etal 1998,
Lanzara \etal 1998) as well as Raman scattering (Yoon \etal 1998,
Bj\"ornsson \etal 2000).

Calculations of the Curie temperature within a model including both
double exchange and electron-phonon interactions (Millis \etal
1996a, 1996c) yielded results in agreement with experimental
findings. The calculated Curie temperature decreases significantly
with the electron-phonon coupling. The resistivity as a function of
temperature displayed a metal-insulator transition at the Curie
temperature for strong electon-phonon coupling. This transition is
accompanied by a large magnetoresistance. $\rho(T)$ in the
paramagnetic phase can be tuned from semiconducting to metallic
behaviour with decreasing electron-phonon coupling strength. The
optical conductivity shows a Drude peak in the ferromagnetic phase,
whereas $\sigma(\omega) \rightarrow 0$ at low frequencies in the
paramagnetic region. Moreover, above the Curie temperature a broad
maximum in $\sigma(\omega)$ appears due to transitions between the
Jahn-Teller split e$_g$ levels. The agreement with experiment is quite
encouraging and indicates that this model contains the essential
physics.

Narimanov and Varma (2001) determined the properties of the double
exchange model coupled to phonons as a function of the
electron-lattice coupling. Here also a competion between the
itinerancy of electrons and localization due to polaron formation is
found. If the elecron-lattice coupling is larger than the bandwidth,
the ferromagnetic transition is first order; in this case the lattice
distortions present in the paramagnetic phase abruptly vanish below
the Curie temperature. In the opposite limit of a small
electron-phonon coupling, the transition is second order and lattice
distortions in the paramagnetic phase are very small. This is
consistent with observations of a diffuse component in neutron
scattering experiments (Lynn \etal 1996). Furthermore, within this
model a strong dependence of the Curie temperature on the ion mass is
found in agreement with the isotope effect.
\paragraph{Spin polarons.}
R\"oder \etal (1996), see also Zang \etal (1996), investigated a
similar model of double exchange
with a coupling of the charge carriers to longitudinal optical phonons
due to the Jahn-Teller effect. Since both the static and dynamic
Jahn-Teller effect split the e$_g$ degeneracy, the model was solved
using mean-field theory in a single orbital approximation valid for
dopings $x < 0.5$. The Curie temperature was found to be considerably
reduced by the electron-phonon coupling in agreement with the results
of Millis \etal (1996a). The effective kinetic energy $K_{eff}$ shows an abrupt
decrease as a function of the electron-phonon coupling constant due to
``self-trapping'' of charge carriers into a small polaronic state. The
small polaron has spin character as evidenced by the temperature
dependence of the numerically calculated spin-distribution. Strictly
speaking, ``self-trapping'' is impossible in a model with 
translational invariance and there should always be polaronic band
conductivity (Friedman 1964); however, small polarons might be localized
by charge fluctuations if the polaronic bandwidth is narrow. Within
this model, the metal-insulator transition is due to an abrupt
unbinding of ``self-trapped'' spin-polarons as the effective kinetic
energy is increased by the onset of ferromagnetic order. This
transition is promoted by an applied magnetic field in qualitative
agreement with experimental data.

The model of R\"oder \etal (1996) is a discrete version of a continuum
model for the thermally induced abrupt shrinking of a donor state in a
ferromagnetic semiconductor as proposed by Emin and coworkers (Emin
\etal 1987, Hillery \etal 1988, Emin and Holstein 1976). These models
were developed in order to understand the sharp metal-insulator
transition observed in ferromagnetic semiconductors such as
non-stoichiometric EuO. In these an impurity band forms and
consequently the charge-carrier density is low. Main ingredients of
the model are an intra-atomic exchange interaction as well as a
short-range electron-phonon coupling. The strength of the
electron-phonon interaction is assumed to be close to the critical
value for ``self-trapping'' such that some additional mechanism like 
spin fluctuations might tip the
balance and lead to the localization of small polarons in the
paramagnetic phase. On cooling through $T_C$ the onset of
ferromagnetic ordering triggers an abrupt expansion of the small
polaronic state. This model is in agreement with data on EuO 
(Oliver \etal 1970, Penney \etal 1972) and might have some relevance
for CMR (Ziese and Srinitiwarawong 1998).
\paragraph{Vibronic states.}
The strong increase in the Curie temperature with tolerance factor
above a critical value $tol_c$, see figure~\ref{phasedia}(b), and
with doping above $x > x_c \sim 1/8$, see figure~\ref{phasedia}(a) was
addressed by Goodenough and co-workers within a vibronic state
model. The basic idea is best illustrated for data as function of
tolerance factor and is as follows (Goodenough 1997, 1999):
below a critical tolerance factor $tol_c$ the perovskite structure is
distorted by static Jahn-Teller displacements of the O ions into an
O'-orthorhombic symmetry (c $<$ a$\sqrt{2}$). The $e_g$ orbitals show
long range order and the manganites are antiferromagnetic or weakly
ferromagnetic insulators in this regime (Zhou and Goodenough
1998). The decisive role of cooperative Jahn-Teller deformations in
stabilizing orbital ordering and the A-type antiferromagnetic
structure was confirmed in a theoretical study (Capone \etal
2000). Above $tol_c$ the structure changes to O-orthorhombic (c $>$ 
a$\sqrt{2}$) and the Curie temperature increases drastically by over
200~K. Goodenough argues that in this region the cooperative oxygen
displacements are dynamic. In this regime the transport properties
change from insulating to metallic. In view of the strong
electron-lattice coupling, this transition might be understood if the
coupling between orbitals and vibrational modes, both of $e_g$
symmetry, is studied in more detail. Double exchange within a
Mn$^{4+}$--O--Mn$^{3+}$ cluster, see equation~(\ref{equilibrium}),
occurs with a hole transfer time $\tau_h$. If there are dynamic
cooperative oxygen displacements with accompanying orbital
reorientations, the holes will not be delocalized over the whole
crystal, since the transitions are only possible along bridges with
occupied $e_g$ orbitals parallel to the Mn$^{4+}$--O--Mn$^{3+}$
bond. If the vibration period is long, $\omega_0^{-1} \gg \tau_h$, a
hole is only delocalized on a Mn pair; this is called a Zener polaron. It
will be itinerant as soon as the molecular orbital reorientation time
$\tau_r$ is of the order of the vibration period. With the estimate
$\tau_r \simeq (\Delta_{JT}/W)\omega_0^{-1}$, where $W$ is the
elctronic bandwidth and $\Delta_{JT}$ the Jahn-Teller-stabilization
energy, one obtains the following conditions for itinerancy
(Goodenough 1999, Goodenough \etal 1961):
\begin{equation}
\Delta_{JT} \ll \hbar\omega_0 \ll W\, .
\label{itinerancy}
\end{equation}
For a $(180^\circ-\phi)$ Mn--O--Mn bond angle the bandwidth is obtained
within a tight-binding model as (Goodenough 1999)
\begin{equation}
W \propto \cos(\phi)\cos(\Theta/2)\, ,
\label{bandwidth}
\end{equation}
where $\Theta$ is the angle between adjacent core spins. Moreover, the
phonon frequency $\omega_0$ will also depend on $\phi$. Thus, within
this vibronic state model, the strong increase of the Curie
temperature and the electron delocalization might be attributed to
both an increase in the bandwidth or a decrease in $\omega_0$. The
isotope effect puts a more vital role on changes in $\omega_0$. A
larger oxygen mass $M_O$ softens the breathing mode $\omega_0 \propto
M_O^{-1/2}$ and drives the system towards static oxygen displacements
and the O'-orthorhombic structure.

Since the transport properties change from insulating to metallic in
this crossover regime, one might use the virial theorem to obtain a
clue towards the interplay between lattice and electronic properties
(Goodenough 1992, Archibald \etal 1996). For spherically symmetric
potentials decaying algebraically as $V(r) \propto r^{-n}$ the virial
theorem relates the mean kinetic energy $\langle T\rangle$ to the mean
potential energy $\langle V\rangle$:
\begin{equation}
2\langle T\rangle + n\langle V\rangle = 0\, .
\label{virial}
\end{equation}
If the kinetic energy is diminished by electron delocalization, the
absolute value of the potential energy $|\langle V\rangle|$ has to
decrease accordingly. Since the relevant electrons occupy antibonding
$e_g$ states, a shortening of the $\langle {\rm Mn}-{\rm O}\rangle$
bond produces such a decrease in $|\langle V\rangle|$. If the
electronic system undergoes a discontinuous metal-insulator
transition, the volume change must be discontinuous and the $\langle
{\rm Mn}-{\rm O}\rangle$ bond distribution shows a double well
potential. Therefore, such a system is likely to undergo phase
separation into hole-rich and hole-depleted regions. Rivadulla \etal
(2001) studied the evolution of magnetization and Jahn-Teller
vibrational anisotropy as a function of Mn--O--Mn bond angle. At a
critical bond angle ($\sim 159^\circ$) the vibrational anisotropy
changes between two degenerate Jahn-Teller modes and the saturation
moment changes drastically. Phase separation might arise from a spatial
variation of the vibrational anisotropy over the sample.

Evidence for this model, especially for dynamic Jahn-Teller
distortions comes from measurements of the pressue dependent
resistivity and thermopower of La$_{1-x}$Sr$_x$MnO$_3$ ($x = 0.12$,
0.15) single crystals (Zhou \etal 1997) and $^{16}$O/$^{18}$O oxygen
isotope exchanged (La$_{1-x}$Nd$_x$)$_{0.7}$Ca$_{0.3}$MnO$_3$ samples
(Zhou and Goodenough 1998) as well as from thermal conductivity
measurements on La$_{1-x}$Sr$_x$MnO$_3$ crystals (Zhou and Goodenough
2001). A discontinuous change in the Curie temperature of a
La$_{0.86}$Sr$_{0.14}$MnO$_3$ crystal as a function of pressure at $p
= 3$~kbar was interpreted as a transition from vibronic to metallic
ferromagnetism (Zhou and Goodenough 2000). Further evidence comes from
resistivity and thermopower measurements on a LaMnO$_3$ single crystal
above the orbital ordering temperature $T_{JT} = 750$~K (Zhou and
Goodenough 1999) and from a comprehensive study of the phase diagram
of La$_{1-x}$Sr$_x$MnO$_3$, $0 \le x \le 0.35$ melt grown samples (Liu
\etal 2001). The vibronic model was originally proposed to explain the
magnetic properties of LaMn$_{1-x}$Ga$_x$O$_3$ (Goodenough \etal 1961)
and was verified by thermal conductivity, magnetization and
ac-susceptibility measurements on this system under pressure (Zhou
\etal 2001).
\paragraph{Localization.}
The models discussed so far stress the
importance of polaronic effects due to the strong
electron-lattice coupling. In another class of models proposed by Kogan
and Auslender (1988, 1998), Kogan \etal (1999), Coey \etal (1995),
Varma (1996), Viret \etal (1997b) and Wagner \etal (1998), Anderson
localization of the 
charge carriers due to spin fluctuations above $T_C$ (and charge
fluctuations due to the doping disorder) is supposed to cause the
semiconducting behaviour in the paramagnetic regime as well as the
metal-insulator transition at the Curie temperature. Kogan and
Auslender (1988, 1998) showed that spin fluctuations localize charge
carriers leading to a shift in the mobility edge as a function of the
fluctuation strength. Accordingly, the metal-insulator transition
occurs, when the mobility edge crosses the Fermi level. Kogan and
Auslender (1988) found for the resistivity the expression
\begin{equation}
\rho = \rho_0
\exp\left\lbrack\frac{1-\langle\vec{S}_0\cdot\vec{S}_1\rangle/S^2}
{1+\langle S_z\rangle/S}\,
\frac{W}{4{\rm k}T}\right\rbrack\, ,
\label{kogan}
\end{equation}
where $\rho_0$ is a constant, $\langle\vec{S}_0\cdot\vec{S}_1\rangle$
the spin-correlator for spins of magnitude $S$, $\langle
S_z\rangle$ the mean spin-component along the magnetic field
direction and $W$ the bandwidth. This model is in
quantitative agreement with data on the ferromagnetic semiconductor
$\rm Cd_{0.99}In_{0.01}Cr_2Se_4$ (Kogan and Auslender 1988). Kogan
and Auslender (1998) and Li \etal (1997a), however, pointed out that such
models are likely to apply only to low density materials. Indeed, it
can be shown that electronic states near the band center of a double
exchange ferromagnet are delocalized irrespective of the disorder
strength. Since the manganites have a metallic carrier density of
about one hole per unit cell (Jakob \etal 1998, Matl \etal 1998,
Asamitsu and Tokura 1998, Ziese and Srinitiwarawong 1999), 
localization models might not explain
the magnetotransport properties of these compounds.
Wang and Zhang (1999) calculated
the temperature dependent resistivity of a nearly half-metallic
ferromagnet with Anderson localized minority carriers. These authors
found a temperature dependence
\begin{equation}
\rho=\rho_0+\rho_{5/2}T^{5/2}
\label{rthalf}
\end{equation}
above some cross-over temperature that was estimated to about
60~K. Below this temperature the resistivity follows a
$T^{3/2}$--law. Equation~(\ref{rthalf}) is also in good agreement with
experimental data below about $T_C/2$ (Ziese 2000b).
\paragraph{Bi-polaron models}
Alexandrov and Bratkovsky (1999a, 1999b, 1999c, 1999d) proposed a model based
on the bi-polaron unbinding transition in order to explain the
magnetotransport properties of the manganites. Based on results of
electron-energy-loss spectroscopy (EELS) measurements on $\rm
La_{1-x}Sr_xMnO_3$ films showing a significant O(2$p$)
hole density (Ju \etal 1997), Alexandrov and Bratkowsky (1999a, 1999b)
concluded that the manganites are charge-transfer-type doped
insulators with O(2$p$) holes. The holes in the O(2$p$) band are supposed
to form bi-polarons immobilized by the strong electron-phonon
interaction. If a pair of holes is localized on a single O ion, a singlet
state forms; the exchange interaction with local Mn($3d^4$) moments
leads to a pair-breaking of these singlets. Accordingly, the polaron
density decreases in the paramagnetic phase on cooling due to a
gradual condensation into bi-polarons. At the Curie temperature the
instability of the bi-polarons caused by the exchange interaction leads
to a sharp polaron density increase. Numerical calculations show a
large effect of an applied magnetic field on the polaron density
consistent with a large magnetoresistance. The competing energy scales
driving the metal-insulator transition in this model are the
bi-polaron binding energy and the exchange interaction. In contrast to
the preceding models which predicted a large mobility increase when
entering the ferromagnetic state, this model predicts a 
carrier-density collapse at the Curie temperature. Hall-effect
measurements usually show small carrier-density variations near the
ferromagnetic transition (Matl \etal 1998, Jakob \etal 1998, Ziese and
Srinitiwarawong 1999) and large mobility variations in
contradiction to the model. Moreover, the EELS measurements are very
sensitive to surface contamination and have to be treated with some
care, since significant oxygen condensation can occur even at quite
low pressures. Zhao \etal (2000a, 2000b) analyzed the temperature
dependence of the resistivity and found evidence for small polaron
transport in the ferromagnetic phase and bi-polaron formation above
$T_C$. The low temperature resistivity can be excellently modelled by
band transport of small polarons with
\begin{equation}
\rho(T) = (\hbar^2/n{\rm e}^2a^2t)\,
(A\omega_0)/\sinh^2(\hbar\omega_0/2{\rm k_B}T)\, .
\label{zhao}
\end{equation}
Here $a$ denotes the lattice constant, $n$ the carrier density, $t$
the hopping integral, $\omega_0$ an optical phonon frequency and $A$
is a constant. It has to be noted, however, that this expression also
holds true for small polarons derived from Mn($3d$) states. In further
work, Alexandrov \etal (2001) presented evidence for a polaronic Fermi
liquid in optimally doped manganites from studies of the resistivity
and thermopower of oxygen isotope-exchanged thin films. Here the
effective carrier mass $m^*$ is renormalized due to the strong
electron-phonon interaction and depends on the ion mass according to
$m^* = m\exp(A/\omega)$, where $m$ denotes the bare band mass. From an
experimental point of view, the polaronic liquid scenario yields a
much better description of the low temperature resistivity behaviour
than any of the power laws found within double exchange or
localization models (Zhao \etal 2000a, Ziese unpublished).
\paragraph{Phase separation and percolation.}
In studies of CMR manganites the issue of phase separation has
emerged, especially in the regime of low dopings $x$ or small
tolerance factor. Phase separation in manganites is discussed in
detail in a review by Dagotto \etal (2001); here the presentation will
be restricted to some general ideas.

The issue of phase separation in manganites is not new. Already the
first extensive neutron study of $\rm La_{1-x}Ca_xMnO_3$ by Wollan and
Koehler (1955) reported the co-existence of ferromagnetic and A-type
antiferromagnetic reflections in non-stoichiometric LaMnO$_3$ (14, 18
and 20\% Mn$^{4+}$) and in $\rm La_{0.89}Ca_{0.11}MnO_3$. The authors
concluded: ``The fact that the magnetic measurements show the (200)
and (220) ferromagnetic reflections to be field sensitive and the
superstructure reflections (100) and (210) to be unaffected by the
field is strong evidence in favour of accounting for the overall
magnetic structure of samples in this composition range as consisting
of an incoherent mixture of ferro- and antiferromagnetic regions or
domains.'' Astonishingly enough, this major experimental work was
published five years before the theoretical work of de Gennes (1960)
who interpreted the spin structure at low dopings as a canted
antiferromagnet. Within de Gennes' model the spontaneous magnetization
at low doping is proportional to the doping (de Gennes 1960). Khomskii
(2000) pointed out that the electron energy in this approximation is
given by $E = E_0 - t^2/J_H x^2$ such that 
$\partial^2E/\partial x^2 < 0$
and accordingly the compressibility is negative. This indicates the
instability of the canted state and a tendency to phase separation;
for more detailed arguments see Kagan \etal (1999) and Arovas and
Guinea (1998). This discussion is far from being settled, see e.g.~the
magnetization data by Geck \etal (2001) that support evidence for a
canted antiferromagnetic state in $\rm La_{0.94}Sr_{0.06}MnO_3$ single
crystals. 

Intimately related to the concept of phase separation is the idea of
percolation of insulating and metallic regions.
A percolation mechanism for the metal-insulator transition was
proposed by Bastiaansen and Knops (1998). Starting from a
half-metallic band-structure, where electrons can propagate only in
magnetic domains with the magnetization parallel to the electron
spin, the metal-insulator transition temperature corresponds to the
percolation threshold for magnetic domains. Above $T_C$ both majority
and minority carriers percolate leading to a resistivity
decrease. Monte-Carlo simulations of resistor networks within the
Ising model including nearest neighbour and next-nearest neighbour
bonds yielded qualitative agreement with experimental data concerning
both the temperature variation of the resistivity and the influence of
a magnetic field (Bastiaansen and Knops 1998). Since this model does
not take into account the gradual spin precession in the rather thick
domain walls, it is unlikely to capture the essential physics of the
manganites. However, in the light of recent results on phase
separation (Moreo \etal 1999), percolation of metallic and
semiconducting regions might occur near $T_C$, especially in compounds
with a low Curie temperature. The manganites have a
tendency to segregate into hole-rich ferromagnetic metallic and
undoped antiferromagnetic insulating regions. These might coexist near
the metal-insulator transition, such that the resistivity below $T_C$
is determined by percolating metallic regions. Scanning-tunnelling
microscopy studies of LCMO single crystals and films revealed the
coexistence of metallic and semiconducting regions on length scales of
several 10~nm (F\"ath \etal 1999). The application of a magnetic field
leads to the reversible growth of metallic regions. The origin of this
phase separation is not fully understood, since a separation into
differently charged regions is energetically unfavourable due to the
long range Coulomb interaction. Therefore, the experimental findings
might be related to structural disorder like twinning, variation of
the oxygen content or cation disorder (F\"ath \etal 1999).
Uehara \etal (1999) reported a percolation transition in 
$\rm La_{5/8-y}Pr_yCa_{3/8}MnO_3$ as evidenced by electron
microscopy. In this sytem phase separation into metallic
ferromagnetic and charge-ordered insulating domains occurs on a
sub-micrometer scale. This phase separation is not of a
charge-segregation type, but is one between the optimally doped
ferromagnetic and $x = 1/2$ charge-ordered states. This explains the
large length scale of about 0.5~$\mu$m observed for the domains.

Mayr \etal (2001) modelled the resistivity of manganites by a random
resistor network, based on the idea of phase separation between
metallic and insulating regions. Near percolation small magnetic
fields induce large changes in the resistivity in agreement with
experiment. Wei{\ss}e \etal (2001) propose a two-phase scenario of
competing ferromagnetic metallic and insulating polaronic phases; the
balance between these two phases can be tuned by the variation of
various parameters. The magnetization exhibits a first order
transition which is consistent with the neutron scattering data of
Lynn \etal (1996) and the magnetization data of Mira \etal (1999) and
Ziese (2001a). Dzero \etal (2000) apply percolation theory to study the
phase diagram at low dopings $x$. Modelling the metallic phase as a
two-band Fermi liquid and the insulating phase as a band insulator,
they arrive at a transition from the antiferromagnetic insulating to
the metallic ferromagnetic state at $ x \simeq 0.16$.
\paragraph{Magnetoimpurity theory.}
The magnetoimpurity theory developed by E.\ L.\ Nagaev (see
Nagaev 2001 and references therein) treats the manganites as specific
examples within the more general class of ferromagnetic
semiconductors. According to this model the mechanism underlying the
resistivity maximum at the Curie temperature and CMR should be
essentially the same in the manganites as in other ferromagnetic
semiconductors such as EuO and NdCr$_2$S$_4$. Within this approach the
manganites are considered as degenerate semiconductors; holes might
move in a band derived from Mn states, but are more likely to move in
an O band (Nagaev 1996). The transport properties are modelled within
a $sd$-model, where the d states refer to the Mn t$_{2g}$ states and the
``s-states'' to the Mn e$_g$ states. The theory explicitly takes
random fluctuations of the acceptor states into account: this leads to
Anderson localization in the band tails enhancing the carrier density
close to charged impurities. At finite temperature the carrier density
fluctuations lead to spatial fluctuations of the magnetization, where
regions close to impurities acquire an excess magnetic moment. This
scatters charge carriers and leads to the temperature dependence of the
resistivity, especially the resistivity maximum. Nagaev (1999) argues that
polaron formation is unimportant, since the $sd$-exchange energy is
much larger than the lattice polarization energy. Furthermore, since
this mechanism is supposed to describe the magnetotransport phenomena
in all ferromagnetic semiconductors, with many systems showing no
Jahn-Teller distortions, Nagaev argues that the Jahn-Teller effect is
unimportant for CMR. This model seems to contradict the isotope effect
found in many manganites. This was reconciled by Nagaev (1998) by
relating the isotope effect to the fact that the thermodynamic
equilibrium densities of oxygen vacancies or excess oxygen
atoms depend on the oxygen mass. Data on the oxygen isotope effect
in $\rm La_{0.8}Ca_{0.2}MnO_3$ partially support this scenario (Franck
\etal 1998).
\subsubsection{Optical properties.}

\begin{figure}[t]
\begin{center}
\vspace*{-0.5cm}
\hspace*{+1.0cm} \epsfysize=12cm \epsfbox{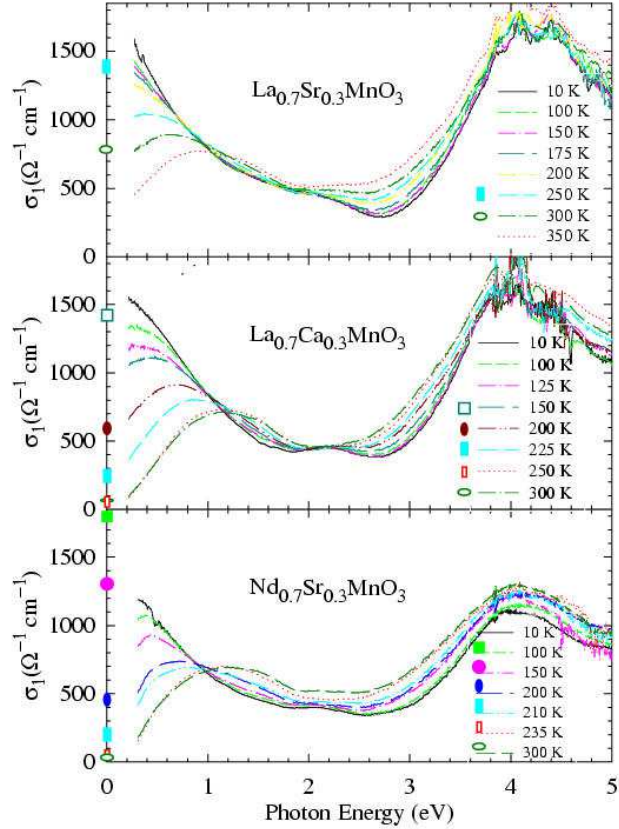}
\vspace*{0.0cm}
\end{center}
\caption{Frequency dependence of the real part of the optical conductivity
$\sigma_1$ of epitaxial manganite films at different temperatures.
Reproduced from Quijada \etal (1998).}
\label{quijada}
\end{figure}
The optical properties of the manganites are still highly controversial.
Here only a very brief overview of recent developments is given. Okimoto \etal (1997)
measured the reflectivity of $\rm La_{1-x}Sr_xMnO_3$ single crystals
with dopings $x = 0.0$, 0.1, 0.175 and 0.3 in a wide range of temperatures
for frequencies from the infrared to the ultraviolet. At high photon energies
three major, doping and temperature independent peaks are observed; 
these were assigned to one intra-atomic transition from La(5$p$)
$\rightarrow$ La(5$d$) at 25~eV and interband transitions from O(2$s$)
$\rightarrow$ Mn(3$d$) at 17~eV and O(2$p$) $\rightarrow$ La(5$d$) at
8~eV, respectively. Below about 3~eV the optical conductivity 
is strongly temperature and frequency dependent. For dopings 0.1, 0.175 and 0.3
a strong transfer of optical weight to lower frequencies is observed on decrease
of temperature. Above $T_C$ an optical gap is seen for all dopings that disappears
on cooling through the ferromagnetic transition for $x = 0.175$ and 0.3
in agreement with the metallic ground state of these compounds.
Quijada \etal (1998) measured the optical conductivity of high quality
epitaxial $\rm La_{0.7}Sr_{0.3}MnO_3$, $\rm La_{0.7}Ca_{0.3}MnO_3$
and $\rm Nd_{0.7}Sr_{0.3}MnO_3$ films for photon energies up to 5~eV.
The three samples show a similar optical conductivity that is in agreement
with the data of Okimoto \etal (1997) for $x = 0.3$. Below photon energies
of 5~eV there are three major features, see figure~\ref{quijada}: 
(i) a strong, temperature independent
peak centered at 4~eV, (ii) a shallow, strongly temperature dependent minimum
developing at low temperatures and (iii) above $T_C$ a small peak near 1~eV
that shifts to lower energy on decrease of temperature and develops into
a Drude peak. The assignment of the features is controversial. Okimoto \etal
(1997) interpreted the maximum near 1~eV as an interband transition 
O(2$p$) $\rightarrow$ Mn(3$d$) (e$_g$). These authors estimated Hund's rule
coupling energy from the peak position of the interband transitions 
and found about 0.9~eV for $x = 0.3$. Quijada \etal (1998), on the other
hand, assigned the temperature independent peak at 4~eV to the interband
transition O(2$p$) $\rightarrow$ Mn(3$d$) (e$_g$), since this transition involves
electrons of both spin directions on the oxygen and is not believed
to be strongly temperature dependent. The features at 1~eV and 3~eV were
interpreted as arising from e$_g$ $\rightarrow$ e$_g$ interband transitions
with the same and parallel spins, respectively. This yields an estimate of the
Hund's rule coupling energy of about 1.5~eV. The interpretation of 
Quijada \etal (1998) was later criticized by Chattopadhyay \etal (2000).
Chattopadhyay \etal (2000) derived the optical conductivity and the
Curie temperature within a double exchange only model; these authors
found that the Curie temperature depends linearly on the change of the
kinetic energy between zero temperature and $T_C$. The kinetic energy
does not depend on details of the band structure; it can be derived from
the optical conductivity and compared with the measured transition temperature.
On the basis of this model, Chattopadhyay \etal (2000) argued that the
strongly temperature dependent feature in the optical conductivity 
at 3~eV does not correspond to transitions into the minority e$_g$ band,
since the derived value of the Hund's rule coupling energy is too small
to reproduce the measured Curie temperature. The e$_g$ $\rightarrow$ e$_g$
transition is expected to be at a higher photon energy above 4~eV. The
analysis of the spectral weight transfer as a function of temperature
in terms of a kinetic energy change leads to the conclusion that
electron-phonon coupling and correlation effects are vital to explain the
physics of the manganites.
\subsection{$\rm Sr_2FeMoO_6$}
$\rm Sr_2FeMoO_6$ belongs to the class of ordered double perovskites
AA'BB'O$_6$. These are known to be ferromagnetic for B' = Cr, Fe, B =
Mo, Re and A = A' an alkaline earth element (Longo and Ward 1961,
Sleight \etal 1962, Patterson \etal 1963, Sleight and Weiher
1972). For the ordered double perovskites a rock-salt structure is
observed. There is a rapidly increasing bulk of research work
on double perovskites; recent research focused mainly on $\rm Sr_2FeMoO_6$ with a
Curie temperature of about 420~K (Kobayashi \etal 1998). Ca and Ba
substitution was found to decrease the Curie temperature to 345~K
($\rm Ca_2FeMoO_6$) and 367~K ($\rm Ba_2FeMoO_6$) (Borges \etal
1999). The highest Curie temperature was reported for $\rm
Ca_2FeReO_6$ with $T_C \sim 538$~K (Longo and Ward 1961).

Band-structure calculations for $\rm Sr_2FeMoO_6$ and $\rm Sr_2FeReO_6$
using the full potential augmented plane-wave (FLAPW)
method based on the generalized gradient approximation (GGA) yielded a
half-metallic band structure (Kobayashi \etal 1998,
Kobayashi \etal 1999). In the majority band an energy gap of about
1~eV was seen at the Fermi level between the occupied Fe e$_g$ and
the unoccupied Re or Mo t$_{2g}$ levels. The minority density of
states is finite at the Fermi level with carriers of hybridized Fe(3$d$)
and Mo(4$d$) or Re(5$d$) character, respectively.

Within an ionic model, $\rm A_2MoFeO_6$ is a ferrimagnet with Fe and
Mo sublattices. Recent neutron-powder diffraction, M\"ossbauer
spectroscopy and X-ray diffraction studies yield the following
consistent picture regarding crystal and magnetic structure. At room
temperature and below the compound is cubic, tetragonal or monoclinic
for Ba, Sr, and Ca substitution, respectively (Borges \etal 1999,
Greneche \etal 2001, Chmaissem \etal 2000, Ritter \etal
2000). $\rm Sr_2FeMoO_6$ shows a crystallographic transition from cubic to
tetragonal on cooling through the Curie temperature (Chmaissem \etal
2000, Ritter \etal 2000). First M\"o{\ss}bauer-investigations on $\rm Ca_2MoFeO_6$
showed a formal Fe$^{3+}$/Mo$^{5+}$ charge configuration
(Pinsard-Gaudart \etal 2000). The Fe$^{3+}$ (3$d^5$) ion is in a high spin
state with $\mu_{\rm Fe} = 5\mu_{\rm B}$ and the Mo$^{5+}$ (4$d^1$) ion
has a magnetic moment $\mu_{\rm Mo} = \mu_{\rm B}$, such that a net
moment of $4\mu_{\rm B}$ results. Neutron diffraction data, however,
indicate reduced magnetic moments between 0..0.5$\mu_{\rm B}$ on the
Mo site coupled antiferromagnetically to Fe moments of magnitude
$\mu_{\rm Fe} = 3.7...4.3\mu_{\rm B}$ (Chmaissem \etal 2000, Ritter
\etal 2000, Garc{\'\i}a-Landa \etal 1999). X-ray absorption
spectroscopy (Ray \etal 2001b) indicates a Mo moment smaller than
0.25$\mu_B$. The measured isomer shift is
rather large and indicates a mixed valence state of the Fe-ion
(Greneche \etal 2001, Balcells \etal 2001, Lind\'en \etal 2000). This
is in agreement with the reduced magnetic moment on the Mo site.
The low temperature magnetic moment as determined from global
magnetization is often found
to be considerably reduced from the ideal value of $4\mu_{\rm B}$ to
about $3-3.5\mu_{\rm B}$ (Garc{\'\i}a-Landa \etal 1999, Borges \etal
1999, Manako \etal 1999, Tomioka \etal 2000, Balcells \etal
2001). This is attributed to cation disorder on the Fe/Mo sites (Ogale
\etal 1999). Balcells \etal (2001) observed a decrease of the
saturation magnetization proportional to the antisite
concentration. The magnetization is reduced by 8$\mu_{\rm B}$ per
antisite in agreement with a simple ionic model. This is also
consistent with the data of Tomioka \etal (2000). From the analysis of
M\"ossbauer spectra Greneche \etal (2001) concluded that antisite
defects predominate in comparison to anphase boundaries.
Mart{\'\i}nez \etal (2000) determined the effective magnetic moment
$\mu_{\rm eff} \simeq 3.4\mu_{\rm B}$ from the high temperature
susceptibility. This value indicates a Fe(3$d^6$)Mo(4$d^0$) state.
The Re-compounds $\rm A_2FeReO_6$ are less well studied; measurements
of isomer shifts, however, also indicate the mixed valence nature of
Fe in these materials; the mixed valent character was observed to
decrease from Ba to Ca substitution (Gopalakrishnan \etal
2000). In conclusion, these data indicate that the double perovskites are
itinerant ferromagnets with a mixed valence of the Fe-ions; the
itinerant carriers are mainly of Mo(4$d$) and Re(5$d$) character.

The resistivity depends sensitively on the preparation conditions,
presumably due to cation disorder, grain-boundary scattering and
oxygen content. In $\rm Sr_2FeMoO_6$ both semiconducting and metallic
behaviour has been observed (Tomioka \etal 2000, Westerburg \etal 2000, Asano \etal
1999, Manako \etal 1999, Chmaissen \etal 2000). Judging from measurements on a single
crystal grown by the floating zone method, the stoichiometric compound
has a metallic resistivity below and above the Curie temperature
(Tomioka \etal 2000). However, Niebieskikwiat \etal (2000) presented
evidence for a strong influence of environmental conditions on the
resistivity. Careful measurements under vaccum show a metal-insulator
transition at the Curie temperature (Niebieskikwiat \etal 2000). The
residual resistivity is quite high with 
$200-300$~$\mu\Omega$cm (Tomioka \etal 2000, Westerburg \etal 2000)
presumably due to cation disorder scattering. The nature of carriers is
electron-like with a density of about $1.1\times10^{28}$~m$^{-3}$
corresponding to one electron per Fe/Mo pair (Tomioka \etal 2000). The
optical conductivity in the ferromagnetic phase shows a Drude
component and two excitation maxima at 0.5~eV and 4~eV,
respectively. These have been interpreted as charge-transfer
transitions from the up spin Fe(e$_{g\uparrow}$) to the Mo(t$_{2g}$) band (0.5~eV)
and the O(2$p$) to Mo/Fe(t$_{2g\downarrow}$) down spin band (4~eV), respectively
(Tomioka \etal 2000). In the Re-compounds there is some evidence that
$\rm Ba_2FeReO_6$ is metallic, whereas $\rm Ca_2FeReO_6$ is insulating
(Gopalakrishnan \etal 2000, Prellier \etal 2000); this correlates to
the decreasing mixed valence character of the Fe ion.

A considerable low field magnetoresistance often appears in
magnetotransport measurements that is likely to be of extrinsic origin
arising from grain-boundary or cation-disorder scattering. At the
Curie temperature a small magnetoresistance maximum of about -5\% in
8~T was observed (Westerburg \etal 2000). Alonso \etal (2000) report a
magnetoresistance increasing with temperature in $\rm
Ca_2FeMoO_6$. The data on the intrinsic 
magnetoresistance are too scarce to allow a further discussion.

The origin of the itinerant ferrimagnetism in FeMo and FeRe double
perovskites seems to be intimately linked to a specific band
structure, namely the strong overlap between the 
Fe t$_{\rm 2g\downarrow}$ and the Mo(4$d$) or Re(5$d$) levels,
respectively. It is striking indeed that Sr$_2$MMoO$_6$ with M = Cr,
Mn, Co (Moritomo \etal 2000) and Sr$_2$FeWO$_6$ (Kobayashi \etal 2000,
Dass and Goodenough 2001, Ray \etal 2001a) are insulating, especially
since W and Mo are iso-electronic. Moreover, Mo is known to have a
small exchange integral and a magnetic moment on the Mo site is
rare. Theoretical models stress the hybridization between the Fe
t$_{\rm 2g\downarrow}$ and Mo(4$d$) bands at the Fermi level that
induces spin-polarization on the Mo site. Sarma \etal (2000) find a
large enhancement of the effective exchange integral that leads to a
negative effective Coulomb interaction strength $U$. This is
consistent with Mo core level spectroscopy (Sarma \etal 2000). Fang
\etal (2001) point out that the strong hybridization leads to both a
gain in kinetic energy and an induced spin-polarization on Mo/Re, both
stabilizing ferrimagnetism. The W($5d$) bands are slightly too high in
energy to allow for sufficient hybridization with the Fe($3d$) states;
consequently, Sr$_2$FeWO$_6$ is an antiferromagnetic insulator
dominated by super-exchange. These models obviously have to be further
developed; they are, however, in agreement with the percolative nature
of charge transport in Sr$_2$FeMo$_x$W$_{1-x}$O$_6$ alloys (Kobayashi
\etal 2000, Dass and Goodenough 2001, Ray \etal 2001a).

Electron doping in the series $\rm Sr_{2-x}La_xFeMoO_6$ raises the
Curie temperature from 420~K to 490~K (Navarro \etal 2001).
\subsection{$\rm Fe_3O_4$}
\begin{figure}[t]
\vspace{-1.0cm}
\begin{center}
\hspace*{2.5cm} \epsfysize=12cm \epsfbox{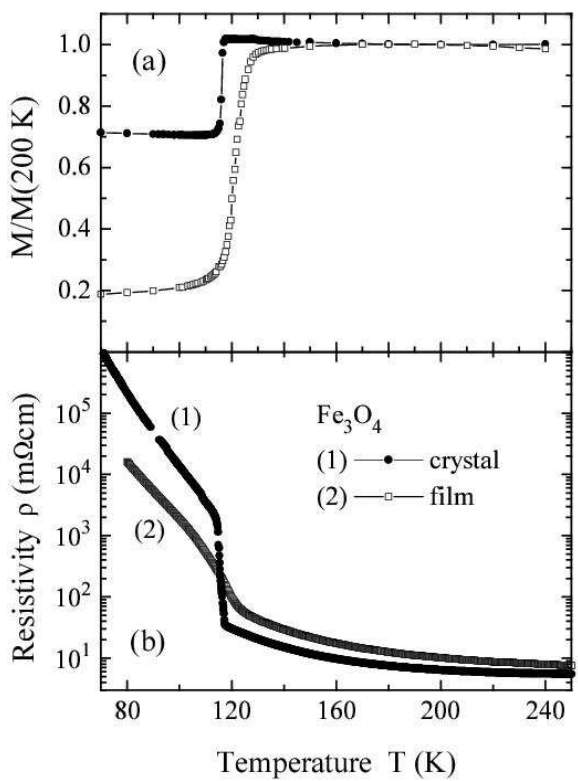}
\end{center}
\caption{(a) Magnetization and (b) resistivity of a magnetite single
crystal and a 200~nm thick magnetite film as a function of temperature.
The magnetization was measured in applied fields of 5~mT along [110]
in the case of the crystal and in 10~mT along [100] in the case of the
film. After Ziese and Blythe (2000).}
\label{feo1}
\end{figure}

Magnetite is a well-studied ferrimagnet and for many details the
reader is referred to the extensive reviews of Brabers, Brabers
and Whall as well as Krupi\v{c}ka and Nov\'ak; here
only some magnetotransport properties will be discussed.
Magnetite shows semiconducting behaviour between the Verwey transition
temperature $T_V$ (Verwey 1939) and 320~K, crossing over to metallic behaviour at higher
temperatures (Todo \etal 1995). At the Verwey transition a jump in the resistivity
is observed; the magnitude of this jump depends on the stoichiometry and
reaches up to a factor of 100. The temperature dependent resistivity of a single crystal
and a 200~nm thick film are shown in figure~\ref{feo1}(b).
The single crystal shows a sharp resistivity
jump of nearly two orders of magnitude at the Verwey temperature,
$T_V = 116.5$~K; the film has a higher Verwey temperature $T_V = 119$~K,
but the resistivity transition is much more gradual. Stoichiometric
magnetite has a Verwey temperature $T_V \simeq 123$~K; this indicates
that the two samples shown in figure~\ref{feo1} are somewhat iron deficient.
The magnetization measured in small magnetic fields shows a sharp 
increase when going through the Verwey temperature.

Magnetite crystallizes in the inverse spinel structure. At room temperature,
in this structure large O ions are located on a close-packed face-centered
cubic lattice, whereas the Fe ions occupy interstitial sites.
There are two kinds of cation sites, namely the tetrahedrally
coordinated A site occupied only by Fe$^{3+}$ ions and the
octahedrally coordinated B site occupied by both Fe$^{2+}$ and Fe$^{3+}$
ions. The A- and B-site sublattices are ferrimagnetically aligned such
that the net moment is equal to the magnetic moment 
$\mu = 4\mu_{\rm B}$ of the Fe$^{2+}$ (3$d^6$) ion.

The Verwey transition is associated with a order-disorder
transition from a charge ordered state of the Fe ions on the
B sites at low temperatures to a statistical
distribution at higher temperatures (Verwey 1939). The
conductivity of magnetite at room temperature which is exceptionally
high among the ferrites is due to electron
transfer between Fe$^{2+}$ and Fe$^{3+}$ ions on the B sites -- or, to
put it in other words, due to the mixed valence nature of the Fe ions
on the B sites. Below the Verwey transition, in the charge ordered state, carrier
transport occurs via electron hopping and it is intuitively clear that
the resistivity shows semiconducting/insulating behaviour. Band-structure
calculations (de Groot and Buschow 1986, P\'enicaud \etal 1992,
Yanase and Siratori 1984, Yanase and Hamada 1999) 
indicate a half-metallic nature with a gap in the majority density of states.
However, this is not in agreement with the experimentally observed
semiconducting resistivity up to about 320~K. Here the strong electron-phonon
interaction is important leading to the formation of small polarons.
In the charge ordered phase, the strong intersite Coulomb interaction
leads to a band-splitting and semiconducting behaviour.
Ihle and Lorenz (1985, 1986) showed that short-range polaronic order 
and band-splitting due to the strong intersite Coulomb interaction 
persist above the Verwey transition up to high temperatures. However,
above $T_V$ a narrow band is formed at the Fermi level and gradually populated
at higher temperatures. This leads to thermally activated polaronic 
band motion above the Verwey transition. 
In parallel to the polaronic band conductivity, polaronic hopping
conductivity becomes important at higher temperatures. If both conduction
processes are taken into account, good agreement between the
calculated and measured conductivity is found, and especially
the conductivity maximum near room temperature is reproduced (Ihle and
Lorenz 1986). The band-splitting is confirmed in LSDA+U band-structure
calculations performed by Anisimov \etal (1996).

Photoemission measurements of the band structure of $\rm Fe_3O_4$
single crystals (Chainanai \etal 1995) and thin films
grown on Pt (111) (Cai \etal 1998) indicate a finite
density of states at the Fermi level and corroborate the predicted
metallic nature in the cubic phase. This is in agreement with the
analysis of optical conductivity data by Degiorgi \etal (1987)
and Park \etal (1998). Park \etal (1998) observe a clear
opening of an optical gap below the Verwey transition. Above the
transition a significant spectral weight transfer in the Fe(3$d$)
intersite transition region occurs in qualitative agreement with
considerable short-range order as assumed in the model of 
Ihle and Lorenz (1985, 1986). A careful analysis of the optical
conductivity by Degiorgi \etal (1987) revealed a small Drude contribution
at 300~K and 130~K in agreement with polaronic band transport
above $T_V$. A broad conductivity maximum near 0.2~eV was attributed
to small polaron hopping (Degiorgi \etal 1987). On the other hand,
Park \etal (1997) conclude from their photoemission and inverse 
photoemission data on a magnetite single crystal that there is no
spectral weight at the Fermi level. On heating through the
transition the single particle gap does not collapse, but
is merely reduced by about 50~meV from $150\pm 30$~meV to $100\pm 30$~meV.
This is consistent with the conductivity jump by a factor of a hundred,
if semiconducting transport is assumed above and below $T_V$.
In conclusion, a unified picture of transport in magnetite has not
yet been found. The model of Ihle and Lorenz (1985, 1986), however, is supported
by a bulk of dc conductivity, optical conductivity and photoemission
spectroscopy data. For further discussions on the transport properties
of magnetite the reader is referred to the review of Brabers (1995) and
to a recent review on metal-insulator transitions by Imada \etal (1998).
\section{Intrinsic magnetoresistance \label{intrinsic}}
In this section a brief overview on the intrinsic
magnetoresistance of SrRuO$_3$, Tl$_2$Mn$_2$O$_7$, CrO$_2$, the
manganites and magnetite is given in order to facilitate comparison
with the extrinsic magnetoresistance that will be discussed in later
sections.
\subsection{SrRuO$_3$}
The magnetoresistance of SrRuO$_3$ films grown on 2$^\circ$ miscut SrTiO$_3$
substrates was measured by Kacedon \etal (1997) and Klein \etal (1998).
The magnetoresistance depends sensitively on either the current and the
magnetic field direction as usually found in single crystals (Campbell and Fert 1982).
Near the Curie temperature a maximum in the magnetoresistance ratio was found
that does not saturate in magnetic fields up to 8~T. The value of the
maximal magnetoresistance depends on the current and field direction
with values between $-2$\% and $-11\%$. Klein \etal (1998) interpreted this
magnetoresistance peak as arising from an increase of the magnetization
and a corresponding reduction of spin-disorder scattering. At lower temperatures
anisotropic magnetoresistance is found and the resistance change is mainly
due to a change in the magnetization direction. Again, the magnetoresistance values
reported by both groups do not agree; Klein \etal (1998) found about $-18$\%
in a field of 6~T for currents along [001] at low temperatures, whereas 
Kacedon \etal (1997) reported only about $-3$\% in a field of 8~T and for
the same current direction. It is not clear, to what extent the magnetoresistance
values are influenced by sample inhomogeneities and extrinsic effects.
\subsection{Tl$_2$Mn$_2$O$_7$}
The pyrochlore $\rm Tl_2Mn_2O_7$ shows a large magnetoresistance near
the Curie temperature that is actually larger than the
magnetoresistance observed in $\rm La_{0.7}Ca_{0.3}MnO_3$ (Ramirez and
Subramanian 1997).
As already indicated in the previous section, the model of Majumdar
and Littlewood (1998a, 1998b) describing the dynamics of a low density
electron gas coupled to spin fluctuations seems to capture the
relevant physics of the magnetotransport phenomena in $\rm
Tl_2Mn_2O_7$. Within this model the large magnetoresistance near the
Curie temperature arises from two mechanisms: in the metallic phase
spin-disorder scattering is anomalously large due to the low carrier
density; above $T_C$ spin-polarons form in an intermediate temperature
range that display a strong field dependence of the binding energy. In
both cases the application of a magnetic field leads to a large
decrease of the resistivity.
\subsection{CrO$_2$}
\begin{figure}[t]
\begin{center}
\vspace*{0.0cm}
\hspace*{0.5cm} \includegraphics[width=0.8\textwidth]{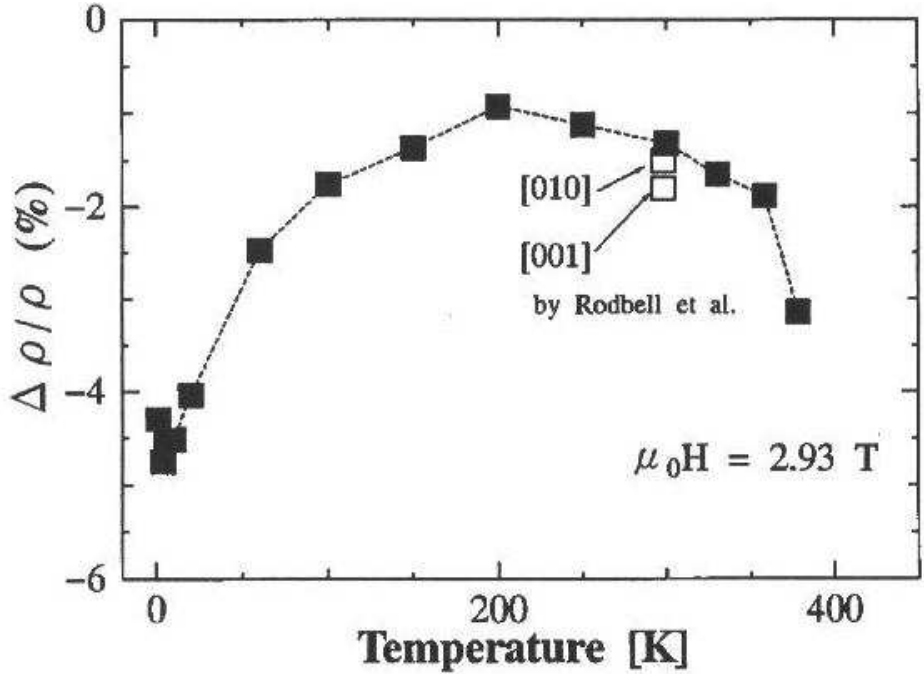}
\vspace*{-9.5cm}
\end{center}
\caption{Longitudinal magnetoresistance ratio at 2.93~T of a textured CrO$_2$ 
film on ZrO$_2$ as a function of temperature. For comparison, data by
Rodbell \etal (1966) are shown. Reproduced from  Suzuki and Tedrow (1998).}
\label{suzuki2}
\end{figure}
The magnetoresistance of textured CrO$_2$ films was
studied by Suzuki and Tedrow (1998, 1999). The longitudinal
magnetoresistance ratio in a magnetic field of 2.93~T is shown
in figure~\ref{suzuki2} as a function of temperature. The magnetoresistance
shows a maximum at low temperatures and an increase at high temperatures
above about 300~K. The Curie temperature of this film was 390~K; thus, one
cannot decide if CrO$_2$ displays a magnetoresistance maximum at $T_C$
as expected from the interpretation as a self-doped double-exchange ferromagnet
(Korotin \etal 1998). Above about 200~K the longitudinal magnetoresistance is linear
in the applied field, whereas a concave non-linear behaviour is seen
below this temperature. The non-linear field dependence was attributed
to cyclotron orbital motion of the conduction electrons (Suzuki and Tedrow 1998).
\subsection{La$_{0.7}$A$_{0.3}$MnO$_3$}
\subsubsection{Colossal magnetoresistance.}
\begin{figure}[t]
\begin{center}
\vspace{-11.0cm}
\hspace*{0.0cm} \includegraphics[width=0.9\textwidth]{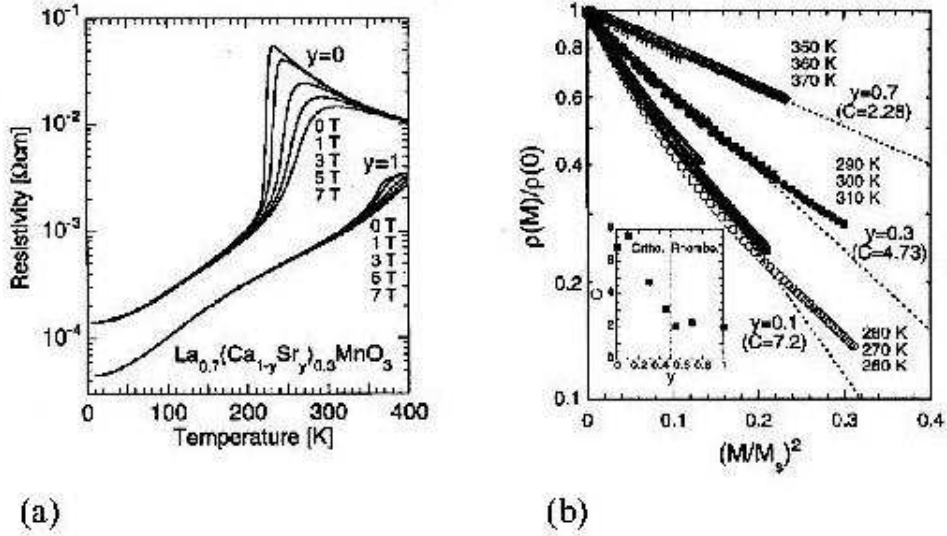}
\vspace{-2.8cm}
\end{center}
\caption{(a) Resistivity versus temperature for  
  $\rm La_{0.7}(Ca_{1-y}Sr_y)_{0.3}MnO_3$ ($y = 0$, $y = 1$) single
  crystals in various applied fields. The magnetoresistance is maximal
  near the metal-insulator transition. (b) Magnetoresistance as a
  function of the normalized magnetization $M/M_S$ for samples with $y
  = 0.1$, 0.3 and 0.7. A scaling $\rho/\rho_0$ with $(M/M_S)^2$ is
  observed. After Tomioka \etal (2001).} 
\label{tokura2}
\end{figure}
The manganites have received intense research interest, since these compounds
display a new kind of magnetoresistance called colossal magnetoresistance.
This magnetoresistance appears as a sharp magnetoresistance peak at the
Curie temperature. In figure~\ref{tokura2}(a) typical resistivity data of
$\rm La_{0.7}(Ca_{1-y}Sr_y)_{0.3}MnO_3$ single crystals in various
magnetic fields are shown. The magnetoresistance is maximal near the
metal-insulator transition leading to a peak in the magnetoresistance ratio
\begin{equation}
\frac{\Delta \rho}{\rho_0} = \frac{\rho(H)-\rho_0}{\rho_0}\, .
\label{ratio}
\end{equation}
The height of this magnetoresistance
peak is seen to decrease with increasing Curie temperature. This is
a general trend in the manganites and magnetoresistance values of nearly
100\% can be found in compounds with low Curie temperatures (McCormack
\etal 1994, Jin \etal 1995a, 1995b, Coey \etal
1999). Figure~\ref{tokura2}(b) shows that the magnetoresistance
depends on temperature and field through the magnetization. Often a
scaling $\Delta\rho/\rho_0 = -C(M/M_S)^2$ is found for small values of
the reduced magnetization (Tokura \etal 1994, Urushibara \etal 1995,
Fontcuberta \etal 1996, O'Donnell \etal 1996). The scaling constant
$C$ lies in the range $1 \le C \le 4$ and depends on doping
(Urushibara \etal 1995). This can be understood within
the Kondo lattice model in the classical spin limit (Furukawa 1994,
1995a, 1995b, Inoue and Maekawa 1995). $C$ depends on the value of the
Hund's rule coupling with $C = 1$ in the weak coupling limit $J_H \ll
1$; in the strong coupling limit $C$ depends on doping and is of the
order $C \sim 5$ in qualitative agreement with experiment (Furukawa
1994). It has to be pointed out, however, that a quadratic dependence
of the resistivity on the magnetization is also present in
conventional ferromagnets such as Ni (Gerlach and Schneiderhan 1930).

\begin{figure}[t]
\begin{minipage}[t]{0.5\textwidth}
\centerline{\mbox{\epsfysize=5.5cm \epsffile{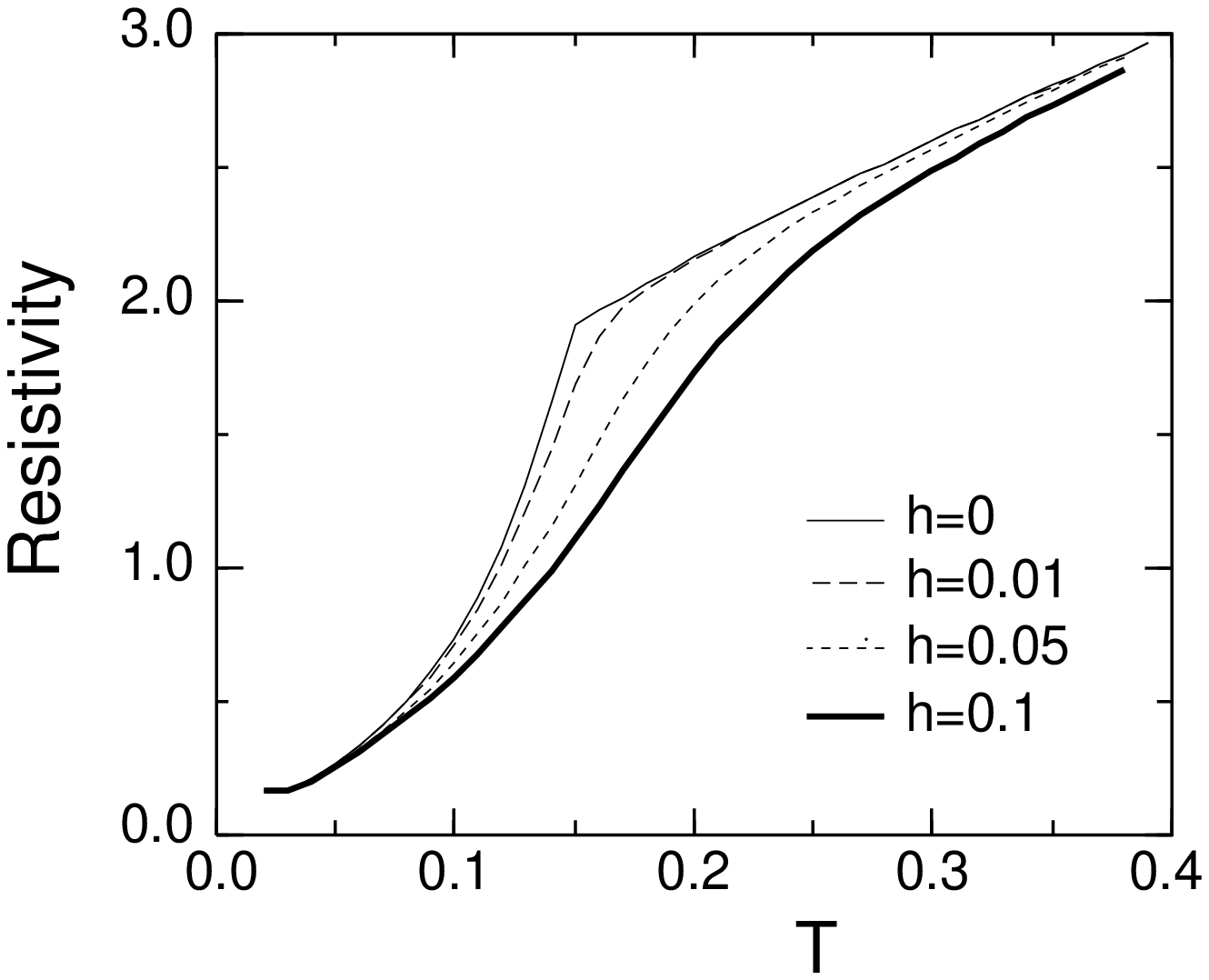}}}
\end{minipage}
\hfill
\begin{minipage}[t]{0.5\textwidth}
\centerline{\mbox{\epsfysize=5.5cm \epsffile{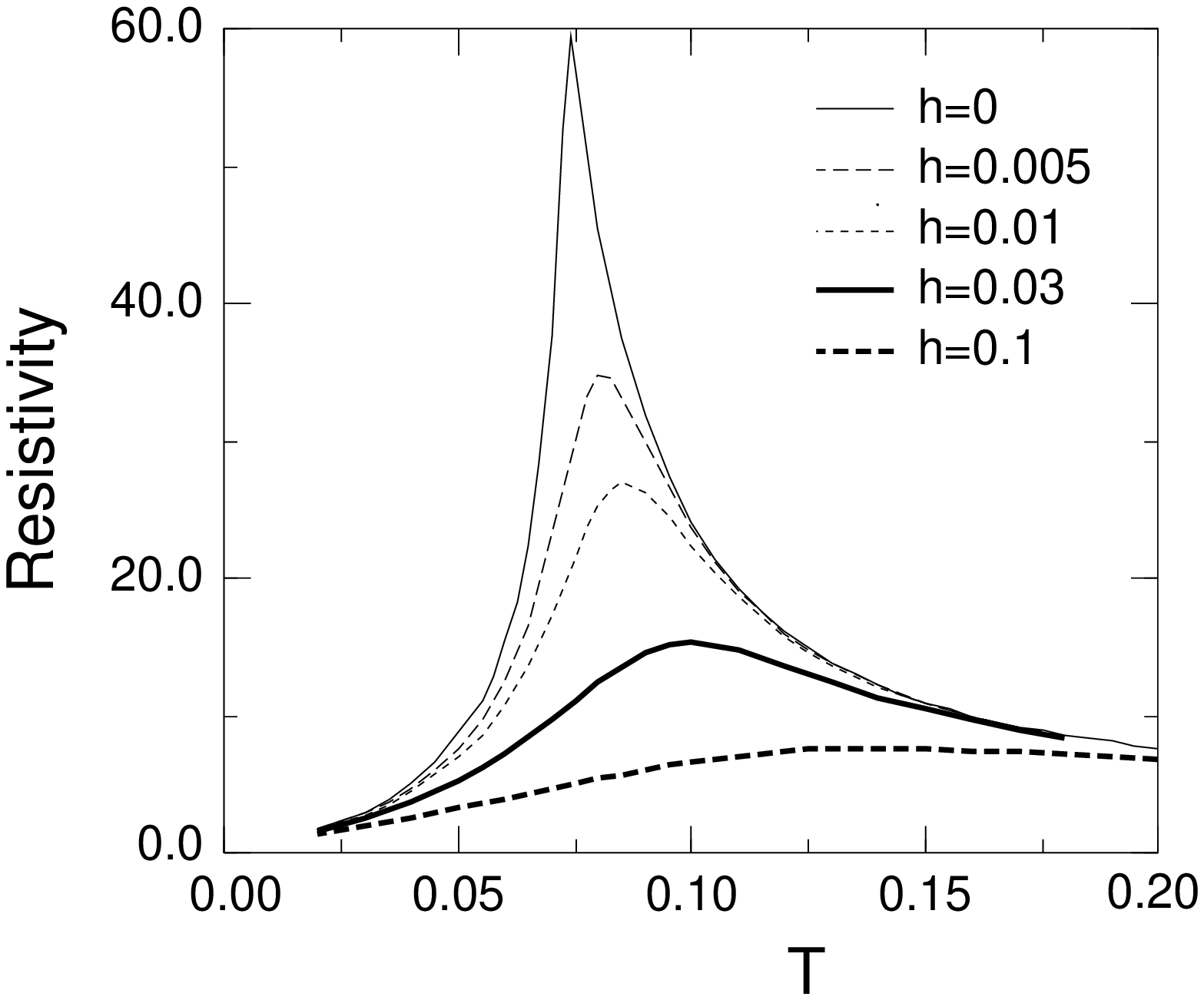}}}
\end{minipage}
\caption{Temperature dependence of the resistivity at different values
  of the normalized magnetic field $h$ for two values of the electron-phonon
  coupling, $\lambda = 0.7$ (left panel) and $\lambda =
  1.12$ (right panel). The parameter $h$ is related to the physical field $H$ by $h
  = g\mu_{\rm B}SH/t$. Using a gyromagnetic ratio $g = 2$, a hopping
  integral $t = 0.6$~eV and the core spin $S = 3/2$ means that $h =
  0.01$ corresponds to $H = 15$~T. The reduced electron-lattice
  coupling strength is given by $\lambda = G^2/kt$, where $G$ denotes
  the electron-phonon coupling constant and $k$ the phonon
  spring-constant. The temperature $T$ is given in units of $t$.
  Reproduced from Millis \etal (1996c).}
\label{millis}
\end{figure}
Colossal magnetoresistance can be qualitatively understood within the
double-exchange model. An applied magnetic field leads to a better alignment
of the core spins and, therefore, to a decrease in conductivity. This effect
is strongest near the Curie temperature, where both spin disorder and 
the susceptibility are large. Accordingly, a maximum in the magnetoresistance
appears near $T_C$. This argument applies to spin-disorder scattering
in ferromagnets in general and does not explain the extraordinary
magnitude of the magnetoresistance in the manganites. As already
discussed in the previous section, many
groups have suggested that double exchange alone is not sufficient in
order to explain the colossal magnetoresistance. The models proposed
evoke a competition between double exchange and another mechanism -- such
as polaron formation due to the strong electron-phonon coupling or
localization by spin fluctuations; this competition is supposed to drive the
metal-insulator transition. The balance between the two competing
mechanisms is very sensitive to an applied magnetic field that
suppresses spin fluctuations and enhances the ferromagnetic order. The
debate on the essential transport mechanism in the manganites has not
yet been decided. However, as indicated in the previous section there
is strong evidence for polaron formation and there seems to be
consensus that the electron-phonon coupling is large. Here theoretical
results by Millis \etal (1996c) are reproduced, see
figure~\ref{millis}(a) and (b). The resistivity was obtained from a
dynamical mean field calculation including double exchange and a coupling
of carriers to phonons. The calculations show that the resistivity
above $T_C$ can be tuned from semiconducting to metallic on decrease
of the electron-phonon coupling strength. This is in qualitative
agreement with measurements on $\rm La_{0.7}Ca_{0.3}MnO_3$ and $\rm
La_{0.7}Sr_{0.3}MnO_3$ single crystals, see figure~\ref{tokura2}(a). An applied
magnetic field leads to a strong resistivity decrease in agreement
with the observation of colossal magnetoresistance. Furthermore, the
calculated magnetoresistance increases with decreasing Curie
temperature as observed in experiments. The field scale for CMR,
however, appears to be too large.
\subsubsection{Anisotropic magnetoresistance.}
\begin{figure}[t]
\begin{center}
\vspace{-0.5cm}
\hspace*{0.0cm} \includegraphics[width=0.9\textwidth]{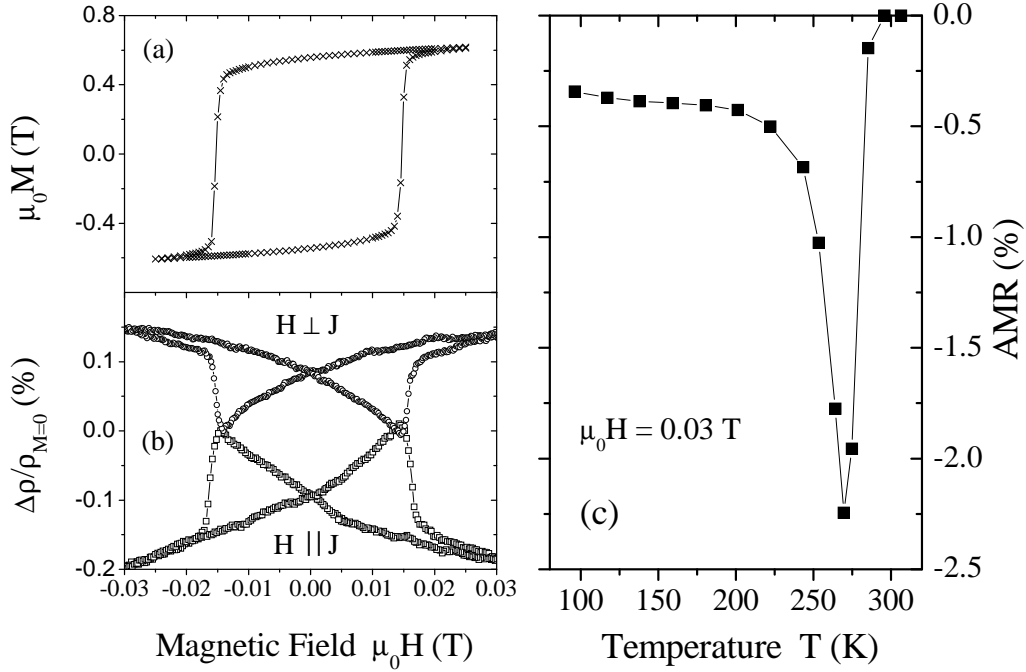}
\end{center}
\caption{(a) Magnetization hysteresis loop of a high quality epitaxial
$\rm La_{0.7}Ca_{0.3}MnO_3$ film at 96~K. (b) Magnetoresistance
ratio of the same film as a function of the applied field at 96~K.
The magnetic field was applied parallel and perpendicular to the current.
The maxima (minima) in the longitudinal (transverse) magnetoresistance
appear at the coercive field. (c) Anisotropic magnetoresistance
$AMR$ at 0.03~T as a function of temperature. After Ziese and Sena (1998).}
\label{lcmo3}
\end{figure}
Anisotropic magnetoresistance (AMR) is present in all ferromagnets and is
defined as the resistivity change as a function of angle between the
current $J$ and the magnetization $M$. If $\rho_\parallel$ and $\rho_\perp$
denote the resistivity in the longitudinal ($J \parallel M$) and
transverse ($J \perp M$) geometries, respectively, then the
anisotropic magnetoresistance is defined by
\begin{equation}
AMR = \frac{\rho_\parallel-\rho_\perp}{\frac{1}{3}\rho_\parallel+\frac{2}{3}\rho_\perp}\, .
\label{amr}
\end{equation}
Anisotropic magnetoresistance in elemental ferromagnets was reviewed by 
Campbell and Fert (1982).

The anisotropic magnetoresistance in the manganites was found to be much
smaller than the colossal magnetoresistance 
(Eckstein \etal 1996, O'Donnell \etal 1997a, 1997b, Li \etal 1997b, Ziese and Sena 1998,
Louren\c{c}o \etal 1999, O'Donnell \etal 2000, Ziese 2000b).
Therefore, only a few studies have been published on anisotropic magnetoresistance
in the manganites. It is a low field effect and is discussed here for
later comparison with extrinsic magnetoresistance effects.

Low field magnetoresistance data of a high quality $\rm La_{0.7}Ca_{0.3}MnO_3$
epitaxial film are shown in figure~\ref{lcmo3}(a). At saturation, the transverse
resistivity is larger than the longitudinal resistivity leading to a
negative anisotropic magnetoresistance. The longitudinal (transverse) resistivity
has maxima (minima) at the coercive field as can be seen from the comparison
with the magnetization-hysteresis loop taken at the same
temperature. Thus, the magnetic field dependence of the resistivity
reflects the magnetic domain structure in the sample.
$AMR$ values determined at 0.03~T are shown
in figure~\ref{lcmo3}(b) as a function of temperature. Whereas the $AMR$
is temperature independent with a value of about $-0.4$\% at low temperatures,
a maximum is seen near the Curie temperature. The temperature
dependent AMR can be decomposed into a normal component $\propto
M_S^2$ and an anomalous component $\propto \rho^2 M_S^2$ that scales
with the anomalous Hall resistivity (Ziese 2001b). Since the low field
magnetoresistance in the manganites depends sensitively on the microstructure,
such a clear anisotropic magnetoresistance can only be found in high
quality epitaxial films.

Anisotropic magnetoresistance is due to the mixing of majority and minority
states by the spin-orbit interaction (Campbell and Fert 1982).
Ziese and Sena (1998) derived an expression for the anisotropic magnetoresistance
within a simple atomic state model following the work of Campbell \etal (1970) and
Malozemoff (1985); this describes the normal AMR component. A
Hamiltonian containing only the crystal field splitting 
$\Delta_{CF}$, the exchange splitting $E_{ex}$ and the spin orbit coupling
$A[L_zS_z+(L_+S_-+L_-S_+)/2]$ was considered. Treating the spin-orbit interaction
as a small perturbation, the eigenfunctions derived from the 3$d$ wave
functions were calculated to second order
in this interaction. Assuming scattering by spherically symmetric impurities,
the anisotropic magnetoresistance can be derived by the analysis of the symmetry
of the scattering matrix elements. This yields
\begin{equation}
\frac{\rho_{\parallel}-\rho_{\perp}}{\frac{1}{3}\rho_\parallel+\frac{2}{3}\rho_\perp} =
-\frac{3}{2}\, \left\lbrack \frac{A^2}{(E_{ex}-\Delta_{CF})^2}
-\frac{A^2}{\Delta_{CF}^2}\right\rbrack\, .
\label{amrtheo}
\end{equation}
This expression contains only the local parameters $\Delta_{CF}$, $E_{ex}$
and $A$. With $\Delta_{CF} \simeq 1.5$~eV, $E_{ex} \simeq 2.0$~eV and
$A \simeq 0.04$~eV a value of $AMR = -0.85$\% is found. Considering the simplicity
of the model, this is in good agreement with the experimental value at 
low temperatures.
\subsection{$\rm Fe_3O_4$}
\begin{figure}[t]
\vspace{-1.0cm}
\begin{center}
\hspace*{2.5cm} \epsfysize=12cm \epsfbox{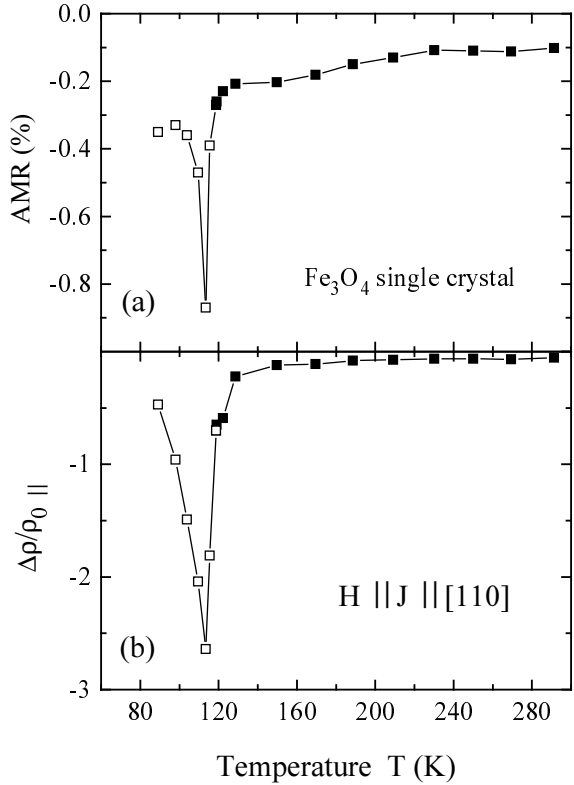}
\end{center}
\caption{(a) Anisotropic magnetoresistance of a magnetite single crystal
for current along [110]. (b) Longitudinal magnetoresistance for current
and applied magnetic field along [110]. The open (solid) symbols show
the magnetoresistance in the monoclinic (cubic) phase below
and above the Vervey transition, respectively. After Ziese and Blythe (2000).}
\label{feo_2}
\end{figure}
The magnetoresistance of magnetite single crystals
(Domenicali 1950, Kostopoulos 1972, Kostopoulos and Alexopoulos 1976,
Shiozaki \etal 1981, Belov \etal 1982, 1983, Gridin \etal 1996, Ziese and Blythe 2000)
and films 
(Feng \etal 1975, Gong \etal 1997, Li \etal 1998a, Ogale \etal 1998,
Ziese and Blythe 2000)
was studied by several authors.
The magnitude of the magnetoresistance in a constant applied
field varies greatly among the studies. Coey \etal (1998a) pointed
out that this is related to the microstructure of the samples;
extended defects like grain boundaries lead to a significant 
magnetoresistance increase, especially an unsaturated high field
magnetoresistance. There are two recent studies on the
magnetoresistance of $\rm Fe_3O_4$ single crystals by Gridin \etal (1996)
and Ziese and Blythe (2000). Both groups report a maximum of a few percent
in the magnetoresistance at the Verwey transition temperature,
presumably due to a magnetic field dependent shift of this
charge-order transition. Typical
magnetoresistance data in an applied field of 1~T 
are shown in figure~\ref{feo_2}(a) and (b). The current flow was along [110].
The longitudinal magnetoresistance has a minimum at $T_V$ of -3\%.
The anisotropic magnetoresistance was derived from measurements with the
applied field along and transverse to [110]. Apart from the
magnetoresistance maximum at the Vervey transition the $\rm Fe_3O_4$
single crystal shows only anisotropic magnetoresistance with
values between -0.2\% and -0.1\% above the transition (Ziese and
Blythe 2000).
\section*{Extrinsic magnetotransport phenomena}
\addcontentsline{toc}{section}
{\protect{Part 2: Extrinsic magnetotransport phenomena}}
\section{Domain-wall scattering \label{dw}}
In 1934 Heaps reported on a resistance discontinuity associated with
the Barkhausen effect. A Ni wire under bending stress showed a large
Barkhausen jump and a sudden resistance decrease of relative magnitude
$\Delta R/R = 6.35\times10^{-5}$ at about the same applied
field. Although Heaps (1934) interpreted this observation in terms of
magnetization rotations, a contemporary interpretation might as well
associate the resistance jump with domain-wall resistance. The concept
of domain-wall scattering emerged in the sixties and has been studied
since, see the experimental work on Fe whiskers by Taylor \etal (1968)
and the theoretical work by Cabrera and Falicov (1974a, 1974b) and
Berger (1978, 1991). More recent work was stimulated by studies into
quantum tunnelling of domain walls using magnetoresistance as a probe 
(Hong and Giordano 1998), the analogy between domain-walls and
metallic multilayer systems showing giant magnetoresistance (Gregg
\etal 1996), the prospect of using a domain-wall switch in
spin-electronic devices as well as the speculation of large
domain-wall scattering in the manganites due to the large
spin-polarization (Zhang and Yang 1996).  

Domain-wall resistance and anisotropic magnetoresistance (AMR) depend sensitively
on the domain configuration; furthermore, both are believed to be of the same 
order of magnitude. Great care has therefore to be devoted to the study
of the micromagnetic state of ferromagnetic samples to distinguish
between these two magnetoresistance mechanisms. At present, a consensus on the
observability of domain-wall scattering has not been reached. Experiments
yield a range of values and agreement has not even been reached on
the sign of the effect. In the following two sections we discuss the state
in elemental ferromagnets and manganites, respectively.
\subsection{Elemental ferromagnets}
Early experiments on iron whiskers showed large magnetoresistive effects
partially attributed to domain-wall scattering (Taylor \etal 1968). The
interpretation relied on an assumed domain structure. The results were
interpreted within models by Cabrera and Falicov (1974a, 1974b) and Berger 
(1978). Interest into domain-wall scattering revived during the nineties.
Gregg \etal (1996) prepared a pattern of stripe domains in a Co film 
with perpendicular anisotropy and 
measured the magnetoresistance in a perpendicular field with an electric 
current perpendicular to the domain walls. They argued that the 
magnetoresistance must arise from domain-wall scattering, since the 
magnetization and current are always perpendicular. This assertion
was later contested by R\"udiger \etal (1999a) on the grounds of
magnetic force microscopy (MFM) studies and micromagnetic modelling
of Co wires. Viret \etal (1996) tried to extract the domain-wall
resistance in Co and Ni films by adding the longitudinal and transverse
magnetoresistance and identifying the non-vanishing contributions with
domain-wall contributions. This method relies on the fact that the 
anisotropic magnetoresistance depends on the square of the cosine
of the angle between magnetization and current, such that the AMR
adds to a constant when measured before and after a field or current
rotation by 90$^\circ$. However, this argument relies on the assumption
of perfect cubic crystal anisotropy and the absence of misalignments between 
current, field and crystal axes. Since the observed effects are very small,
these assumptions are doubtful. Both Gregg \etal (1996) and Viret \etal (1996)
found a positive domain-wall resistivity. Hong and Giordano (1998) measured
the magnetoresistance of 30~nm thin Ni wires and found a negative contribution
to the resistivity in the presence of domain walls. However, this result
also relies on an assumed domain structure and has to be treated with
care. Wegrowe \etal (1999) compared experimental magnetoresistance
data of Ni and Co wires with a model for the magnetization reversal
and found discrepancies in the case of Co wires presumably due to
domain-wall nucleation.
It has to be clear that any interpretation of magnetoresistance data has 
to be based on the actual domain structure of the sample. 
Such comprehensive investigations have been conducted by 
Kent \etal (Kent \etal 1999, R\"udiger \etal 1998a, 1998b, 1999a, 1999b).
Therefore their results are reviewed here in more detail.

\begin{figure}[t]
\vspace*{0.0cm}
\begin{center}
\hspace*{1.0cm} \includegraphics[width=0.8\textwidth]{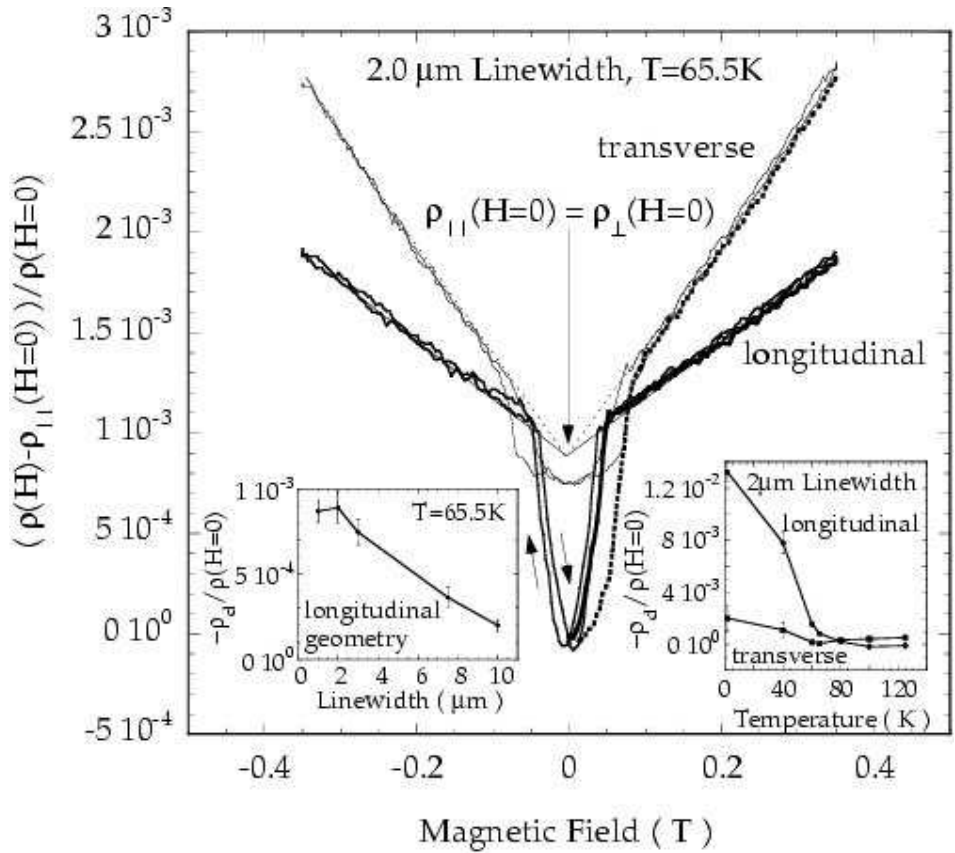}
\end{center}
\vspace*{0.0cm}
\caption{Magnetoresistance of a 2~$\mu$m Fe wire at 65.5~K. The extrapolation
of the high field MR data in transverse (dotted line) and longitudinal
(solid line) geometry shows that $\rho_\perp(H = 0) = \rho_\parallel(H=0)$.
The resistivity with walls present, $\rho(H = 0)$, is smaller than this 
extrapolation and indicates that domain walls lower the wire resistivity.
The left-hand inset shows this negative domain-wall contribution as a function
of Fe wire linewidth at this compensation temperature in the longitudinal
geometry. The right-hand inset shows the domain-wall contribution
as a function of temperature deduced using the model described in the text.
Reproduced from R\"udiger \etal (1998a).}
\label{kent}
\end{figure}
Kent \etal studied the magnetoresistance of Fe and Co wires of
width in the range 0.65 to 20~$\mu$m and thickness in the range
25 to 200~nm. The domain structures in these wires were characterized using
MFM and micromagnetic modelling. The Co wires have a perpendicular
anisotropy such that the magnetization within a domain is oriented
along the surface normal. MFM and micromagnetic modelling, however,
show that flux-closure domains exist; the fraction of the flux-closure
caps depends on the film thickness varying between 35\% and 17\%
for thicknesses between 50 and 200~nm. The existence of these
flux-closure domains was neglected by Gregg \etal (1996). 
The magnetoresistance in these wires is due to normal
state (Lorentz) magnetoresistance, anisotropic magnetoresistance as well as
a domain-wall contribution. The anisotropic magnetoresistance in Fe and Co
is positive, i.e.\ $\rho_\parallel > \rho_\perp$. In contrast, the Lorentz
magnetoresistance shows the opposite behaviour, i.e.\ the transverse
MR is larger than the longitudinal MR. Since the Lorentz magnetoresistance
depends on the local induction, a considerable normal state magnetoresistance
due to the large saturation magnetization in Fe ($\mu_0M_S = 2.2$~T)
and Co ($\mu_0 M_S = 1.8$~T) is present even at zero applied field.
This leads  to the existence of a compensation temperature with
equal resistivities of magnetic domains oriented perpendicular or parallel
to the current. These resistivities are obtained by extrapolation of the normal
state resistivity above saturation. The compensation temperatures are about 
65~K for Fe and 85~K for Co, respectively. In the absence of domain wall
effects, at the compensation temperature
the resistivity measured in zero field should agree with that
determined by extrapolation. The experiment, however, shows a
negative resistivity contribution in Fe films for thicknesses
below 100~nm due to domain walls, see figure~\ref{kent}. This negative
contribution increases with domain-wall density and decreases with
film thickness. At other temperatures, the domain-wall resistivity
can be estimated by the difference in the measured resistivity and
the resistivity estimated from the high field extrapolation combined with the
measured domain configuration. This yields a negative domain-wall
resistivity in Fe wires persisting up to temperatures of about 80~K.
In Co the situation is different. The measured resistivities are
always very close to the effective resistivities calculated from
the extrapolation values and the measured domain configuration.
However, at the compensation temperature of about 85~K a small positive
contribution to the magnetoresistance ratio
of $9\times 10^{-4}$ is found. 
This is consistent with an additional contribution due to domain-wall 
scattering. 

A report on domain-wall resistance
in Co zig-zag wires found a negative domain-wall contribution
of $-6\times 10^{-4}$ at 5~K (Taniyama \etal 1999). This persists
up to 200~K. An investigation by Ebels \etal (2000) on 35~nm
thin Co wires identified a steplike resistance increase of about
$1\times 10^{-3}$-$3\times10^{-3}$. In these thin wires the c-axis
grows parallel to the wire axis and both shape and crystal anisotropy
lead to magnetic domains being oriented along the wire. During a
hysteresis loop in a parallel magnetic field head-to-head domain walls
can be induced that supposedly lead to the resistance increase.

This discussion shows that the experimental situation is
controversial; the status of theoretical studies is also far from
being clearcut. The analogy between a domain
wall and a metallic multilayer system intuitively leads to the idea
of a positive resistivity contribution of the domain wall. This
idea was advanced by Gregg \etal (1996) and later confirmed by
Levy and Zhang (1997) by a more rigorous calculation. These treatments
visualize an itinerant electron traversing the domain wall and adiabatically
tracking the rotating exchange field. Majority and minority
carriers in the bulk conduct in parallel with different
resistivities $\rho^\uparrow$, $\rho^\downarrow$. Levy and Zhang (1997) 
showed that the carrier wave functions in the domain wall contain 
contributions of both spin orientations, thus facilitating a spin mixing
that leads to an additional resistivity due to the different resistivities
for majority and minority carriers. The domain-wall resistivity was
found to be anisotropic with respect to the current direction relative
to the domain wall. This anisotropy was investigated by Viret \etal
(2000) in FePd films and was found to be consistent with the
predictions. Levy and Zhang (1997) obtained
a domain-wall resistivity proportional to $\delta^{-2}$, where
$\delta$ denotes the domain-wall width. van Hoof \etal (1999) and
Brataas \etal (1999b) calculated the domain-wall resistivity
in Ni, Fe and Co due to spin-flip scattering both within a 
two-band model and using a realistic 
band-structure. Within the two-band model these authors recovered
Levy and Zhang's (1997) result of a domain-wall resistivity proportional
to $\delta^{-2}$. The first-principles calculation including a realistic
bandstructure, however, yielded a considerably larger domain-wall
magnetoresistance inversely proportional to $\delta$ in contrast
to the two-band result.

On the other hand, Tatara and Fukuyama
(1997) and Lyanda-Geller \etal (1998) pointed out that domain walls
might contribute to the de-coherence of electrons, thus leading
to a resistivity decrease due to the suppression of weak
localization effects. Although this is in qualitative agreement
with the results of Kent \etal (1999) and Taniyama \etal (1999), Kent \etal (1999)
pointed out that the negative MR contribution persists to much higher
temperatures than expected for quantum interference effects. Indeed,
R\"udiger \etal (1998a) estimated the maximum temperature for the
observation of weak localization effects by equating the wall de-coherence
time to the inelastic scattering time and found a value of 7~K being an order
of magnitude lower than observed. Finally, based on the observation
that the negative domain-wall contribution vanishes for large
film thicknesses, R\"udiger \etal (1999b) argued that surface scattering might 
be important; in their model the negative magnetoresistance arises
from the deflection of charge carriers away from the surface that is
being mediated by the presence of domain walls.
A recent calculation by van Gorkom \etal (1999a) showed that a
negative domain-wall resistance can also arise from the reduced
magnetization in a domain wall, if the ratio of the spin-dependent
relaxation times is appropriate. This can be seen as follows. The
Drude resistivity of a single domain ferromagnet is given within the
two-band Stoner model by (van Gorkom \etal 1999a)
\begin{equation}
\rho = \frac{{\rm m}}{{\rm e}^2}\, \frac{1}
{n_\uparrow\tau_\uparrow+n_\downarrow\tau_\downarrow}\, ,
\label{gorkom1}
\end{equation}
where $n_\uparrow$ ($n_\downarrow$) denote the majority (minority)
density of states and $\tau_\uparrow$ ($\tau_\downarrow$) the
relaxation time for majority (minority) electrons. A redistribution of
the electrons in a domain wall with $n_{\uparrow(\downarrow)} = 
n_{0\uparrow(\downarrow)}+\delta n_{\uparrow(\downarrow)}$, 
$\delta n_\uparrow = -\delta n_\downarrow$ modifies the resistivity
according to
\begin{equation}
\delta\rho = -\rho^2\, \frac{{\rm e}^2}{{\rm m}}\, \delta n_\uparrow\, 
\left(\tau_\uparrow-\tau_\downarrow\right)\, .
\label{gorkom2}
\end{equation}
Depending on the ratio of the spin-dependent relaxation times, this
resistivity change can be positive or negative. van Gorkom \etal (1999a)
calculated $\delta n_\uparrow$ and $\delta \rho$ self-consistently for
a semiclassical domain-wall model and indeed found a negative
domain-wall resistance for $\tau_\uparrow > \tau_\downarrow$. Since
the relaxation times depend on the type of impurity (Campbell and Fert
1982), this model can be experimentally checked by measuring the
domain-wall resistivities of a series of alloys.

Despite the experimental and theoretical debate on the relation
between domain-wall magnetoresistance and domain-wall width, it
is generally accepted that a significant magnetoresistance will only
arise in the case of a narrow wall. It is therefore promising to investigate
nanocontacts with geometrically constrained domain walls. Indeed, Bruno (1999)
showed that the domain-wall width of such a geometrically constrained
wall is of the order of the width of the constriction. Garc{\'\i}a \etal (1999)
investigated the magnetoresistance of Ni nanocontacts and found values of
up to 75\% (with respect to the zero field value) at room temperature
for contacts with a conductance of only 
a few conductance units $2{\rm e}^2/{\rm h}$. This was explained by
Tatara \etal (1999) by scattering at a narrow domain wall located at
the nanocontact. The agreement between theory and experimental values,
however, is only qualitative (Tatara \etal 1999); the theory shows that
the magnetoresistance is enhanced at small contact sizes corresponding
to conductances of only a few conductance units. A similar experiment
was performed by Wegrowe \etal (2000). These authors investigated the
magnetoresistance of Co nanowires exchange biased by a GdCo$_{1.6}$ layer.
In this configuration, a domain wall is expected to form at the interface
and to be compressed by an applied magnetic field. Wegrowe \etal (2000)
observed a strong decrease of the domain-wall resistance as a function of 
the domain-wall width in the range 5~nm to 10~nm. The limited range
of domain-wall thicknesses accessible in this experiment did not allow
to distinguish between the various predictions for the domain-wall
width dependence. The resistance contribution of partial domain walls
located on either side of permalloy/permalloy, Co/permalloy, Co/Co,
Ni/Ni and Co/Cu point contacts was found to be negative (van Gorkom
\etal 1999b, Theeuwen \etal 2001). Following the theoretical results
of van Gorkom \etal (1999a) this could be interpreted as arising from
boundary scattering in combination with an appropriate ratio of the
spin dependent relaxation times.
\subsection{Magnetic oxides}
Within the double-exchange model, the transfer integral $t\cos(\Theta/2)$
for the e$_g$ electrons depends sensitively on the angle $\Theta$ between 
the core spins of the adjacent Mn$^{3+}$ and Mn$^{4+}$ sites. Therefore,
it is expected that a narrow domain wall has a
considerable influence on the conductance of the manganites. 
The scattering by domain walls in double-exchange
ferromagnets was treated by Zhang and Yang (1996), Yamanaka and
Nagaosa (1996), Gehring (1997) and Brey (1999). Whereas Zhang and 
Yang (1996) calculated the temperature
and field dependence of the resistivity due to the temperature and field
dependence of the average domain size, Yamanaka and Nagaosa (1996), as well
as Gehring (1997), calculated the conductance of a single domain wall as a function
of the domain-wall width $\delta$. Both assumed a 180$^\circ$ N\'eel wall
with a constant spin-rotation angle in the wall, $\Theta = \pi a/\delta$, 
where $a$ denotes the lattice parameter. Within a tight binding model
the dispersion relation is given by
\begin{equation}
\epsilon = 2t\cos(\Theta)\left[1-\cos(ka)\right]\, ,
\label{tightbind}
\end{equation}
where $k$ denotes the wave vector perpendicular to the wall.
Since $\Theta = 0$ outside the wall and $\Theta = \pi a/\delta$ inside the
domain wall, the band width in the wall is reduced. Thus, depending on
the wave vector of the electrons incident under an angle $\phi$
with respect to the wall normal,
\begin{enumerate}
\item
the electrons will be scattered regardless of incoming angle $\phi$,
if $k_f a < \Theta/2$,
\item
for $\Theta/2 < k_f a < \pi-\Theta/2$ only the electrons propagating
at an angle $\phi$ such that $k_f a \cos(\phi) < \Theta/2$
will be scattered,
\item
for $k_f a> \pi-\Theta/2$ electrons satisfying $k_f a\cos(\phi) <\Theta/2$
and $k_f a\cos(\phi)>\pi-\Theta/2$ will be scattered.
\end{enumerate}
Gehring (1997) estimated $k_f \sim 0.7 \pi/a$ such that case (ii) is likely to apply.
Thus, electrons incident under an angle $\phi_0 < \phi < \pi/2$ with
$\cos(\phi_0) = \Theta/(2k_fa)$, will be scattered through an angle 
$\pi-2\phi$ giving a contribution to the resistivity proportional to 
$k_f[1-\cos(\pi-2\phi)] = 2k_f\cos^2(\phi)$, whereas electrons incident
at smaller angles pass the wall nearly undisturbed. The magnetoconductivity
of a single wall is then given by
\begin{equation}
\frac{\Delta G_{wall}}{G_0} = \frac{\int_{\phi_0}^{\pi/2}d\phi\sin\phi
\left[1-\cos(\pi-2\phi)\right]}{\int_0^{\pi/2}d\phi\sin\phi}
= \frac{2}{3}\, \left(\frac{\pi}{2k_f\delta}\right)^3\, ,
\label{gehring}
\end{equation}
and decays very strongly with the domain-wall thickness. Surprisingly,
according to this calculation, the domain-wall resistance in the manganites 
decays much more strongly with wall thickness than in elemental ferromagnets.

Yamanaka and Nagaosa (1996) numerically calculated the conductivity as a 
function of the energy $\epsilon$ of the incident electron using Landauer's formula. 
In the case of a thick wall, a one-dimensional continuum model can be 
applied. An effective potential
\begin{equation}
V(x) = \left\lbrace
\begin{array}{l@{\quad : \quad}l} 
V = 2t\left[1-\cos\left(\frac{\pi a}{2\delta}\right)\right] & {\rm in\, the\, wall}\\
0 & {\rm elsewhere}
\end{array}\right.
\end{equation}
was used in the one-dimensional Schr\"odinger equation leading to a transmission coefficient
\begin{equation}
G_{s}(\epsilon) = \frac{1}{1+\frac{V^2}{4\epsilon(V-\epsilon)}\, 
\sinh^2\left(\frac{\delta}{a}\sqrt{\frac{V-\epsilon}{t}}\right)}
\label{yamanaka1}
\end{equation}
with $\epsilon = 2t[1-\cos(ka)]$. For large domain-wall thicknesses
this can be approximated by 
\begin{equation}
G_s \sim \left\lbrace
\begin{array}{l@{\quad : \quad}l}
\left[\frac{\delta}{a}\right]^2\, \frac{\epsilon}{t} 
& \epsilon < t\, \left[\frac{a}{\delta}\right]^2\\
1
& \epsilon > t\, \left[\frac{a}{\delta}\right]^2\, . 
\end{array}\right.
\label{yamanaka2}
\end{equation}
Integrating over all wave vectors up to $k_f$ yields
\begin{equation}
\frac{\Delta G_{wall}}{G_0} = \frac{2}{5}\, \left(\frac{1}{k_f\delta}\right)^3\, ,
\label{yamanaka3}
\end{equation}
which, apart from a numerical factor, agrees with Gehring's result.

Brey (1999) calculated the magnetoresistance of a domain wall in the
manganites numerically as a function of domain-wall thickness taking
into account the modulation of the hopping amplitude due to the spin
canting inside the domain wall as well as the shift of the chemical
potential of the Mn ion levels. He obtained a decreasing
magnetoresistance with increasing wall thickness; for a wall thickness
$\delta = 10a$ he found a value $\Delta G/G = 0.01$.

Experimentally the issue of domain-wall scattering in the manganites
was addressed by Wolfman \etal (1998), Wang and Li (1998), Mathur
\etal (1999), Wu \etal (1999), Ziese \etal (1999b) and Suzuki \etal (2000). 
Mathur \etal (1999) measured the resistivity of a patterned LCMO track.
This track had narrow constrictions separating regions with wider
linewidth (``bellies''); small permanent magnets were placed near
every second belly. After magnetizing to saturation in an in-plane field,
the magnetic field was slowly reversed; the unbiased ``belly'' regions are
believed to reverse the magnetization at small negative fields, thus creating
domain walls near the constrictions. Mathur \etal (1999) report a step-like
behaviour of the resistivity hysteresis below about 130~K. This was attributed
to the switching of individual ``belly'' regions and, accordingly, to the creation
of single domain walls. The areal resistivity was found to be large about
$8\times 10^{-14}$~$\Omega$m$^2$ at 77~K. Mathur \etal (1999) estimated
an areal domain-wall resistivity of $1.6\times 10^{-18}$~$\Omega$m$^2$
within a Born approximation. Using $\rho\delta (\Delta G/G)$ with 
$\rho = 100$~$\mu\Omega$cm, a domain-wall width $\delta = 30$~nm and
$\Delta G/G$ according to equation~(\ref{yamanaka3}) yields an even smaller
value of about $7.6\times 10^{-20}$~$\Omega$m$^2$. This discrepancy might indicate
some internal structure of the domain walls (Mathur \etal 1999).

Wu \etal (1999) and Suzuki \etal (2000) investigated the magnetoresistance of compressively
strained LSMO films on LaAlO$_3$. These films have a perpendicular
magnetic anisotropy leading to the formation of perpendicular magnetic
domains on a scale of about 200~nm. If the magnetoresistance is
measured after sweeping an in-plane or an out-of-plane field
perpendicular to the current, the
resistivity should return to the same value in zero field apart from
domain-wall scattering effects due to the different domain
configurations. The measurement on a 80~nm LSMO film at 300~K yielded
a deviation of about 0.1\% in the two field configurations. This
corresponds to a domain-wall areal resistivity of 
$10^{-15}$~$\Omega$m$^2$. This value is significantly smaller than the
value obtained by Mathur \etal (1999), possibly due to the higher
measurement temperature. The calculation of Brey (1999) yielded an
areal resistivity of $3\times 10^{-16}$~$\Omega$m$^2$ smaller than the
value observed by Mathur \etal (1999), but close to the value found by
Wu \etal (1999). This shows that the precision of the experimental
values is not sufficient at the moment in order to test different theories.

\begin{figure}[t]
\vspace{-0.5cm}
\begin{center}
\hspace*{0.5cm} \includegraphics[width=0.8\textwidth]{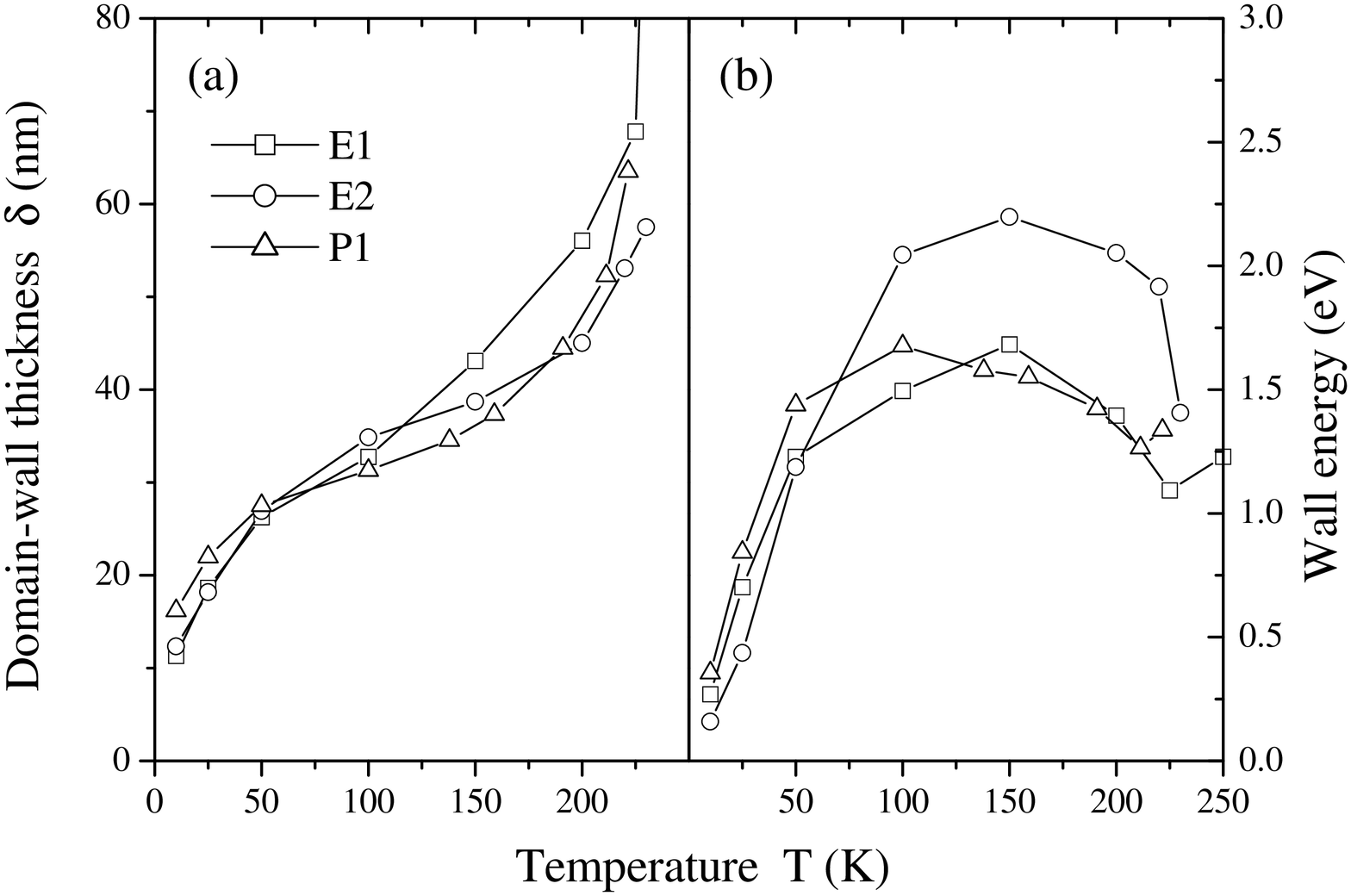}
\end{center}
\vspace{-0.5cm}
\caption{(a) Domain-wall thickness $\delta$ and (b) domain-wall
energy as determined from magnetic viscosity measurements on
a 9~nm thick annealed LCMO film on LaAlO$_3$ (E1), 
a 120~nm thick as-deposited LCMO film on LaAlO$_3$ (E2) and 
a 200~nm thick polycrystalline LCMO film on Si (P1). After Ziese \etal (1999b).}
\label{domain}
\end{figure}

Ziese \etal (1999b) determined the domain-wall width in various
LCMO films, i.e.\ a polycrystalline film on Si (200~nm), an as-deposited 
film on LaAlO$_3$ (120~nm) and an annealed film on LaAlO$_3$ (9~nm)
using magnetic viscosity measurements. The magnetic viscosity
$S = dM/d\ln(t)$ is related to the average activation volume $v$ by
\begin{equation}
S = \frac{{\rm k}T}{vM_S}\, \chi_{irr}\, ,
\label{magvis}
\end{equation}
where $M_S$ denotes the saturation magnetization and $\chi_{irr}$
the irreversible susceptibility. From measurements of $S$, $M_S$ and
$\chi_{irr}$ the average activation volume can be determined. This
is related to the domain-wall thickness $\delta$ via $v = \delta^3$.
Ziese \etal (1999b) found that the domain-wall thickness did not
significantly depend on the microstructure, see figure~\ref{domain}. 
This is surprising, since it was believed that domain walls located 
near grain boundaries were relatively narrow due to the weakened 
double-exchange coupling near the defect. Gehring (1997) speculated that
the large low field magnetoresistance observed in polycrystalline materials
is due to electron scattering by narrow domain walls near the
grain boundaries. Experimentally, this idea could not
be confirmed, since the magnetoresistance of the epitaxial and polycrystalline
films investigated by Ziese \etal (1999b) varied by one order of magnitude, 
whereas the measured domain-wall thickness was identical.
Furthermore, the observed domain-wall thickness is very large
compared to the lattice parameter being in 
agreement with the calculated Bloch-wall thickness using
typical values for the magnetocrystalline anisotropy and Curie temperature.
The domain-wall energy, also shown in figure~\ref{domain},
can be calculated from the measured magnetic viscosity, the coercive
and fluctuation fields.

The magnetoresistance of $\rm Pr_{0.67}Sr_{0.33}MnO_3$ films grown
under compressive strain on LaAlO$_3$ substrates was investigated by
Wang and Li (1998) and Wolfman \etal (1998). In these samples a large
magnetoresistance arises when the magnetic field is applied
perpendicular to the film. It is known from magnetic anisotropy
investigations that manganite films under compressive strain exhibit a
perpendicular anisotropy (O'Donnell \etal 1998). Since the observed
magnetoresistance is not correlated to specific crystallographic
defects, the authors suggested that it arises from domain-wall
scattering at domain walls separating out-of-plane domains.

Domain-wall scattering was also reported in SrRuO$_3$ films at 
low temperature (Klein \etal 1998). SrRuO$_3$ grows on SrTiO$_3$
with the c-axis in-plane and the a- and b-axes at 45$^\circ$ with
respect to the substrate normal. Lorentz microscopy imaging at
low temperatures revealed a stripe domain pattern with domain walls
along [1$\overline{1}$0]. Resistance measurements at 5~K after
zero-field cooling with the current along [001] and the magnetic field
along [1$\overline{1}$0] showed a large irreversible drop in magnetic
field below about 0.3~T. This effect was not observed with the current
along [1$\overline{1}$0]. In 0.3~T the magnetization of the film was 
technically saturated and the stripe domains had been driven out of the film;
the stripe domain pattern was not recovered
after removal of the magnetic field, implying that the irreversible
resistivity drop was related to this particular domain structure.
Klein \etal (1998) interpreted these results as evidence for domain-wall
scattering. The magnitude of the areal domain-wall resistivity was found to
be large of the order of $2\times 10^{-15}$~$\Omega$m$^2$. As already
noted in the beginning of this section, this result has to be treated
with care, since the domain structure during the complete hysteresis cycle
is unknown and AMR effects cannot be completely ruled out.

Versluijs \etal (2001) studied the magnetoresistance of Fe$_3$O$_4$
nanocontacts in an experimental arrangement similar to that of
Garc{\'\i}a \etal (1999). For conductances smaller than the conductance
quantum a large magnetoresistance up to 90\% at 7~mT was determined; this was
found to decay strongly with increasing conductance similar to the
results of Garc{\'\i}a \etal (1999) for Ni nanocontacts.
The contacts have non-linear I--V characteristics with $I = GV+cV^3$,
$c = G^{0.3\pm0.1}$. The authors favour an interpretation of the data
within a model of spin scattering at a constricted domain-wall. The
``magnetic ballon effect'', i.e.\ a deplacement of the domain wall
from the constriction due to the spin pressure exerted by the electron
flow qualitatively accounts for the non-linear I--V curves as well as
the decrease of the magnetoresistance with applied voltage. 
\section{Surface and interface properties \label{LCMOmetal}}
Since the main application of half-metallic magnets will be in
integrated devices, interfacial properties are of crucial
importance. The investigation of interfacial properties is in its
early stages and few quantitative results have yet been
established. Here we report on first investigations of oxide-vacuum,
oxide-metal and oxide-superconductor interfaces.

\subsection{Oxide-vacuum interface}

\begin{figure}[t]
\begin{center}
\vspace*{-0.7cm}
\hspace*{1.8cm} \includegraphics[width=0.9\textwidth]{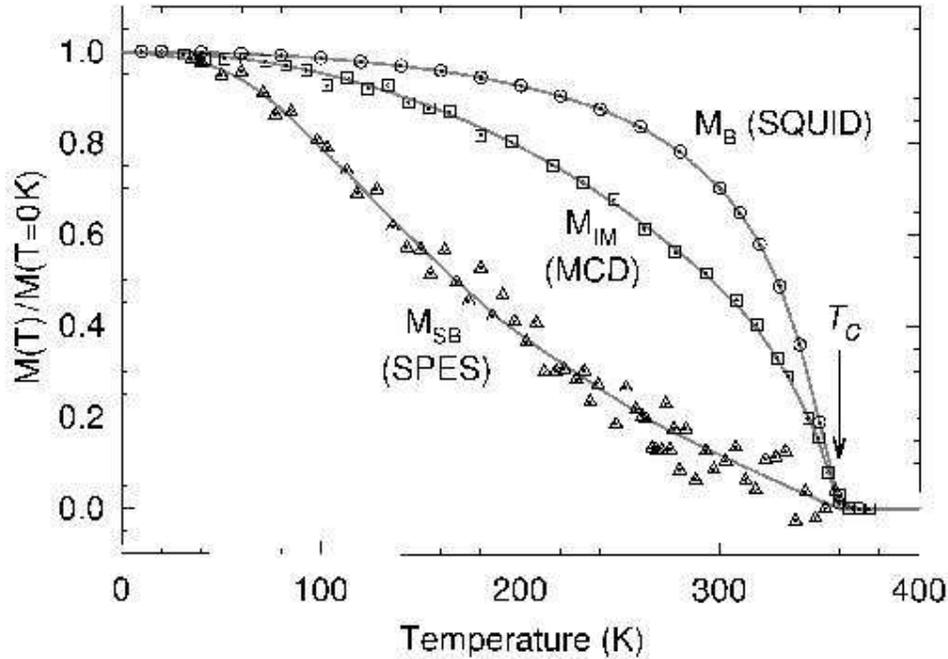}
\vspace*{-11.5cm}
\end{center}
\caption{Temperature dependence of the magnetization measured
on different length scales. $M_B$, $M_{IM}$ and $M_{SB}$ denote
the bulk, the intermediate length scale ($\sim 5$~nm) and
the surface boundary ($\sim0.5$~nm) magnetization, respectively.
Reproduced from Park \etal (1998a).}
\label{park}
\end{figure}

The magnetic properties near the surface of a 
$\rm La_{0.7}Sr_{0.3}MnO_3$ film were investigated by Park \etal
(1998a, 1998b)
using SQUID magnetometry, magnetic circular dichroism (MCD) at the Mn L-edge
and spin-resolved photoemission spectroscopy (SPES). These techniques
probe the magnetization in the bulk and in surface layers of thickness
$\sim 5$~nm (MCD) and $\sim 0.5$~nm (SPES), respectively. The film investigated
was cleaned in an UHV-chamber by a sequence of annealing processes
that resulted in a large surface roughness of $\sim 2$~nm. The main result
is reproduced in figure~\ref{park}, showing the normalized magnetization
measured on different length scales. The spin-polarization near the surface
is strongly reduced and recovers on a length scale of more than 5~nm. Since
numerous applications depend on the interfacial spin-polarization, this result
might have serious implications for possible applications at room temperature.
However, since the roughness was considerably larger than the penetration depth
for SPES measurements, the intrinsic nature of this result is not
obvious.

Alvarado (1979) compared the surfacial magnetization of a Fe$_3$O$_4$
crystal as determined by SPES to the bulk magnetization and found a
significant reduction of the former. Data were collected between 4~K
and 500~K and in this temperature range the surfacial magnetization
decreases approximately linearly with temperature. This behaviour can
be successfully modelled taking into account the reduction of nearest
neighbour magnetic ions and the surface reconstruction
(Srinitiwarawong and Gehring 2001).

Wei \etal (1997, 1998) investigated the surface electronic properties
of $\rm La_{0.7}Ca_{0.3}MnO_3$ and $\rm Fe_3O_4$ using scanning tunnelling
microscopy. At 77~K both compounds show a density of states compatible with
a half-metallic state.

Choi \etal (1999a, 1999b) investigated the surfaces of crystalline $\rm La_{1-x}Ca_xMnO_3$
films with dopings $x = 0.1$ and $x = 0.35$ using angle-resolved
core-level photoemission. From the ratio of the Mn, La and Ca core-level
intensities it was concluded that $\rm La_{0.65}Ca_{0.35}MnO_3$ films
are terminated by a Mn--O layer, whereas $\rm La_{0.9}Ca_{0.1}MnO_3$ films
have a La/Ca--O terminal layer. Moreover, a significant surface segregation 
was found with the Ca content being strongly enhanced near the surface;
this was also found to depend on doping with a surface Ca fraction of
0.9 for $x = 0.1$ and 0.6 for $x = 0.35$. The authors conclude that the
surface could be fundamentally different from the bulk and that any 
measurement of bulk properties using surface sensitive techniques such as 
angle-resolved photoemission has to be treated with care. These
investigations were extended with two studies by Dulli \etal
(2000a, 2000b) on crystalline films with composition $\rm
La_{0.65}Sr_{0.35}MnO_3$. In line with the results on the Ca-doped
compound an appreciable amount of Sr segregation near the surface was
detected and the formation of a Ruddlesden-Popper phase
(La,Sr)$_2$MnO$_4$ was suggested (Dulli \etal 2000b). The same authors
argue that a surface electronic phase transition occurs at 240~K that
is different from the bulk ferromagnetic transition at
370~K. Measurements of the O(1$s$) core level binding energy and the
density of states as a function of temperature show that the surface
becomes insulating below 240~K (Dulli \etal 2000a). This result is
consistent with transport measurements on ultrathin films described in
the next paragraph.  

\begin{figure}[t]
\begin{center}
\vspace{-1.0cm}
\hspace{2.5cm} \epsfysize=12cm \epsfbox{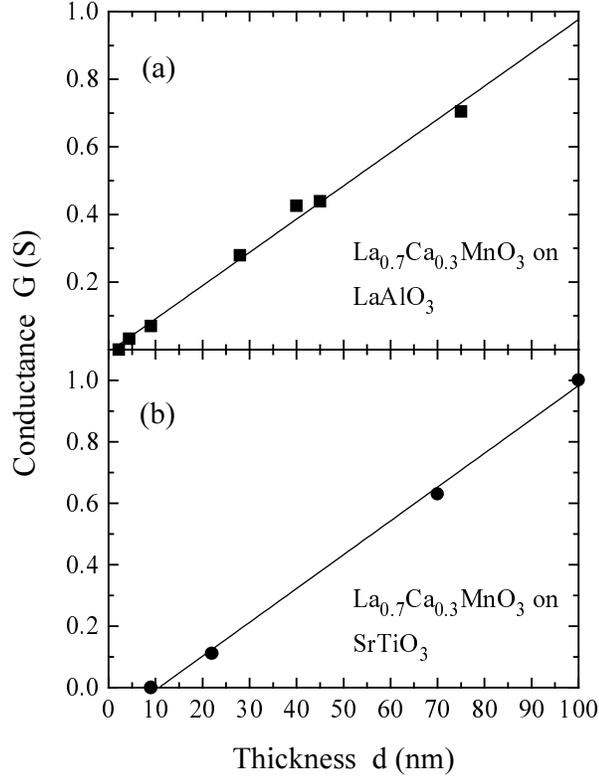}
\end{center}
\vspace{-0.5cm}
\caption{Conductance measured at 77~K in zero field
for annealed LCMO films grown on (a) LaAlO$_3$ and (b) SrTiO$_3$.
The lines were calculated according to equation~({\protect{\ref{Gsmall}}}).
Reproduced from Ziese \etal (1999c).}
\label{thickness}
\end{figure}
An indirect approach to investigate the interface properties of manganite
films was suggested by Sun \etal (1999) and Ziese \etal (1999c),
measuring the thickness dependence of the conductance of
$\rm La_{0.67}Sr_{0.33}MnO_3$ films on LaAlO$_3$ and NdGaO$_3$ and
$\rm La_{0.7}Ca_{0.3}MnO_3$ films on LaAlO$_3$, 
$\rm (LaAlO_3)_{0.3}(Sr_2AlTaO_6)_{0.7}$ (LSAT) and SrTiO$_3$,
respectively. The conductance was found to depend linearly on film
thickness, see figure~\ref{thickness}; the extrapolation
to zero conductance yielded finite values that were interpreted
as electrically dead layers (Sun \etal 1999). The dead layer thickness
depends on the substrate with $d_{dead} = 3$~nm (LSMO on NdGaO$_3$, 14~K),
5~nm (LSMO on LaAlO$_3$, 14~K), 1~nm (LCMO on LaAlO$_3$ and LSAT, 77~K)
and 11~nm (LCMO on SrTiO$_3$, 77~K), respectively. 
However, in the case of thin films surface scattering has to be taken
into account and deviations from a linear dependence of the conductance
on the thickness are to be expected for very thin films.
The conductance of a film is given by (MacDonald 1956)
\begin{equation}
G = \sigma_b d \bigg[1-3\ell_b/8d+
(3\ell_b/2d)\int_1^\infty\left(x^{-3}-x^{-5}\right)\exp(-dx/\ell_b)dx\bigg],
\label{G}
\end{equation}
where $\sigma_b$ denotes the bulk conductivity and $\ell_b$ the mean free path 
in the bulk. In the limits of thick and thin films
this reduces to (MacDonald 1956)
\begin{eqnarray}
G & = & \sigma_b(d-3\ell_b/8)\qquad \ell_b \ll d
\label{Gsmall}\\
G & = & \sigma_b(3d^2/4\ell_b)\ln(\ell_b/d) \qquad \ell_b \gg d\, .
\label{Glarge}
\end{eqnarray}
The bulk mean free path at 77~K can be estimated from the resistivity of 
thick films using the Drude formula with a carrier density 
(Ziese and Srinitiwarawong 1999)
$n = 5.2\times10^{-27}$~m$^{-3}$; this yields a value $\ell_b = 1.7$~nm.
Accordingly, the majority of the films investigated satisfy the condition
$d \gg \ell_b$ and Eq.~(\ref{Gsmall}) may be used. This predicts a linear
dependence of the conductance on film thickness with a finite abscissa
as found experimentally, see figure~\ref{thickness}.
The value of the dead layer of about 1~nm found
for LCMO films on LaAlO$_3$ and LSAT is of the same magnitude
as the bulk mean free path. Thus, annealed films on LaAlO$_3$ and LSAT
do not show a dead layer. The situation for annealed films on 
SrTiO$_3$ is different, since here an insulating layer 
of about 10~nm is found. This might be related to coherent film growth
of LCMO on SrTiO$_3$ resulting in large strains at small film thickness.
Indeed, high resolution microscopy (HREM) studies of thin (6~nm -
12~nm) $\rm La_{0.73}Ca_{0.27}MnO_3$ films grown on SrTiO$_3$ showed a
change in structure in those films as compared to the bulk due to the
in-plane tensile strain (Zandbergen \etal 1999). An elongation of the
in-plane oxygen square was found which is a Jahn-Teller-like
distortion induced by the substrate. The resulting misfit energy is
relaxed via twin boundaries; the twin-boundary density was found to
increase strongly with decreasing film thickness. This
Jahn-Teller-like distortion is supposed to lead to a ferromagnetic
insulating state of these thin films. It was predicted that the Curie
temperature is very sensitive to bi-axial strain (Millis \etal 1998)
in qualitative agreement with the experimental observation. In
ultrathin films, however, the structural change seems to stabilize the
Curie temperature at a rather high value.

\subsection{Theoretical results}

\begin{figure}[t]
\begin{center}
\includegraphics[width=0.3\textwidth]{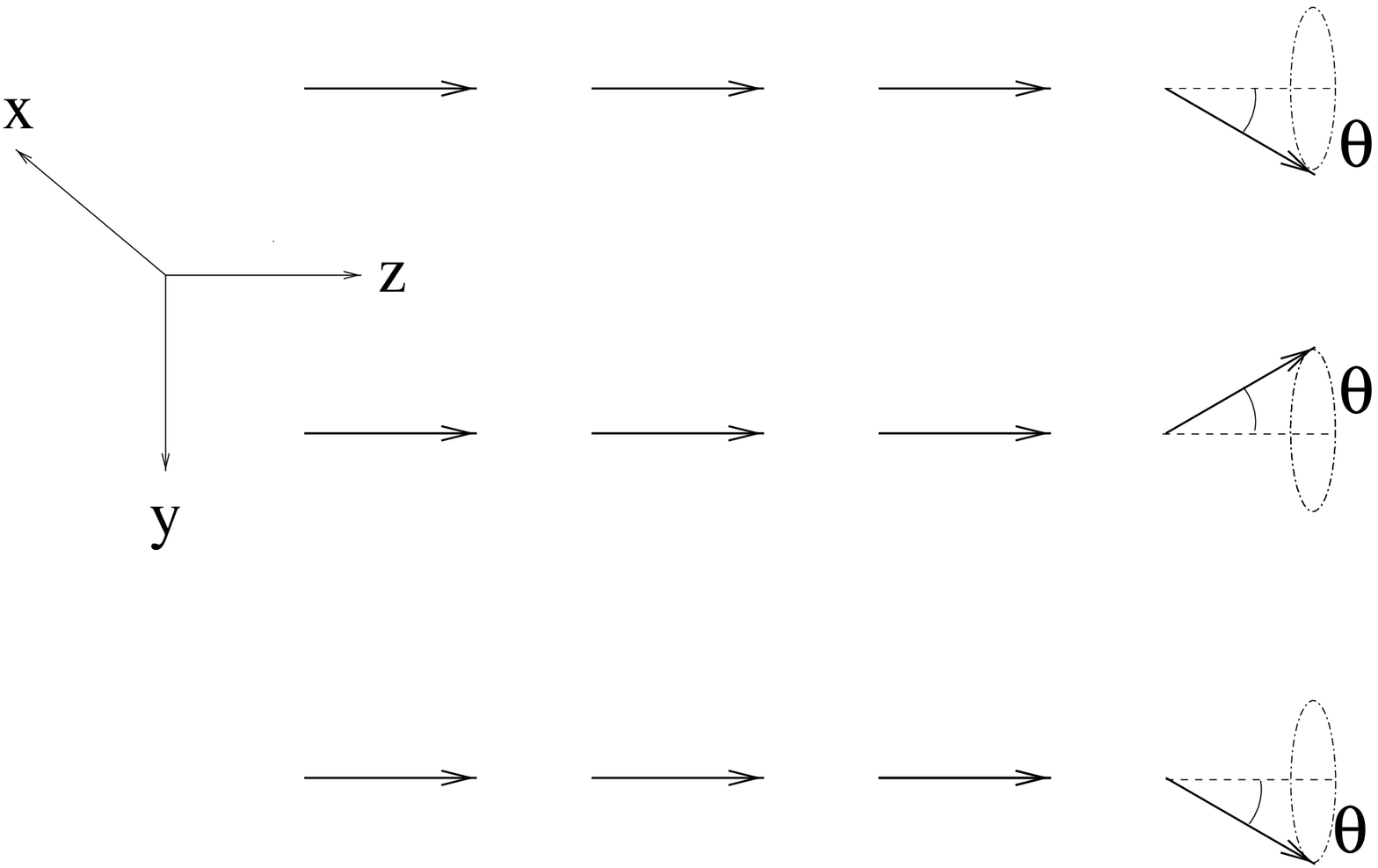}
\includegraphics[width=0.6\textwidth]{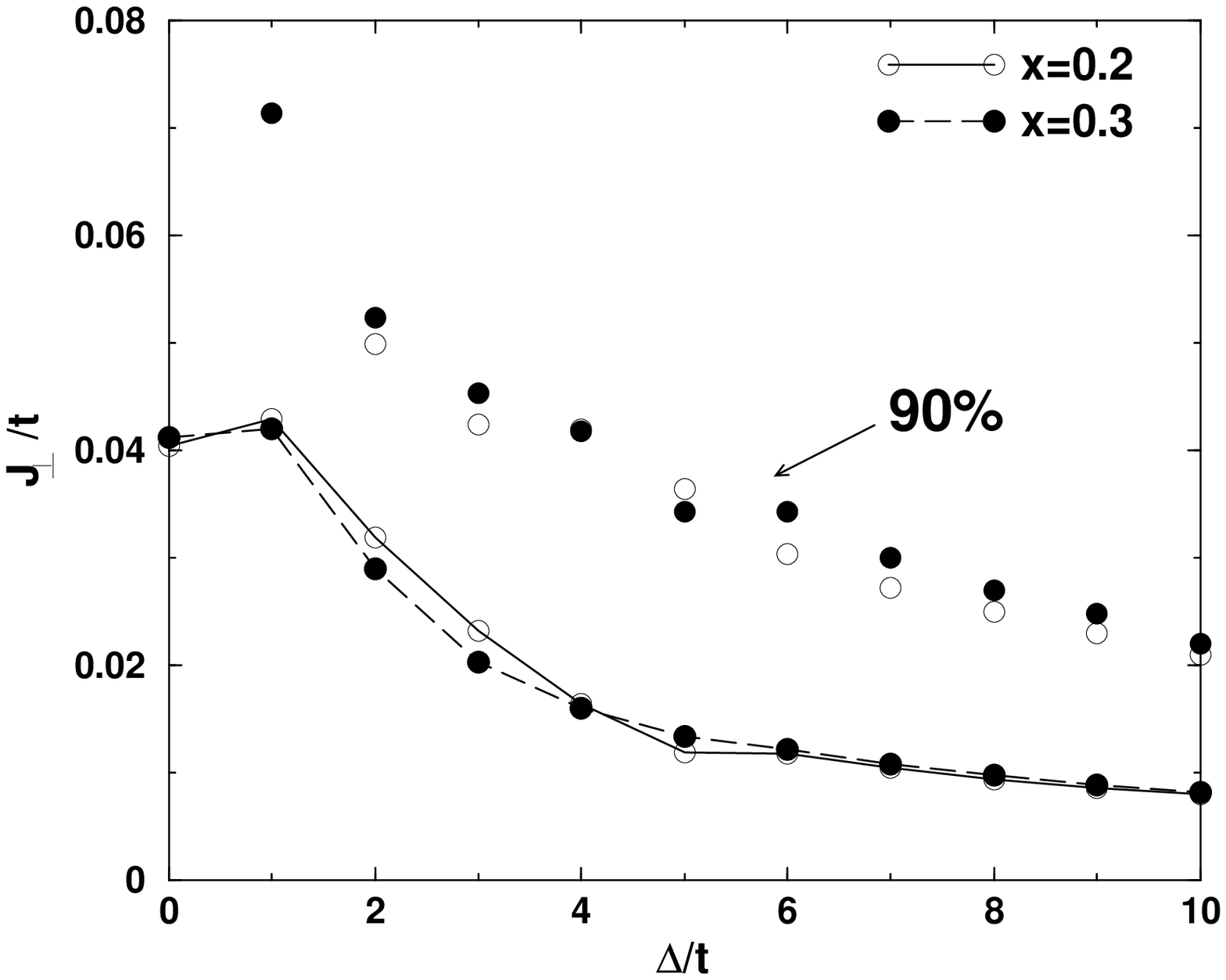}
\end{center}
\caption{Left panel: Magnetic structure with canting angle $\Theta$ considered
in the theoretical calculation. Right panel: Surface magnetic phase diagram
of a doped manganite. The lines separate the fully ferromagnetic and 
the canted regions. The symbols indicate that the order is 90\% antiferromagnetic,
i.e.\ $\Theta = 81^\circ$. Solid and open dots are the results for hole 
concentrations $x = 0.3$ and $x = 0.2$, respectively. $\Delta$ denotes
the surface e$_g$ level energy splitting, $t$ the hopping integral and $J_\perp$
the antiferromagnetic coupling constant. Reproduced from Calderon \etal (1999b).}
\label{calderon1}
\end{figure}
\begin{figure}[t]
\begin{center}
\centerline{\hspace{1.5cm} \mbox{\epsfysize=10cm \epsffile{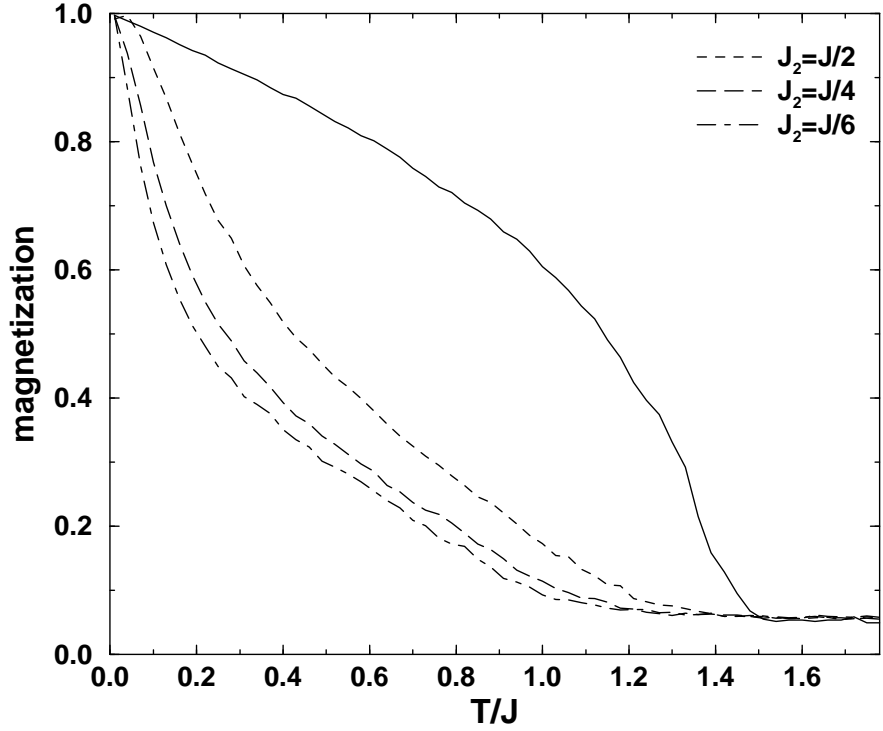}}}
\end{center}
\vspace{-1.0cm}
\caption{Normalized magnetization of the surface layer as a function of
temperature for various ferromagnetic coupling constants $J_2$ between
the surface spins and the spins of the first innermost layer. The solid line
is the result for the bulk magnetization. $J$ denotes the bulk
ferromagnetic coupling constant. Reproduced from Calderon \etal (1999b).}
\label{calderon2}
\end{figure}
Two models for the description of the manganite-vacuum surface have
been proposed by Calderon \etal (1999b) and Filippetti and Pickett
(1999, 2000), respectively. These models arrive at seemingly opposite
conclusions and the main results are summarized below.

The (001) surface electronic structure and surface magnetization of manganites
were theoretically investigated by Calderon \etal (1999b). At the surface,
the oxygen octahedra surrounding each manganese ion are incomplete
resulting in a tetragonal distortion and a splitting of the e$_g$ levels
by an amount $\Delta$. Since the d$_{3z^2-r^2}$ orbitals point towards the surface,
these are shifted to lower energy levels than the d$_{x^2-y^2}$ orbitals.
The calculation shows that for energy splittings $\Delta > t$, where $t$
is the hopping matrix element, the d$_{3z^2-r^2}$ levels at the surface
are fully occupied, whereas the d$_{x^2-y^2}$ levels are empty. This
results in a suppression of the double-exchange coupling at the surface;
the surface-spin structure is therefore determined by the antiferromagnetic
super-exchange coupling between Mn ions with coupling constant $J_\perp$.
Calderon \etal (1999b) considered the case of a simple spin-canting
by an angle $\Theta$ in the surface layer, see figure~\ref{calderon1}(a).
The canting angle $\Theta$ was then calculated as a function of the
e$_g$ level surface splitting $\Delta$ and the antiferromagnetic
exchange constant $J_\perp$. The resulting magnetic phase diagram
is shown in figure~\ref{calderon1}(b). For realistic values of the
antiferromagnetic coupling constant, $J_\perp \sim 0.02t$ (Perring \etal 1997),
and values $t \sim 0.1-0.3$~eV as well as $\Delta \sim 0.5-1.5$~eV,
the surface spin-structure is nearly antiferromagnetic. The magnetization
of the surface layer was calculated within an effective Heisenberg
model with bulk ferromagnetic coupling $J$, surface antiferromagnetic 
coupling $J_1 \sim -J/100$ and ferromagnetic coupling $J_2 \sim J/6-J/2$
between the surface layer and the first innermost layer. The surface
magnetization for various couplings $J_2$ is compared to the bulk magnetization
in figure~\ref{calderon2}. The surface magnetization decays much
more strongly with temperature than the bulk magnetization due
to a reduced spin stiffness at the surface. The theoretical results are
in qualitative agreement with the measured spin-polarization,
see figure~\ref{park}.

\begin{figure}[h]
\begin{center}
\includegraphics[width=0.6\textwidth]{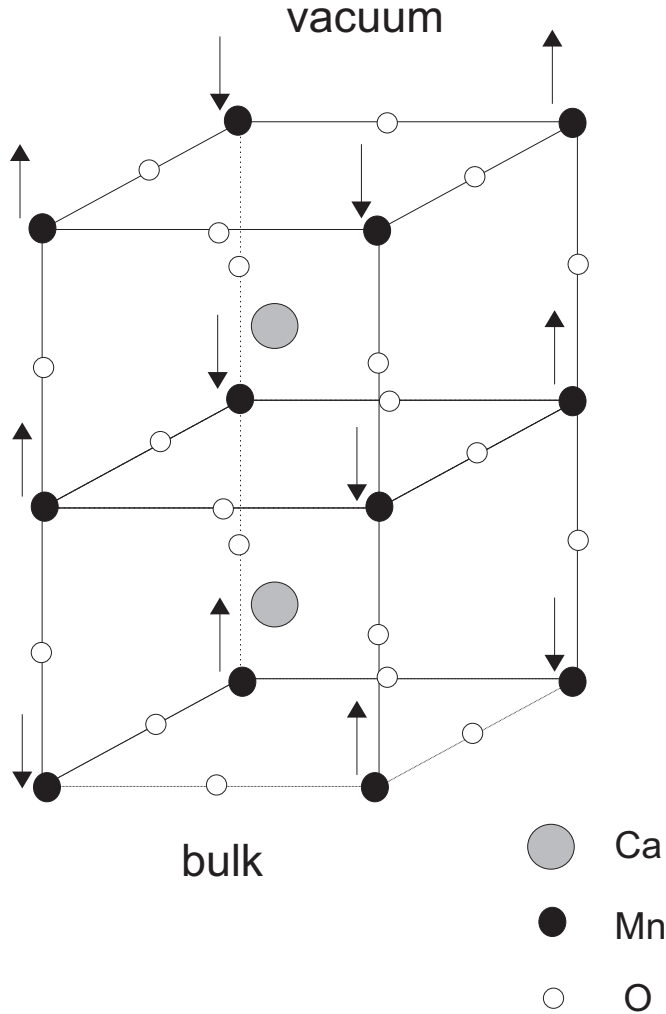}
\caption{Structure of cubic CaMnO$_3$. The arrows near the Mn-ions
  indicate the core spins. These have a G-type antiferromagnetic order
  in the bulk. The surface layer, however, is found to be
  ferromagnetically coupled to the sub-surface layer while retaining
  the antiferromagnetic in-plane order. After Filippetti and Pickett (2000).} 
\label{pickett}
\end{center}
\end{figure}
Filippetti and Pickett performed first principles calculations using
local spin-density functional theory in order to investigate magnetic
reconstructions at the (001) surface of CaMnO$_3$ (Filippetti and Pickett
1999) and  of $\rm La_{0.5}Ca_{0.5}MnO_3$ (Filippetti and Pickett
2000). At these dopings a bulk antiferromagnetic structure prevails
with G-type and A-type antiferromagnetic ordering for CaMnO$_3$ and
$\rm La_{0.5}Ca_{0.5}MnO_3$, respectively. The striking result of
these studies is a ferromagnetic coupling of the surface spins to the
sub-surface layer while retaining the antiferromagnetic in-plane
structure. This result does not depend on the Ca-doping and is thought
to apply also in the ferromagnetic phase. The resulting spin structure
is sketched in Fig.~\ref{pickett} for CaMnO$_3$.
This spin-flip process is driven by the formation of a deep surface
state with $d_{z^2}$ orbital character at the Fermi level. This state
promotes a double-exchange like coupling between the spins in the
surface and sub-surface layer, a scenario in stark contrast to the
assumption of weakened double exchange at the surface made by Calderon
\etal (1999b).

The experimental data collected to date do not allow for a distinction
between these two models.

\subsection{Oxide-metal interface}

The resistance of oxide/metal interfaces was systematically
investigated by Mieville \etal (1998) for $\rm La_{0.67}Sr_{0.33}MnO_3$,
SrRuO$_3$ and $\rm La_{0.5}Sr_{0.5}CoO_3$ as conducting ferromagnetic
oxides and the metals Au, Pd, Nb and Al. It was found that the interface
resistance increased with the oxygen affinity of the metal, presumably
due to both the formation of an oxygen depleted layer in the ferromagnetic
oxide as well as an oxide layer in the metal. Pd yielded a particularly
low interface resistance. RF Ar ion plasma cleaning was found to increase
the interface resistance. Values for the areal resistivity $RA$ are given
in table~\ref{metalresist}. Ziese (1999) investigated LCMO/Cr and
LCMO/Ti contacts and also found the formation of an interfacial layer
with a high resistance. Values are listed in table~\ref{metalresist}.
On the other hand, SrRuO$_3$ showed a remarkable compositional
stability evidenced by very low interface resistivities with all metals
investigated: Au, Pd, Al, Nb. The high corrosion resistance was also
shown by lifting a SrRuO$_3$ film from its SrTiO$_3$ substrate by
selective wet chemical etching with an acid 
(50\% HF: 70\% HNO$_3$: H$_2$O = 1:1:1) (Gan \etal 1998). Magnetic and
electrical measurements on the free-standing film in comparison
to strained films on SrTiO$_3$ substrates showed an enhancement of
the magnetocrystalline anisotropy due to strain and reductions
of the Curie temperature and the resistivity in the strained state.
\begin{table}[t]
\caption{Interface resistance $RA$ of various manganite-metal
contacts.}
\begin{indented}
\item[]\begin{tabular}{@{}lllll}
\br
Manganite & Metal  & $T$ (K) & $RA$ ($\Omega$cm$^2$) & Ref. \\
\mr
LSMO & Pd & 4.2 & $<2 \times 10^{-8}$      & Mieville (1998)\\
LSMO & Au & 4.2 & $(0.64-1)\times 10^{-6}$ & Mieville (1998)\\
LSMO & Nb & 4.2 & $4 \times 10^{-4}$       & Mieville (1998)\\
LSMO & Al & 4.2 & $10-20$                 & Mieville (1998)\\
LCMO & Ti & 77  & $0.5$                   & Ziese (1999)   \\
LCMO & Cr & 10  & $3.0$                   & Ziese (1999)   \\
LCMO & Cr & 77  & $3.3$                   & Ziese (1999)   \\
\br
\end{tabular}
\end{indented}
\label{metalresist}
\end{table}

Coombes and Gehring (1998) theoretically investigated the perturbation 
produced on manganites by a metallic interface. These authors considered
a nickel-LCMO interface modelled as a perfect epitaxy within a tight binding
approximation. The magnetic moment on a Mn lattice site was calculated
as a function of the distance from the interface assuming a full
spin-polarization of the nickel ion at the interface. The perturbation
of the Mn spin was found to decay to the bulk values within about
five lattice sites from the interface. Below $T_C$ a drop in the magnetic
moment of the first Mn ions was observed, whereas above $T_C$ a magnetization
is induced on the Mn site. Ziese \etal (1998a) and Gibbs \etal (1998)
investigated LCMO/Ni heterostructures in out-of-plane geometry in
order to investigate the induced magnetization. However, the transport
properties of those heterostructures turned out to be dominated by a
large interface resistance due to oxidation.

\subsection{Ferromagnet-superconductor hybrids}

Since the manganites have the perovskite structure and a similar lattice
constant as high temperature superconductors, 
attempts have been made to grow hetero-epitaxial 
ferromagnetic/superconducting structures to investigate magnetotransport
properties of such devices. These investigations fall into three classes:
the growth and investigation of multilayers, the investigation
of spin-injection devices and the direct study of interface properties,
especially Andreev reflection. These studies are briefly reviewed
in the following.

Jakob \etal (1995) and Przyslupski \etal (1997) investigated
$\rm La_{0.7}Ba_{0.3}MnO_3$/\-$\rm YBa_2Cu_3O_7$ and 
$\rm Nd_{0.67}Sr_{0.33}MnO_3$/\-$\rm YBa_2Cu_3O_7$ multilayers, respectively.
Jakob \etal (1995) reported hetero-epitaxial growth of multilayers
with a Curie temperature of about 220~K and a reduced superconducting 
temperature of about 50~K. The multilayers show the coexistence of
superconductivity and ferromagnetism, especially a colossal
magnetoresistance effect. The YBCO layers are decoupled by the
LBMO layers and show quasi-two-dimensional behaviour as derived
from the scaling of the resistance with the magnetic field
component perpendicular to the layers. Przyslupski \etal (1997)
also reported hetero-epitaxial multilayer growth, a Curie temperature of 
about 150~K and a critical temperature of about 75~K. Superconductivity 
and ferromagnetism also coexist in these NSMO/YBCO-multilayers.

\begin{figure}[t]
\vspace*{-4.0cm}
\begin{center}
\hspace*{2.5cm} \epsfysize=12cm \epsfbox{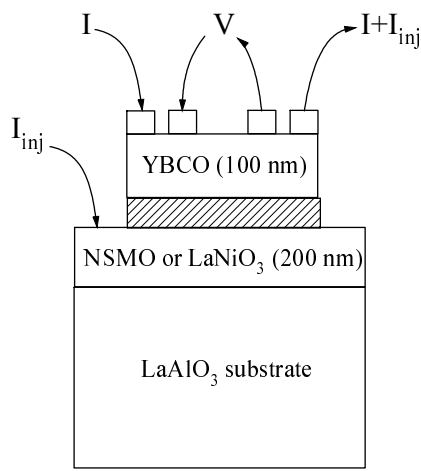}
\end{center}
\vspace{-2.0cm}
\caption{Schematic drawing of a spin-injection device as used
by Dong \etal (1997).}
\label{dong1}
\end{figure}
\begin{figure}[t]
\begin{center}
\vspace*{-0.0cm}
\hspace*{0.0cm} \includegraphics[width=0.7\textwidth]{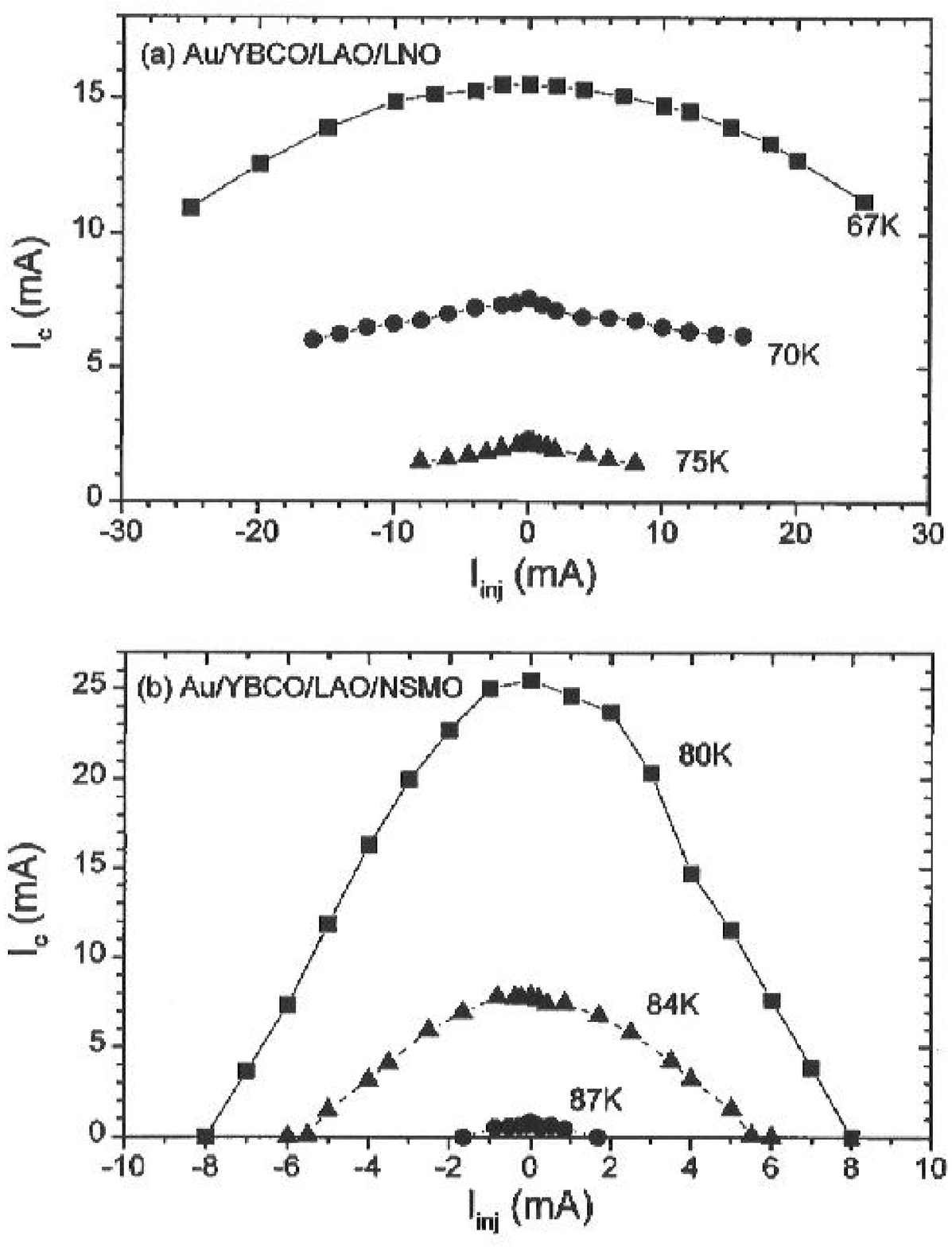}
\vspace*{-2.7cm}
\end{center}
\caption{Critical current of the YBCO layer shown in 
figure~\protect{\ref{dong1}} as a function of the current
injected from (a) a LaNiO$_3$ and (b) a $\rm Nd_{0.7}Sr_{0.3}MnO_3$
layer. The injection of a spin-polarized current apparently leads to a
much stronger decay of the critical current. Reproduced from Dong \etal (1997).}
\label{dong2}
\end{figure}
Considerable work has been devoted to the study of spin-injection
devices based on ferromagnetic manganites and high temperature superconductors,
see Dong \etal (1997, 1998), Yeh \etal (1999) 
and Vas'ko \etal (1997). Various geometries
were used to study spin-polarized carrier injection; the set-up
used by Dong \etal (1997) is sketched in figure~\ref{dong1}.
The critical current of a YBCO layer was measured with current
injection through a NSMO or a LaNiO$_3$ layer. The critical current
was found to decrease much stronger with injection of spin-polarized
quasi-particles from the NSMO layer than with injection of unpolarized
quasi-particles from the LaNiO$_3$ layer, see figure~\ref{dong2}. However,
since the YBCO film grown on LaNiO$_3$ shows degraded properties, 
especially a lower critical temperature, pinning effects from
the changed microstructure have to be taken into account. The comparison
between the two results for spin-polarized and unpolarized carrier injection
seems to rule out heating effects 
being responsible for the decrease of the critical current.
Dong \etal (1997) interpreted their results within a semi-quantitative
non-equilibrium thermodynamic model of quasi-particle injection at high energies.
Yeh \etal (1999) investigated spin-injection in
LCMO/YSZ/YBa$_2$Cu$_3$O$_7$ and LSMO/SrTiO$_3$/YBa$_2$Cu$_3$O$_7$
heterostructures with two thicknesses (2~nm and 10~nm) of the
SrTiO$_3$ barrier. YSZ denotes yttrium-stabilized zirconia. A large
difference in the critical current densities determined from
continuous and pulsed current measurements was observed. This
indicates significant Joule heating in the dc measurements. For 2~nm
thick SrTiO$_3$ and 1.3~nm thick YSZ barriers a strong decrease of the
critical current with the injected spin-polarized current was
found. Surprisingly, for the 10~nm thick SrTiO$_3$ barrier, a
variation of the critical current with the injected spin-polarized
current could not be detected. A
LaNiO$_3$/SrTiO$_3$/YBa$_2$Cu$_3$O$_7$ control sample also did not
show any suppression of the critical current on quasiparticle
injection. Yeh \etal (1999) argued that the suppression of the
critical current density in the heterostructures with thin barriers is
due to the pair-breaking effect of spin-polarized quasiparticles. The
thicker SrTiO$_3$ barrier is supposed to provide a larger hypothetical
interface impedance, thereby suppressing spin-polarized quasiparticle
injection. 

In a related study Mikheenko \etal (2001) investigated the
effect of spin-injection on the relaxation from the Bean critical
state established in a YBa$_2$Cu$_3$O$_7$ film; a large change of the
magnetic moment was detected. In view of the experimental data
obtained so far the large variation of device parameters seems to
preclude the observation of any reliable trend. Therefore, in order to
decide on the mechanism of spin-injection and pair-breaking more
experimental data are clearly necessary.

\begin{figure}[t]
\begin{center}
\vspace*{-0.2cm}
\hspace*{1.0cm} \includegraphics[width=0.8\textwidth]{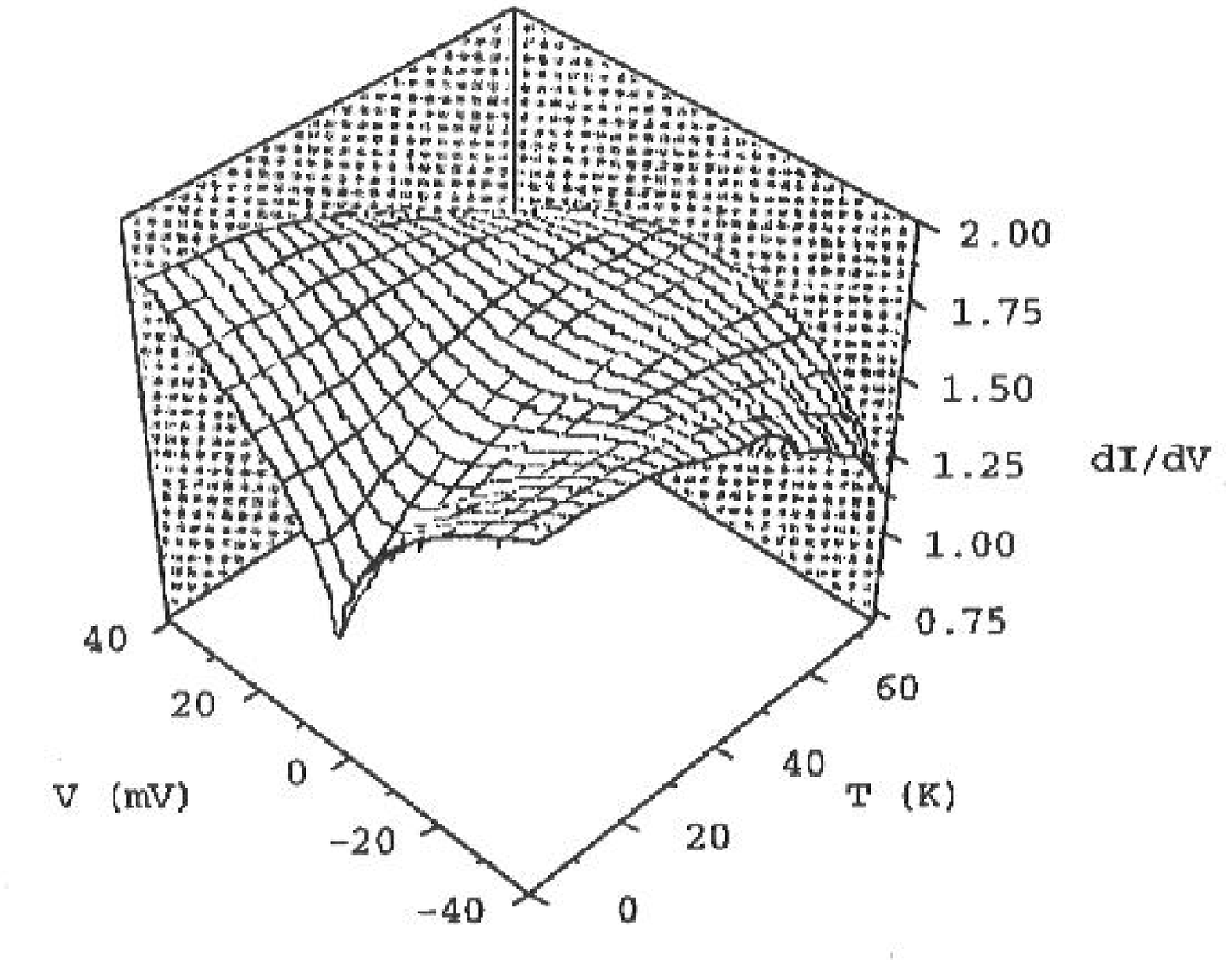}
\vspace*{-8.7cm}
\end{center}
\caption{Temperature and voltage-bias dependence of the differential conductance
of the interface between the ferromagnet $\rm La_{0.67}Ba_{0.33}MnO_3$
and the superconductor $\rm DyBa_2Cu_3O_7$. Reproduced from
Vas'ko \etal (1998).}
\label{vasko}
\end{figure}
A detailed study of the transport properties of 
$\rm La_{0.67}Ba_{0.33}MnO_3/DyBa_2Cu_3O_7$-interfaces was performed
by Vas'ko \etal (1998). The Curie temperature was 305~K and the critical
temperature 90~K. The differential conductance of the 
ferromagnetic/superconducting interface was measured
as a function of temperature, bias voltage and magnetic field.
The main finding was a dip in the differential conductance near
zero bias at temperatures below about 60~K, see figure~\ref{vasko}. 
At high temperatures the conductance decreases at large bias voltages,
presumably due to heating effects. The voltage dependence of the
differential conductance at low temperatures does not depend on the applied
magnetic field for fields up to 12~T. The conductance dip was
interpreted as being due to Andreev reflection at a clean ferromagnetic/superconducting
interface. Since LBMO has a high spin-polarization, Andreev reflection
is suppressed. The disappearance of the conductance dip at about 60~K might
indicate that the superconductor is in the orthorombic II phase
near the interface. This study shows that the fabrication of fairly clean
superconductor/ferromagnet interfaces is feasible using manganites and
high temperature superconductors. The measurement of Andreev reflection
at these interfaces might prove to be a powerful tool for the investigation
of symmetries of the superconducting order parameter. Calculations of
conductance spectra for s- and d-wave superconductors have already been 
performed (Zhu \etal 1999, Merrill and Si 1999, Bhattacharjee and
Sardar 2000).

The electronic structure and magnetism of a LSMO interface buried
under YBa$_2$Cu$_3$O$_7$ capping layers of various thicknesses $\le
8$~nm was investigated by Stadler \etal (1999, 2000) using x-ray magnetic
circular dichroism (XMCD) and x-ray absorption spectroscopy (XAS) at
the Mn L$_{2,3}$ edge. From a comparison of the data with spectra
measured on a $\rm La_{1-x}Sr_xMnO_3$ series it was concluded that the
cation stoichiometry at the interface is changed. La ions are
presumably replaced by Ba ions from the YBa$_2$Cu$_3$O$_7$ layer
leading to a decreased La concentration near the interface. The XMCD
signal is consistent with a progressive replacement of the
ferromagnetic order by an antiferromagnetic structure. A comparison
with Al capped LSMO films shows that the spectral changes are not due
to de-oxygenation. The antiferromagnetic structure at the
interface is likely to decrease the spin-polarization.

\subsection{Exchange biasing, multilayers and GMR}

In view of possible applications of the magnetic oxides in tunnelling
junctions that generally have a pinned magnetic electrode, the study 
of exchange biasing of the magnetic layer by an antiferromagnet is 
important. Exchange biasing of magnetite layers by antiferromagnetic
layers such as NiO (Berry \etal 1993, Lind \etal 1995, Ball \etal
1996, van der Heijden \etal 1999) and CoO (Ijiri \etal 1998a, 1998b, 
Kleint \etal 1998) has been achieved.
In this case, the coupling mechanism seems to be similar to
exchange coupling in elemental ferromagnets. In the manganites the
situation might be different, since the double-exchange mechanism
that provides the magnetic coupling is short-ranged and very sensitive
to structural disorder. So far, only two groups reported studies on
the exchange-biasing of manganite layers: Niarchos and coworkers
(Panagiotopoulos \etal 1999a, 1999b, Moutis \etal 2001) and Nikolaev
\etal (2000a) studied exchange coupling in 
$\rm La_{2/3}Ca_{1/3}MnO_3/La_{1/3}Ca_{2/3}MnO_3$ multilayers.
Panagiotopoulos \etal (1999a, 1999b) grew
multilayers with various bilayer thicknesses between 2 and 32~nm
using pulsed laser deposition. The magnetization hysteresis
loops were found to be shifted after field cooling in a field of 1~T with
respect to the hysteresis loops obtained after zero field cooling.
The exchange biasing field measured at 10~K displays a marked dependence on
the bilayer thickness with a maximum exchange field at a bilayer
thickness of 10~nm. The temperature dependence of the magnetization,
however, indicates superparamagnetic behaviour of the multilayers
with a blocking temperature of about 70~K. The exchange-biasing field
vanishes at the blocking temperature. The superparamagnetic behaviour 
indicates an unfavourable structural morphology of the samples and 
possibly very rough interfaces. Nikolaev \etal (2000) fabricated 
$\rm
La_{1/3}Ca_{2/3}MnO_3/La_{2/3}Ca_{1/3}MnO_3/La_{1/3}Ca_{2/3}MnO_3$
trilayers using ozone-assisted molecular beam epitaxy. Even very thin
(2.3~nm -- 4.6~nm) manganite films showed metallic conductivity and
ferromagnetism below comparatively high Curie temperatures of
200~K. The exchange field was found to be about 800~G at 5~K. 
Shifted hysteresis loops were also
observed in NSMO ceramics (Baszy\'nski \etal 1999) 
and $\rm La_{0.5}Ca_{0.5}MnO_3$ films (Roy \etal 1999).

There is substantial work on the properties of magnetite multilayers
that will not be reviewed here. It is more interesting to look at the
studies of manganite multilayers, since is is clear from the
theoretical work discussed above that these samples might exihibit
interesting physical properties. First studies of multilayers formed
from thin ferromagnetic manganite layers sandwiched inbetween
insulating layers report an enhancement of the magnetoresistance (Kwon
\etal 1997, Venimadhav \etal 2000, Pietambaram \etal 2001); the
mechanism leading to this enhancement is unclear but might simply be
due to structural and chemical inhomogeneity. Studies on high
quality samples reveal antiferromagnetic interlayer coupling and giant
magnetoresistance (Nikolaev \etal 1999, Nikolaev \etal 2000b,
Krivorotov \etal 2001). Nikolaev \etal (1999, 2000b) showed that
ferromagnetic $\rm La_{0.67}Ba_{0.33}MnO_3$ layers coupled via
LaNiO$_3$ spacers showed oscillatory exchange coupling as a function
of the spacer thickness. At antiferromagnetic coupling a small
positive magnetoresistance was seen. This is in contrast to the usual
giant magnetoresistance in metallic multilayers which is negative. The
result, however, is in agreement with the observations on $\rm
Fe_3O_4/TiN$ multilayers also showing positive magnetoresistance
(Orozco \etal 1999). The explanation of this phenomenon is still
qualitative: the positive magnetoresistance is attributed to quantum
confinement enhanced by the half-metallic nature of the ferro- and
ferrimagnetic constituent layers.

\subsection{Oxide-semiconductor interfaces}

To the best of the author's knowledge, no data are available on 
ferromagnetic oxide/semiconductor interfaces. It is possible
to grow manganite films directly on Si or GaAs substrates
resulting in polycrystalline films due to the large thermal
expansion coefficient of the semiconductor substrate and strong
inter-diffusion. Several groups reported the heteroepitaxial
growth of manganite films on YSZ buffered Si substrates 
(Trajanovic \etal 1996, Fontcuberta \etal 1999, Gillman \etal 1998).
However, the film quality seems to be worse than for films
grown on lattice matched substrates. Due to the presence
of the thick insulating buffer layer, spin-injection from the 
ferromagnetic oxide into the semiconductor cannot be facilitated.
\section{Ferromagnetic tunnelling junctions \label{tunnelling}}
\subsection{Basic theory \label{theorytunnel}}
\begin{figure}[t]
\vspace*{-0.5cm}
\begin{center}
\hspace{2.0cm} \epsfysize=10cm \epsfbox{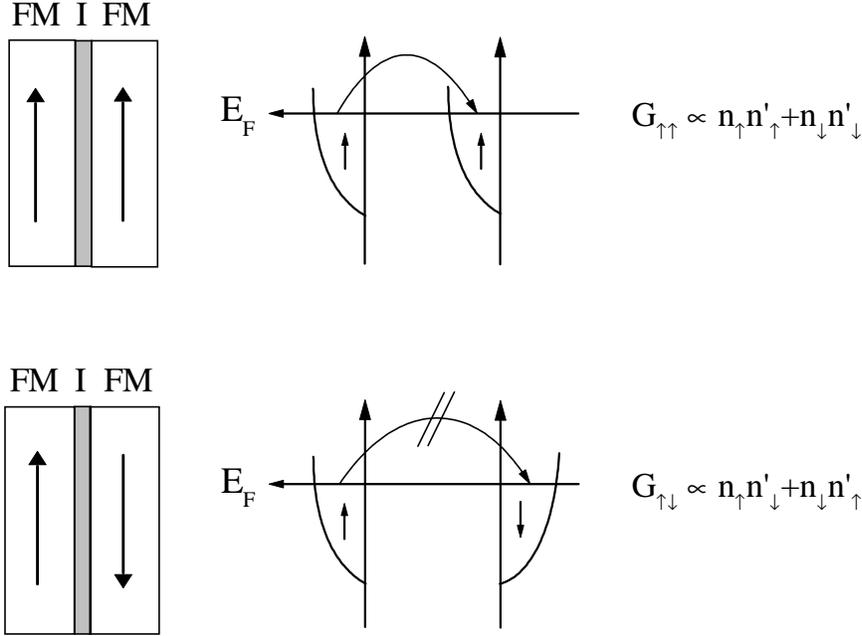}
\end{center}
\vspace{-1.0cm}
\caption{Schematic drawing of a ferromagnetic tunnelling junction
and of the corresponding density of states. FM indicates the
ferromagnetic electrodes and I the insulating barrier. For simplicity
the density of states of a half-metallic ferromagnet are shown. The
conductivities in the parallel and antiparallel state,
$G_{\uparrow\uparrow}$ and $G_{\uparrow\downarrow}$ are proportional
to the densities of states at the Fermi level, $n_\uparrow$,
$n_\downarrow$ (left electrode) and $n'_\uparrow$, $n'_\downarrow$
(right electrode), if spin-flip processes in the barrier can be
neglected.}
\label{tunprin}
\end{figure}

The magneto-conductance of a ferromagnet-insulator-ferromagnet
tunnelling junction was first derived by Julliere (1975). 
A schematic drawing of a ferromagnetic tunnelling junction is shown in
figure~\ref{tunprin}: two ferromagnetic electrodes are separated by
a thin insulating layer. The parallel and anti-parallel orientation
of the ferromagnetic electrodes leads to different conductances
in these two states. A schematic density of states for the tunnelling
junction is indicated in the figure assuming identical, half-metallic
ferromagnets. If the spin of the carrier is conserved in the tunnelling
process, tunnelling is only possible for parallel electrode orientation.
A simple expression for the tunnelling magnetoconductance can be
derived as follows (Julliere 1975). Let $P$ and $P'$ denote the
spin-polarization of the ferromagnetic electrodes. If the carrier 
spin is conserved during tunnelling, the conductance is proportional 
to the sum of the products of the spin-polarized densities of states.
One obtains in the parallel ($\uparrow\uparrow$) and antiparallel
($\uparrow\downarrow$) states:
\begin{eqnarray}
G_{\uparrow\uparrow} \propto \left(1+P\right)\left(1+P'\right)
+\left(1-P\right)\left(1-P'\right) \propto 1+PP'
\label{Gparallel}\\
G_{\uparrow\downarrow} \propto \left(1+P\right)\left(1-P'\right)
+\left(1-P\right)\left(1+P'\right) \propto 1-PP'\, .
\label{Gantiparallel}
\end{eqnarray}
This yields for the magnetoconductance and magnetoresistance (TMR), respectively
\begin{eqnarray}
\frac{\Delta G}{G} \equiv \frac{G_{\uparrow\uparrow}-G_{\uparrow\downarrow}}{G_{\uparrow\uparrow}}
= \frac{2PP'}{1+PP'}
\label{tuncond}\\
\frac{\Delta R}{R} \equiv \frac{R_{\uparrow\downarrow}-R_{\uparrow\uparrow}}{R_{\uparrow\uparrow}}
= \frac{2PP'}{1-PP'}\, .
\label{tunresist}
\end{eqnarray}
In the case of two identical ferromagnets, the magnetoresistance is always
negative; it diverges for two half-metallic electrodes. In the general case,
the magnetoresistance can be both positive and negative.

Slonczewski (1989) calculated the tunnelling conductance within a free
electron model using the Landauer-B\"uttiker formula (Landauer 1957,
B\"uttiker 1988). The exchange fields in the ferromagnetic
electrodes were assumed to span an angle $\Theta$.
The transmission coefficient $T(k_\parallel)$ of an incoming
electron of defined spin is calculated. $k_\parallel$ denotes
the wave vector component parallel to the barrier.
Assuming the absence of diffuse scattering in the barrier, 
the tunnelling conductance is given by
\begin{equation}
G = \frac{{\rm e}^2}{(2\pi)^2{\rm h}}\, \int d^2k_\parallel T(k_\parallel)\, .
\label{landauer}
\end{equation}
Slonczewski evaluated equation~(\ref{landauer}) by integrating over $k_\parallel$
and keeping only the leading term in $1/d$. The conductance
is found to be of the form
\begin{equation}
G \propto \left\lbrack1+P_{fb}^2\cos(\Theta)\right\rbrack
\label{slonc1}
\end{equation}
with an effective polarization $P_{fb}$ depending on the band splitting.
In the one-band case $P_{fb} = 1$ resulting in a complete spin-valve
effect, in the two-band case
\begin{equation}
P_{fb} = \frac{(k_\uparrow-k_\downarrow)}{(k_\uparrow+k_\downarrow)}\, 
\frac{(\kappa^2-k_\uparrow k_\downarrow)}{(\kappa^2+k_\uparrow k_\downarrow)}
\equiv P\, A_{fb}
\label{slonc2}
\end{equation}
with the wave-vectors $k_\uparrow$, $k_\downarrow$ 
of the $\uparrow$ and $\downarrow$ electrons and the inverse of the 
decay length in the barrier $\kappa = \sqrt{2(U_0-E_F)}$. $U_0$ denotes the barrier
height and $E_F$ the Fermi energy. For free electrons, the first factor
in equation~(\ref{slonc2}) simply reduces to the standard definition
of the spin-polarization $P = (n_\uparrow-n_\downarrow)/(n_\uparrow+n_\downarrow)$.
The second factor $A_{fb}$, however, is related to interface properties and 
has the range $-1 < A_{fb} < 1$. It changes sign as a function of barrier
height, indicating that the sign of the apparent spin-polarization
can be modified by the appropriate choice of the barrier material.

\begin{figure}[t]
\begin{center}
\vspace*{0.2cm}
\hspace*{1.0cm} \includegraphics[width=0.8\textwidth]{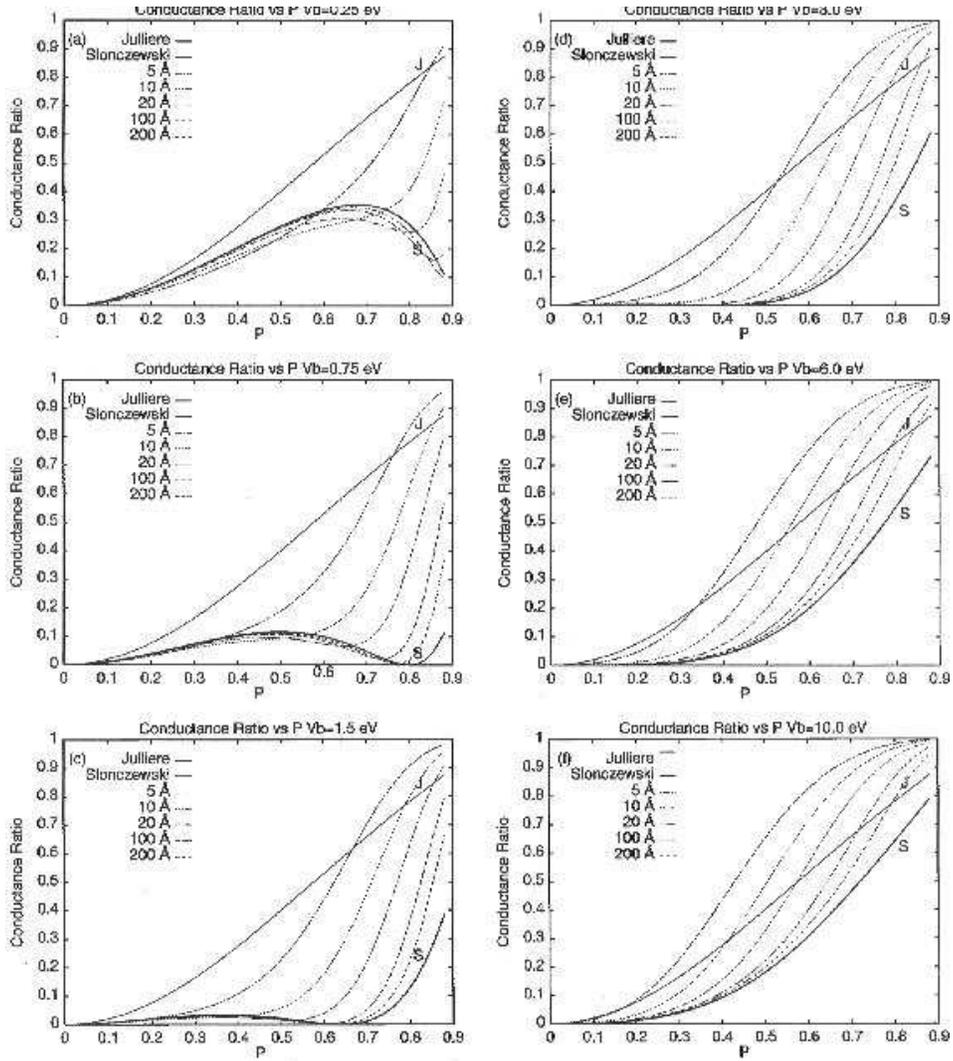}
\vspace*{-4.5cm}
\end{center}
\caption{Conductance ratio $\Delta G/G$ for free electron spin-dependent
tunnelling for various barrier heights (a) 0.25~eV, (b) 0.75~eV, (c) 1.5~eV,
(d) 3.0~eV, (e) 6.0~eV and (f) 10.0~eV. In each panel, barrier widths
of 0.5, 1, 2, 10 and 20~nm are shown along with the Julliere and Slonczewski 
results labelled by (J) and (S), respectively. Reproduced from Maclaren \etal (1997).}
\label{maclaren}
\end{figure}
The tunnelling conductance in the free-electron model was numerically 
calculated by MacLaren \etal (1997) in order to investigate
the validity of Slonczewski's and Julliere's results. In figure~\ref{maclaren}
the conductance ratio $\Delta G/G$ calculated by numerically integrating
Eq.~(\ref{landauer}) is shown as a function of the polarization $P$
for various barrier heights and thicknesses. For comparison, Julliere's
expression $\Delta G/G = 2P^2/(1+P^2)$ and Slonczewski's approximation
are also shown. Slonczewski's expression for $\Delta G/G$ is a good
approximation for large barrier thicknesses and small barrier heights,
whereas Julliere's expression fails to reproduce any dependence
on barrier height and thickness.

MacLaren \etal (1997) further calculated the tunnelling conductance
taking into account the band-structure of the iron electrodes.
These results also confirmed a considerable dependence of the tunnelling 
magnetoconductance on the barrier height and a small variation with 
barrier thickness. Therefore, it has to be concluded that Julliere's model
fails to incorporate the relevant physics of spin-polarized tunnelling.
As will be seen later, however, this model yields TMR values in
surprisingly good agreement with experiment, if the spin-polarization
is defined in an appropriate way.

The bias-dependence of the tunnelling conductance is generally
quite complicated. Since the carriers tunnel directly between the
electrodes, a quadratic voltage dependence is expected at low
bias according to the Simmons' model for free electron tunnelling
(Simmons 1963). Guinea (1998)
investigated the bias dependence due to the excitation of bulk and
interface magnons.
He found that, for a barrier of thickness $d$, bulk magnons with
wavelengths larger than $d$ can be created with roughly equal
probability. At zero temperature magnons are created at finite voltage
bias yielding a conductance
\begin{equation}
G(V) \propto \left\lbrace \begin{array}{r@{\quad: \quad}l}
\left(V/J \right)^{3/2} & V \ll J\, \left(a^2/d^2\right)
\label{guinea1}\\
\left(a/d \right)^3 & V \gg J\, \left(a^2/d^2\right)\, .
\label{guinea2}
\end{array}\right.
\end{equation}
$a$ denotes the lattice parameter and $J$ the exchange constant.

The contribution to the bias dependence by interface antiferromagnons
was found to be
\begin{equation}
G(V) \propto \left\lbrace \begin{array}{r@{\quad: \quad}l}
\left(V/J_{AF} \right)^2 & V \ll J_{AF}\, \left(a/d\right)
\label{guinea3}\\
\left(a/d \right)^2 & V \gg J_{AF}\, \left(a/d\right)\, .
\label{guinea4}
\end{array}\right.
\end{equation}
$J_{AF}$ denotes the interface exchange constant.
\subsection{Model systems}
In this section some tunnelling systems using the elemental ferromagnets
Fe, Co, Ni and their alloys Ni$_{80}$Fe$_{20}$ and Fe$_{50}$Co$_{50}$
will be discussed in order to give a brief overview of the current
status of magnetic tunnelling junctions before turning to oxide tunnelling 
junctions. Various barrier materials such as Ge, GeO, NiO, Al$_2$O$_3$,
AlN, MgO and HfO$_2$ have been used; Al$_2$O$_3$ is the most popular
one.

Experimental work on ferromagnetic tunnelling junctions was initiated
by Julliere as early as 1975. However, interest in these devices was small, 
since the tunnelling magnetoresistance at room temperature was only a 
fraction of a percent. This situation changed after magnetoresistance
values in excess of 10\% were reported by Moodera \etal in 1995.
Meanwhile it is possible to fabricate ferromagnetic tunnelling 
junctions reproducibly with a magnetoresistance of more than 20\% at room temperature.
Moodera \etal (1998) reported TMR values for a Co/Al$_2$O$_3$/Ni$_{80}$Fe$_{20}$ 
junction of 20.2\%, 27.1\% and 27.3\% at 295~K, 77~K and 4.2~K, respectively.
As already described in section~\ref{materials}, the spin-polarization of the
conduction electrons can be measured using a superconducting counter-electrode 
as a spin detector. This yields $P_{\rm Co} = 35$\%
and $P_{\rm NiFe} = 45$\%. Within Julliere's model, a tunnelling
magnetoresistance of 27.2\% is found, being in perfect agreement with
the measured value. Thus, Julliere's model apparently yields a
good description of the tunnelling magnetoresistance. The spin-polarization
has to be defined via measurements of ferromagnet/insulator/superconductor
tunnelling using the same barrier material.

In order to achieve a large tunnelling magnetoresistance it is necessary
to (i) grow smooth and clean FM/I-interfaces, (ii) have defect free barriers
of large height and (c) have well defined and separated coercive fields
to fully realize the antiferromagnetic state in the field range
$H_{c1} < H < H_{c2}$.

An exciting development in the field of tunnelling junctions is
the recent investigation of spin-polarized tunnelling in double tunnelling
junctions (Schelp \etal 1997, Br\"uckl \etal 1998, Montaigne \etal 1998,
Takahashi and Maekawa 1998, Barn\'as and Fert 1998, Brataas \etal 1999a).
A double tunnelling junction consists of two ferromagnetic electrodes
separated by a small ferromagnetic island. If the electrode dimensions
are very small such that the island capacitance $C$ is also very small, the junction
enters the Coulomb blockade regime for temperatures smaller
than the Coulomb energy $E_c = {\rm e}^2/2C$. Br\"uckl \etal (1998) observed
an anomalous increase of the magnetoresistance in 
permalloy/Al$_2$O$_3$/Co/Al$_2$O$_3$/permalloy tunnelling junctions at
temperatures below 5~K. The lateral dimensions of the junctions were of order
$100\times 200$~nm$^2$. This was interpreted as arising from spin-polarized tunnelling
in the Coulomb-blockade regime. Takahashi and Maekawa (1998) calculated the
tunnelling magnetoresistance of a double tunnelling junction in the Coulomb
blockade regime. At low temperatures charging of the intermediate island
is unfavourable and tunnelling proceeds via a co-tunnelling process, namely
by the simultaneous tunnelling of a charge carrier from one electrode to the
island and from the island to the other electrode. Since this is
a second order process, the magnetoresistance is found to be
\begin{equation}
\frac{\Delta R}{R_{\uparrow\uparrow}} = \left[\frac{1+P^2}{1-P^2}\right]^2-1\, ,
\label{takahashi}
\end{equation}
being larger than for a single tunnelling junction. Barn\'as and Fert (1998)
analyzed double tunnelling junctions using a different approach and
found oscillations in the magnetoresistance as a function of the applied
voltage. These are due to discrete charging effects. A quantitative
experimental verification of these predictions in microfabricated
tunnelling junctions has not yet been carried out. This is
due to the challenging microfabrication requirements to achieve small
structures with small capacitances. However, higher-order tunnelling
processes have been successfully studied in insulating Co--Al--O films
(Mitani \etal 1998a, 1998b). These films show a striking enhancement
of the magnetoresistance in the Coulomb blockade-regime. Surprisingly,
the magnetoresistance in this regime is independent of the bias voltage,
although the resistivity is a strong function of the applied voltage bias.
These results can be understood quantitatively within a model of spin-polarized
tunnelling between large grains via several small grains (Mitani \etal
1998a), see discussion in the following section.
\subsection{Temperature and voltage dependence}
It is generally observed that the tunnelling magnetoresistance decreases
with temperature as well as with bias voltage. A number of studies
were devoted to the voltage dependence and we briefly review this
field of research at the end of this section. First we discuss the
temperature dependence that was studied in great detail by Shang \etal
(1998).
\begin{figure}[t]
\begin{center}
\vspace*{-0.0cm}
\hspace*{1.0cm} \includegraphics[width=0.8\textwidth]{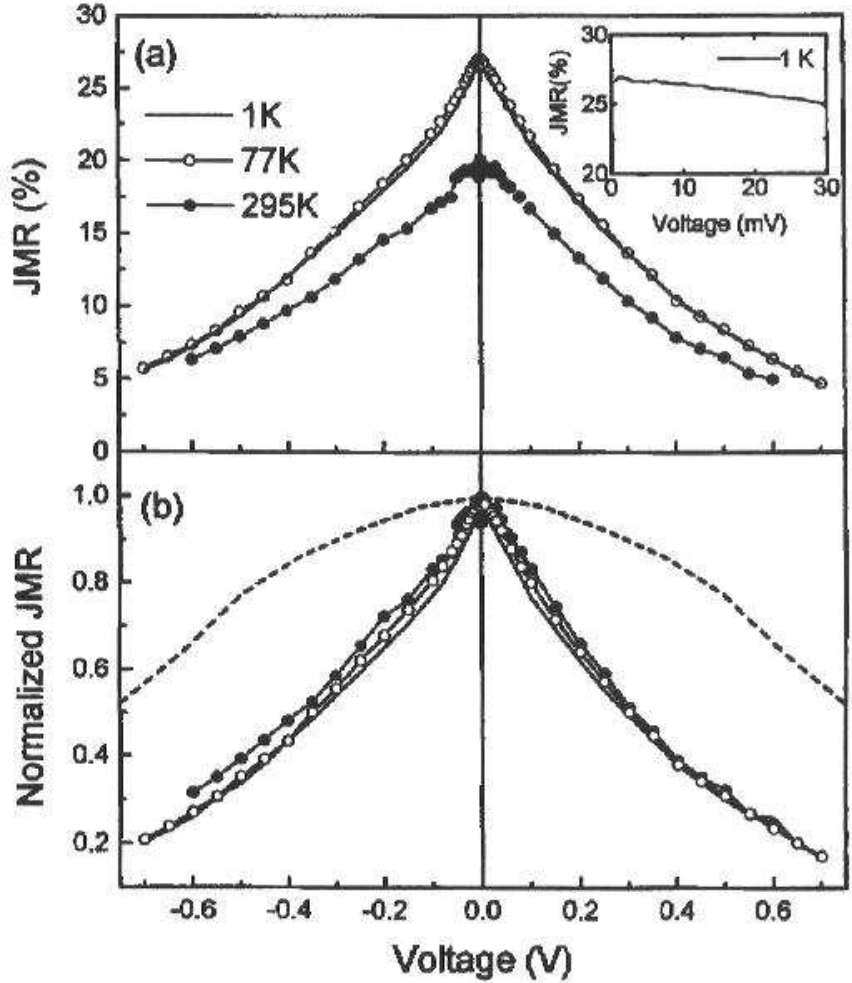}
\vspace*{-5.5cm}
\end{center}
\caption{TMR versus bias voltage at three temperatures for
a Co/Al$_2$O$_3$/Ni$_{80}$Fe$_{20}$ junction. Data shown are
(a) the actual percentages and (b) normalized at zero bias.
The inset shows the TMR in the low bias region displaying near
constancy of TMR. The dashed line in (b) is the theoretically
expected variation for a Fe-Al$_2$O$_3$-Fe junction with a 3~eV 
barrier height (from Fig.~1 of Bratkovsky (1997)). Reproduced from
Moodera \etal (1998).}
\label{moodera1}
\end{figure}
\begin{figure}[t]
\begin{center}
\vspace*{-0.0cm}
\hspace*{0.7cm} \includegraphics[width=0.75\textwidth]{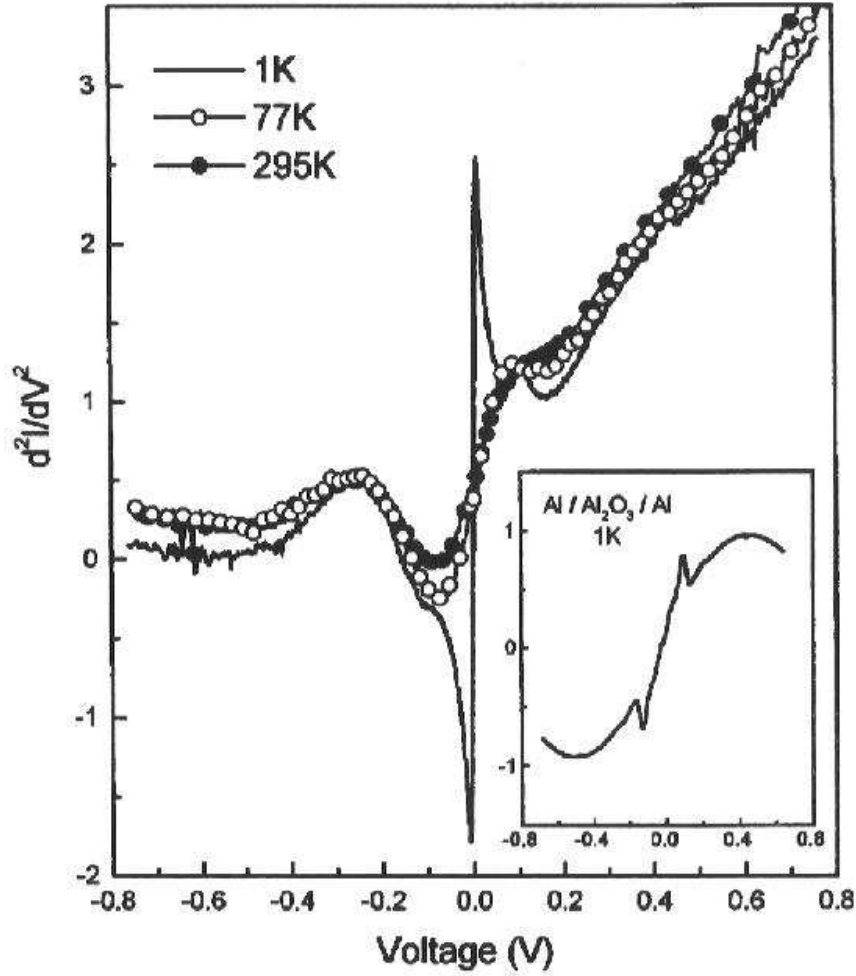}
\vspace*{-4.0cm}
\end{center}
\caption{Inelastic tunnelling spectroscopy spectra at three temperatures
for the same junction as in Fig.~\protect{\ref{moodera1}}, measured at $B = 0$.
Similar spectra are seen for junctions where one electrode is a ferromagnet
and the other electrode is Al. The inset shows an inelastic tunnelling
spectrum of an Al/Al$_2$O$_3$/Al reference junction for comparison.
The dip near 100~mV was interpreted as arising from the excitation of
an Al--O stretching mode, compare with the inset that shows the same
feature in a non-magnetic Al/Al$_2$O$_3$/Al junction, whereas the sharp
features near 17~mV were assigned to magnon excitation. 
Reproduced from Moodera \etal (1998).}
\label{moodera2}
\end{figure}

Shang \etal (1998) based their investigation of the temperature
dependent conductance and magnetoconductance on Julliere's model, since this
reproduces the correct magnitude of the TMR at low temperatures.
The total conductance is written as
\begin{equation}
G(\Theta) = G_T\left[1+P_1P_2\cos(\Theta)\right]+G_{SI}\, ,
\label{shang1}
\end{equation}
where the first term on the right hand side is the usual spin-polarized
tunnelling conductance and the second term $G_{SI}$
is an inelastic contribution independent of the magnetization. This
inelastic contribution might arise from inelastic tunnelling via
localized states in the barrier or from conduction through pin-holes. The 
temperature dependent pre-factor is given by
\begin{equation}
G_T = G_0\frac{CT}{\sin(CT)}\, ,
\label{shang2}
\end{equation}
where $G_0$ is a constant and $C = 1.387\times 10^{-3}d/\sqrt{\Phi}$
with the barrier width $d$ in nm and the barrier height $\Phi$ in eV.
For typical parameters, $G_T$ at room temperature is only a few percent 
higher than at 4.2~K. The spin-polarization is thought to be proportional
to the surface magnetization. The magnetization far below the Curie 
temperature is reduced by the excitation of spin waves producing a term
proportional to $T^{3/2}$. This temperature dependence was confirmed
in bulk samples, thin films and for the surface magnetization. The surface
spin waves, however, showed a largely reduced spin stiffness. Therefore,
for the spin-polarization at low temperatures one can write
\begin{equation}
P(T) = P_0\left(1-\alpha T^{3/2}\right)
\label{shang3}
\end{equation}
with the zero temperature spin-polarization $P_0$ and
a material dependent parameter $\alpha$.

Shang \etal (1998) analyzed the measured tunnelling resistance and TMR
of Co/Al$_2$O$_3$/Ni$_{80}$Fe$_{20}$, Co/Al$_2$O$_3$/Ni$_{80}$Fe$_{20}$/NiO
and Co/Al$_2$O$_3$/Co/NiO tunnelling junctions. They found good
agreement between theory and experimental data in the temperature
range 77~K $\le T \le 400$~K using the parameters
$\alpha_{\rm Co} = 1-6\times 10^{-6}T^{-3/2}$ and 
$\alpha_{\rm NiFe} = 3-5\times 10^{-5}K^{-3/2}$
as well as an inelastic conductance $G_{SI} \propto T^{1.35\pm 0.15}$.
The $\alpha$ values obtained are comparable to values extracted
from magnetization measurements; these show a rough scaling with
the Curie temperature ($T_C = 1360$~K for Co and 850~K for Ni$_{80}$Fe$_{20}$).
The temperature dependence of $G_{SI}$ is in good agreement with
theoretical predictions of inelastic tunnelling through an amorphous barrier
via pairs of localized states that yields a temperature dependence 
$\propto T^{4/3}$ (Glazman and Matveev 1988). 
This indicates the existence of defect states in the barrier.
Although this analysis was performed on the basis of the simple Julliere
model, it indicates that the magnetoresistance in high quality tunnelling 
junctions is reduced by both the excitation of interface spin waves as
well as inelastic processes via defect states in the barrier material.

The tunnelling resistances in the parallel and antiparallel states show
different bias voltage dependences leading to a bias voltage dependence
of the tunnelling magnetoconductance. The TMR is found to be
voltage independent at very low voltages, $V < 10$~mV and is then seen
to decrease strongly on a voltage scale of about 500~mV, see figure~\ref{moodera1}.
The junction resistance as a function of bias voltage usually
shows a cusp at zero voltage and at low temperatures. This cusp was
termed ``zero bias anomaly''.
Slonczewski's model contains a bias voltage dependence through the
parameter $\kappa$ being related to the effective barrier height.
This voltage dependence, however, is much too small to explain
the experimental data. Zhang \etal (1997a) and Bratkovsky (1997) calculated the TMR
at high electron energies and found a quenching of the TMR by
hot electrons. The tunnelling conductance was found to be linear
in bias voltage at low voltages $< 200$~mV in qualitative agreement 
with theory. Moodera \etal (1998), however, observed in high quality junctions
a stronger voltage dependence than predicted by theory, see
figure~\ref{moodera1}, and interpreted this result as arising from
magnon scattering. Indeed, inelastic tunnelling spectroscopy spectra
of a Co/Al$_2$O$_3$/Ni$_{80}$Fe$_{20}$ junction showed features
at about $\pm 100$~mV as well as 17~mV, see figure~\ref{moodera2}. 
Although the feature near 100~mV has been identified as being due to an Al--O
stretching mode, it is believed to also have some magnon contribution; the 
sharp feature emerging at 1~K at 17~mV is also attributed to a magnon excitation.
According to the results of Guinea (1998) magnon scattering should
lead to a voltage dependence $\propto V^{3/2}$, see equation~(\ref{guinea1}),
in qualitative agreement with experiment.

Zhang and White (1998) suggested that extrinsic factors might cause
the considerable voltage dependence of the magnetoresistance. This
idea was based on a correlation between the maximal magnetoresistance
value and the non-linear $I$--$V$-curves of permalloy/Al$_2$O$_3$/Co
(CoFe) junctions showing that a small magnetoresistance ratio is
related with a strong non-linearity. The junctions with unfavourable
characteristics are likely to have numerous defect states in the
barrier leading to inelastic tunnelling processes that are
spin-independent. Zhang and White (1998) proposed a phenomenological
model taking into accoung the inelastic tunnelling current $I_i$ and
derived the magnetoresistance ratio in this case:
\begin{equation}
\frac{\Delta G}{G} = \frac{(\Delta G/G)_{\rm
    Julliere}}{1+I_i/I_{\uparrow\downarrow}}\, ,
\label{white}
\end{equation}
where $I_{\uparrow\downarrow}$ denotes the direct tunnelling current
in the case of antiparallel electrode magnetization. Assuming a
spatially and energetically uniform distribution of the defect states,
the calculated $I$--$V$-curves agree qualitatively with the measured
characteristics of high and low quality junctions, lending support to
this model.
\subsection{Oxide tunnelling junctions \label{oxidetunnelling}}
Ferromagnetic oxide tunnelling junctions based on manganite electrodes
were fabricated and investigated by the IBM group (Sun \etal 1996,
Lu \etal 1996, Li \etal 1997c, Sun \etal 1997, 1998, Sun 1998),
the Orsay group (Viret \etal 1997a), the Cambridge group (Moon-Ho Jo
\etal 2000a, 2000b) and further by Obata \etal (1999), Yin \etal
(2000) and Noh \etal (2001).
Kwon \etal (1998) investigated LSMO ramp-edge junctions. Heteroepitaxial
magnetite tunnelling junctions were investigated by the IBM group
(Li \etal 1998b) and the Eindhoven group (van der Zaag \etal 2000). 
A ``mixed'' junction based on a LSMO and a magnetite
electrode was studied by Ghosh \etal (1998). This research was motivated
by the large spin-polarization of the manganites and of magnetite. Indeed,
TMR values in excess of 100\% at 4.2~K have been reported 
by Sun \etal (1998), Viret \etal (1997a) and Obata \etal (1999). 
In this section the fabrication technique is reviewed,
typical results for resistance, magnetoresistance and non-linear
conductance are discussed and an analysis of limiting factors and future 
trends is given.

LSMO/SrTiO$_3$/LSMO trilayer structures are usually grown by
pulsed laser deposition on LaAlO$_3$ or SrTiO$_3$ substrates. Whereas
Sun \etal (1998) report a peak-to-peak roughness of a single 
$\rm La_{0.7}Sr_{0.3}MnO_3$ layer of 1.5~nm, Obata \etal (1999) achieved
the growth of atomically flat $\rm La_{0.8}Sr_{0.2}MnO_3$ films.
Noh \etal (2001) report on the fabrication of LSMO/SrTiO$_3$/LSMO
trilayer structures by 90$^\circ$ off-axis sputtering and found
improved junction uniformity compared to junctions made by pulsed
laser deposition.
SrTiO$_3$ (STO) layers with thicknesses between 1.6~nm and 5~nm have
been used as tunnelling barriers. 
High resolution transmission electron micrographs of cross-sectional
images obtained from LSMO/STO/LSMO trilayers indicate heteroepitaxial
growth (Lu \etal 1996).
Typically, a self-aligned photolithographic
process is employed to define the junctions. The trilayer structure
is patterned using ion-beam milling to define the base electrode.
After applying photoresist and developing, the top electrode 
is defined by ion-beam milling the top manganite layer; the ion milling
is timed to stop at the bottom electrode. Next the 
junctions are coated with a SiO$_2$ film by sputtering. The photoresist
left after the second ion-milling step is used as the lift-off stencil
to open self-aligned contact holes to the top electrodes.
Finally a Au or Cu metallization layer is deposited and patterned to
make electrical contacts to the bottom and top electrodes.
\begin{figure}[t]
\vspace*{+2.0cm}
\begin{center}
\hspace*{0.3cm} \epsfysize=20cm \epsfbox{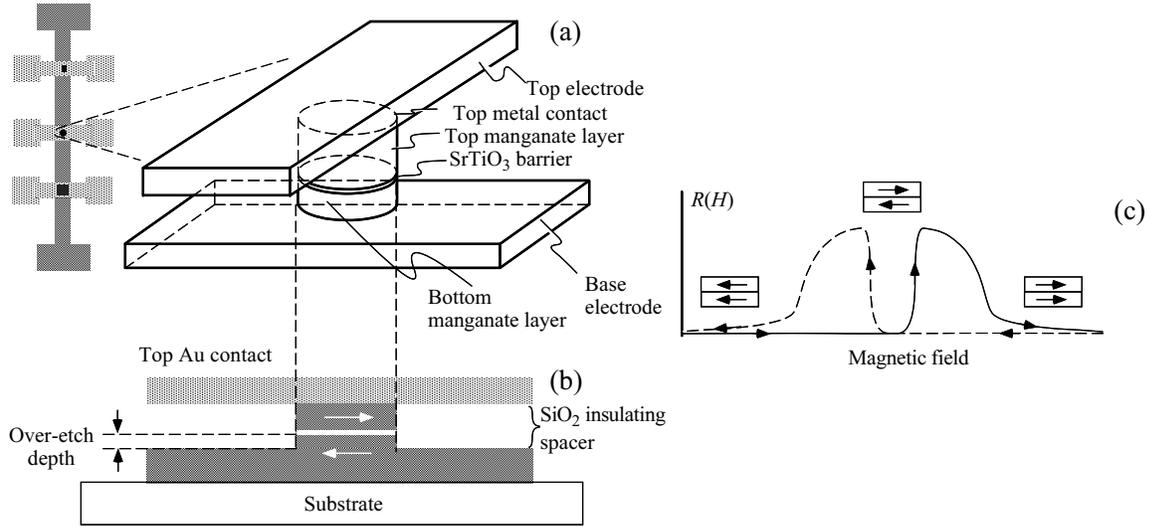}
\end{center}
\vspace*{-14.5cm}
\caption{A schematic view of a LSMO/STO/LSMO trilayer
thin film junction structure. (a) Left: top-view of the device;
right: 3-dimensional illustration of the current-perpendicular
pillar structure. (b): Side view of the structure, showing the over-etch
steps which add additional magnetic coupling between the top
and bottom ferromagnetic electrodes. (c) Schematic junction resistance
as a function of sweeping magnetic field, showing the transitions from
parallel to anti-parallel to parallel state of the magnetic moment
alignments of the electrodes. Reproduced from Sun (1998).}
\label{sun1}
\end{figure}
\begin{figure}[t]
\begin{center}
\hspace{2.5cm} \epsfysize=12cm \epsfbox{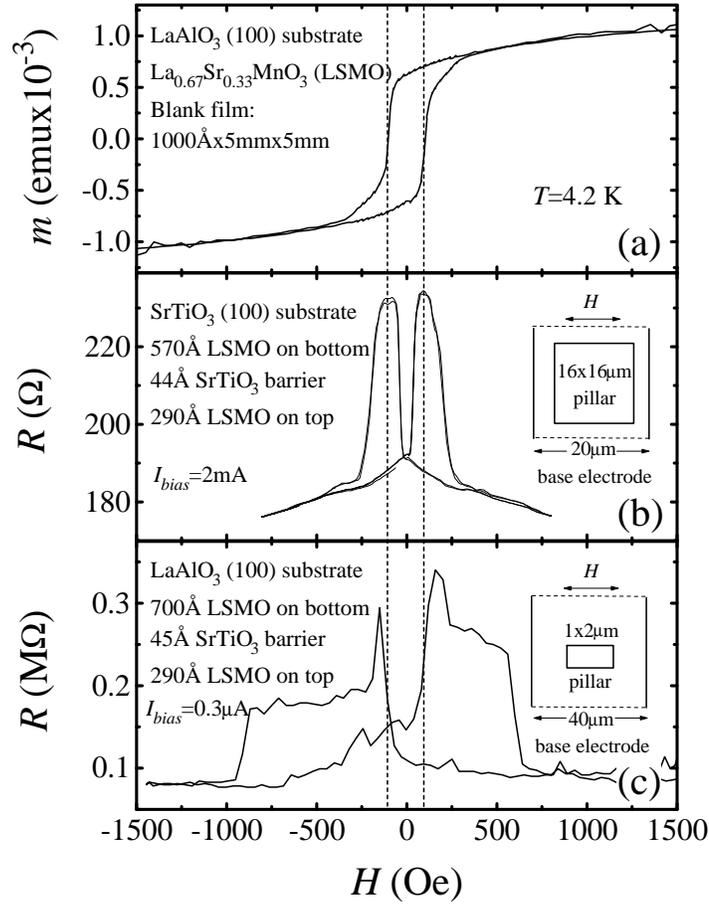}
\end{center}
\caption{Comparing the $R(H)$ curves of devices to 
magnetic hysteresis from a blank film. (a) Magnetic hysteresis
loop of a blank film. (b) $R(H)$ loop of a low resistance junction
showing similar switching field as the blank film's coercive
field $H_c$. (c) High resistance
junction. The lower switching field corresponds well to the blank film's $H_c$, 
whereas the upper switching field is well above $H_c$, indicating an 
additional magnetic interaction present for magnetic states within the 
pillar. The insets in (b) and (c) show the geometry of the electrodes for 
the particular junctions and the relative field orientation in each case. 
Reproduced from Sun (1998).}
\label{sun2}
\end{figure}
\begin{figure}[t]
\begin{center}
\vspace*{-0.3cm}
\hspace*{1.0cm} \includegraphics[width=0.9\textwidth]{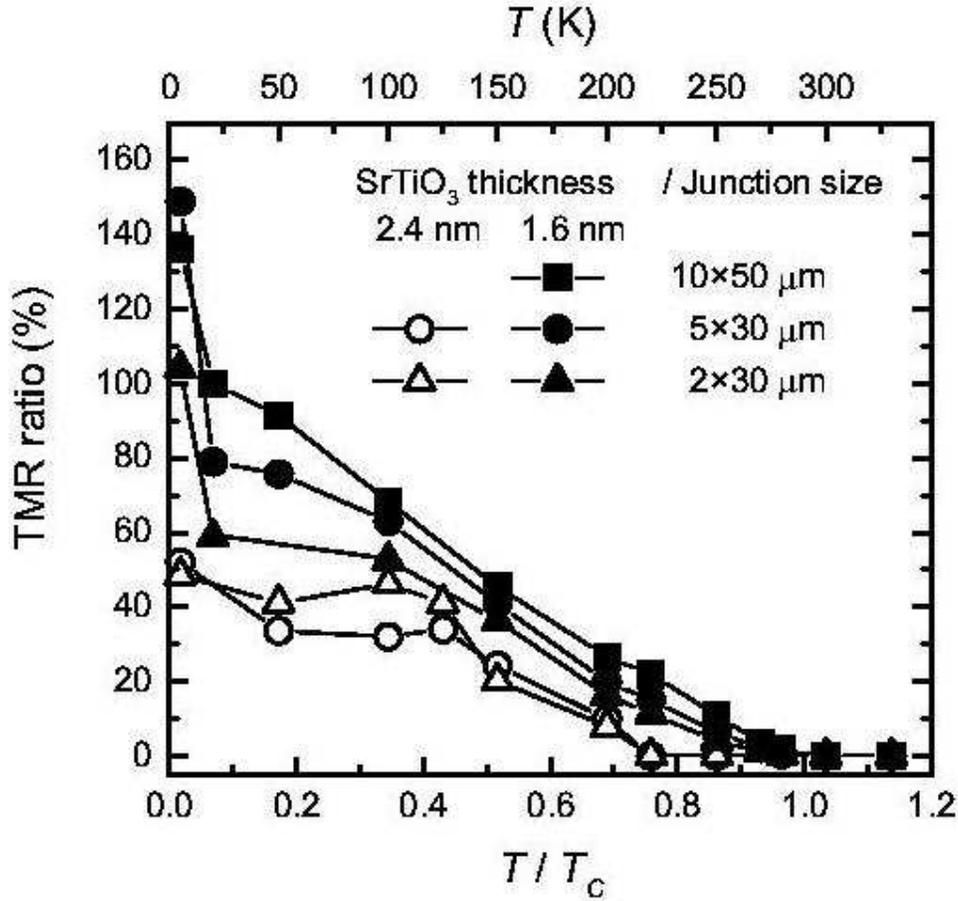}
\vspace*{-8.5cm}
\end{center}
\caption{Temperature dependence of maximum TMR ratios 
for $\rm La_{0.8}Sr_{0.2}MnO_3/SrTiO_3/La_{0.8}Sr_{0.2}MnO_3$
junctions with different areas and SrTiO$_3$ thicknesses. Reproduced
from Obata \etal (1999).}
\label{obata1}
\end{figure}
\begin{figure}[t]
\vspace*{0.2cm}
\begin{center}
\hspace{2cm} \epsfysize=9cm \epsfbox{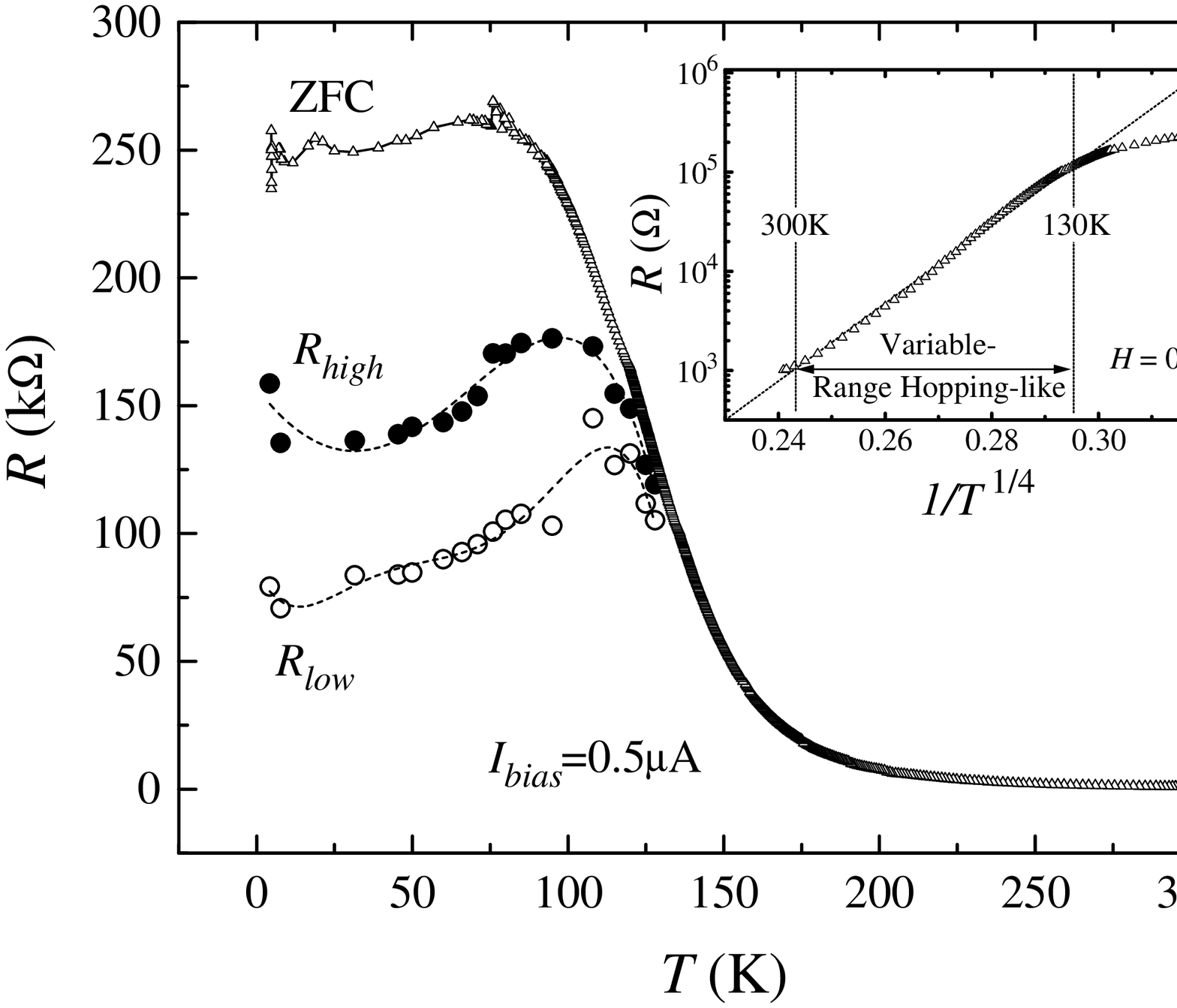}
\end{center}
\caption{Temperature dependence of a LSMO/STO/LSMO junction.
The initial zero-field-cooled trace of this junction gives a higher 
resistance value at low temperatures than observed in subsequent measurement
cycles. Solid and open circles: Resistive--high and --low
states as measured from individual $R(H)$ loops.
Inset: junction resistance plotted versus $T^{-1/4}$ in order to show
the variable range hopping in the high temperature region between 130~K and 
room temperature. Reproduced from Sun (1998).}
\label{sun3}
\end{figure}

A schematic view of a LSMO/STO/LSMO trilayer thin film junction
structure is shown in figure~\ref{sun1}. These structures show
non-linear conductance curves; the non-linear conductance is often found to
be quadratic in the bias voltage at low voltages which is taken as
an indication of tunnelling as the main transport mechanism.
However, more complex bias voltage dependences of the conductance
have been reported that are not understood at present, see Sun (1998).
Figure~\ref{sun2} shows (a) the magnetization hysteresis of a single
LSMO layer indicating the coercive field and (b), (c) the
resistance hysteresis at 4.2~K of two LSMO/STO/LSMO tunnelling
junctions. Both junctions show typical TMR behaviour:
a resistance maximum for antiparallel alignment of the electrode
magnetizations and a nearly field independent resistance for
parallel alignment. The resistance hysteresis of the larger
tunnelling junction in figure~\ref{sun2}b is very similar to
corresponding curves of conventional ferromagnetic tunnelling
junctions. The smaller junction in figure~\ref{sun2}c shows
large noise, presumably due to complicated domain patterns
in the ferromagnetic electrodes. The magnetoresistance of the 
junction in (c) reaches about 200\%, whereas the larger junction in (b)
has a magnetoresistance of only 20\%. Very large values
of 870\% (Sun \etal 1998) and 450\% (Viret \etal 1997a) have been reported
at low temperatures. However, the junction characteristics are
not reproducible.

Several groups have investigated the TMR ratio as a function
of temperature. Lu \etal (1996) reported a strong decrease of the magnetoresistance
with temperature in LSMO/STO/LSMO junctions; the TMR ratio is found to vanish 
above about 200~K, whereas the Curie temperature of the electrodes
is much higher with $T_C = 347$~K. Obata \etal (1999) report TMR
values extending up to the Curie temperature in 
LSMO/STO/LSMO junctions with very smooth interfaces and thin STO barriers
with thickness of 1.6~nm, see 
figure~\ref{obata1}. Yin \etal (2000) obtained a similar result in
LSMO/La$_{0.85}$Sr$_{0.15}$MnO$_3$/LSMO junctions. However, even in
this case the TMR ratio decreases much faster with temperature than
the bulk magnetization. 
This behaviour might not be surprising, since the
spin-polarization is controlled by the interfacial magnetization that
shows a much stronger decay with temperature than the bulk
magnetization. Indeed, as discussed in section~\ref{LCMOmetal},
measurements of the surface spin-polarization by Park \etal (1998a)
indicate that the surface spin-polarization 
is strongly reduced in comparison to the bulk magnetization in 
qualitative agreement with the tunnelling magnetoresistance. The strong
temperature dependence, however, might to some extent still be of extrinsic
origin related to non-stoichiometry of the interface region and
interface roughness. Viret \etal (1997a) observed a maximum in the
tunnelling resistance for parallel magnetization at about 200~K.
In view of the dependence of the Curie temperature on oxygen
content, this might be interpreted as due to an oxygen deficient LSMO layer
near the interface. This is corroborated by the results of Gilabert
\etal (2001) who observed a strong photoconductivity in a 
La$_{0.81}$MnO$_3$/Al$_2$O$_3$/Nb junction below
95~K. This finding was interpreted as arising from photoinduced
charge-carrier generation in an oxygen-depleted region near the
interface, thus effectively reducing the tunnelling barrier thickness.
Sun \etal (1997) found indications of variable range 
hopping in the junction resistance above about 130~K, where the TMR vanishes,
see figure~\ref{sun3}.
This was interpreted as indicating a high defect density in the
SrTiO$_3$ barrier leading to defect-state mediated tunnelling through
the barrier at higher temperatures. Fits of the Simmons model
to the non-linear conductance yield small tunnelling barriers
with values 0.5-0.7~eV (Sun \etal 1997) and 0.1-0.2~eV (Obata \etal 1999).
These are much smaller than the SrTiO$_3$ band gap of about 3~eV
and also indicate the presence of defect states. 
Sun \etal (1998) reported a zero bias dip in the conductance
for temperatures below 130~K and bias voltages below 200~meV.
Since this zero bias dip was also present in LSMO/Al$_2$O$_3$/permalloy
junctions, it was related to the properties of the LSMO bottom electrode.
Sun \etal (1998) suggested this feature to arise from
a Coulomb gap effect due to metallic inclusions of 
approximate size 1.5~nm at the interface.
The nature of the metallic inclusions is unknown; it was speculated
that these are related to the localization length of the
spin-polarized carriers. Moreover, the different
switching fields of the electrodes are only induced by shape anisotropy.
This might not be sufficient to guarantee the full development
of the antiferromagnetic state at intermediate fields and might
lead to a strong reduction of the magnetoresistance.
\begin{figure}[t]
\begin{center}
\vspace*{0.5cm}
\hspace*{2.5cm} \includegraphics[width=0.85\textwidth]{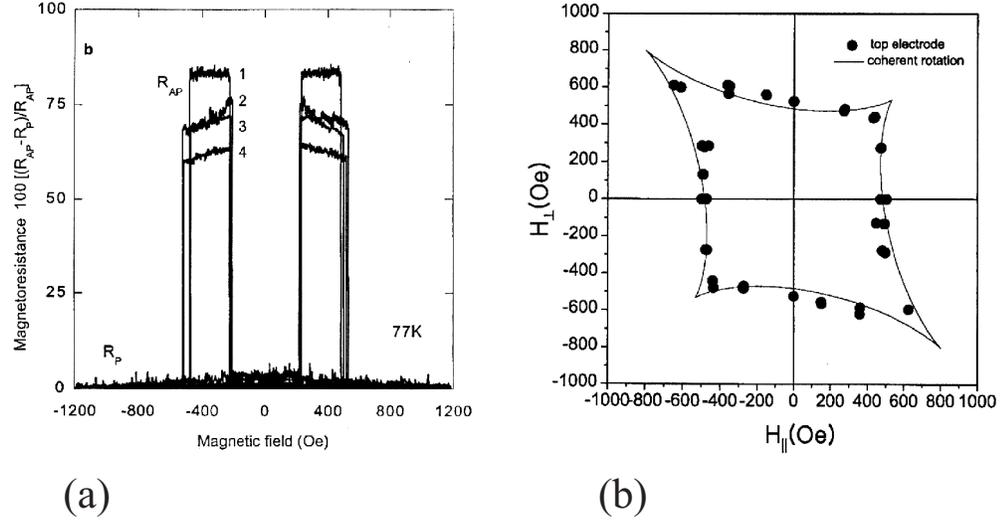}
\vspace*{-13.5cm}
\end{center}
\caption{(a) Resistance of LCMO/NdGaO$_3$/LCMO trilayer junctions at
  77~K. Here the magnetoresistance is defined with respect to
  $R_{\uparrow\downarrow}$. The junctions show coherent sharp
  switching at well defined switching fields and a rather high
  magnetoresistance. This corresponds to a spin-polarization of
  86\%. Adapted from Moon-Ho Jo \etal (2000a). (b) Switching field of
  a junction as shown in (a) plotted in the $(H_\parallel,H_\perp)$
  plane. $H_\parallel$ and $H_\perp$ are the field components defined
  with respect to the magnetic easy axis. The solid lines show the
  result of a calculation within a single domain model with an
  anisotropy consisting of a twofold and a fourfold term; this is
  consistent with the observed angular dependence of the switching
  field. The agreement between calculation and measured values
  strongly suggests a coherent rotation mechanism. Adapted from
  Moon-Ho Jo \etal (2000b).}
\label{moonhojo}
\end{figure}

Moon-Ho Jo \etal (2000a, 2000b) reported the fabrication of trilayer
tunnelling junctions made from LCMO with NdGaO$_3$ as a tunnelling
barrier. Due to the small lattice mismatch between LCMO and NdGaO$_3$
strain effects are minimized and these junctions show excellent
reproducibility and very sharp resistance switching between bistable
resistance states. Magnetoresistance curves are shown in
figure~\ref{moonhojo}(a). The magnetization reversal mechanism in the
bottom and top electrodes was investigated by angular dependent
magnetotransport measurements (Moon-Ho Jo \etal 2000b). The bottom electrode
shows a twofold symmetry consistent with magnetocrystalline
anisotropy; the top electrode shows an additional fourfold
contribution. The analysis of the reversal mechanism within
Stoner-Wohlfarth single domain theory strongly indicates coherent
magnetization reversal in the top electrode. This is illustrated in
figure~\ref{moonhojo}(b) showing the observed switching fields in the
$(H_\parallel,H_\perp)$ plane compared to the calculation for coherent
rotation. The agreement is remarkable. Thus one might assume that
extrinsic effects such as incomplete magnetization reversal do not
influence the magnetotunnelling values obtained for these junctions.
Indeed, Moon-Ho Jo \etal observe a spin-polarization of 86\% at 77~K (2000a)
that is higher than the value measured directly by Andreev reflection
(Soulen \etal 1998). However, the effective spin-polarization in these
junctions does also strongly decrease with temperature and was seen to
vanish above 150~K. As in the data of Sun \etal (1997) the junction
resistivity indicates an activated non-tunnelling conductance. Moon-Ho
Jo \etal (2000a), however, argue that this is insufficient to explain
the drastic decrease in tunnelling magnetoresistance above 120~K. They
propose a model based on percolative phase separation at the
interface: within this model the decrease of the spin-polarization is
caused by the growth of paramagnetic regions nucleated near interface
defects at the expense of ferromagnetic regions. This model is
consistent with the mesoscopic results of F\"ath \etal (1999), see
section~\ref{optical}.

Direct evidence for phase separation near 
$\rm La_{0.67}Ca_{0.33}MnO_3$/SrTiO$_3$ interfaces has been obtained
from a NMR study as a function of thickness, 6~nm $\le t\le$ 108~nm of
the manganite layer (Bibes \etal 2001). $^{55}$Mn NMR spectra show the
occurrence of two lines: a dominant line of mixed-valent character and
a small contribution from Mn$^{4+}$. The NMR enhancement factors of
both lines are large, thus indicating ferromagnetic states. These NMR
signals are related to ferromagnetic metallic and ferromagnetic
insulating regions. Since the Mn$^{4+}$ line becomes more pronounced
at small film thickness, the ferromagnetic insulating region is
located near the interface. The total spectrum intensity normalized to
the layer thickness decreases with decreasing film thickness; this
indicates that some fraction of the sample is non-ferromagnetic and
presumably insulating.

From this discussion it becomes clear that an understanding of manganite
tunnelling junctions is far from complete. Calculations of
the tunnelling conductance using the full manganite band structure
might be of help. Experimentally the interface quality, the barrier
material as well as the oxygenation state have to be improved.
It might prove helpful to investigate junctions of mixed type
with a well defined Al$_2$O$_3$ barrier and a conventional
ferromagnetic electrode of known properties.

\begin{figure}[t]
\begin{center}
\vspace*{-0.0cm}
\hspace*{0.0cm} \includegraphics[width=0.8\textwidth]{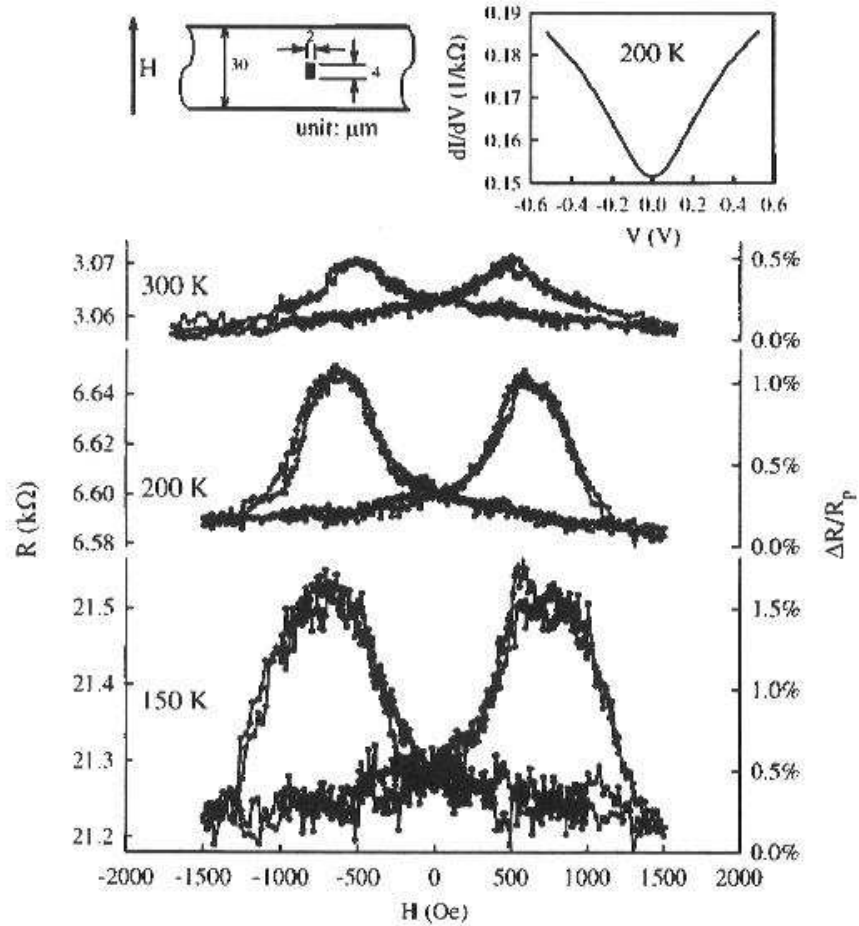}
\vspace*{-5.7cm}
\end{center}
\caption{Resistance and magnetoresistance versus magnetic field
at various temperatures for a $\rm Fe_3O_4/MgO/Fe_3O_4$ tunnelling junction
with a rectangular $2\times 4$~$\mu$m$^2$ top electrode. The insets
show the junction geometry and the dynamic conductance as function of bias
voltage at 200~K. Reproduced from Li \etal (1998b).}
\label{li1}
\end{figure}
Li \etal (1998b) fabricated tunnelling junctions from heteroepitaxially
grown $\rm Fe_3O_4/MgO/Fe_3O_4$ trilayers using a similar process
as for the fabrication of manganite junctions. Since magnetite
is believed to be a half-metallic ferrimagnet with a high Curie temperature,
the junctions have potential to be used at room temperature. Different
coercive fields were achieved by growing on a CoCr$_2$O$_4$ buffer layer
that yields a magnetically softer bottom electrode. At room temperature, 
however, the difference between the two coercitvities becomes quite small.
The tunnelling magnetoresistance is disappointingly small, see figure~\ref{li1}.
At room temperature a magnetoresistance of about 0.5\% increasing
to 1.5\% at 150~K was found. This is in agreement with results by van
der Zaag (2000). Although a strong decay of the surface magnetization
with temperature was found by Alvarado (1979) in magnetite crystals,
this effect should not place a severe limit on the spin-polarization
at 150~K or 300~K due to the high Curie temperature. At present, this small
magnetoresistance in magnetite tunnelling junctions is not understood.

Barry \etal (2001) studied a tunnelling junction farbicated from a
CrO$_2$ bottom electrode, a native oxide layer, probably composed of
antiferromagnetic Cr$_2$O$_3$, and a Co top electrode. The
current-voltage characteristics are non-linear and an analysis within
Simmons' model yields a barrier thickness of about 2~nm and a barrier
height of 0.76~eV. The tunnelling magnetoresistance is 1\% at 77~K.

Ghosh \etal (1998) investigated LSMO/STO/Fe$_3$O$_4$ tunnelling junctions.
A positive magnetoresistance was observed in large applied fields
consistent with the opposite spin-polarization in the manganites 
and magnetite. However, the junction resistance did not show any 
apparent correlation with magnetic properties, especially the coercive
fields. Therefore, the results on these junctions are at present 
not very well understood.
\subsection{Barriers and band-structure}
In the foregoing sections, it has been shown that Julliere's model
provides a consistent description of spin-polarized tunnelling, if the
spin-polarization as determined from
ferromagnet-insulator-superconductor tunnelling is used. The meaning
of these spin-polarization values, however, is somewhat unclear,
since these contradict basic ideas on the band-structure of
ferromagnetic metals. Indeed, the spin-polarization of Ni, Co and
LSMO determined from tunnelling measurements into Al is {\em
positive}, see table~\ref{spinpol}. For LSMO this result is expected,
since in the simplest band picture only the majority e$_g$ band lies
at the Fermi level and the spin-polarization should be $P = +1$. In
the case of the elemental ferromagnets, however, physical intuition
leads one to expect a negative spin-polarization, since the minority
density of states significantly exceeds the majority one. Gadzuk
(1969) studied band-structure effects on the tunnelling electron
distribution for non-magnetic metals. The band-structure was modelled
as consisting of a wide, free-electron-like $s$-band with parabolic
dispersion and a narrow tight-binding $d$-band with a linear
$k$-dependence. Gadzuk (1969) found that the tunnelling probability of
$d$ electrons is drastically reduced compared to that of $s$
electrons, i.e.\ ``tight-binding electrons are not only tightly bound
with respect to conduction processes but also with respect to
tunneling.'' Similar methods were used to treat the spin-polarization
of ferromagnetic metals observed in tunnelling (Hertz and Aoi 1973,
Stearns 1977) and in field emission (Politzer and Cutler 1972,
Chazalviel and Yafet 1977). Main conclusion from this theoretical work
is that electrons from bands of low-effective mass contribute most to
tunnelling. The spin-polarization of these $s$ electrons is induced by
the exchange splitting of the $d$ bands through $s-d$ hybridization
and generally is of the opposite sign as the spin-polarization derived
from band structure. Within a rigid Stoner band-model Stearns (1977)
obtained a relationship between the spin-polarization and the magnetic
moment of the conduction electrons $\mu_c$:
\begin{equation}
P = \mu_c\, f(\Delta/E_F)\, ,
\label{stearns}
\end{equation}
where $f(\Delta/E_F)$ denotes a slowly varying function of the ratio
of band splitting $\Delta$ and Fermi energy $E_F$ with $f(0) =
1/3$. This expression is in agreement with the results of
Paraskevopoulos \etal (1977) and Meservey \etal (1980). A second
solution to explain the experimentally observed spin-polarizations was
proposed by Fulde \etal (1973). These authors studied the electronic
structure of a Ni surface and found a spin-dependent surface resonance
state in the ferromagnetic region. A majority spin resonance
significantly changes the local surface density of states and might
lead to a spin-polarization drastically different from the bulk
value. The classical data of Meservey and Tedrow did not settle the
controversy on the physical mechanisms underlying spin-polarized
tunnelling, see discussion in Meservey \etal (1980). Here recent work
will be reviewed in more detail, since it makes essential
contributions to the understanding of this problem.

When comparing data on ferromagnet-insulator-ferromagnet and
ferromagnet-insulator-superconductor tunnelling, it is immediately
clear that the choice of barrier is quite limited. Actually most of
the experiments involving elemental ferromagnets have been performed
with the use of Al$_2$O$_3$ as a barrier, although some experiments with
AlN (Plaskett \etal 1997), GdO$_x$ (Nowak and Rauluszkiewicz 1992),
NiO (Nowak and Rauluszkiewicz 1992), MgO (Platt \etal 1997) and
HfO$_2$ (Platt \etal 1997) barriers have been reported. Among
these Al$_2$O$_3$ has proven to be the most successful barrier for
spin-polarized tunnelling junctions. This might be attributed to the
excellent wetting properties of Al and its ability to oxidize
readily. The barrier material of choice when working with manganites is
SrTiO$_3$, although experiments with PrBa$_2$Cu$_{2.8}$Ga$_{0.2}$O$_7$
and CeO$_2$ barriers have been reported (Viret \etal 1999). Since a positive
spin-polarization of LSMO in conjunction with a SrTiO$_3$ barrier is
found in tunnelling experiments with an Al counterelectrode (Worledge
and Geballe 2000) and is expected from band-structure, LSMO might be
used as a spin-analyzer.

\begin{figure}[t]
\begin{center}
\vspace*{-0.0cm}
\hspace*{-1.5cm} \includegraphics[width=1.3\textwidth]{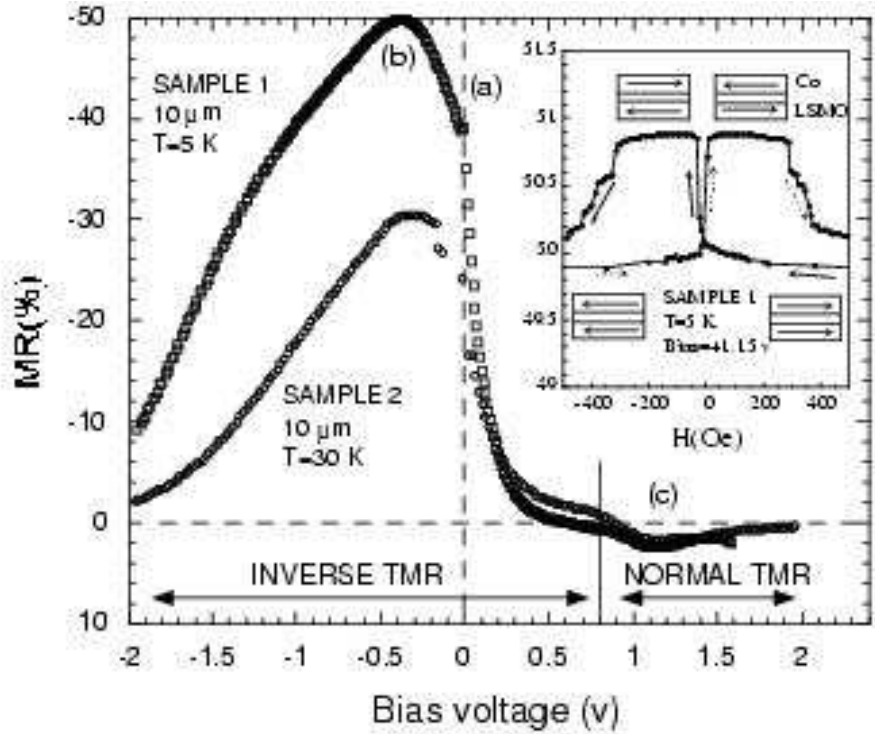}
\vspace*{-20.5cm}
\end{center}
\caption{TMR ratio as a function of the applied dc bias for 10~$\mu$m
  Co/SrTiO$_3$/LSMO junctions. The inverse TMR is maximum at about
  $-0.4$~V, and reaches -50\% and -30\% for samples 1 and 2,
  respectively. At positive bias the TMR decreases rapidly and a
  normal TMR of, respectively, 1.5\% and 1\% is measured at $+1.15$~V
  for samples 1 and 2. The inset shows the normal TMR measured at 5~K
  on sample 1 for a positive bias of $+1.15$~V. The different TMR values
  of the two samples are likely to be related to the LSMO/SrTiO$_3$
  interface quality. Reproduced from de Teresa \etal (1999a).}
\label{fert1}
\end{figure}
\begin{figure}[t]
\begin{center}
\vspace*{0.2cm}
\hspace*{0.0cm} \includegraphics[width=1.0\textwidth]{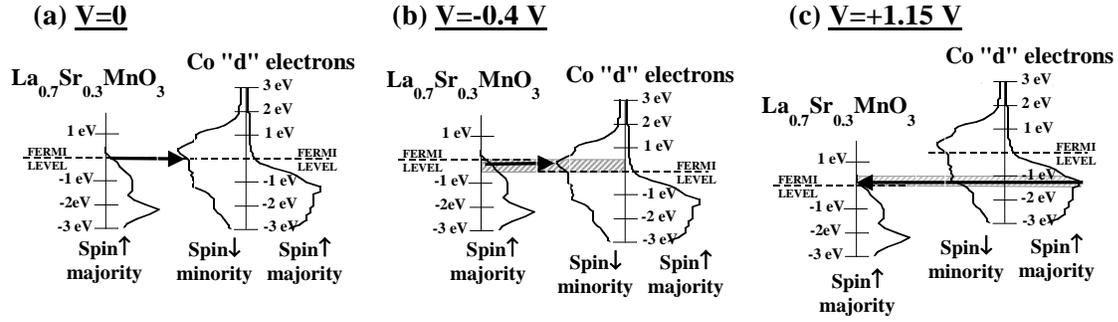}
\vspace*{-7.8cm}
\end{center}
\caption{Relative positions of the $d$-band DOS in Co and LSMO for (a) a
  bias around zero, (b) a negative bias of $-0.4$~V, and (c) a
  positive bias of $+1.15$~V. In each case, arrows indicate the route
  of the higher tunnelling rate which occurs between majority states
  of LSMO and minority states of Co in the antiparallel configuration
  [(a),(b)] or between majority states of LSMO and majority states of
  Co in the parallel configuration (c). Reproduced from de Teresa \etal (1999a).}
\label{fert2}
\end{figure}
A convincing experiment demonstrating the influence of both barrier and
band-structure on spin-polarized tunnelling was performed by de Teresa
\etal (1999a, 1999b). This group fabricated LSMO/SrTiO$_3$/Co
tunnelling junctions using a combined pulsed laser
ablation/sputtering process and conventional UV lithography. The
junctions showed a clear magnetotunnelling effect and, depending on
the bias voltage, an inverse tunnelling magnetoresistance with the
resistance lower in the antiparallel state was observed. Since LSMO
has a positive spin-polarization, this clearly indicates a {\em
negative} spin-polarization of the Co electrode in agreement with the
band-structure. The bias dependence of the tunnelling
magnetoresistance of this LSMO/SrTiO$_3$/Co structure is shown in
figure~\ref{fert1}. For bias voltages between -2~V and 0.8~V an
inverse tunnelling magnetoresistance is seen with a broad maximum near
-0.4~V; above 0.8~V the magnetoresistance becomes normal. This
behaviour can be understood considering the schematic density of
states shown in figure~\ref{fert2}. At zero bias, majority-spin
electrons can tunnel from the LSMO electrode into the minority
$d$-states of the Co electrode, when the magnetizations are
antiparallel; this leads to an inverse tunnelling
magnetoresistance. At a bias voltage of -0.4~V tunnelling into the
minority DOS peak of the Co electrode is possible and the TMR is
maximum. On the other hand, at positive bias voltages the majority
density of states in the Co electrode becomes slightly larger than the
minority DOS and a small, but normal TMR is found. In this way,
spin-polarized tunnelling provides a direct map of the spin-integrated
band-structure of Co.

This experiment shows that Co has a negative spin-polarization when
adjacent to a SrTiO$_3$ barrier, whereas the spin-polarization is
positive when adjacent to a Al$_2$O$_3$ barrier. In a related
experiment, Sharma \etal (1999) showed that permalloy
(Ni$_{80}$Fe$_{20}$) has a positive spin-polarization when adjacent to
Al$_2$O$_3$ but a negative one when adjacent to Ta$_2$O$_5$. In this
experiment tunnelling junctions were grown with two permalloy
electrodes and a mixed Al$_2$O$_3$/Ta$_2$O$_5$ barrier. If a positive
bias was applied with respect to the Al$_2$O$_3$ side of the barrier, a
positive tunnelling magnetoresistance was observed, whereas an inverse
TMR appeared when a negative bias was applied. This clearly shows that
the apparent spin-polarization of permalloy depends on the barrier
material.

An influence of the barrier was already predicted by Slonczewski
within the free electron model, see equation~(\ref{slonc2}). A more
detailed model by Tsymbal and Pettifor (1997) considers the character of the
metal-insulator bonding. Tsymbal and Pettifor (1997) calculate the tunnelling
conductance within the ballistic regime using the Kubo formula and
taking into account a realistic band-structure for the ferromagnetic
electrodes (Co and Fe). In order to investigate the influence of the
metal-insulator bonding, three models were investigated. In the first
case only $ss\sigma$ bonding was taken into account, whereas
$sp\sigma$ and $sd\sigma$ bonding was considered in the second and
third calculations. The $ss\sigma$, $sp\sigma$ and $sd\sigma$ hopping
integrals were chosen to be the same as for the bulk ferromagnet.

In the case of Co, the total density of states near the interface is
asymmetric giving a much larger weight to the minority states, as
expected. The conductance taking only $ss\sigma$ bonding into account,
however, is larger in the majority channel and leads to a positive
spin-polarization of 34\% in excellent agreement with measurements on
Co/Al$_2$O$_3$. If $sp\sigma$ coupling is switched on, the result
remains unchanged. This change of sign in the spin-polarization is due to
the fact that only $s$-electrons from the ferromagnet are coupled into
the insulator such that the conductance is dominated by the $s$-electron
partial DOS. The $s$-electron partial DOS, however, is reduced in the
minority channel as compared to the majority channel as a consequence
of the strong $s$-$d$ hybridization; the conductance is therefore
quite similar to the $s$-electron partial DOS. When $sd\sigma$ bonding
is switched on, $d$ electrons participate strongly in the tunnelling
process and the spin-polarization is found to be negative, $-11$\%, as
expected. In the case of Fe electrodes a similar picture is found with
a spin-polarization of 45\% in the case of $ss\sigma$ bonding and of
8\% in the case of combined $ss\sigma$, $sp\sigma$ and $sd\sigma$
bonding. Again the calculated spin-polarization of 45\% is in good
agreement with measurements on Fe/$\rm Al_2O_3$.

These calculations indicate that there is strong bonding between the
Co $d$-orbitals and the Al $sp$-orbitals at a Co/Al$_2$O$_3$ interface
leading to a positive spin-polarization due to propagating
$s$-electrons. In the case of a Co/SrTiO$_3$ interface, the bonding seems
to be mainly of $d$-$d$ character between Co and Sr/Ti. Although
these ideas give a qualitative explanation of the trends seen in the
experiments, microscopic understanding can only be achieved through
detailed calculations in comparison to investigations of the
respective interfaces. The dependence of spin-polarized tunnelling on
the barrier material presents an additional degree of freedom to the
physicist and requires numerical techniques for a controlled engineering.

Seneor \etal (1999) reported magnetotunneling measurements on
Co/Al$_2$O$_3$/Fe$_3$O$_4$ junctions at 4.2~K. Here a sharp gap in the
magnetoresistance was seen below a bias voltage of 10~meV. This might
be related to a gap in the magnetite band-structure due to correlation
effects in the charge-ordered state. The spin-polarization observed in
this experiment and in the Fe$_3$O$_4$/MgO/Fe$_3$O$_4$ tunnelling
junctions of Li \etal (1998b) was much smaller than the predicted full
spin-polarization. Srinitiwarawong and Gehring (2001) suggested that
this might be a feature of the more than half-filled band
structure. Since the dwell time of an electron on a Fe$^{2+}$ ion is
comparatively long, the wave function relaxes to an atomic state wave
function. An electron tunnelling from such an ion is described by the
angular momentum coupling relation $|2;2\, \rangle =
\sqrt{5/6}\, {|5/2;5/2\downarrow\rangle}-\sqrt{1/6}\, {|5/2;3/2\uparrow\rangle}$;
it follows that the spin-polarization of this slow process is $P =
(\sqrt{5/6})^2-(\sqrt{1/6})^2 = 2/3$. Srinitiwarawong and Gehring
(2001) further calculated the tunnelling probabilities between
Fe$^{2+}$ and Fe$^{3+}$ ions as a function of the magnetic quantum
number.
\subsection{Current-induced switching of the magnetization}
Recent interest has focused on switching of magnetic nanoparticles or
thin magnetic layers by a spin-polarized current. This concept was
originally proposed by Slonczewski (1996). Although the theoretical
description is not settled yet (Bazaliy \etal 1998, Sun 2000, Heide
\etal 2001), convincing evidence of current-driven magnetization
switching has been presented for the switching of thin Co layers in
Co/Cu/Co-pillars (Katine \etal 2000, Albert \etal 2000, Grollier \etal
2001). Here the work of Sun (1999) on the switching of magnetic
nanoparticles in manganite trilayer junctions is reviewed, which was
the first report on the occurrence of current-induced magnetic
switching and is the only report of this phenomenon in oxide
magnets. The interpretation of the data follows the model of 
Slonczewski (1996).

\begin{figure}[t]
\begin{center}
\vspace*{-0.5cm}
\hspace*{0.0cm} \includegraphics[width=0.75\textwidth]{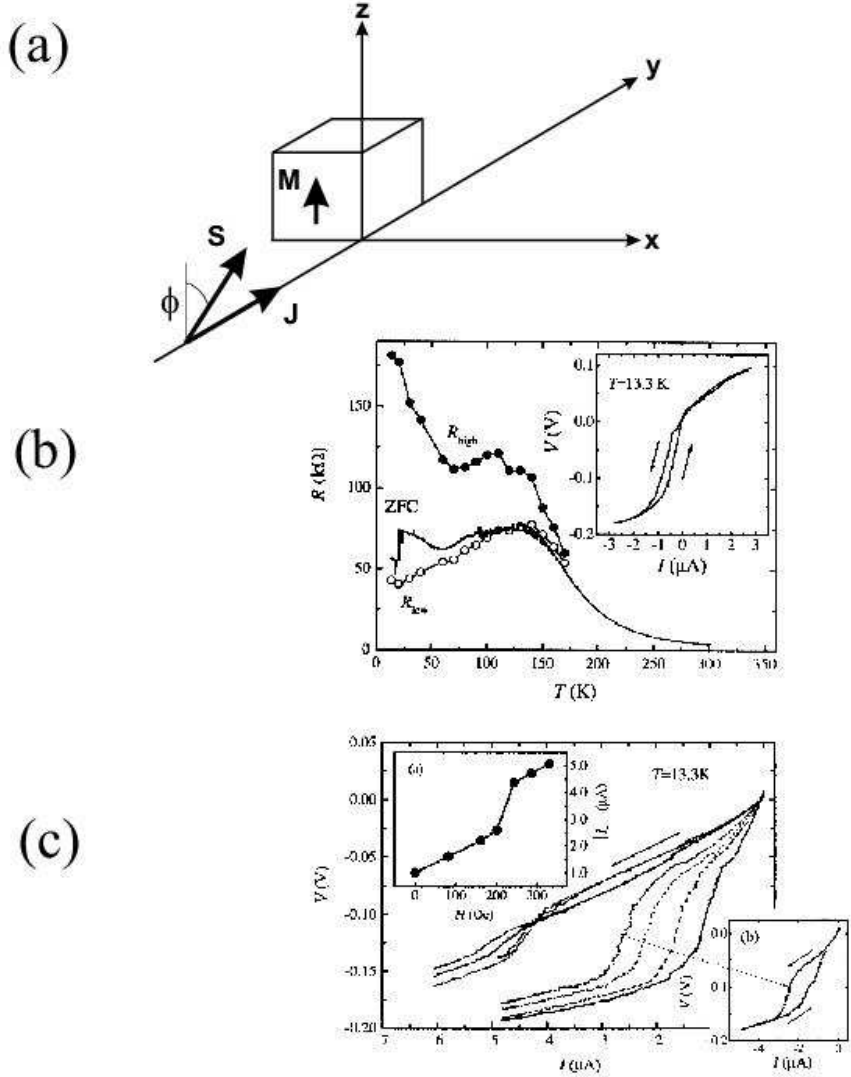}
\vspace*{-2.5cm}
\end{center}
\caption{(a) Schematical setup of a current-driven switching
  experiment. An electric current $\vec{J}$ is injected into a
  ferromagnetic nanoparticle, here modelled as a cubic
  Stoner-Wohlfarth particle with magnetization $\vec{M}$. The injected
  spin-current density $\vec{S}$ is oriented under an angle $\phi$
  with respect to the z-axis. (b) Temperature dependence of a
  manganite trilayer junction in the zero field cooled (ZFC),
  resistance low (parallel electrode magnetization) and resistance
  high state. The inset shows a current-voltage characteristic
  measured at 13.3~K in zero field. A hysteretic jump near $I_c \sim
  -1$~$\mu$A is clearly visible. (c) Current-voltage characteristics
  taken on the same sample at 13.3 K in various applied fields. As
  shown in the right inset these characteristics are hysteretic; for
  clarity only curves taken for one sweep direction are shown in the
  main panel. The left inset shows the critical current as a function
  of the applied field. At about 200~K one of the electrodes switches
  and induces a step in the $I_c(H)$ curve. (b) and (c) after Sun (1999).}
\label{cswitch}
\end{figure}
Figure~\ref{cswitch}(a) shows the principal design of such a switching
experiment. A ferromagnetic nanoparticle, here approximated by a cubic
Stoner-Wohlfarth particle of side length $a$, has a magnetization
$\vec{M}$, $|\vec{M}| = M_S$, making an angle $\theta$ with the
z-axis. A spin-polarized 
electric current $\vec{J}$ flows through the nanoparticle; the
spin-current density $\vec{S}$ makes an angle $\phi$ with the
z-axis. For a spin-polarization $P$ the spin-current density is given
by $S = (\hbar/2{\rm e})PJ$. Within a spin-diffusion length
$\lambda_s$ the direction of the spin-current relaxes towards
$\vec{M}$. The angular momentum is transferred to the particle's
magnetization which induces a torque density on the magnetization of
magnitude (Sun 1999) 
\begin{equation}
\vec{\tau_s} = \vec{M}\times(\vec{S}\times\vec{M})/\lambda_sM_S^2\, .
\label{torque1}
\end{equation}
If for simplicity one assumes uniaxial anisotropy with an anisotropy
constant $K_u$ and an easy axis along $\hat{z}$ and if a magnetic
field $\vec{H}$ is applied along $\hat{z}$, one obtains a second
torque density
\begin{equation}
\vec{\tau_a} = \mu_0(H_A\cos(\theta)+H)\vec{M}\times\hat{z}\, ,
\label{torque2}
\end{equation}
with the anisotropy field $H_A = 2K_u/M$. The dynamic evolution of the 
magnetization is governed by the Landau-Lifshitz-Gilbert equation:
\begin{equation}
\frac{d\vec{M}}{dt} =
\gamma(\vec{\tau}_1+\vec{\tau}_2)-(\alpha/M_S)\vec{M}\times\frac{d\vec{M}}{dt}
\label{LLG}
\end{equation}
with gyromagnetic ratio $\gamma$ and damping parameter $\alpha$.

If the spin-current density exceeds a critical value, the
magnetization might be flipped. The critical current is given by (Sun
1999)
\begin{equation}
I_c = \frac{2{\rm e}\alpha}{P\hbar}\,
\frac{a^2\lambda_sMH_A}{|\cos(\phi)|}\left[1+\frac{H}{H_A}\right]\, .
\label{critical}
\end{equation}
Assuming the nanoparticle to be superparamagnetic, the anisotropy
energy can be related to the blocking temperature $T_B$ via
$a^2\lambda_sMH_A \simeq 40T_B$. Thus the theory yields two
predictions: (1) the magnitude of the critical current is determined
by the blocking temperature and (2) the critical current is
proportional to the applied field.

The experiment uses manganite trilayer junctions as already discussed
in section~\ref{oxidetunnelling}. Since these junctions were grown by
laser ablation, there is a certain number of particulates distributed
between the electrodes. Typical areal densities are $10^6$~cm$^{-2}$
for particles larger than 100~nm and $10^8-10^9$~cm$^{-2}$ for small
particles in the range 10--50~nm. These particles form grain boundaries
of low conductivity and weak magnetic coupling, see next section, with the
electrodes. Accordingly, bottom and top electrode will be connected by
some particles; the local current density at these shunts is likely to
be high.

Detailed magnetotransport measurements of trilayer junctions revealed
three distinctive features shown in figures~\ref{cswitch}(b) and (c):
(1) the resistance after zero field cooling as well as in the high
resistance state shows a minimum around 60~K. This is identified with
the blocking temperature $T_B$ of the particulates in the film. (2)
Current-voltage characteristics are asymmetric and show a hysteretic
step-like feature. This is related to switching of a magnetic
nanoparticle in the barrier. The critical current is defined at the
location of the step. (3) If a magnetic field is applied parallel to
the trilayer junction, this critical current increases linearly with
field below $H < 200$~Oe. Above this field one of the electrodes
switches influencing the magnetic coupling between the nanoparticle
and the electrodes and yielding a step in the $I_c(H)$ curve.
From the blocking temperature a typical particle dimension of 17~nm
was estimated which is reasonable. The anisotropy field $H_A \simeq
120$~Oe was determined from the field dependence of the critical
current. With $P = 1$, the damping coefficient can be estimated to
$\alpha \simeq 0.01$ in agreement with measurements (Lofland \etal 1995).
\subsection{Intrinsic tunnelling in layered manganites}
\begin{figure}[t]
\vspace*{0.3cm}
\begin{center}
\hspace*{1.5cm} \epsfysize=14cm \epsfbox{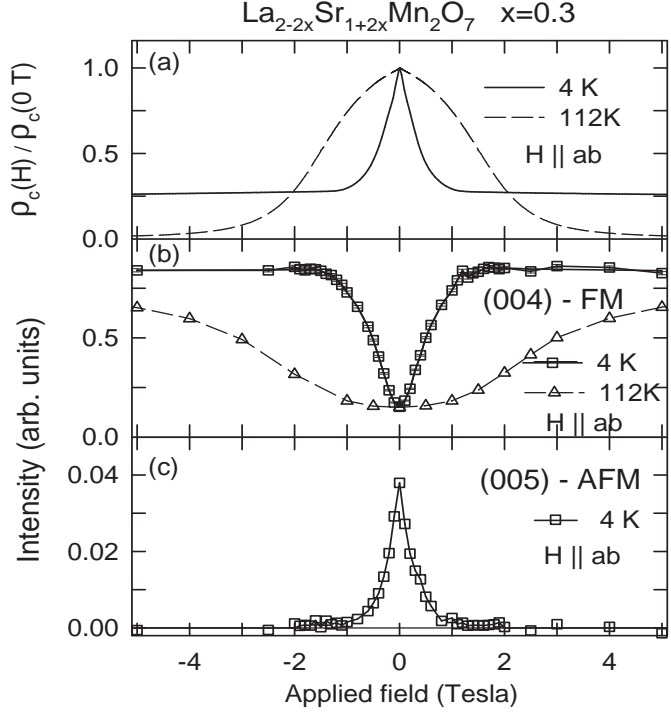}
\end{center}
\vspace*{-4.8cm}
\caption{(a) Field-dependent switching of the resistivity, (b)
ferromagnetic, and (c) antiferromagnetic Bragg intensities in the
tunnelling ($T = 4$~K) and colossal ($T = 112$~K) magnetoresistance
regimes. The simultaneous disappearance of the antiferromagnetic
superstructure peaks (see (c)) and the saturation of the magnetoresistance
(see (a)) was interpreted as arising from intrinsic spin-polarized
tunnelling between ferromagnetically ordered sheets. Reproduced from
Perring \etal (1998).}
\label{perring98}
\end{figure}
The resistivity and magnetoresistance of $\rm La_{2-2x}Sr_{1+2x}Mn_2O_7$
single crystals for the dopings $x = 0.3$ and $x = 0.4$ were found to
differ significantly (Moritomo \etal 1996, Kimura \etal 1996). Whereas
crystals for both dopings showed a large resistivity anisotropy, the
magnetoresistance in the ordered phase was found to be isotropic
for $x = 0.4$, but strongly anisotropic for $x = 0.3$. The ordering
temperatures are 90~K ($x = 0.3$) and 121~K ($x = 0.3$), respectively.
The resistivity along the c-axis displays insulating behaviour above
the magnetic ordering temperature. Perring \etal (1998) performed
a neutron scattering study on single crystals with dopings $x = 0.3, 0.4$
in order to determine the magnetic structure. Whereas the $x = 0.4$
compound appears to be ferromagnetically ordered below 121~K, the
$x = 0.3$ compound was found to order ferromagnetically in the ab-planes
with these ferromagnetic sheets stacked antiferromagnetically along
the c-axis. A magnetic field of 1.5~T applied parallel to the
ab-planes was found to switch the magnetic order from antiferromagnetic
to ferromagnetic. Simultaneous measurements of the antiferromagnetic
superstructure peaks and the c-axis magnetoresistance revealed
a similar field dependence; the magnetoresistance saturates, when the
antiferromagnetic order is destroyed, see figure~\ref{perring98}. 
Perring \etal (1998) argued that
the semiconducting behaviour of the c-axis resistivity above the magnetic
ordering temperature indicates the insulating nature of the
$\rm (La, Sr)_2O_2$ spacing layers. The MnO layers are metallic ferromagnets
due to the double-exchange interaction. Accordingly, the layered
manganites can be viewed as an intrinsic stack of tunnelling junctions
such as $\rm Bi_2Sr_2CaCu_2O_8$ single crystals are stacks of
intrinsic Joesphson junctions. At 4~K a magnetoresistance $\Delta R/R = 300$\%
is measured, see figure~\ref{perring98}. Neglecting shunting currents,
the spin-polarization can be calculated from the magnetoresistance
ratio within Julliere's model, equation~(\ref{tunresist}), yielding
$P = 0.77$.
\section{Grain-boundary junctions \label{JMR}}
As already discussed in section~\ref{optical},
the double exchange mechanism is extraordinarily sensitive to
distortions of the Mn--O--Mn bond. This leads to the formation of
insulating regions in the strain fields of extended defects such as
grain boundaries. Therefore, polycrystalline manganite samples do
actually behave as granular metals, although the variation in
stoichiometry across grain boundaries is likely to be negligible. An
unexpectedly large low field magnetoresistance was discovered in
polycrystalline manganite bulk and thin film samples (Hwang \etal 1996,
Gupta \etal 1996). This finding initiated numerous studies on
extrinsic magnetoresistance effects in magnetic oxides; investigations
were extended to other systems such as CrO$_2$, $\rm Tl_2Mn_2O_7$, the
double perovskite $\rm Sr_2MoFeO_6$ as well as SrRuO$_3$. The
characteristics of the extrinsic magnetoresistance in these materials
will be compared at the end of this section. The bulk of the
experimental data, however, is related to the manganites.
\subsection{Basic theory: granular metals \label{theorypoly}}
With few exceptions, models for grain-boundary magnetoresistance are
based on spin-polarized tunnelling.\footnote{A notable exception is
  the model of Evetts \etal (1998) that relates the grain-boundary
  magnetoresistance to a magnetization enhancement in the
  grain-boundary region induced by the alignment of the adjacent,
  magnetically soft grains.}
This implies that the grain
boundary strongly hinders charge transport leading to the formation of
an insulating region; simultaneously the adjacent grains are
only weakly magnetically coupled across this interfacial region. Thus,
polycrystalline manganite samples are similar to granular
ferromagnetic metals; at this point a brief review of the properties
of granular metals seems appropriate.

A typical granular metal consists of small grains surrounded by
insulating material. Usually the values of both grain diameter $d$ as
well as grain separation $s$ follow a broad distribution. Carrier
transport involves charging of grains that costs a charging energy
$E_{\rm c} = {\rm e}^2/(4\pi\epsilon_0d)F(s/d)$, where $\epsilon_0$
denotes the vacuum permittivity and the function $F(s/d)$ depends on
grain shape. The conductivity is proportional to the tunnelling matrix
element $\exp[-2\chi s]$ with $\chi = (2m\Phi/\hbar^2)^{1/2}$ (Simmons
1963) and the population of a single grain $\exp[-E_{\rm c}/2{\rm
  k_B}T]$; these terms determine the essential temperature dependence
of the conductivity.
\begin{equation}
\sigma \propto \exp\left[-2\chi s-E_{\rm c}/(2{\rm k_B}T)\right]
\label{granular1}
\end{equation}
$\Phi$ denotes the barrier height and $m$ the effective mass.
In order to obtain the conductivity, this expression has to be averaged
over the distributions for grain diameter and separation.

Granular metals are often fabricated by a co-deposition process of a
metal and a dielectric, e.g.\ the co-sputtering of Ni, Co, Fe and
SiO$_2$ (Gittleman \etal 1972, Barzilai \etal 1981, Honda \etal
1997). In this case the metal grains are formed by diffusion
of metal atoms or clusters and, for a given composition, there exists
a definite relationship between grain diameter $d$ and grain
separation $s$ such that $s/d$ is constant (Sheng \etal
1973). Therefore, $sE_{\rm c}$ is a constant: $s\chi E_{\rm c} = C$.
Under this constraint the averaging of the conductivity can be easily
carried out. Tunnelling occurs mainly between grains with an optimum
charging energy $E_{\rm c}^0 = 2(C{\rm k_B}T)^{1/2}$ such that the
conductivity in equation~(\ref{granular1}) is maximized (Sheng \etal
1973). This results in a typical temperature dependence of the
conductivity given by
\begin{equation}
\sigma \propto \exp\left[-2(C/{\rm k_B}T)^{1/2}\right]\, .
\label{granular2}
\end{equation}

Whereas this model for the temperature dependence of the conductivity
is broadly accepted, the theoretical situation for the
magnetotransport properties of granular metals is less clear. In a
pioneering work Helman and Abeles (1976) proposed a model for the
magnetoconductivity. They assumed the existence of a magnetic energy
term $E_{\rm M}$ such that the hopping probability of a charge carrier
from a grain is reduced (enhanced) if its spin is parallel
(antiparallel) to the grain magnetization:
\begin{eqnarray}
\sigma & \propto & \exp(-2\chi s)\left\lbrace\frac{1}{2}(1+P)
\exp\left[-(E_{\rm c}+E_{\rm M})/(2{\rm k_B}T)\right]\right.\nonumber\\
& & \left. +\frac{1}{2}(1-P)
\exp\left[-(E_{\rm c}-E_{\rm M})/(2{\rm k_B}T)\right]\right\rbrace
\label{granular3}
\end{eqnarray}
The magnetic energy term is expressed in terms of the spin correlation
function of neighbouring grains: $E_{\rm M} = J[1-\langle
\vec{S}_1\cdot\vec{S}_2\rangle/S^2]/2$; $J$ denotes the exchange
coupling constant. In a linear approximation the magnetoresistance is
given by
\begin{equation}
\Delta\rho/\rho_0 = -\, \frac{JP}{4{\rm
    k_B}T}\frac{\left[M^2(H)-M^2(0)\right]}{M_{\rm S}^2}
\label{granular4}
\end{equation}
with the magnetization $M(H)$ and the saturation magnetization 
$M_{\rm S}$. This expression yields semiquantitative agreement with
experimental data on Co--SiO$_2$ and Ni--SiO$_2$ granular metals at
temperatures above 50~K; especially the ferromagnetic to
superparamagnetic transition can be successfully modelled (Helman and
Abeles 1976, Barzilai \etal 1981). It is surprising, however, that the
spin polarization $P$ appears linearly in the expression for the
magnetoresistance. Thus, the magnetoresistance should be positive for
Fe and negative for Ni grains which is not seen experimentally. 
There seems to be some inconsistency in the work of Helman and Abeles
(1976) in that it relates the spin dependence of the hopping
probability to one grain, whereas the magnetic energy is an intergrain
property. This issue is somehow resolved by Wagner \etal (1998) within
a different context, namely hopping between spin polarons in single
crystalline material. If the existence of a magnetic energy term
$E_{\rm M} = (J/2)[1-\vec{S_1}\cdot\vec{S_2}/S^2]$ is accepted, the
conductivity is given by 
$\sigma \propto \exp[-2\chi s-(E_{\rm c}+E_{\rm M})/({\rm k_B}T)]$ and
the magnetoresistance by 
$\Delta \rho/\rho_0 = -[1-\exp[-J(M/M_{\rm S})^2/2{\rm k_B}T]]$. 
It is reasonable that two mechanisms contribute to the conductivity,
namely tunnelling through the barrier and thermal hopping over the
barrier; these lead to two contributions to the magnetoresistance, a
temperature independent one due to tunnelling and a temperature
dependent one due to hopping. A systematic investigation of these
effects, however, has not yet been carried out.

Data on the magnetoresistance of granular metals, see e.g.\ Mitani
\etal (1997), Honda \etal (1997), show that the magnetoresistance in
granular metals is often independent of or weakly dependent on
temperature. This lead Inoue and Maekawa (1996) to contest
equation~(\ref{granular4}) and propose another mechanism for the
magnetoresistance in granular materials. Since the tunnelling
probability depends on the relative directions of the grain
magnetizations, see equation~(\ref{slonc1}), the field dependence of the
conductivity arises mainly from the pre-exponential factor, $\sigma
\propto (1+P^2\cos(\Theta))\exp(-2\chi s)$. Averaging over $\Theta$
and the grain separation $s$ yields a temperature dependence of the
conductivity as in equation~(\ref{granular2}) and a magnetoconductivity
\begin{equation}
\left\langle\frac{\Delta G}{G}\right\rangle = 
\left\langle1-\frac{G(\Theta)}{G_{\uparrow\uparrow}}\right\rangle
= \frac{m^2P^2}{1+m^2P^2}
\label{Gpoly}
\end{equation}
with $m = M(H)/M_{\rm S}$. $\langle...\rangle$ denotes the angular
average. At saturation this is just half of the magnetoconductivity of
a single ferromagnetic tunnelling junction. Coey \etal (1998b)
suggested to replace the factor $(1-P^2)/(1+P^2)$ by $2S/(2S+1)^2$
where $S$ denotes the core spin.

The model of Inoue and Maekawa was refined by Mitani \etal (1998a) by
taking into account co-tunnelling processes between large and small
grains. These authors consider the simultaneous tunnelling of charge
carriers from a large grain with charging energy $E_{\rm c}/n$ via $n$
small grains with charging energy $E_{\rm c}$ to another large
grain. In this case the conductivity is given as a sum over all these
higher order processes:
\begin{equation}
\sigma \propto \sum_{n=1}^\infty \, \exp\left[-\, \frac{E_{\rm
      c}}{2n{\rm k_B}T}\right]\, \left[(1+P^2m^2)\exp\left[-2\chi
      s\right]\right]^nf(n)
\label{granular5}
\end{equation}
The function $f(n)$ includes the influence of a distribution of
conduction paths as well as temperature. The exponential in the sum is
strongly peaked as a function of the order number $n$. Following an
analysis similar to that above, the conductivity is found to be of a
form
\begin{equation}
\sigma \propto (1+P^2m^2)^{n^*+1}\,
\exp\left[-2\sqrt{\frac{2\chi sE_{\rm c}}{{\rm k_B}T}}\right]
\label{granular6}
\end{equation}
with $n^* = (E_{\rm c}/8\chi s{\rm k_B}T)^{1/2}$. The
magnetoresistance is found to be
\begin{equation}
\frac{\Delta\rho}{\rho_0} = (1+P^2m^2)^{-(n^*+1)}-1 \simeq
-P^2m^2\left[1+\sqrt{\frac{E_{\rm c}}{8\chi s{\rm k_B}T}}\right]\, ,
\label{granular7}
\end{equation}
where the last approximation is valid for $P^2 \ll 1$. In conclusion,
higher order tunnelling processes become important at low temperatures
and lead to a gradual increase of the magnetoresistance. This is in
agreement with experiments on granular CoAlO films (Mitani \etal 1998a).

Equation~(\ref{Gpoly}) was derived under the assumption that the
grains within a specific percolation path are connected in
parallel. In reality a granular material is a complicated conduction
network. In the case of a narrow distribution of grain separations 
as it might occur in polycrystalline films with tunnelling barriers
defined by grain boundaries, a multitude of conduction paths
contribute to the global resistance. This can be approximated by a
resistor network. As an illustration the magnetoresistance of a
one-dimensional array of resistors connected in series is calculated
here. Assuming a random distribution of grain magnetization at zero
field and a resistance $R = 1/(G_0(1+P^2\cos(\Theta)))$ the following
magnetoresistance is obtained: 
\begin{equation}
\left\langle\frac{\Delta R}{R_0}\right\rangle
= \left\langle\frac{R(\Theta)-R_{\uparrow\uparrow}}{R(\Theta)}\right\rangle 
= 1-\frac{P^2}{(1+P^2){\rm atanh}(P^2)}\, .
\label{Rpoly}
\end{equation}
For spin-polarizations $P > 0.5$ this result differs significantly
from equation~(\ref{Gpoly}) and demonstrates the influence of the
conduction network. Therefore, care has to be taken, whenever spin
polarization values are derived from measurements on granular samples.

The dynamic conductance has already been discussed in
section~\ref{theorytunnel}. The mechanisms listed there, most
importantly bulk and interface magnon scattering, also
contribute to the tunnelling
processes in grain-boundary junctions. However, it has been experimentally
found that often inelastic tunnelling processes are observed in grain-boundary
junctions. Inelastic tunnelling through a barrier with a constant
density of states was investigated by Glazman and Matveev (1988).
They calculated the contributions to the conductance from tunnelling 
via $n$ impurities; these are generally bias dependent and proportional
to $V^{n-2/(n+1)}$. 
\subsection{Experimental data on grain-boundary junctions}
Various grain-boundary systems such as polycrystalline samples,
pressed powders, bi-crystal junctions, step-edge junctions and
laser-patterened junctions have been investigated. The main
experimental data are summarized in this section.
\subsubsection{Are polycrystalline manganite samples classical granular metals?}
Typical polycrystalline manganite samples have grain sizes in the
range 100~nm...10~$\mu$m being too large for charging effects to
dominate. Therefore, the temperature dependence of the resistivity
does not follow equation~(\ref{granular2}), but is rather determined by the
specific transport mechanisms in the barrier. Moreover, the grains are
coupled ferromagnetically.

A systematic study of the magnetotransport and magnetic properties of
$\rm La_{2/3}Sr_{1/3}MnO_3$ ceramics was reported by Balcells \etal
(2000). At small grain sizes below about 40~nm charging effects
become significant. The charging energy was found to be reciprocal to
the grain diameter as expected. This regime can be understood using
the theory of granular metals as outlined above. The intergranular
magnetoresistance to be discussed in the following sections involves
effects beyond this classical model; within the discussion a theoretical
description will evolve.
\subsubsection{Polycrystalline materials.}
The effects of grain boundaries on the resistivity and magnetoresistance
of polycrystalline manganite compounds were reported very early
(Volger 1953, van den Brom and Volger 1967). The recent research was initiated
by the work of Hwang \etal (1996) and Gupta \etal (1996). These authors
compared the magnetoresistance and magnetization of 
$\rm La_{0.67}Sr_{0.33}MnO_3$ single crystals
and polycrystalline ceramics (Hwang \etal 1996) and 
$\rm La_{0.67}Ca_{0.33}MnO_3$ and $\rm La_{0.67}Sr_{0.33}MnO_3$
epitaxial and polycrystalline films (Gupta \etal 1996), respectively.
Both investigations found that the resistivity and magnetoresistance
depended sensitively on the microstructure, whereas the magnetization 
was hardly affected by it.

\begin{figure}[t]
\begin{center}
\vspace*{0.0cm}
\hspace*{0.0cm} \includegraphics[width=0.8\textwidth]{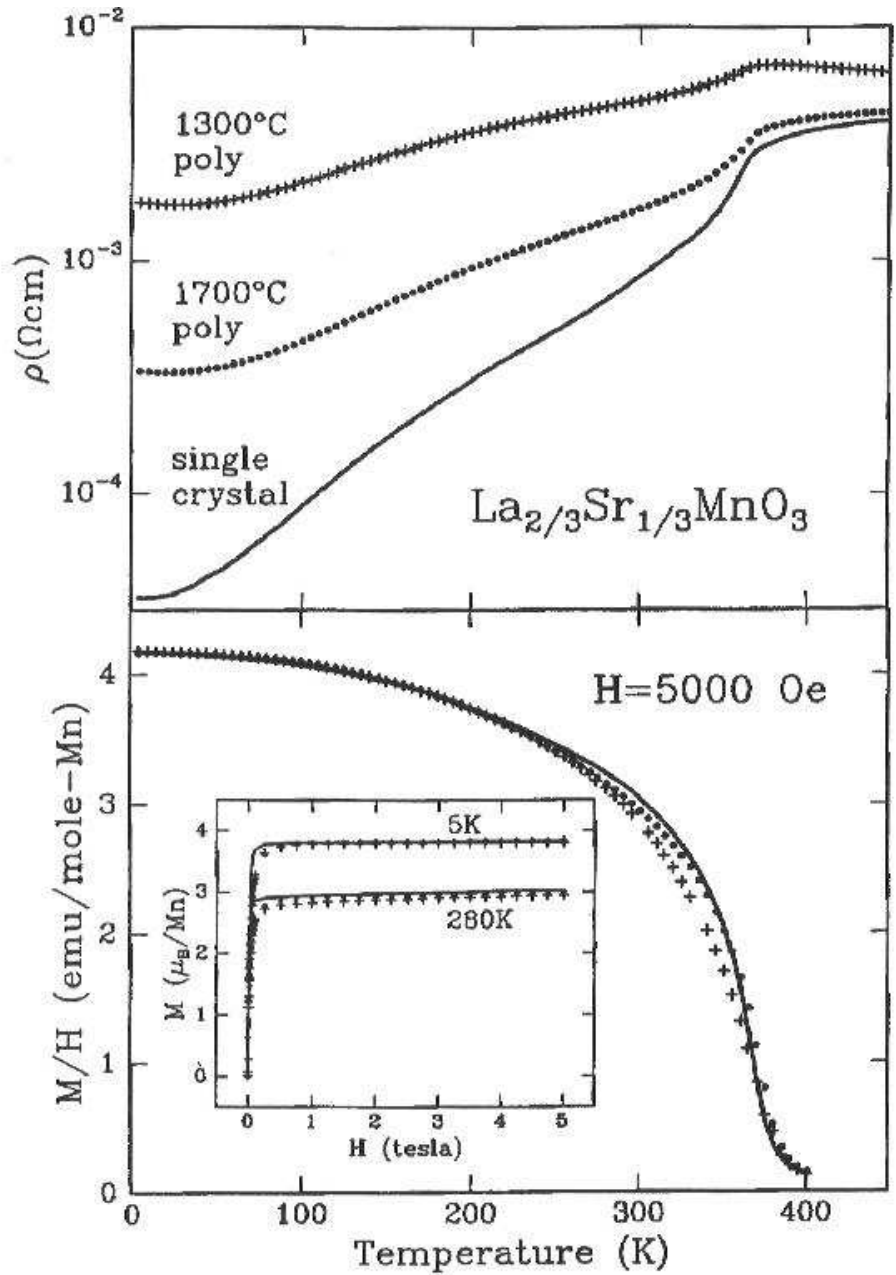}
\vspace*{-1.6cm}
\end{center}
\caption{Top panel: zero field resistivity of $\rm La_{0.67}Sr_{0.33}MnO_3$
single crystal and polycrystals as a function of temperature.
Bottom panel: magnetization of the samples as a function of temperature
measured at $B = 0.5$~T. The inset shows the field dependent
magnetization at 5~K and 280~K. Reproduced from Hwang \etal (1996).}
\label{hwang1}
\end{figure}

\begin{figure}[t]
\begin{center}
\vspace*{0.0cm}
\hspace*{0.0cm} \includegraphics[width=0.8\textwidth]{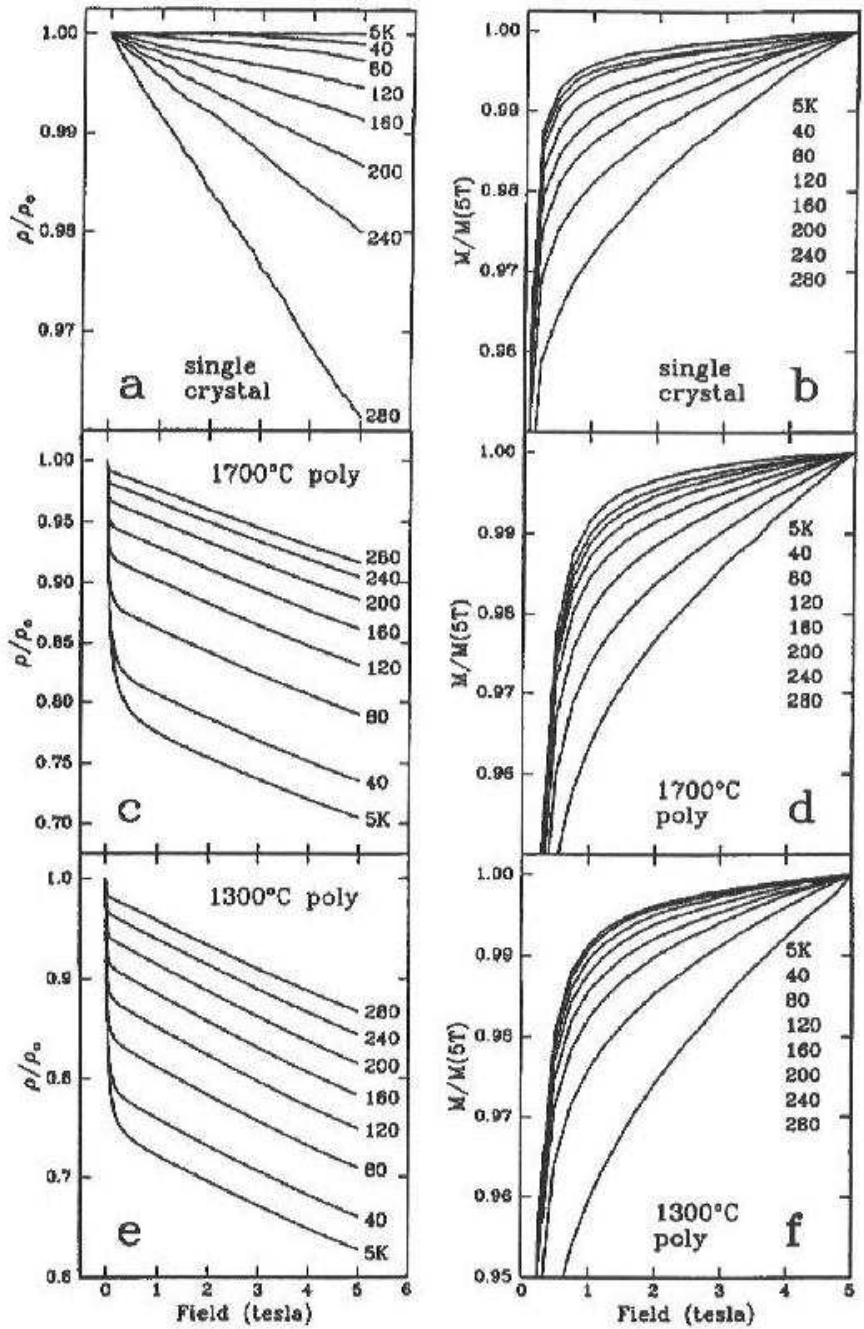}
\vspace*{-0.6cm}
\end{center}
\caption{Magnetoresistance data of the samples of
  figure~{\protect{\ref{hwang1}}}. Panels a, c and e: normalized
  resistivity $\rho/\rho_0$ as a function of magnetic field. $\rho_0$
  denotes the zero field resistivity. Panels b, d and f: magnetic
  field dependence of the normalized magnetization. Reproduced from
  Hwang \etal 1996.}
\label{hwang2}
\end{figure}

Hwang \etal (1996) investigated a LSMO single crystal and two
LSMO ceramic samples sintered at 1300$^\circ$C and 1700$^\circ$C,
respectively. The sample sintered at the higher temperature
had the larger grain size.
The data of Hwang \etal (1996) are reproduced in figures~\ref{hwang1}
and \ref{hwang2}.
Figure~\ref{hwang1} shows the zero field resistivity and the magnetization
of the samples as a function of temperature. Whereas the low
temperature resistivity
depends strongly on the microstructure, 
the magnetization of the three samples is virtually identical. The
effect of the grain boundaries on the magnetoresistance is even more
dramatic. Figure~\ref{hwang2} shows the field dependent resistivity and
magnetization of the samples investigated. Whereas the single crystal
shows a magnetoresistance linear in magnetic field, the polycrystalline 
samples show a sharp drop at low magnetic fields followed by a
linear dependence at higher fields. Again the field dependence
of the magnetization is virtually identical for the three samples.
The magnitude of the low field magnetoresistance increases with
decreasing temperature in contrast to the intrinsic magnetoresistance
that has a maximum near the Curie temperature and decreases with decreasing
temperature.

These results cannot be explained by the intrinsic magnetoresistance
alone, since the intrinsic magnetoresistance is only a function of the
magnetization. Hwang \etal (1996) suggested that the low field magnetoresistance
in polycrystalline samples is due to spin-polarized tunnelling between 
misaligned grains. It was shown by Wang \etal (1998) that, phenomenologically, one has
to distinguish weak and strong links between the grains. Only weak links
give rise to a considerable low field magnetoresistance. Whereas
the microstructural characteristics of the two types of links
are not clear, the formation of weak or strong links can be controlled
by the fabrication conditions.

The results of Gupta \etal (1996) on epitaxial and polycrystalline
films are in full agreement with the work on polycrystalline ceramics.
However, by growing manganite films on polycrystalline SrTiO$_3$
substrates with controlled grain size, Gupta \etal (1996) were able
to determine the grain-size dependence. If the polycrystalline films
are idealized to consist of low resistivity grains ($\rho_g$)
of size $l_g$ and thin ($l_{gb} \ll l_g$) high resistivity grain boundaries
($\rho_{gb}$), then the resistivity is expected to follow 
$\rho = \rho_g + (l_{gb}/l_g)\rho_{gb}$. A linear dependence of the
resistivity measured
at 10~K on the inverse grain diameter was indeed observed and
the interface resistivity was derived as 
$l_{gb}\rho_{gb} \sim 6\times 10^{-5}$~$\Omega$cm$^2$.
Versluijs \etal (1999) used a scanning tunnelling microscope in order
to simultaneously image the surface topography and map the potential
distribution of $\rm La_{0.7}Sr_{0.3}MnO_3$ films deposited on single
crystal and polycrystalline MgO. Near the crystallite boundaries
potential steps were found. At room temperature grain-boundary areal
resistivities were found to be in the range $3\times 10^{-7}$--$3\times
10^{-5}$~$\Omega$cm$^2$ with a typical value of $6\times
10^{-6}$~$\Omega$cm$^2$. This is one order of magnitude smaller than
the value deduced by Gupta \etal (1996) and might be related to the
higher measurement temperature. Scanning tunnelling potentiometry
measurements performed by Gr\'evin \etal (2000) on epitaxial LSMO
films on MgO revealed occasional potential steps; these have low areal
resistivities in the range $0.3\dots0.8\times10^{-7}$~$\Omega$cm$^2$
possibly related to the good crystallinity of the film.

The resistivity and magnetoresistance of manganite, magnetite
and CrO$_2$ powder compacts was investigated by Coey (1998c, 1999),
Coey \etal (1998a, 1998b) and Manoharan \etal (1998). Coey \etal
(1998b) investigated a CrO$_2$ and a diluted 25\% CrO$_2$/75\% Cr$_2$O$_3$
powder compact. Both samples show an increasing low field magnetoresistance
with decreasing temperature; the diluted powder compact has a low temperature
magnetoresistance of roughly 50\%. Using equation~(\ref{Gpoly}) and
a spin-polarization of 100\% one obtains $\langle\Delta G/G\rangle =
50$\%, in agreement with the experimentally observed value.
This is a rare example of a system that attains the theoretically
expected magnetoresistance ratio at low temperature; it is more often
observed that the measured magnetoresistance is considerably 
smaller than expected from Julliere's model. Coey \etal (1998b)
report dynamic conductance data at low temperatures that were attributed
to Coulomb-gap effects.

The dependence of the intergranular magnetoresistance of LSMO and LCMO
ceramics on the grain size was investigated by Balcells \etal (1998b)
and Hueso \etal (1999), respectively. The intergranular
magnetoresistance was found to increase with decreasing grain
size. Balcells \etal (1998b) reported a saturation of the low field
magnetoresistance at about 30\%. The logarithm of the resistivity at
constant temperature varied inversely proportional to the grain size
indicating that the tunnelling barrier thickness increased with
decreasing grain size. For submicronic grains an intergranular Coulomb
gap with a charging energy $E_C < 50$~K was found. Walter \etal (1999)
also reported a saturation of the low field magnetoresistance of LSMO
films with various degrees of texture at about 34\%.

Balcells \etal (1999) and Petrov \etal (1999) reported the
magnetoresistance of granular LSMO/CeO$_2$ and LCMO/SrTiO$_3$
composites as a function of the manganite fraction. An enhanced low
field magnetoresistance was found near the percolation threshold.
The low field magnetoresistance was also found to be enhanced in
LSMO/glass composites (Gupta \etal 2001) and LSMO/$\rm
Pr_{0.5}Sr_{0.5}MnO_3$ composites (Liu et al 2001).
\subsubsection{Bi-crystal junctions.}
Bi-crystal junctions have been investigated by Steenbeck \etal (1997, 1998),
Mathur \etal (1997, 1999), Isaac \etal (1998), Evetts \etal (1998), 
Klein \etal (1999), Westerburg \etal (1999), Miller \etal (2000),
Philipp \etal (2000) and Mathieu \etal (2001a, 2001b).
Generally these junctions are fabricated as follows.
Manganite thin films are deposited on a bi-crystal substrate with a
misalignment between the crystallographic directions of 24$^\circ$, 36.8$^\circ$ or
45$^\circ$. After deposition the manganite films are patterned
into a meander-like track crossing the grain-boundary several times.
Thus the investigation of a single grain boundary is possible.
45$^\circ$ grain boundaries were also fabricated on SrTiO$_3$
substrates using MgO and CeO$_2$ buffer layers in a chess board
pattern (Mathieu \etal 2000); these junctions yielded results similar
to bi-crystal junctions.

\begin{figure}[t]
\vspace*{0.0cm}
\begin{center}
\hspace*{1.0cm} \includegraphics[width=0.8\textwidth]{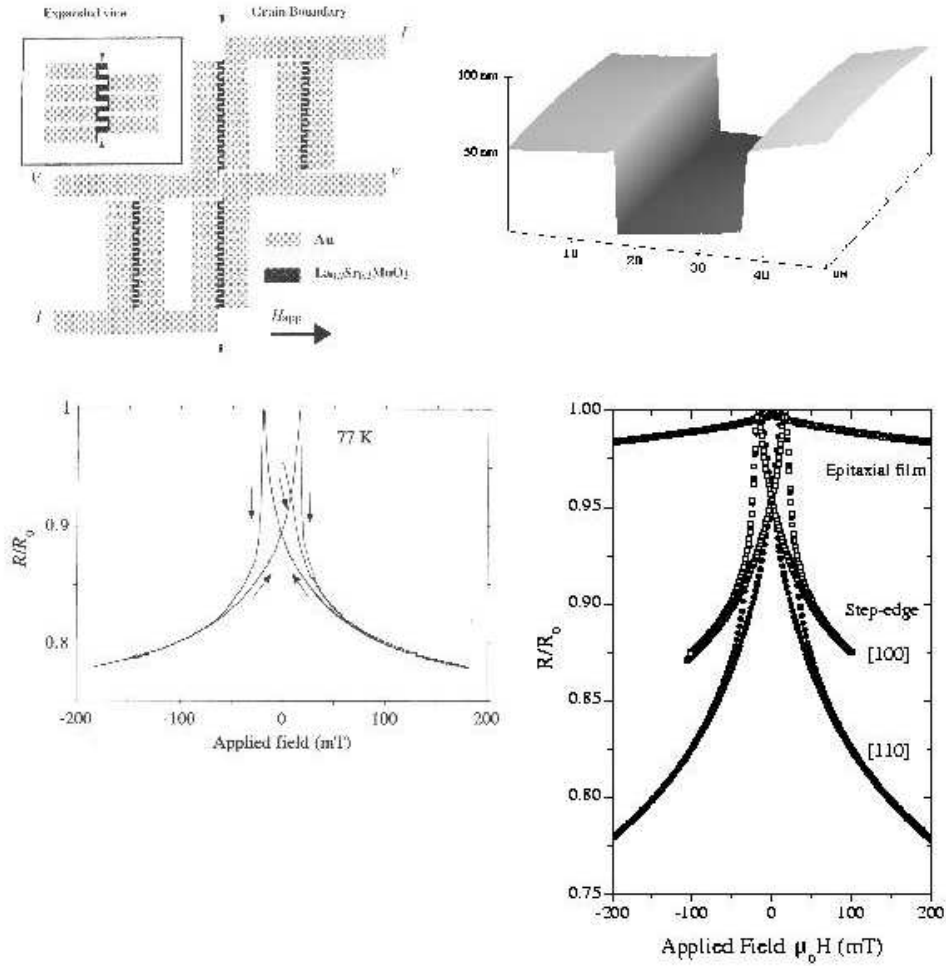}
\end{center}
\vspace*{-4.0cm}
\caption{(a) General layout and magnetoresistance response of a
bi-crystal junction. Adapted from Mathur \etal (1997). (b) Atomic force
microscopy picture of a step-edge junction. Magnetoresistance ratio at
100~K of an epitaxial film and step-edge arrays along [100] and [110],
respectively. For the step-edge arrays the electrical current flows
across the steps. After Ziese \etal (1999a).} 
\label{stepmr}
\end{figure}
The characteristics of resistivity and magnetoresistance seen in the
bi-crystal junctions are similar to polycrystalline samples, see
figure~\ref{stepmr}a. The resistivity shows a maximum far below the
Curie temperature; a sharp decrease is found in the resistivity for
small applied magnetic fields. This low field magnetoresistance
increases with decreasing temperature with values up to nearly 100\%
at low temperatures. Steenbeck \etal (1998), Westerburg \etal (1999)
and Philipp \etal (2000) report field dependencies of the
magnetoresistance similar to tunnel junctions for magnetic fields
applied parallel to the junction, i.e.\ a large two-level
magnetoresistance effect. The areal resistivity of the grain boundary
increases with tilt angle (Isaac \etal 1998). However, the various
groups report a range of values, see table~\ref{interresist}. 

\begin{table}
\caption{Interface resistance $RA$ of various bi-crystal
  junctions. For comparison, the interface resistance measured in
  polycrystalline films is also shown.}
\begin{indented}
\item[]\begin{tabular}{@{}lllll}
\br
Temperature $T$ (K) & Tilt angle & Material & $RA$ ($\Omega$cm$^2$) & Ref. \\
\mr
32 & $36.8^\circ$ & LSMO & $4.1\times 10^{-6}$ & Steenbeck (1997)\\
32 & $36.8^\circ$ & LSMO & $16 \times 10^{-6}$ & Steenbeck (1998)\\
77 & $4^\circ$    & LCMO & $0.2\times 10^{-8}$ & Isaac     (1998)\\
77 & $24^\circ$   & LCMO & $20 \times 10^{-8}$ & Isaac     (1998)\\
77 & $36.8^\circ$ & LCMO & $10 \times 10^{-8}$ & Isaac     (1998)\\
77 & $45^\circ$   & LCMO & $25 \times 10^{-8}$ & Isaac     (1998)\\
10 & $24^\circ$   & LCMO & $10^{-2}-1$         & Klein     (1999)\\
4.2& $45^\circ$   & LCMO & $170 \times 10^{-6}$& Westerburg (1999)\\
\hline
10  & poly        & LCMO & $6\times 10^{-5}$   & Gupta (1996)\\
300 & poly        & LSMO & $6\times 10^{-6}$   & Versluijs (1999)\\
300 & epitaxial   & LSMO & $5\times 10^{-8}$   & Gr\'evin (2000)\\
\br
\end{tabular}
\end{indented}
\label{interresist}
\end{table}
\subsubsection{Step-edge junctions.}
Ziese \etal (1999a) investigated step-edge junctions made from LCMO
films. LaAlO$_3$ substrates were patterned prior to film deposition
by chemically assisted ion-beam etching such that an array of
steps along [100] or [110] was formed. The steps were 100~nm to 200~nm
high and 20~$\mu$m apart; the substrates contained 150 [100]
or 200 [110] steps, respectively.
25~nm thick LCMO films were deposited on the patterned substrates
using pulsed laser deposition. These films show a large resistance 
anisotropy, the resistance showing intrinsic behaviour
for electric currents flowing along the steps and typical grain-boundary
behaviour for currents across the steps. This resistance anisotropy
can be related to disordered regions near the step edges.
In comparison to epitaxial films, the magnetoresistance is strongly enhanced, 
see figure~\ref{stepmr}b. The magnetoresistance value at fixed field
and temperature seems to be determined by the local defect structure and varies
between different samples. After annealing the film deposited on [100] step edges
at 950$^\circ$C for 2~h in flowing oxygen, the resistance and 
magnetoresistance resumed the typical behaviour of epitaxial films.
Similar results were obtained on ``scratch'' junctions by
Srinitiwarawong and Ziese (1998).
\subsubsection{Laser-patterned junctions.}
Bibes \etal (1999a, 1999b) reported on the temperature and magnetic field
dependence of the magnetotransport properties of laser-patterned
planar junctions. A 248~nm KrF Excimer laser with a fluence of about
2.5~J/m$^2$ was used to define tracks of 10~$\mu$m and 40~$\mu$m width
on SrTiO$_3$ substrates. These tracks consisted of overlapping disks
of molten material, about 0.1-0.2~$\mu$m deep. Microcrack formation
was observed within these disk regions. LSMO films were deposited by
pulsed laser deposition on these patterned substrates. A strongly
enhanced resistance was only found for the 40~$\mu$m wide
tracks. However, both 10~$\mu$m and 40~$\mu$m tracks lead to a
significantly enhanced low field magnetoresistance with the
characteristic magnetic field and temperature dependence.
\subsection{General characteristics and models}
Since considerable research efforts have been focused on the investigation of 
grain-boundary junctions, a lot of systems have been studied and certain
general features have emerged. In this section the general characteristics
will be summarized and the current status of models for grain-boundary
transport will be reviewed.

When discussing magnetoresistance of ferromagnets, it is useful to distinguish
the low field magnetoresistance from the high field behaviour. The magnetic
field scale of the low field magnetoresistance is the coercive field. 
Grain-boundary junctions generally show a large low field magnetoresistance
crossing over to a much more gradual decrease of the resistance, see
figure~\ref{hwang2}. There is consensus that the low field magnetoresistance
is due to spin-polarized tunnelling between grains with different orientations.
These grains are aligned in magnetic fields of the order of the
coercive field, thus explaining the steep decrease in resistance
at low fields. This is consistent with noise measurements (Mathieu
\etal 2001a) that prove the importance of magnetic domain fluctuations
in this field regime. The observed magnetoresistance, however, is always
considerably smaller than the value derived from equation~(\ref{Rpoly}).
Furthermore, Julliere's model predicts no magnetoresistance for parallel
orientation of the grain magnetization, in contradiction with the ubiquitous
high field magnetoresistance slope. These deviations from the basic theory
will be discussed in the following. Only few exceptions have been
found to this general picture, see e.g.\ the large two-level
magnetoresistance in a bi-crystal junction reported by Philipp \etal
(2000). In this case the tunnelling barrier is apparently of high
quality and, correspondingly, the high field magnetoresistance is
absent and the spin polarization derived from the resistance
switching is high. These exceptions, however, enforce the general
conclusions.

\begin{figure}[t]
\vspace*{-1.0cm}
\begin{center}
\hspace*{0.5cm} \epsfysize=10cm \epsfbox{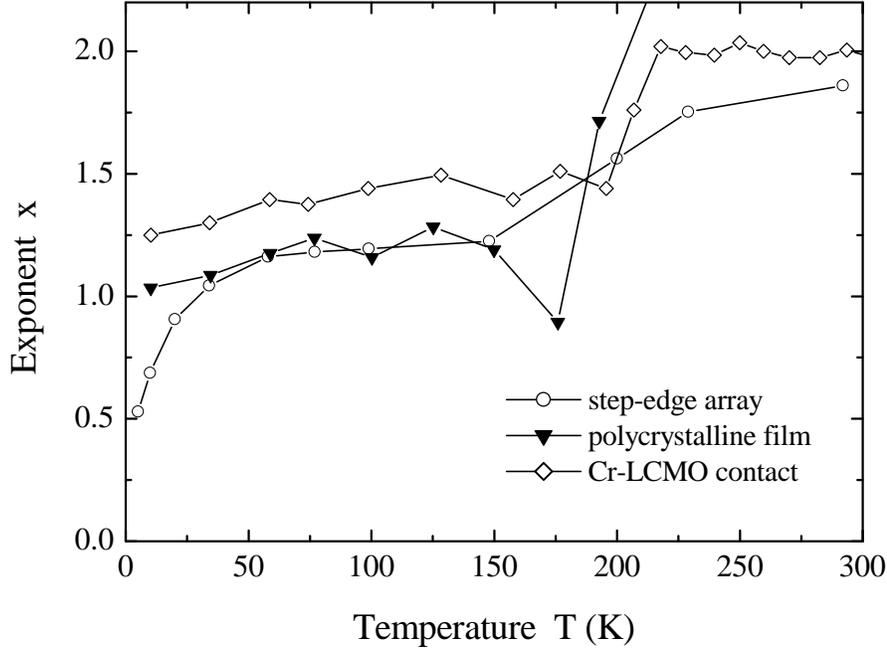}
\end{center}
\vspace*{0.0cm}
\caption{Conductance exponent $x$ as a function of temperature
for a step-edge array, a polycrystalline LCMO film on MgO and 
a Cr-LCMO contact. Reproduced from Ziese (1999).}
\label{expx}
\end{figure}

Some non-linear conductance measurements have been performed to assess the
transport mechanism. Whereas measurements on polycrystalline ceramics
are usually performed at low voltages in the linear region, the non-linear
conductance of bi-crystal junctions (Steenbeck \etal 1998, Mathur \etal 1999,
Klein \etal 1999, Westerburg \etal 1999, H\"ofener \etal 2000),
step-edges junctions (Ziese \etal 1999a, Ziese 1999) and mechanically induced
grain boundaries (Srinitiwarawong and Ziese 1998, Ziese 1999) has been measured.
Whereas the non-linear conductance depends somewhat on the microstructure,
measurements on several systems indicate an inelastic tunnelling
process. Steenbeck \etal (1998) reported
a quadratic voltage dependence of the conductance in annealed
bi-crystal junctions, whereas Klein \etal (1999), Westerburg \etal
(1999) and H\"ofener \etal (2000)
observe clear deviations from a quadratic dependence. Klein \etal
(1999), Ziese (1999) and H\"ofener \etal (2000) measured the
non-linear conductance at various temperatures 
and studied the evolution of the non-linearity with temperature.
These investigations found clear evidence for inelastic tunnelling via
localized states. Ziese (1999) analyzed the voltage dependence of
the conductance using the general form
\begin{equation}
G = \frac{dI}{dV} = G_0 (1+g_xV^x),
\label{condx}
\end{equation}
where $G_0$ denotes the zero bias conductance and $g_x$ the non-linear 
conductance. This ansatz was motivated by the results of Glazman and
Matveev (1988) on inelastic tunnelling via $n$ localized states; in
this case $x = n-2/(n+1)$. Moreover, the analysis of the junction
resistivity of conventional ferromagnetic tunnelling junctions
by Shang \etal (1998), see previous section, indicated the presence of
inelastic tunnelling processes via two localized states in the
barrier. The conductance exponent $x$ of a step-edge array,
a polycrystalline LCMO film on MgO and a Cr-LCMO contact
(see section~\ref{LCMOmetal}) are shown in figure~\ref{expx}
as a function of temperature. 
The conductance exponent shows a crossover from a value $x \sim 1.2-1.4$
below the Curie temperature to $x \simeq 2$ above $T_{\rm C}$. Accordingly,
the non-linear conductance $g_x$ decreases sharply at the Curie
temperature indicating that inelastic tunnelling processes are only
present below $T_{\rm C}$,
whereas a small elastic tunnelling component persists in the 
paramagnetic phase. These results are in agreement with the investigation
on bi-crystal junctions by Klein \etal (1999) and H\"ofener \etal
(2000). From these non-linear conductance 
measurements it might be concluded that tunnelling between ferromagnetic grains
occurs mainly via one or two localized states. This qualitatively explains
the experimentally observed reduction of the magnetoresistance
in comparison to Julliere's model, since a spin-polarization loss
results during the inelastic tunnelling process. This idea was further
developed by H\"ofener \etal (2000): the bi-crystal
magnetoresistance was observed to decrease drastically with increasing
voltage bias, since the junction is shunted by inelastic tunnelling
processes that do not conserve the spin. H\"ofener \etal (2000) proposed an
extension of Julliere's model to a three-current model by taking
inelastic tunnelling into account; this is the same approach used by
Zhang and White (1998). This leads to a voltage dependent
magnetoresistance
\begin{equation}
\frac{\Delta R}{R}(V,T) =
\frac{I_{\uparrow\downarrow}}{I_{\uparrow\downarrow}+I_{\rm i}}(V,T)\,
\left(\frac{\Delta R}{R}\right)_{\rm Julliere}\, .
\label{hoefener}
\end{equation}
Here $I_{\rm i}$ denotes the current due to inelastic tunnelling and
$I_{\uparrow\downarrow}$ the current through the junction in the
antiparallel magnetization state due to direct tunnelling. Independent
measurements of the voltage dependent magnetoresistance and the
elastic to inelastic current ratio showed significant correlation,
thus corroborating the importance of inelastic tunnelling processes.
Lee \etal (1999) argued
that spin-polarized tunnelling in manganite ceramics proceeds mainly
via one localized state. Within this model these authors derived
a ``universal'' magnetoresistance value of $\Delta R/R \simeq m^2/3$.
$m$ denotes the magnetization of the grains normalized by the
saturation magnetization. At low temperatures a value
of $1/3$ is found in rough agreement with experiment (Lee \etal 1999). However,
the assumption of tunnelling via one localized state as well as the
derived temperature dependence proportional to the square of the magnetization
are in contradiction to experimental results. Coey \etal (1998b)
observed a non-linear conductance in CrO$_2$ pressed powders at
low temperature and interpreted this within a Coulomb-blockade model.
Versluijs \etal (2000) obtained non-linear $I$--$V$--curves in
nanocontacts between LSMO single crystals that might also arise from
inelastic tunnelling.

\begin{figure}[t]
\begin{center}
\vspace*{0.0cm}
\hspace*{0.2cm} \includegraphics[width=0.75\textwidth]{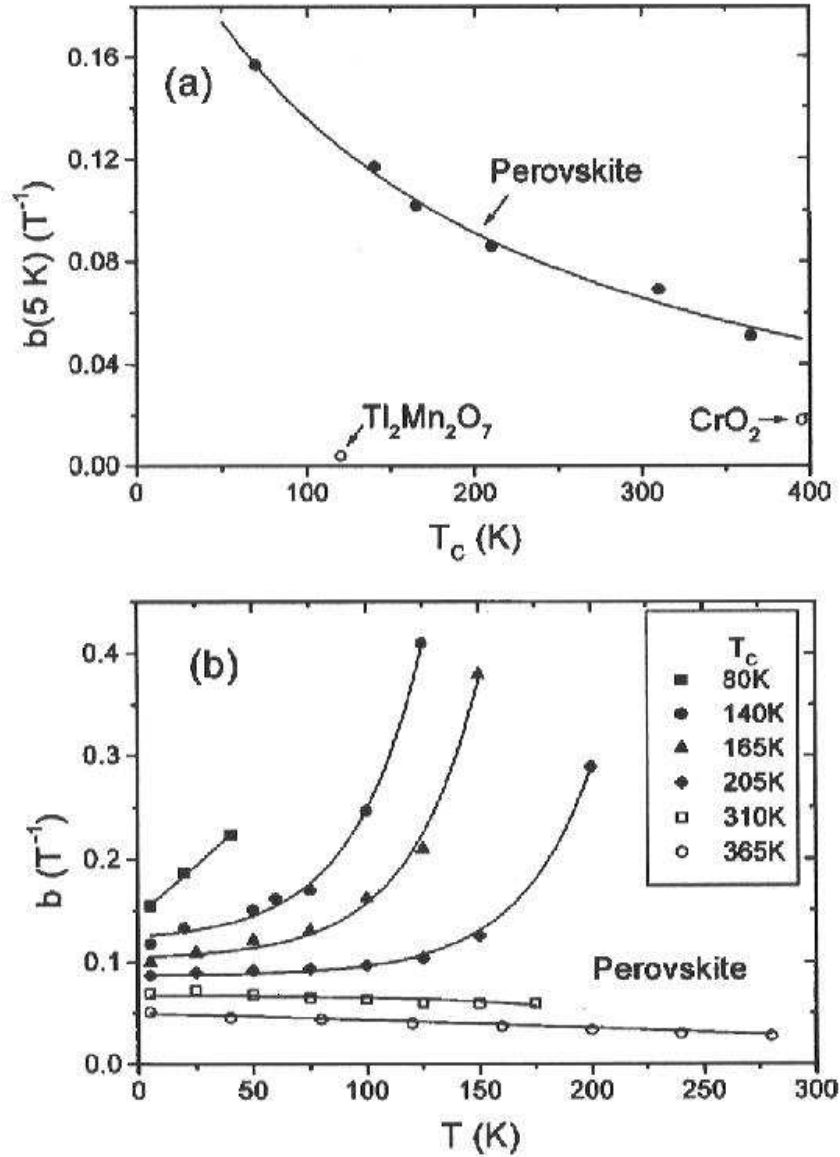}
\vspace*{-2.0cm}
\end{center}
\caption{Top panel: slope $b$ of the high field magnetoconductance as a
  function of the Curie temperature for various manganites. Bottom
  panel: high field magnetoconductivity slope as a function of
  temperature. $b$ is proportional to the grain-boundary susceptibility
  $\chi_{\rm GB}$. Reproduced from Lee \etal (1999).} 
\label{chigb}
\end{figure}
\begin{figure}[t]
\vspace*{-1.0cm}
\begin{center}
\hspace*{2.5cm} \epsfysize=12cm \epsfbox{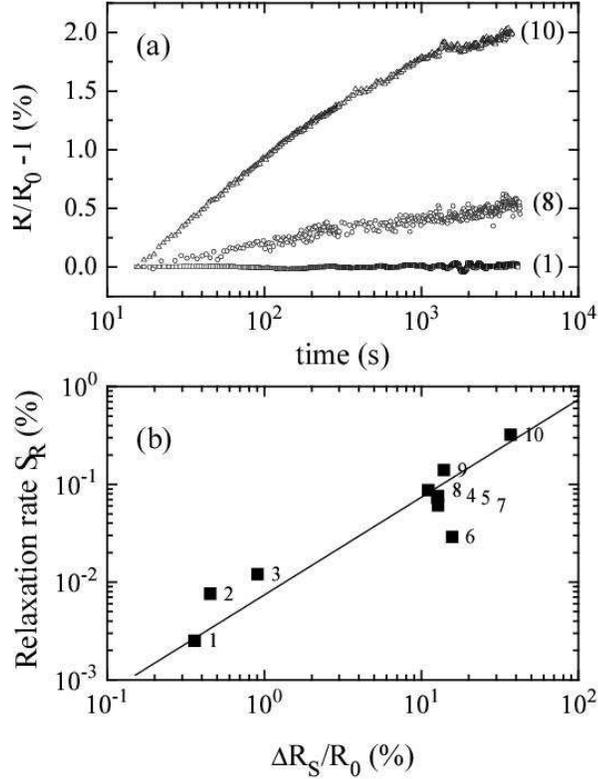}
\end{center}
\vspace*{-0.5cm}
\caption{(a) Resistance relaxation after a sudden field
change from 1~T to 7~mT of (1) an annealed
LCMO film on $\rm (LaAlO_3)_{0.3}(Sr_2AlTaO_6)_{0.7}$ (LSAT), (8) a
mechanically induced grain boundary, 
and (10) a Ti/Ni/LCMO/Ti heterostructure.
(b) Resistance relaxation $S_R$ versus low-field
magnetoresistance $\Delta R_{\rm S}/R_0$ for various 
LCMO structures:
(1) 250~nm thin annealed film on LSAT,
(2) 120~nm thin as-deposited film on LaAlO$_3$,
(3) 20~nm thin as-deposited film on LaAlO$_3$,
(4) annealed film on Si,
(5) as-deposited film on Si,
(6) as-deposited film on MgO,
(7) step-edge array,
(8) mechanically induced grain boundary,
(9) Ti/LCMO/Ti heterostructure,
(10) Ti/Ni/LCMO/Ti heterostructure.
The solid line is a fit of a linear law 
$S \propto \Delta R_{\rm S}/R_0$ to the data.
Reproduced from Ziese \etal (1998b).}
\label{relax}
\end{figure}

The high field magnetoresistance slope has been consistently interpreted
as arising from the barrier material (Guinea 1998, Evetts \etal 1998, Ziese 1999,
Lee \etal 1999). In a limited field range, the magnetoresistance, as well
as the magnetoconductance, are linear at high fields. However,
Lee \etal (1999) showed that the magnetoconductance of manganite polycrystals
is linear in magnetic field to a very good approximation.
It is generally argued that the high field slope $d(\sigma/\sigma_0)/dB$
is proportional to the grain-boundary susceptibility $\chi_{\rm GB}$
(Guinea 1998, Evetts \etal 1998, Ziese 1999, Lee \etal 1999). Here
$\sigma$ denotes the conductivity and $\sigma_0$ the zero field
conductivity. Such a relationship
was derived by Guinea (1998) within a model including tunnelling via
paramagnetic impurities which leads to a high field slope being
proportional to the Curie susceptibility. However, the temperature dependence of 
$\chi_{\rm GB}$, see figure~\ref{chigb} for data on various manganite
samples, indicates some magnetic ordering of the grain-boundary region.
Moreover, the high field magnetoresistance slope extends to very high fields
$> 8$~T, whereas a paramagnetic grain-boundary region containing
Mn$^{4+}$-ions is expected to saturate at about 1~T at 4.2 K.
Some evidence for the magnetic state of the grain-boundary region
comes from the work of Fontcuberta (Fontcuberta \etal 1998, Balcells
\etal 1998a, 1998b, Mart{\'\i}nez \etal 1998), Ziese (Ziese \etal
1998b, Ziese \etal 1999a, Ziese 1999), Zhang \etal (1997) and Zhu (Zhu
\etal 2001). Fontcuberta \etal studied LSMO ceramics with various 
grain sizes and observed an increase of the thickness of the
tunnelling barrier as well as a reduction of the saturation
magnetization with decreasing grain size. This indicates that the
surface layer of the grains is in a magnetically disordered
state. This magnetically frustrated interface region is presumably
insulating and serves as the tunnelling barrier between the
ferromagnetic grains. This interpretation is corroborated by studies
of various rare-earth substitutions on the magnetotransport properties
(Fontcuberta \etal 1998, Zhou \etal 1999). There seem to be
ideal cation substitutions maximizing the low field
magnetoresistance. This might be related to enhanced spin disorder
induced by the competition of double exchange and super-exchange
interactions. Zhang \etal (1997) related the tunnelling barrier
to the energy difference between the bulk and surface double-exchange
sytems. Within this approach resistivity versus temperature curves
could be successfully modelled. Zhu \etal (2001) directly observed a
spin-freezing transition in LSMO nanoparticles with a mean grain size
below 50~nm: at temperatures below 45~K the field cooled magnetization
suddenly increases. This transition indicates the alignment of the
surface magnetic moments with the moments in the nanoparticle core.
Ziese \etal investigated the grain-boundary magnetoresistance
as a function of angle between the applied field and the current and
found typical anisotropic magnetoresistance (see also Coey 1999)
of the same order of magnitude as in epitaxial films
(Ziese \etal 1999a, Ziese 1999). This was interpreted by 
Ziese (1999) to indicate tunnelling via manganese ions in the barrier, 
since the anisotropic magnetoresistance is mainly determined by the
local environment of the magnetic ion, see section~\ref{intrinsic}. It
is therefore likely that electrons tunnel between different grains via
magnetically coupled, frustrated
manganese ions in the grain-boundary region. Since this region has a 
negligible volume compared to the rest of the film, a direct magnetic
investigation has not been possible so far. However, indirect information
on the magnetic state of the grain boundary stems from resistance
relaxation experiments (Ziese \etal 1998b). In these
experiments the resistance relaxation of various manganite samples
was measured after a field step from 1~T to the remanence field of the magnet
of about 7~mT. The relaxation was found to be logarithmic in time,
see figure~\ref{relax}a. The relaxation rate
\begin{equation}
S_R = \frac{1}{R_0}\, \frac{dR}{d\ln t}
\label{SR}
\end{equation}
scales with the low field magnetoresistance, see figure~\ref{relax}b 
proving that the relaxation
is due to magnetization processes in the grain-boundary region.
This was corroborated by measurements of the magnetic viscosity of a 
polycrystalline LCMO film on Si (Ziese \etal 1999b). The resistance
relaxation expected from magnetic viscosity of the grains is one
order of magnitude smaller than the resistance relaxation due to the
barrier spins.
The logarithmic time decay indicates a frustrated magnetic state
of the barrier spins.

Calculations of the surface structure of double-exchange ferromagnets
indicate an antiferromagnetic ordering (Calderon \etal 1999b). Assuming
an antiferromagnetically ordered interface between two ferromagnetic grains,
Calderon \etal (1999b) calculated the high field magnetoresistance
and found a linear resistivity decrease in agreement with experiment.
However, it might be argued that the linear high field magnetoresistance is 
a signature of any frustrated spin structure in the barrier, since
it arises from the action of the applied field against the exchange
field.

\begin{figure}[t]
\vspace*{0.0cm}
\begin{center}
\hspace*{0.0cm} \includegraphics[width=0.8\textwidth]{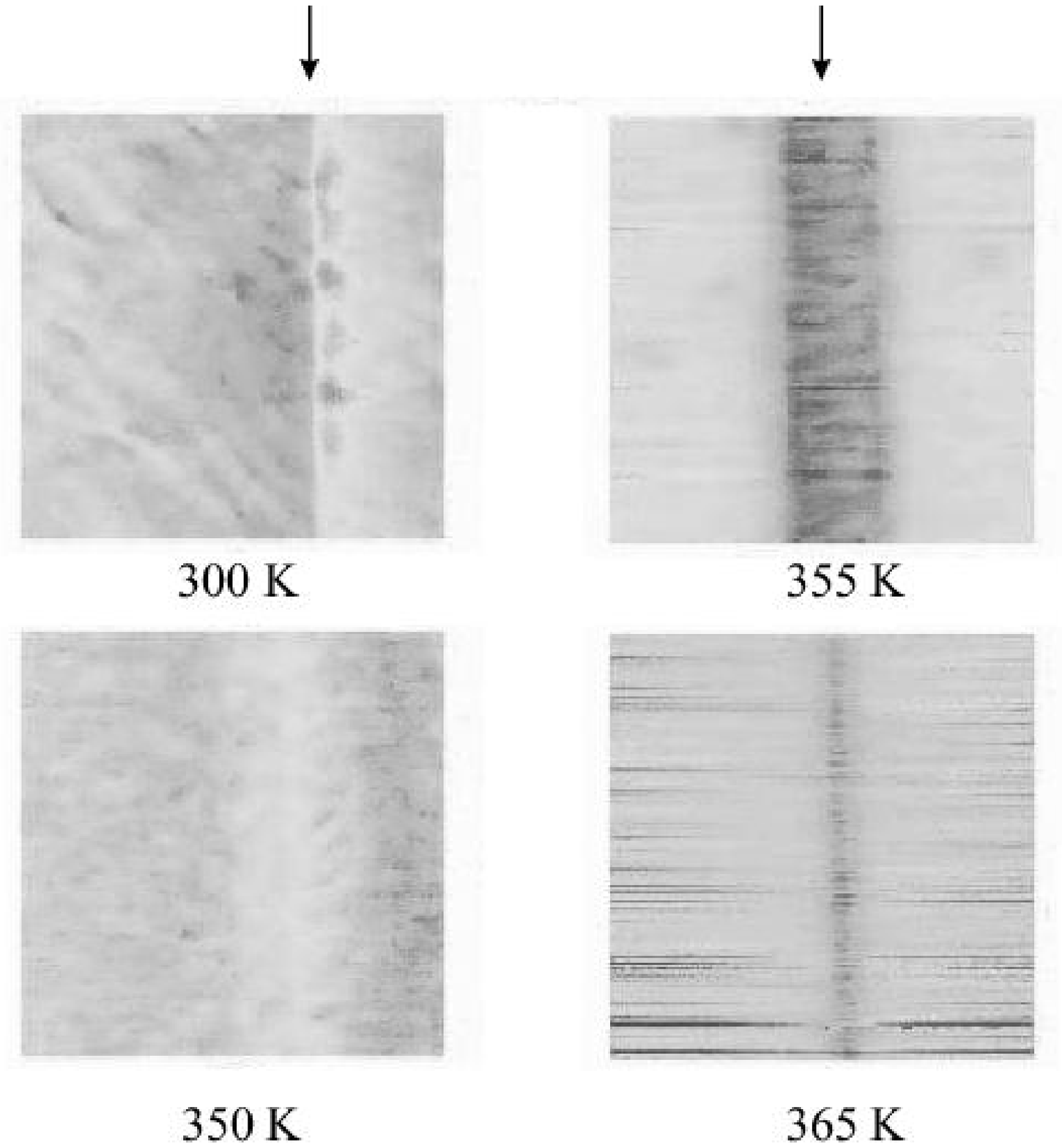}
\end{center}
\vspace*{-7.0cm}
\caption{MFM-images showing the evolution of the magnetic structure
near a grain boundary in a $\rm La_{0.7}Sr_{0.3}MnO_3$ film grown on a
SrTiO$_3$ bi-crystal; the location of the grain boundary is marked by
arrows. At 300~K a domain wall located at the grain
boundary is observed. Above 350~K the grain magnetization vanishes; at
355~K, however, a clear magnetic signal is obtained in the vicinity of
the grain boundary. The grain boundary magnetization vanishes above
370~K. Adapted from Soh \etal (2001).}
\label{mfm}
\end{figure}
The magnetic properties of the grain boundary play a central role in
the understanding of the magnetotransport properties in
polycrystalline manganites. Local magnetic information has been
obtained by magneto-optical imaging (Miller \etal 2000) and magnetic
force microscopy (MFM) (Soh \etal 2000, 2001) of bi-crystal
junctions. Miller \etal (2000) detected a reorientation of the
grain-boundary magnetization pointing out-of-plane, whereas the grain
magnetization lies in plane. This magnetization rotation might
contribute to the low field magnetoresistance. More astonishing are
the results of MFM studies on the temperature dependence of the
grain-boundary magnetization: this was found to have a higher Curie
temperature than bulk material (Soh \etal 2000, 2001)!
Figure~\ref{mfm} shows MFM-scans near a grain boundary grown on a
bi-crystal substrate at temperatures between 300~K and 365~K. At 300~K
a domain-wall located at the grain boundary is seen. Above the bulk
Curie temperature of 300~K the grain magnetization vanishes; a clear
magnetic signal is recorded, however, at 355~K, localized near the
grain boundary. This magnetic state is also found near natural
defects. It vanishes above 370~K. From these studies it follows that
the grain-boundary region is magnetically ordered; moreover, the size
of this mesoscale magnetic region varies with temperature and the
nature of the underlying defect. This provides additional evidence for
the mechanism of tunnelling through a magnetic barrier.

The temperature dependence of the grain-boundary resistance is not
very well understood. In all samples a broad resistance maximum is
seen well below the Curie temperature. At low temperatures $< 20$~K, a
resistance upturn is observed. Phenomenologically, such a temperature
dependence can be reproduced by considering a model of parallel
conduction channels (de Andr\'es \etal 1999). Since the resistivity is
dominated by the least resistive paths, parallel conduction through
well linked grains and intergranular regions can be considered. In
such a model, the intergranular regions are assumed to be
semiconducting, whereas the grains are metallic. Introducing effective
cross sections for both channels, a satisfactory fit to the data can
be made (de Andr\'es \etal 1999). This model, however, is not fully
satisfying, since it does not provide information on the microscopic
transport mechanism.
Ziese and Srinitiwarawong (1998) showed that the resistivity of
polycrystalline LCMO and LBMO films above the Curie temperature
followed a variable range hopping law $\rho \propto
\exp\lbrack -(T_0/T)^{1/4}\rbrack$, wheras epitaxial films are better described by
polaron transport in the adiabatic limit, $\rho \propto T\exp\lbrack -U/{\rm
  k}T\rbrack$. This result is consistent with the idea of intergranular
transport via impurities.
Balcells \etal (1998b) found a $\rho \propto \exp[(E_C/T)^{1/2}]$
variation of the resistivity of ceramic LSMO samples at low
temperature and interpreted this behaviour as arising from a Coulomb
gap. In contrast to this result, Raychaudhuri \etal (1999) observed
variable range hopping at low temperatures in polycrystalline LSMO.
The broad resistance maximum at intermediate temperatures might be an
indication of a mainly antiferromagnetic spin structure at the
interface. The ubiquitous resistance minimum at low temperatures was
interpreted by Rozenberg \etal (2000) as due to the thermal
disordering of a mainly antiferromagnetically aligned spin structure.

In conclusion, a consensus seems to emerge on the nature of
spin-polarized transport in polycrystalline magnetic oxides. It was realized 
that charge-carrier transport occurs via inelastic tunnelling processes.
This implies that the magnetotransport depends strongly on the barrier
characteristics. The grain-boundary region shows magnetic order,
presumably a frustrated magnetic state or a mainly antiferromagnetic
state with some frustration, causing an approximately
linear high field magnetoresistance extending to very large fields.
The inelastic tunnelling process is detrimental to the low field
magnetoresistance, since the spin-polarization is reduced during
the transit through the magnetically disordered barrier. This is
confirmed by calculations of the apparent spin-polarization of
ferromagnetic tunnelling junctions with a disordered barrier (Tsymbal
and Pettifor 1998). The great future
challenge is to produce grain-boundary junctions with non-magnetic
or magnetically ordered tunnelling barriers in order to improve the effective
spin-polarization. Annealed bi-crystal junctions seem to be promising
candidates for such a development (Steenbeck \etal 1998).
\subsection{Other ferromagnetic oxides}
\subsubsection{Polycrystalline material.}
\begin{figure}[t]
\begin{center}
\vspace*{0.0cm}
\hspace*{0.5cm} \includegraphics[width=0.8\textwidth]{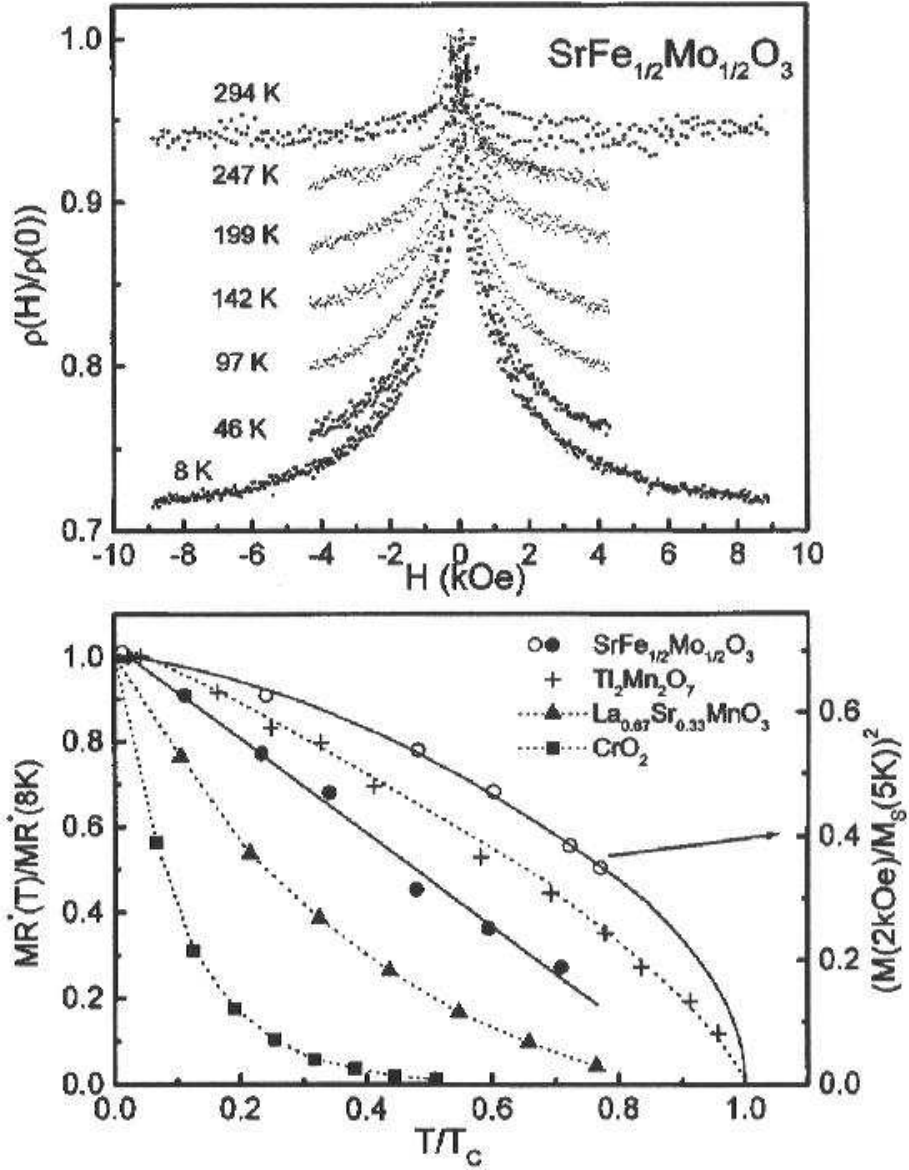}
\vspace*{-3.0cm}
\end{center}
\caption{Top panel: the magnetic field dependence of the normalized resistance
of $\rm Sr_2FeMoO_6$ at various temperatures. Bottom panel: the normalized
low field magnetoresistance of $\rm Sr_2FeMoO_6$, defined as 
$MR^* = [\rho(0)-\rho(2{\rm kOe})]/\rho(0)$, plotted as a function
of the reduced temperature $T/T_C$ with those of $\rm Tl_2Mn_2O_3$,
CrO$_2$ and $\rm La_{0.67}Sr_{0.33}MnO_3$. Reproduced from Kim \etal (1999).}
\label{kim}
\end{figure}

Apart from the low field magnetoresistance studies of manganite
ceramics discussed above, the extrinsic magnetoresistance of
polycrystalline material of CrO$_2$ (Hwang and Cheong 1997a, Coey
\etal 1998a, Manoharan \etal 1998, Dai \etal 2000, Dai and Tang 2000a,
2000b), $\rm Tl_2Mn_2O_7$ (Hwang and Cheong
1997b), $\rm Sr_2MoFeO_6$ (Kim \etal 1999, Yuan \etal 1999) and $\rm
La_{1.2}Sr_{1.8}Mn_2O_7$ (D\"orr \etal 1999) was studied. These  
investigations usually show a large low field magnetoresistance between
about 20\% and 60\% at low temperature. This is consistent with
spin-polarized tunnelling between (nearly) half-metallic
ferromagnets. The magnetoresistance of magnetite was found to be small
of only a few percent. 
Typical data of a $\rm Sr_2MoFeO_6$ polycrystal are
shown in figure~\ref{kim}a. The temperature dependence of the
tunnelling magnetoresistance, however, varies strongly among these
compounds. This is illustrated in figure~\ref{kim}b comparing the
normalized low field magnetoresistance as a function of the reduced
temperature, $T/T_C$, for CrO$_2$, $\rm La_{2/3}Sr_{1/3}MnO_3$, $\rm
Tl_2Mn_2O_7$ and $\rm Sr_2MoFeO_6$. A clear trend emerges: the decay
of the tunnelling magnetoresistance with temperature becomes smaller
along this series. This was corroborated by Lee \etal (1999). For $\rm
Sr_2MoFeO_6$ the magnetoresistance is
proportional to $M^2$ as expected for spin-polarized tunnelling. The
interpretation of these data is not fully clear. The different
temperature dependences seem to be related to both the interfacial
magnetism and the tunnelling barrier properties. One might speculate
that $\rm Sr_2MoFeO_6$ has the most robust interfacial magnetization
of the four compounds compared. At the same time the tunnelling
barrier might contain less magnetically active localized states. On
a microscopic scale the distinction between grains and barrier might
not be suitable, since the transition between those is supposed to be
gradual. The spin structure of the itinerant and super-exchange
ferromagnets $\rm Sr_2MoFeO_6$ and $\rm Tl_2Mn_2O_7$ might be less
sensitive to structural disorder than in the double exchange systems
$\rm La_{2/3}Sr_{1/3}MnO_3$ and CrO$_2$, since the latter show a
competition between ferromagnetic double exchange and
antiferromagnetic super-exchange. The double exchange mechanism is
supposed to be weakenend near an interface due to the reduced carrier
mobility. Scanning tunnelling microscopic and spectroscopic
investigations on manganite polycrystals indicated a semiconducting
nature of the intergranular layers with a band gap of 0.3-0.45~eV;
there might also be some band bending in the adjacent grains (Kar \etal
1998). These microscopic investigations, however, are in the early
stages and further studies on well characterized systems are clearly
desirable.

A promising candidate for room temperature applications is $\rm
Fe_3O_4$ with a Curie temperature of 858~K and the predicted
half-metallic character. It has, however, been notoriously difficult
to obtain a significant extrinsic magnetoresistance even at low
temperatures, see the studies by Coey \etal (1998a), Ziese \etal
(1998b), Li \etal (1998a), Nishimura \etal (2000), Kitamoto \etal
(2000), Chen and Du (2000), Uotani \etal (2000) and Taniyama
\etal (2001). Two recent studies along different routes report the
successfull observation of giant room temperature
magnetoresistance in magnetite. Versluijs \etal (2001) used
nanocontacts between magnetite crystals, see section~\ref{dw}. The
authors interpreted their results within a model of domain-wall
scattering at a nano-constricted wall. The non-linear $I$--$V$ curves,
however, are also in agreement with a tunnelling mechanism. Chen \etal
(2001) investigated the magnetoresistance of Zn-doped magnetite
polycrystals and observed a spectacular room temperature
magnetoresistance of more than 50\% at a doping concentration of
40\%. At this doping level insulating $\alpha$-Fe$_2$O$_3$
precipitates form between grains; these appear to be good tunnelling
barriers minimizing spin-polarization loss near the grain-boundary as
well as inelastic tunnelling effects. The spin-polarization vanishes
sharply at the Curie temperature of 318~K. If the interpretation
of the large magnetoresistance due to spin-polarized tunnelling is
correct, this work opens the way to efficiently engineer room
temperature devices based on magnetite.
\subsubsection{Controlled defect structures.}
Some investigations focused on the controlled fabrication of grain
boundaries in other ferro- and ferrimagnetic oxides. This research was
driven by the realization that the strong magnetoresistance decay as a
function of temperature observed in the manganites and in CrO$_2$
precluded room temperature applications. Possible remedies are
magnetic oxides with a high Curie temperature or with a robust
interfacial spin structure as indicated by the encouraging results on
polycrystalline material. Grain-boundary junctions introduced in
SrRuO$_3$ films (Bibes \etal 1999b), $\rm Sr_2MoFeO_6$ films (Yin
\etal 1999) and $\rm Fe_3O_4$ films (Ziese \etal 1998b) were
investigated. All these experiments showed only a small extrinsic 
magnetoresistance. 24$^\circ$ bi-crystal junctions had no effect on the
magnetoresistance of SrRuO$_3$ films, whereas laser-patterned
junctions enhanced the high field magnetoresistance, but did not
induce a low field magnetoresistance characteristic of spin-polarized
tunnelling (Bibes \etal 1999b). 24$^\circ$ bi-crystal junctions in
$\rm Sr_2MoFeO_6$ films showed a small magnetoresistance of about 2\%
at 2~kG and 20~K, twice the value of the magnetoresistance of a virgin
film (Yin \etal 1999). Ziese \etal (1998b) did not observe any
significant enhancement of the magnetoresistance of $\rm Fe_3O_4$
``scratch'' junctions compared to epitaxial films. This is consistent
with the small extrinsic magnetoresistance in $\rm Fe_3O_4$
polycrystalline films and pressed powders (Coey \etal 1998a). 
It has to be noted, however, that magnetite films show a high field
extrinsic magnetoresistance due to antiphase boundaries (Ziese and
Blythe 2000) that might mask any contribution from artificially
introduced defects.

In conclusion, these results indicate that the large low field
magnetoresistance observed in manganite bi-crystal junctions might be
specific to this double-exchange system. It seems that it is much more
difficult to introduce insulating regions in the itinerant ferromagnets
SrRuO$_3$, $\rm Sr_2MoFeO_6$ and the ferrimagnet $\rm Fe_3O_4$ and
thus to decouple the magnetic electrodes. This, apart from the lower
spin-polarization in SrRuO$_3$, seems to be the limiting factor in the
device performance of those junctions. One might state this conclusion
in another way: the instability of the interfacial spin structure of
the manganites toward antiferromagnetic ordering or magnetic
frustration is a pre-requisite to readily fabricate grain-boundary
junctions. The large low field magnetoresistance observed in
polycrystalline samples does not contradict this conclusion, since
strong magnetic disorder is often induced in small
particles. Disordered magnetism at grain boundaries was also reported
for pure nanocrystalline iron (Bonetti \etal 1999).
\section{Summary, Conclusions and Outlook \label{summary}}
In this work extrinsic magnetoresistance phenomena in magnetic oxides
were reviewed. Besides domain-wall scattering, the most important
effects are spin-polarized tunnelling in ferromagnetic junctions and
grain-boundary magnetoresistance. One indicator for the potential of a
magnetic material in tunnelling structures is the degree of
spin-polarization. Here the magnetic oxides are almost unique is
having spin-polarizations approaching 100\%. Great progress has been
made in recent years in both the fabrication of tunnelling junctions and the
understanding of the transport mechanisms in grain-boundary junctions.
A huge magnetoresistance was seen in 
$\rm La_{0.7}Sr_{0.3}MnO_3/insulator/La_{0.7}Sr_{0.3}MnO_3$ tunnelling junctions
at $4.2$~K in agreement with the large spin-polarization. The
magnetoresistance ratio, however, decays strongly with increasing temperature
rendering these junctions useless for room-temperature applications.
This strong decay appears to be related to a reduced interfacial 
spin-polarization. At present, it is not clear whether this is mainly an intrinsic
effect related to surface reconstruction or is more a problem of producing
high quality atomically flat interfaces. A similar problem was found
in grain-boundary junctions. Here, insulating connections between different 
crystallites in polycrystalline material are formed leading to
spin-polarized tunnelling between ferromagnetic grains. This process
is very similar to the tunnelling process in ferromagnetic tunnelling
junctions. The insulating layer between the grains, however, has many
defect states leading to inelastic tunnelling processes and an
apparent spin-polarization decrease.

There are some indications that domain-wall scattering might be
important in magnetic oxides. A rigorous experimental proof including
a precise measurement of its magnitude, however, has not yet been
given. The use of nano-constricted domain walls certainly opens the
most promising perspectives for further research.

In summary, the physics of spin-polarized 
transport in both ferromagnetic tunnel junctions and grain-boundary 
junctions has led to great challenges to both theoretical as well as
experimental physics and many interesting discoveries have been made.
The initial aim of fabricating magnetic field sensors operating at room
temperature, however, has not yet been achieved. This is in part due
to the sensitive surface chemistry of the materials and furthermore
due to the relatively low Curie temperatures. Here, the investigation
of $\rm Sr_2FeMoO_6$ that has a $T_C$ above 400~K and a seemingly
robust interfacial magnetization might lead to a breakthrough. Magnetite has a high
Curie temperature, but has not been successfully used as electrode
material for tunnelling junctions.

An issue that has not been addressed in this review but that is of
particular importance for applications is the noise level in magnetic
oxides. In the oxides studied so far -- mixed-valence manganites,
CrO$_2$ and $\rm Fe_3O_4$ (see Alers \etal 1996, Rajeswari \etal 1996,
Hardner \etal 1997 and Raquet \etal 1999) -- a large $1/f$ noise
exceeding the noise level in elemental metals by three to four orders of
magnitude was found. The $1/f$ dependence can be understood within the
Dutta-Dimon-Horn model as arising from a very broad distribution of
thermally activated fluctuators coupling to the resistivity (Dutta
\etal 1979). This, however, is a very formal treatment analogous to
the analysis of magnetic after-effects by a distribution of thermally
activated processes or the modelling of the low temperature properties
of glasses with a distribution of two-level systems. The physical
nature of the processes is usually difficult to reveal. Two candidates
causing the large noise level in magnetic oxides are thermally
activated motion of oxygen defects as well as electronic excitations
of reversed spin in a nearly half-metallic band-structure. Further
work is clearly necessary in order to understand the noise properties
of magnetic oxides and to limit their negative influence on device
performance.

The investigation of spin-polarized tunnelling in ferromagnetic oxides
constitutes a research area within the emerging field of spin-electronics
(also called magneto-electronics, Prinz 1998).
Spin-electronics denotes a novel field of solid-state electronics that
strives to realize electronic devices based on the differential manipulation
of currents with well-defined spin direction. The simplest device is a magnetic 
field sensor and such devices have been realized on the basis of oxides
working at low temperatures (ferromagnetic tunnelling junctions) and utilizing
colossal magnetoresistance at room temperature 
(thick film sensors, Balcells \etal 1996) or at 360~K (bridge sensors,
Steinbei{\ss} and Steenbeck 1998). There are three-terminal devices,
so-called spin transistors, made from elemental ferromagnets (Johnson 1993a, 1993b)
or semiconductor/ferromagnet hybrids (Monsma \etal 1995), but such devices have not
yet been realized using ferromagnetic oxides. However, since magnetic
oxides are almost unique in having a half-metallic band structure,
research in this direction will certainly have a huge potential.

Many challenging problems, e.g.\ the intrinsic transport mechanism in
mixed-valence manganites, the realization and investigation of
double tunnelling junctions, the relation between tunnelling current
and band-structure, the study of surface and interface magnetism, the
investigation of bonding effects between ferromagnets and insulating
barriers as well as a microscopic description of grain-boundary
transport, remain and will certainly lead to the discovery of much
more interesting physics.
\ack
This work was supported by the European Union TMR ``OXSEN''
network and by the DFG under DFG ES 86/7-1 within the Forschergruppe
``Oxidische Grenzfl\"achen''. I thank Prof.\ Gillian Gehring,
University of Sheffield, for valuable discussions and a critical
reading of the manuscript.
\clearpage
\addcontentsline{toc}{section}{References}
\References
\item
Abe M, Nakagawa T and Nomura S
1973 {\it J.\ Phys.\ Soc.\ Japan} {\bf 35} 1360.
\item
Adams C P, Lynn J W, Mukovskii Y M, Arsenov A A and Shulyatev D A
2000 {\it Phys.\ Rev.\ Lett.} {\bf 85} 3954.
\item
Ahn K, Felser C, Seshadri R, Kremer R K and Simon A
2000 {\it J.\ Alloys Compounds} {\bf 303-304} 252.
\item
Albert F J, Katine J A, Buhrmann R A and Ralph D C
2000 {\it Appl.\ Phys.\ Lett.} {\bf 77} 3809.
\item
Alers G B, Ramirez A P and Jin S
1996 {\it Appl.\ Phys.\ Lett.} {\bf 68} 3644.
\item
Alexandrov A S and Bratkovsky A M
1999a {\it Phys.\ Rev.\ Lett.} {\bf 82} 141.
\item
Alexandrov A S and Bratkowsky A M
1999b {\it J.\ Phys.: Condens.\ Matter} {\bf 11} 1989.
\item
Alexandrov A S and Bratkowsky A M
1999c {\it Phys.\ Rev.\ B} {\bf 60} 6215.
\item
Alexandrov A S and Bratkowsky A M
1999d {\it J.\ Phys.: Condens.\ Matter} {\bf 11} L531.
\item
Alexandrov A S, Zhao G-M, Keller H, Lorenz B, Wang Y S and Chu C W
2001 {\it Phys.\ Rev.\ B} {\bf 64} 140404(R).
\item
Allen P B, Berger H, Chauvet O, Forro L, Jarlborg T, Junod A, Revaz B
and Santi G
1996 {\it Phys.\ Rev.\ B} {\bf 53} 4393.
\item
Allodi G, De Renzi R and Guidi G
1998 {\it Phys.\ Rev.\ B} {\bf 57} 1024.
\item
Alonso J A, Mart{\'\i}nez-Lope M J, Casais M T and Fern\'andez-D{\'\i}az M T
1999 {\it Phys.\ Rev.\ Lett.} {\bf 82} 189.
\item
Alonso J A, Casais M T, Mart{\'\i}nez-Lope M J, Mart{\'\i}nez J L, Velasco
P, Mu\~{n}oz A and Fern\'andez-D{\'\i}az M T
2000 {\it Chem.\ Mater.} {\bf 12} 161.
\item
Alonso J L, Fern\'andez L A, Guinea F, Laliena V and Mart{\'\i}n-Mayor V
2001 {\it Phys.\ Rev.\ B} {\bf 63} 064416.
\item
Alvarado S F, Eib W, Meier F, Pierce D T, Sattler K, Siegmann H C and
Remeika J P
1975 {\it Phys.\ Rev.\ Lett.} {\bf 34} 319.
\item
Alvarado S F
1979 {\it Z.\ Phys.\ B} {\bf 33} 51.
\item
Anderson P W and Hasegawa H
1955 {\it Phys.\ Rev.} {\bf 100} 675.
\item
Anisimov V I, Elfimov I S, Hamada N and Terakura K
1996 {\it Phys.\ Rev.\ B} {\bf 54} 4387.
\item
Ansermet J-Ph
1998 {\it J.\ Phys.: Condens.\ Matter} {\bf 10} 6027.
\item
Archibald W, Zhou J-S and Goodenough J B
1996 {\it Phys.\ Rev.\ B} {\bf 53} 14445.
\item
Arovas D and Guinea F
1998 {\it Phys.\ Rev.\ B} {\bf 58} 9150.
\item
Asamitsu A and Tokura Y
1998 {\it Phys.\ Rev.\ B} {\bf 58} 47.
\item
Asano H, Ogale S B, Garrison J, Orozco A, Li Y H, Smolyaninova V, Galley C,
Downes M, Rajeswari M, Ramesh R and Venkatesan T
1999 {\it Appl.\ Phys.\ Lett.} {\bf 74} 3696.
\item
Babushkina N A, Belova L M, Gorbenko O Yu, Kaul A R, Bosak A A,
Ozhogin V I and Kugel K I
1998 Nature (London) {\bf 391} 159.
\item
Balcells Ll, Enrich R, Mora J, Calleja A, Fontcuberta J and Obradors X
1996 {\it Appl.\ Phys.\ Lett.} {\bf 69} 1486.
\item
Balcells Ll, Fontcuberta J, Mart{\'\i}nez B and Obradors X
1998a {\it J.\ Phys.: Condens.\ Matter} {\bf 10} 1883.
\item
Balcells Ll, Fontcuberta J, Mar{\'\i}nez B and Obradors X
1998b {\it Phys.\ Rev. B} {\bf 58} R14697.
\item
Balcells Ll, Carrillo A E, Mart{\'\i}nez B and Fontcuberta J
1999 {\it Appl.\ Phys.\ Lett.} {\bf 74} 4014.
\item
Balcells Ll, Mart{\'\i}nez B, Sandiumenge F and Fontcuberta J
2000 {\it J.\ Phys.: Condens. Matter} {\bf 12} 3013.
\item
Balcells Ll, Navarro J, Bibes M, Roig A, Mart{\'\i}nez B and Fontcuberta
J
2001 {\it Appl.\ Phys.\ Lett.} {\bf 78} 781.
\item
Ball A R, Leenaers A J G, van der Zaag P J, Shaw K A, Singer B, Lind D
M, Frederikze H and Rekveldt M Th
1996 {\it Appl.\ Phys.\ Lett.} {\bf 69} 583.
\item
Barn\'as J and Fert A
1998 {\it Phys.\ Rev.\ Lett.} {\bf 80} 1058.
\item
Barry A, Coey J M D and Viret M
2001 {\it J.\ Phys.: Condens. Matter} {\bf 12} L173.
\item
Barry A, Coey J M D, Ranno L and Ounadjela K
1998 {\it J.\ Appl.\ Phys.} {\bf 83} 7166.
\item
Barzilai S, Goldstein Y, Balberg I and Helman J S
1981 {\it Phys.\ Rev.\ B} {\bf 23} 1809.
\item
Bastiaansen P J M and Knops H J F
1998 {\it J.\ Phys.\ Chem.\ Solids} {\bf 59} 297.
\item
Baszy\'nski J, Kova\v{c} J and Kowalczyk A
1999 {\it J.\ Magn.\ Magn.\ Mater.} {\bf 195} 93.
\item
Battle P D, Blundell S J, Green M A, Hayer W, Honold M, Klehe A K,
Laskey N S, Millburn J E, Murphy L, Rosseinsky M J, Samarin N A,
Singleton J, Sluchanko N E, Sullivan S P and Vente J F
1996 {\it J.\ Phys.: Condens. Matter} {\bf 8} L427.
\item
Battle P D, Kasmir N, Millburn J E, Rosseinsky M J, Patel R T, Spring L E,
Vente J F, Blundell S J, Hayes W, Klehe A K, Mihut A and Singleton J
1998 {\it J.\ Appl.\ Phys.} {\bf 83} 6379.
\item
Bazaliy Ya B, Jones B A and Zhang S-C
1998 {\it Phys.\ Rev.\ B} {\bf 57} R3113.
\item
Belov K P, Goryaga A N, Pronin V N and Skipetrova L A
1982 {\it Pis'ma Zh.\ Eksp.\ Teor.\ Fiz.} {\bf 36} 118
[1982 {\it JETP Lett.} {\bf 36} 146].
\item
Belov K P, Goryaga A N, Pronin V N and Skipetrova L A
1983 {\it Pis'ma Zh.\ Eksp.\ Teor.\ Fiz.} {\bf 37} 392
[1983 {\it JETP Lett.} {\bf 37} 464].
\item
Berger L
1978 {\it J.\ Appl.\ Phys.} {\bf 49} 2156.
\item
Berger L
1991 {\it J.\ Appl.\ Phys.} {\bf 69} 1550.
\item
Berry S D, Lind D M, Chern G and Mathias H
1993 {\it J.\ Magn.\ Magn.\ Mater.} {\bf 123} 126.
\item
Bhattacharjee S and Sardar M
2000 {\it Phys.\ Rev.\ B} {\bf 62} R6139.
\item
Bibes M, Mart{\'\i}nez B, Fontcuberta J, Trtik V, Ben{\'\i}tez F, S\'anchez
F and Varela M
1999a {\it Appl.\ Phys.\ Lett.} {\bf 75} 2120.
\item
Bibes M, Mart{\'\i}nez B, Fontcuberta J, Trtik V, Benitez F, Ferrater C,
S\'anchez F and Varela M
1999b {\it Phys.\ Rev.\ B} {\bf 60} 9579.
\item
Bibes M, Balcells Ll, Valencia S, Fontcuberta J, Wojcik M, Jedryka E
and Nadolski S
2001 {\it Phys.\ Rev.\ Lett.} {\bf 87} 067210.
\item
Bj\"ornsson P, R\"ubhausen M, B\"ackstr\"om J, K\"all M, Eriksson S,
Eriksen J and B\"orjesson L
2000 {\it Phys.\ Rev.\ B} {\bf 61} 1193.
\item
Bonetti E, Del Bianco L, Fiorani D, Rinaldi D, Caciuffo R and Hernando
A
1999 {\it Phys.\ Rev.\ Lett.} {\bf 83} 2829.
\item
Booth C H, Bridges F, Kwei G H, Lawrence J M, Cornelius A L and
Neumeier J J
1998 {\it Phys.\ Rev.\ Lett.} {\bf 80} 853.
\item
Borges R P, Thomas R M, Cullinan C, Coey J M D, Suryanarayanan R,
Ben-Dor L, Pinsard-Gaudart L and Revcolevschi A
1999 {\it J.\ Phys.: Condens.\ Matter} {\bf 11} L445.
\item
Brabers V A M
in {\it Handbook of Magnetic Materials, Vol.\ 8} 
edited by Buschow K H J (North-Holland Publishing Company, Amsterdam) p.\ 189.
\item
Brabers V A M and Whall T E
in {\it Landolt-B\"ornstein} New Series III/27d p.~17.
\item
Brataas A, Nazarov Yu V, Inoue J and Bauer G E W
1999a {\it Phys.\ Rev.\ B} {\bf 59} 93.
\item
Brataas A, Tatara G and Bauer G E W
1999b {\it Phys.\ Rev.\ B} {\bf 60} 3406.
\item
Bratkovsky A M
1997 {\it Phys.\ Rev.\ B} {\bf 56} 2344.
\item
Brener N E, Tyler J M, Callaway J, Bagayoko D and Zhao G L
2000 {\it Phys.\ Rev.\ B} {\bf 61} 16582.
\item
Brey L
1999 {\it cond-mat/9905209} preprint.
\item
Bruno P
1999 {\it Phys.\ Rev.\ Lett.} {\bf 83} 2425.
\item
Br\"uckl H, Reiss G, Vinzelberg H, Bertram M, M\"onch I and Schumann J
1998 {\it Phys.\ Rev.\ B} {\bf 58} R8893.
\item
Busch G, Campagna M, Cotti P and Siegmann H Ch
1969 {\it Phys.\ Rev.\ Lett.} {\bf 22} 597.
\item
B\"uttiker M
1988 {\it IBM J.\ Res.\ Dev.} {\bf 32} 317.
\item
Cabrera G G and Falicov L M
1974a {\it phys.\ stat.\ sol.\ (b)} {\bf 61} 539.
\item
Cabrera G G and Falicov L M
1974b {\it phys.\ stat.\ sol.\ (b)} {\bf 62} 217.
\item
Cai Y Q, Ritter M, Weiss W and Bradshaw A M
1998 {\it Phys.\ Rev.\ B} {\bf 58} 5043.
\item
Calder\'on M J, Verg\'es J A and Brey L
1999a {\it Phys.\ Rev.\ Lett.} {\bf 59} 4170.
\item
Calder\'on M J, Brey L and Guinea F
1999b {\it Phys.\ Rev. B} {\bf 60} 6698.
\item
Calder\'on M J and Brey L
2001 {\it Phys.\ Rev.\ B} {\bf 64} 140403.
\item
Callaghan A, Moeller C W and Ward R
1966 {\it Inorg.\ Chem.} {\bf 5} 1572.
\item
Campbell I A, Fert A and Jaoul O
1970 {\it J.\ Phys.\ C: Solid State Phys.} {\bf 3} S95.
\item Campbell I A and Fert A
1982 in {\it Ferromagnetic Materials, Vol.\ 3} 
edited by Wohlfarth E P (North-Holland Publishing Company, Amsterdam) p.\ 751.
\item
Cao G, McCall S, Shepard M, Crow J E and Guertin R P
1997 {\it Phys.\ Rev.\ B} {\bf 56} 321.
\item
Capone M, Feinberg D and Grilli M
2000 {\it Eur.\ Phys.\ J.\ B} {\bf 17} 103.
\item
Chainani A, Yokoya T, Moromoto T, Takahashi T and Todo S
1995 {\it Phys.\ Rev.\ B} {\bf 51} 17976.
\item
Chattopadhyay A, Millis A J and Das Sarma S
2000 {\it Phys.\ Rev.\ B} {\bf 61} 10738.
\item
Chazalviel J-N and Yafet Y
1977 {\it Phys.\ Rev.\ B} {\bf 15} 1062.
\item
Chen P and Du Y W
2000 {\it J.\ Magn.\ Magn.\ Mater.} {\bf 219} 265.
\item
Chen P, Xing D Y, Du Y W, Zhu J M and Feng D
2001 {\it Phys.\ Rev.\ Lett.} {\bf 87} 107202.
\item
Chikazumi S
1997 {\it Physics of Ferromagnetism}
(Clarendon Press, Oxford) p.\ 163ff.
\item
Chmaissem O, Kruk R, Dabrowski B, Brown D E, Xiong X, Kolesnik S,
Jorgensen J D and Kimball C W,
2000 {\it Phys.\ Rev.\ B} {\bf 62} 14197.
\item
Choi J, Zhang J, Liou S-H, Dowben P A and Plummer E W
1999a {\it Phys.\ Rev.\ B} {\bf 59} 13453.
\item
Choi J, Dulli H, Liou S-H, Dowben P A and Langell M A
1999b {\it phys.\ stat.\ sol.\ (b)} {\bf 214} 45.
\item
Chuprakov I S and Dahmen K H
1998 {\it Appl.\ Phys.\ Lett.} {\bf 72} 2165.
\item 
Coey J M D, Viret M, Ranno L and Ounadjela K
1995 {\it Phys.\ Rev.\ Lett.} {\bf 75} 3910.
\item
Coey J M D, Berkowitz A E, Balcells Ll, Putris F F and Parker F T
1998a {\it Appl.\ Phys.\ Lett.} {\bf 72} 734.
\item
Coey J M D, Berkowitz A E, Balcells Ll, Putris F F and Barry A
1998b {\it Phys.\ Rev.\ Lett.} {\bf 80} 3815.
\item
Coey J M D
1998 {\it Phil.\ Trans.\ R.\ Soc.\ Lond.\ A} {\bf 356} 1519.
\item
Coey J M D
1999 {\it J.\ Appl.\ Phys.} {\bf 85} 5576.
\item
Coey J M D, Viret M and von Moln\'ar S
1999 {\it Adv.\ Phys.} {\bf 48} 167.
\item 
Coey J M D
``Materials for Spin Electronics'', in {\it Spin Electronics} edited by
Ziese M and Thornton M J (Springer, Heidelberg, 2001).
\item
Coombes D J and Gehring G A
1998 {\it J.\ Magn.\ Magn.\ Mater.} {\bf 177-181} 862.
\item
Dagotto E, Hotta T and Moreo A
2001 {\it Phys.\ Rep.} {\bf 344} 1.
\item
Dai J, Tang J, Xu H, Spinu L, Wang W, Wang K, Kumbhar A, Li M and
Diebold U
2000 {\it Appl.\ Phys.\ Lett.} {\bf 77} 2840.
\item
Dai J and Tang J
2001a {\it Phys.\ Rev.\ B} {\bf 63} 054434.
\item
Dai J and Tang J
2001b {\it Phys.\ Rev.\ B} {\bf 63} 064410.
\item
Dai P, Fernandez-Baca J A, Wakabayashi N, Plummer E W, Tomioka Y and
Tokura Y
2000 {\it Phys.\ Rev.\ Lett.} {\bf 85} 2553.
\item
Dass R I and Goodenough J B
2001 {\it Phys.\ Rev.\ B} {\bf 63} 064417.
\item
de Andr\'es A, Garc{\'\i}a-Hern\'andez M and Mart{\'\i}nez J L
1999 {\it Phys.\ Rev.\ B} {\bf 60} 7328.
\item
de Boer P K, van Leuken H, de Groot R A, Rojo T and Barberis G E
1997 {\it Solid State Commun.} {\bf 102} 621.
\item
Degiorgi L, Wachter P and Ihle D
1987 {\it Phys.\ Rev.\ B} {\bf 35} 9259.
\item
de Gennes P-G
1960 {\it Phys.\ Rev.} {\bf 118} 141.
\item
de Groot R A, Mueller F M, van Engen P G and Buschow K H J
1983 {\it Phys.\ Rev.\ Lett.} {\bf 50} 2024.
\item
de Groot R A and Buschow K H J
1986 {\it J.\ Magn.\ Magn.\ Mater.} {\bf 54-57} 1377.
\item
de Teresa J M, Ibarra M R, Algarabel P A, Ritter C, Marquina C, Blasco J,
Garc{\'\i}a J del Moral A and Arnold Z
1997 {\it Nature (London)} {\bf 387} 256.
\item
de Teresa J M, Barth\'el\'emy A, Fert A, Contour J P, Lyonnet R,
Montaigne F, Seneor P and Vaur\`es A
1999a {\it Phys.\ Rev.\ Lett.} {\bf 82} 4288.
\item
de Teresa J M, Barth\'el\'emy A, Fert A, Contour J P, Montaigne F and
Seneor P
1999b {\it Science} {\bf 286} 507.
\item
Dodge J S, Weber C P, Corson J, Orenstein J, Schlesinger Z, Reiner J W
and Beasley M R
2000 {\it Phys.\ Rev.\ Lett.} {\bf 85} 4932.
\item
Domenicali C A
1950 {\it Phys.\ Rev.} {\bf 78} 458.
\item
Dong Z W, Ramesh R, Venkatesan T, Johnson M, Chen Z Y, Pai S P,
Talyansky V, Sharma R P, Shreekala R, Lobb C J and Greene R L
1997 {\it Appl.\ Phys.\ Lett.} {\bf 71} 1718.
\item
Dong Z W, Pai S P, Ramesh R, Venkatesan T, Johnson M, Chen Z Y,
Cavanaugh A, Zhao Y G, Jiang X L, Sharma R P, Ogale S and Greene R L
1998 {\it J.\ Appl.\ Phys.} {\bf 83} 6780.
\item
D\"orr K, M\"uller K-H, Ruck K, Krabbes G and Schultz L
1999 {\it J.\ Appl.\ Phys.} {\bf 85} 5420.
\item
Dulli H, Plummer E W, Dowben P A, Choi J and Liou S-H
2000a {\it Appl.\ Phys.\ Lett.} {\bf 77} 570.
\item
Dulli H, Dowben P A, Liou S-H and Plummer E W
2000b {\it Phys.\ Rev.\ B} {\bf 62} R14629.
\item
Dutta P, Dimon P and Horn P M
1979 {\it Phys.\ Rev.\ Lett.} {\bf 43} 646.
\item
Dzero M O, Gor'kov L P and Kresin V Z
2000 {\it Eur.\ Phys.\ J B} {\bf 14} 459.
\item
Ebels U, Radulescu A, Henry Y, Piraux L and Ounadjela K
2000 {\it Phys.\ Rev.\ Lett.} {\bf 84} 983.
\item
Eckstein J N, Bozovic I, O'Donnell J, Onellion M and Rzchowski M S
1996 {\it Appl.\ Phys.\ Lett.} {\bf 69} 1312.
\item
Eib W and Alvarado S F
1976 {\it Phys.\ Rev.\ Lett.} {\bf 37} 444.
\item
Emery V J and Kivelson S A
1995 {\it Phys.\ Rev.\ Lett.} {\bf 74} 3253.
\item
Emin D and Holstein T
1976 {\it Phys.\ Rev.\ Lett.} {\bf 36} 323.
\item
Emin D, Hillery M S and Liu N-L H
1987 {\it Phys.\ Rev.\ B} {\bf 35} 641.
\item
Evetts J E, Blamire M G, Mathur N D, Isaac S P, Teo B-S, Cohen L F
and MacManus-Driscoll J L
1998 {\it Phil.\ Trans.\ R.\ Soc.\ Lond.\  A} {\bf 356} 1593.
\item
Fang Z, Terakura K and Kanamori J
2001 {\it cond-mat/0103189} preprint.
\item
F\"ath M, Freisem S, Menovsky A A, Tomioka Y, Aarts J and Mydosh J A
1999 {\it Science} {\bf 285} 1540.
\item
Feng J S-Y, Pashley R D and Nicolet M-A
1975 {\it J.\ Phys.\ C: Solid State Phys.} {\bf 8} 1010.
\item
Fert A and Campbell I A
1968 {\it Phys.\ Rev.\ Lett.} {\bf 21} 1190.
\item
Filippetti A and Pickett W E
1999 {\it Phys.\ Rev.\ Lett.} {\bf 20} 4184.
\item
Filippetti A and Pickett W E
2000 {\it Phys.\ Rev. B} {\bf 62} 11571.
\item
Fisk Z and Webb G W
1976 {\it Phys.\ Rev.\ Lett.} {\bf 36} 1084.
\item
Fontcuberta J, Mart{\'\i}nez B, Seffar A, Pi\~nol S, Garc{\'\i}a-Mu\~noz J L
and Obradors X
1996 {\it Phys.\ Rev.\ Lett.} {\bf 76} 1122.
\item
Fontcuberta J, Mart{\'\i}nez B, Laukhin V, Balcells Ll, Obradors X,
Cohenca C H and Jardim R F
1998 {\it Phil.\ Trans.\ R.\ Soc.\ Lond.\ A} {\bf 356} 1577.
\item
Fontcuberta J, Bibes M, Mart{\'\i}nez B, Trtik V, Ferrater C, S\'anchez F
and Varela M
1999 {\it J.\ Appl.\ Phys.} {\bf 85} 4800.
\item
Fontcuberta J
1999 {\it Physics World} {\bf 12} 33.
\item
Franck J P, Isaac I, Chen W, Chrzanowski J and Irwin J C
1998 {\it Phys.\ Rev.\ B} {\bf 58} 5189.
\item
Friedman L
1964 {\it Phys.\ Rev.} {\bf 135} A233.
\item
Fujioka K, Okamato J, Mizokawa T, Fujimori A, Hase I, Abbate M, Lin H J, Chen C T,
Takeda Y and Takano M
1997 {\it Phys.\ Rev.\ B} {\bf 56} 6380.
\item
Fulde P, Luther A and Watson R E
1973 {\it Phys.\ Rev.\ B} {\bf 8} 440.
\item
Furukawa N
1994 {\it J.\ Phys.\ Soc.\ Japan} {\bf 63} 3214.
\item
Furukawa N
1995a {\it J.\ Phys.\ Soc.\ Japan} {\bf 64} 2734.
\item
Furukawa N
1995b {\it J.\ Phys.\ Soc.\ Japan} {\bf 64} 3164.
\item
Furukawa N
1998 {\it cond-mat/9812066} preprint.
\item
Furukawa N
2000 {\it J.\ Phys.\ Soc.\ Japan} {\bf 69} 1954.
\item
Gadzuk J W
1969 {\it Phys.\ Rev.} {\bf 182} 416.
\item
Gan Q, Rao R A, Eom C B, Garrett J L and Lee M
1998 {\it Appl.\ Phys.\ Lett.} {\bf 72} 978.
\item
Garc{\'\i}a N, Mu\~noz M and Zhao Y-W
1999 {\it Phys.\ Rev.\ Lett.} {\bf 82} 2923.
\item
Garc{\'\i}a-Landa B, Ritter C, Ibarra M R, Blasco J, Algarabel P A,
Mahendiran R and Garc{\'\i}a J
1999 {\it Solid State Commun.} {\bf 110} 435.
\item
Geck J, B\"uchner B, H\"ucker M, Klingeler R, Gross R, Pinsard-Gaudart
L and Revcolevschi A
2001 {\it Phys.\ Rev.\ B} {\bf 64} 144430.
\item
Gehring G A
1997 {\it unpublished}.
\item
Gerlach W and Schneiderhan K
1930 {\it Ann. Physik} {\bf 6} 772.
\item
Ghosh K, Ogale S B, Pai S P, Robson M, Li E, Jin I, Dong Z W,
Greene R L, Ramesh R and Venkatesan T
1998 {\it Appl.\ Phys.\ Lett.} {\bf 73} 689.
\item
Gibbs M R J, Ziese M, Gehring G A, Blythe H J, Coombes D J, Sena S P
and Shearwood C
1998 {\it Phil.\ Trans.\ R.\ Soc.\ Lond.\ A} {\bf 356} 1681.
\item
Gilabert A, Plecenik A, Fr\"ohlich K, Ga\v{z}i \v{S}, Pripko M,
Mozolov\'a \v{Z}, Machajd{\'\i}k D, Be\v{n}a\v{c}ka \v{S}, Medici M G,
Grajcar M and K\'u\v{s} P
2001 {\it Appl.\ Phys.\ Lett.} {\bf 78} 1712.
\item
Gillman E S, Li M and Dahmen K-H
1998 {\it J.\ Appl.\ Phys.} {\bf 84} 6217.
\item
Gittleman J I, Goldstein Y and Bozowski S
1972 {\it Phys.\ Rev.\ B} {\bf 5} 3609.
\item
Glazman L I and Matveev K A
1988 {\it Zh.\ Eksp.\ Teor.} {\bf 94} 332
[{\it Sov.\ Phys.\ JETP} {\bf 67} 1276].
\item
Gong G Q, Gupta A, Xiao G, Qian W and Dravid V P
1997 {\it Phys.\ Rev.\ B} {\bf 56} 5096.
\item
Goodenough J B and Longo M
in {\it Landolt-B\"ornstein} III/4, p.~126.
\item
Goodenough J B
1955 {\it Phys.\ Rev.} {\bf 100} 564.
\item
Goodenough J B
1958 {\it J.\ Phys.\ Chem.\ Solids} {\bf 10} 287.
\item
Goodenough J B, Wold A, Arnott R J and Menyuk N
1961 {\it Phys.\ Rev.} {\bf 124} 373.
\item
Goodenough J B
1992 {\it Ferroelectrics} {\bf 130} 77.
\item
Goodenough J B
1997 {\it J.\ Appl.\ Phys.} {\bf 81} 5330.
\item
Goodenough J B
1999 {\it Aust.\ J.\ Phys} {\bf 52} 155.
\item
Gopalakrishnan J, Chattopadhyay A, Ogale S B, Venkatesan T, Greene R
L, Millis A J, Ramesha K, Hannoyer B and Marest G
2000 Phys.\ Rev.\ B{\bf 61} 9538.
\item
Gregg J F, Allen W, Ounadjela K, Viret M, Hehn M, Thompson S M and Coey J M D
1996 {\it Phys.\ Rev.\ Lett.} {\bf 77} 1580.
\item
Greneche J M, Venkatesan M, Suryanarayanan R and Coey J M D
2001 {\it Phys.\ Rev.\ B} {\bf 63} 174403.
\item
Gr\'evin B, Maggio-Aprile I, Bentzen A, Ranno L, Llobet A and Fischer {\O}
2000 {\it Phys.\ Rev.\ B} {\bf 62} 8596.
\item
Gridin V V, Hearne G R and Honig J M
1996 {\it Phys.\ Rev.\ B} {\bf 53} 15518.
\item
Grollier J, Cros V, Hamzic A, George J M, Jaffr\`es H, Fert A, Faini
G, Ben Youssef J and Legall H
2001 {\it Appl.\ Phys.\ Lett.} {\bf 78} 3663.
\item
Guinea F
1998 {\it Phys.\ Rev.\ B} {\bf 58} 9212.
\item
Gupta A, Gong G Q, Xiao G, Duncombe P R, Lecoeur P, Trouilloud P,
Wang Y Y, Dravid V P and Sun J Z
1996 {\it Phys.\ Rev.\ B} {\bf 54} R15629.
\item
Gupta A and Sun J Z
1999 {\it J.\ Magn.\ Magn.\ Mater.} {\bf 200} 24.
\item
Gupta S, Ranjit R, Mitra C, Raychaudhuri P and Pinto R
2001 {\it Appl. Phys.\ Lett.} {\bf 78} 362.
\item
Hardner H T, Weissmann M B, Jaime M, Treece R E, Dorsey P C, Horwitz J
S and Chrisey D B
1997 {\it J.\ Appl.\ Phys.} {\bf 81} 272.
\item
Heaps C W
1934 {\it Phys.\ Rev.} {\bf 45} 320.
\item
Heide C, Zilberman P E and Elliott R J
2001 {\it Phys.\ Rev.\ B} {\bf 63} 064424.
\item
Helman J S and Abeles B
1976 {\it Phys.\ Rev.\ Lett.} {\bf 37} 1429.
\item
Hertz J A and Aoi K
1973 {\it Phys.\ Rev.\ B} {\bf 8} 3252.
\item
Hillery M S, Emin D and Liu N-L H
1988 {\it Phys.\ Rev.\ B} {\bf 38} 9771.
\item
H\"ofener C, Philipp J B, Klein J, Alff L, Marx A, B\"uchner B and
Gross R
2000 {\it Europhys.\ Lett.} {\bf 50} 681.
\item
Holstein T
1959 {\it Ann.\ Phys.\ (N.Y.)} {\bf 8} 325.
\item
Honda S, Okada T, Nawate M and Tokumoto M
1997 {\it Phys.\ Rev. B} {\bf 56} 14566.
\item
Hong K and Giordano N
1998 {\it J.\ Phys.: Condens.\ Matter} {\bf 10} L401.
\item
Hueso L E, Rivas J, Rivadulla F and L\'opez-Quintela M A
1999 {\it J.\ Appl.\ Phys.} {\bf 86} 3881.
\item
Hwang H Y, Cheong S-W, Radaelli P G, Marezio M and Batlogg B
1995 {\it Phys.\ Rev.\ Lett.} {\bf 75} 914.
\item
Hwang H Y, Cheong S W, Ong N P and Batlogg B
1996 {\it Phys.\ Rev.\ Lett.} {\bf 77} 2041.
\item
Hwang H Y and Cheong S-W
1997a {\it Science} {\bf 278} 1607.
\item
Hwang H Y and Cheong S-W
1997b {\it Nature (London)} {\bf 389} 942.
\item
Ihle D and Lorenz B
1985 {\it J.\ Phys.\ C: Solid State Phys.} {\bf 18} L647.
\item
Ihle D and Lorenz B
1986 {\it J.\ Phys.\ C: Solid State Phys.} {\bf 19} 5239.
\item
Ijiri Y, Borchers J A, Erwin R W, Lee S-H, van der Zaag P J 
and Wolf R M
1998a {\it Phys.\ Rev.\ Lett.} {\bf 80} 608.
\item
Ijiri Y, Borchers J A, Erwin R W, Lee S-H, van der Zaag P J 
and Wolf R M
1998b {\it J.\ Appl.\ Phys.} {\bf 83} 6882.
\item
Imada M, Fujimori A and Tokura Y
1998 {\it Rev.\ Mod.\ Phys.} {\bf 70} 1039.
\item
Imai H, Shimakawa Y, Sushko Yu V and Kubo Y
2000 {\it Phys.\ Rev.\ B} {\bf 62} 12190.
\item
Inoue J and Maekawa S
1995 {\it Phys.\ Rev.\ Lett.} {\bf 74} 3407.
\item
Inoue J and Maekawa S
1996 {\it Phys.\ Rev.\ B} {\bf 53} R11927.
\item
Ioffe A F and Regel A R
1960 {\it Prog.\ Semicond.} {\bf 4} 237.
\item
Irkhin V Yu and Katsnel'son M I
1994 {\it Usp.\ Fiz.\ Nauk} {\bf 164} 705
[1994 {\it Physics-Uspekhi} {\bf 37} 659].
\item
Isaac S P, Mathur N D, Evetts J E and Blamire M G
1998 {\it Appl.\ Phys.\ Lett.} {\bf 72} 2038.
\item
Ishizaka S and Ishibara S
1999 {\it Phys.\ Rev.\ B} {\bf 59} 8375.
\item
Jaime M, Salamon M B, Rubinstein M, Treece R E, Horwitz J S and Chrisey D B
1996 {\it Phys.\ Rev.\ B} {\bf 54} 11914.
\item
Jaime M, Hardner H T, Salamon M B, Rubinstein M, Dorsey P and Emin D
1997 {\it Phys.\ Rev.\ Lett.} {\bf 78} 951.
\item
Jaime M, Lin P, Salamon M B and Han P D
1998 {\it Phys.\ Rev.\ B} {\bf 58} R5901.
\item
Jaime M and Salamon M B
1999 {\it cond-mat/9902284} preprint.
\item
Jakob G, Moshchalkov V V and Bruynseraede Y
1995 {\it Appl.\ Phys.\ Lett.} {\bf 66} 2564.
\item
Jakob G, Martin F, Westerburg W and Adrian H
1998 {\it Phys.\ Rev.\ B} {\bf 57} 10252.
\item
Ji Y, Strijkers G J, Yang F Y, Chien C L, Byers J M, Anguelouch A,
Xiao G and Gupta A
2001 {\it Phys.\ Rev.\ Lett.} {\bf 86} 5585.
\item
Jin S, O'Bryan H M, Tiefel T H, McCormack M and Rhodes W W
1995a {\it Appl.\ Phys.\ Lett.} {\bf 66} 382.
\item
Jin S, Tiefel T H, McCormack M, O'Bryan H M, Chen L H, Ramesh R and
Schurig D
1995b {\it Appl.\ Phys.\ Lett.} {\bf 67} 557.
\item
Johnson M
1993a {\it Phys.\ Rev.\ Lett.} {\bf 70} 2142.
\item
Johnson M
1993b {\it Science} {\bf 260} 320.
\item
Jonker G and van Santen J
1950 {\it Physica} {\bf 16} 599.
\item
Ju H L, Sohn H-C and Krishnan K M
1997 {\it Phys.\ Rev.\ Lett.} {\bf 79} 3230.
\item
Julliere M
1975 {\it Phys.\ Lett.} {\bf 54A} 225.
\item
Kacedon D B, Rao R A and Eom C B
1997 {\it Appl.\ Phys.\ Lett.} {\bf 71} 1724.
\item
Kagan M Yu, Khomskii D I and Mostovoy M V
1999 {\it Eur.\ Phys.\ J. B} {\bf 12} 217.
\item
Kanamori J
1958 {\it J.\ Phys.\ Chem.\ Solids} {\bf 10} 87.
\item
K\"amper K P, Schmitt W P, G\"untherodt G, Gambino R J and Ruf R
1987 {\it Phys.\ Rev.\ Lett.} {\bf 59} 2788.
\item
Kapusta Cz, Riedi P C, Kocemba W, Tomka G J, Ibarra M R, de Teresa J
M, Viret M and Coey J M D
1999 {\it J.\ Phys.: Condens.\ Matter} {\bf 11} 4079.
\item
Kar A K, Dhar A, Ray S K, Mathur B K, Bhattacharya D and Chopra K L
1998 {\it J.\ Phys.: Condens.\ Matter} {\bf 10} 10795.
\item
Katine J A, Albert F J, Buhrman R A, Myers E B and Ralph D C
2000 {\it Phys.\ Rev.\ Lett.} {\bf 84} 3149.
\item
Kent A D, R\"udiger U, Yu J, Thomas L and Parkin S S P
1999 {\it J.\ Appl.\ Phys.} {\bf 85} 5243.
\item
Khomskii D
2000 Physica B {\bf 280} 325.
\item
Kim T H, Uehara M, Cheong S-W and Lee S
1999 {\it Appl.\ Phys.\ Lett.} {\bf 74} 1737.
\item
Kimura T, Tomioka Y, Kuwahara H, Asamitsu A, Tamura M and Tokura Y
1996 {\it Science} {\bf 274} 1698.
\item
Kitamoto Y, Nakayama Y and Abe M
2000 {\it J.\ Appl.\ Phys.} {\bf 87} 7130.
\item
Klein L, Dodge J S, Ahn C H, Snyder G J, Geballe T H, Beasley M R and Kapitulnik A
1996 {\it Phys.\ Rev.\ Lett.} {\bf 77} 2774.
\item
Klein L, Marshall A F, Reiner J W, Ahn C H, Geballe T H, Beasley M R and Kapitulnik A
1998 {\it J.\ Magn.\ Magn.\ Mater.} {\bf 188} 319.
\item
Klein J, H\"ofener C, Uhlenbruck S, Alff L, B\"uchner B and Gross R
1999 {\it Europhys.\ Lett.} {\bf 47} 371.
\item
Kleint C A, Krause M K, H\"ohne R, Walter T, Semmelhack H C,
Lorenz M and Esquinazi P
1998 {\it J.\ Appl.\ Phys.} {\bf 84} 5097.
\item
Kobayashi K-I, Kimura T, Sawada H, Terakura K and Tokura Y
1998 {\it Nature (London)} {\bf 395} 677.
\item
Kobayashi K-I, Kimura T, Tomioka Y, Sawada H, Terakura K and Tokura Y
1999 {\it Phys.\ Rev.\ B} {\bf 59} 11159.
\item
Kobayashi K-I, Okuda T, Tomioka Y, Kimura T and Tokura Y
2000 {\it J.\ Magn.\ Magn.\ Mater.} {\bf 218} 17.
\item
Kogan E and Auslender M
1988 {\it phys.\ stat.\ sol.\ (b)} {\bf 147} 613.
\item
Kogan E and Auslender M
1998 {\it cond-mat/9807069} preprint.
\item
Kogan E, Auslender M and Kaveh M
1999 {\it Eur.\ Phys.\ J.\ B} {\bf 9} 373.
\item
Korotin M A, Anisimov V I, Khomskii D I and Sawatzky G A
1998 {\it Phys.\ Rev.\ Lett.} {\bf 80} 4305.
\item
Kostic P, Okada Y, Collins N C, Schlesinger Z, Reiner J W, Klein L, 
Kapitulnik A, Geballe T H and Beasley M R
1998 {\it Phys.\ Rev.\ Lett.} {\bf 81} 2498.
\item
Kostopoulos D 
1972 {\it phys.\ stat.\ sol.\ (a)} {\bf 9} 523.
\item
Kostopoulos D and Alexopoulos K
1976 {\it J.\ Appl.\ Phys.} {\bf 47} 1714;
\item
Krivorotov I N, Nikolaev K R, Dobin A Yu, Goldman A M and Dahlberg E D
2001 {\it Phys.\ Rev.\ Lett.} {\bf 86} 5779.
\item
Krupi\v{c}ka S and Nov\'ak P
in {\it Handbook of Magnetic Materials, Vol.\ 3} 
edited by Wohlfarth E P (North-Holland Publishing Company, Amsterdam) p.\ 189.
\item
Kubo K and Ohata N
1972 {\it J.\ Phys.\ Soc.\ Japan} {\bf 33} 21.
\item
Kwei G H, Booth C H, Bridges F and Subramanian M A
1997 {\it Phys.\ Rev.\ B} {\bf 55} R688.
\item
Kwon C, Kim K-C, Robson M C, Gu J Y, Rajeswari M and Venkatesan T
1997 {\it J.\ Appl.\ Phys.} {\bf 81} 4950.
\item
Kwon C, Jia Q X, Fan Y, Hundley M F, Reagor D W, Coulter J Y and Peterson D E
1998 {\it Appl.\ Phys.\ Lett.} {\bf 72} 486.
\item
Landauer R
1957 {\it IBM J.\ Res.\ Dev.} {\bf 1} 223.
\item
Lanzara A, Saini N L, Brunelli M, Natali F, Bianconi A, Radaelli P G
and Cheong S-W
1998 {\it Phys.\ Rev.\ Lett.} {\bf 81} 878.
\item
Lee S, Hwang H Y, Shraiman B I, Ratcliff II W D and Cheong S-W
1999 {\it Phys.\ Rev.\ Lett.} {\bf 82} 4508.
\item
Leung L K, Morrish A H and Searle C W
1969 {\it Can.\ J.\ Phys.} {\bf 47} 2697.
\item
Levy P M and Zhang S
1997 {\it Phys.\ Rev.\ Lett.} {\bf 79} 5110.
\item
Lewis S P, Allen P B and Sasaki T
1997 {\it Phys.\ Rev.\ B} {\bf 55} 10253.
\item 
Li Q, Zang J, Bishop A R and Soukoulis C M
1997a {\it Phys.\ Rev.\ B} {\bf 56} 4541.
\item
Li X W, Gupta A, Xiao G and Gong G Q
1997b {\it Appl.\ Phys.\ Lett.} {\bf 71} 1124.
\item
Li X W, Lu Y, Gong G Q, Xiao G, Gupta A, Lecoeur P, Sun J Z, Wang Y Y and Dravid V P
1997c {\it J.\ Appl.\ Phys.} {\bf 81} 5509.
\item
Li X W, Gupta A, Xiao G and Gong G Q
1998a {\it J.\ Appl.\ Phys.} {\bf 83} 7049.
\item
Li X W, Gupta A, Xiao G, Qian W and Dravid V P
1998b {\it Appl.\ Phys.\ Lett.} {\bf 73} 3282.
\item
Lind D M, Berry S D, Borchers J A, Erwin R W, Lochner E, Stoyonov P,
Shaw K A and Dibari R C
1995 {\it J.\ Magn.\ Magn.\ Mater.} {\bf 148} 44.
\item
Lind\'en J, Yamamoto T, Karppinen M, Yamauchi H and Pietari T
2000 {\it Appl.\ Phys.\ Lett.} {\bf 76} 2925.
\item
Liu G-L, Zhou J-S and Goodenough J B
2001 {\it Phys.\ Rev.\ B} {\bf 64} 144414.
\item
Liu J-M, Yuan G L, Sang H, Wu Z C, Chen X Y, Liu Z G, Du Y W, Huang Q
and Ong C K
2001 {\it Appl.\ Phys.\ Lett.} {\bf 78} 1110.
\item
Livesay E A, West R N, Dugdale S B, Santi G and Jarlborg T
1998 {\it cond-mat/9812308} preprint.
\item
Livesay E A, West R N, Dugdale S B, Santi G and Jarlborg T
1999 {\it J.\ Phys.: Condens.\ Matter} {\bf 11} L279.
\item
Lofland S E, Bhagat S M, Kwon C, Robson M C, Sharma R P, Ramesh R and
Venkatesan T
1995 {\it Phys.\ Lett.} {\bf 209A} 246.
\item
Longo J M and Ward R
1961 {\it J.\ Am.\ Chem.\ Soc.} {\bf 83} 1088.
\item
Longo J M, Raccah P M and Goodenough J B
1968 {\it J.\ Appl.\ Phys.} {\bf 39} 1327.
\item
Louren\c{c}o A A C S, Ar\'aujo J P, Amaral V S, Tavares P B, Sousa J B,
Vieira J M, Alves E, da Silva M F and Soares J C
1999 {\it J.\ Magn.\ Magn.\ Mater.} {\bf 196-197} 495.
\item
Lu Y, Li W, Gong G Q, Xiao G, Gupta A, Lecoeur P, Sun J Z, Wang Y Y and Dravid V P
1996 {\it Phys.\ Rev.\ B} {\bf 54} R8357.
\item
Lyanda-Geller Y, Aleiner I L and Goldbart P M
1998 {\it Phys.\ Rev.\ Lett.} {\bf 81} 3215.
\item
Lyanda-Geller Y, Chun S H, Salamon M B, Goldbart P M, Han P D, 
Tomioka Y, Asamitsu A and Tokura Y
2001 {\it Phys.\ Rev.\ B} {\bf 63} 184426.
\item
Lynn J W, Erwin R W, Borchers J A, Huang Q, Santoro A, Peng J-L and Li
Z Y
1996 {\it Phys.\ Rev.\ Lett.} {\bf 76} 4046.
\item
MacDonald D K C
1956 {\it Handbuch der Physik} {\bf 14} (Springer Verlag, Berlin) p.~137.
\item
Mackenzie A P, Reiner J W, Tyler A W, Galvin L M, Julian S R, Beasley M R,
Geballe T H and Kapitulnik A
1998 {\it Phys.\ Rev.\ B} {\bf 58} R13318.
\item
MacLaren J M, Zhang X G and Butler W H
1997 {\it Phys.\ Rev.\ B} {\bf 56} 11827.
\item
Majumdar P and Littlewood P B
1998a {\it Phys.\ Rev.\ Lett.} {\bf 81} 1314.
\item
Majumdar P and Littlewood P B
1998b {\it Nature (London)} {\bf 395} 479.
\item
Mallik R, Sampathkumaran E V and Paulose P L
1997 {\it Appl.\ Phys.\ Lett.} {\bf 71} 2385.
\item
Malozemoff A P
1986 {\it Phys.\ Rev.\ B} {\bf 34} 1853.
\item
Manako T, Izumi M, Konishi Y, Kobayashi K-I, Kawasaki M and Tokura Y
1999 {\it Appl.\ Phys.\ Lett.} {\bf 74} 2215.
\item
Manoharan S S, Elefant D, Reiss G and Goodenough J B
1998 {\it Appl.\ Phys.\ Lett.} {\bf 72} 984.
\item
Mart{\'\i}nez B, Balcells Ll, Fontcuberta J, Obradors X, Cohenca C H
and Jardim R F
1998 {\it J.\ Appl.\ Phys.} {\bf 83} 7058.
\item
Mart{\'\i}nez B, Senis R, Fontcuberta J, Obradors X, Cheikh-Rouhou W,
Strobel P, Bougerol-Chaillout C and Pernet M
1999 {\it Phys.\ Rev.\ Lett.} {\bf 83} 2022.
\item
Mart{\'\i}nez B, Navarro J, Balcells Ll and Fontcuberta J
2000 {\it J.\ Phys.: Condens.\ Matter} {\bf 12} 10515.
\item
Mathieu R, Svedlindh P, Chakalov R A and Ivanov Z G
2000 {\it Phys.\ Rev.\ B} {\bf 62} 3333.
\item
Mathieu R, Svedlindh P, Gunnarsson R and Ivanov Z G
2001a {\it Phys.\ Rev.\ B} {\bf 63} 132407.
\item
Mathieu R, Svedlindh P, Chakalov R and Ivanov Z G
2001b {\it cond-mat/0109052} preprint.
\item
Mathur N D, Burnell G, Isaac S P, Jackson T J, Teo B-S,
MacManus-Driscoll J, Cohen L F, Evetts J E and Blamire M G
1997 {\it Nature (London)} {\bf 387} 266.
\item
Mathur N D, Littlewood P B, Todd N K, Isaac S P, Teo B-S, Kang D-J,
Tarte E J, Barber Z H, Evetts J E and Blamire M G
1999 {\it J.\ Appl.\ Phys.} {\bf 86} 6287.
\item
Matl P, Ong N P, Yan Y F, Li Y Q, Studebaker D, Baum T and Doubinina G
1998 {\it Phys.\ Rev.\ B} {\bf 57} 10248.
\item
Mayr M, Moreo A, Verg\'es J A, Arispe J, Feiguin A and Dagotto E
2001 {\it Phys.\ Rev.\ Lett.} {\bf 86} 135.
\item
Mazin I I and Singh D J
1997 {\it Phys.\ Rev.\ B} {\bf 56} 2556.
\item
Mazin I I
1999 {\it Phys.\ Rev.\ Lett.} {\bf 83} 1427.
\item
Mazin I I, Singh D J and Ambrosch-Draxl C
1999 {\it Phys.\ Rev.\ B} {\bf 59} 411.
\item
McCormack M, Jin S, Tiefel T H, Fleming R M, Phillips J M and Ramesh R
1994 {\it Appl.\ Phys.\ Lett.} {\bf 64} 3045.
\item
Merrill R L and Si Q
1999 {\it Phys.\ Rev.\ Lett.} {\bf 83} 5326.
\item
Meservey R, Paraskevopoulos D and Tedrow P M
1976 {\it Phys.\ Rev.\ Lett.} {\bf 37} 858.
\item
Meservey R, Paraskevopoulos D and Tedrow P M
1980 {\it Phys.\ Rev.\ B} {\bf 22} 1331.
\item
Meservey R and Tedrow P M
1994 {\it Phys.\ Rep.} {\bf 238} 173.
\item
Mieville L, Worledge D, Geballe T H, Contreras R and Char K
1998 {\it Appl.\ Phys.\ Lett.} {\bf 73} 1736.
\item
Mikheenko P, Colclough M S, Severac C, Chakalov R, Welhoffer F and
Muirhead C M
2001 {\it Appl.\ Phys.\ Lett.} {\bf 78} 356.
\item
Miller D J, Lin Y K, Vlasko-Vlasov V and Welp U
2000 {\it J.\ Appl.\ Phys.} {\bf 87} 6758.
\item
Millis A J, Littlewood P B and Shraiman B I
1995 {\it Phys.\ Rev.\ Lett.} {\bf 74} 5144.
\item
Millis A J, Shraiman B I and Mueller R
1996a {\it Phys.\ Rev.\ Lett.} {\bf 77} 175.
\item
Millis A J, Mueller R and Shraiman B I
1996b {\it Phys.\ Rev.\ B} {\bf 54} 5389.
\item
Millis A J, Mueller R and Shraiman B I
1996c {\it Phys.\ Rev.\ B} {\bf 54} 5405.
\item
Millis A J, Darling T and Migliori A
1998 {\it J.\ Appl.\ Phys.} {\bf 83} 1588.
\item
Millis A J, Hu J and Das Sarma S
1999 {\it Phys.\ Rev.\ Lett.} {\bf 82} 2354.
\item
Mira J, Rivas J, Rivadulla F, V\'azquez-V\'azquez C and
L\'opez-Quintela M A
1999 {\it Phys.\ Rev.\ B} {\bf 60} 2998.
\item
Mitani S, Fujimori H and Ohnuma S
1997 {\it J.\ Magn.\ Magn.\ Mater.} {\bf 165} 141.
\item
Mitani S, Takahashi S, Takanashi K, Yakushiji K, Maekawa S and Fujimori H,
1998a {\it Phys.\ Rev.\ Lett.} {\bf 81} 2799.
\item
Mitani S, Takanashi K, Yakushiji K and Fujimori H
1998b {\it J.\ Appl.\ Phys.} {\bf 83} 6524.
\item
Miyazaki T, Tezuka N, Kumagai S, Ando Y, Kubota H, Murai J, Watabe T 
and Yokota M
1998 {\it J.\ Phys.\ D: Appl.\ Phys.} {\bf 31} 630.
\item
Monsma D J, Lodder J C, Popma Th J A and Dieny B
1995 {\it Phys.\ Rev.\ Lett.} {\bf 74} 5260.
\item
Monsma D J and Parkin S S P
2000a {\it Appl. Phys.\ Lett.} {\bf 77} 720.
\item
Monsma D J and Parkin S S P
2000b {\it Appl. Phys.\ Lett.} {\bf 77} 883.
\item
Montaigne F, Nassar J, Vaur\`es A, Nguyen Van Dau F, Petroff F, Schuhl A and Fert A
1998 {\it Appl.\ Phys.\ Lett.} {\bf 73} 2829.
\item
Moodera J S, Kinder L R, Wong T M and Meservey R
1995 {\it Phys.\ Rev.\ Lett.} {\bf 74} 3273.
\item
Moodera J S, Nowak J and van de Veerdonk R J M
1998 {\it Phys.\ Rev.\ Lett.} {\bf 80} 2941.
\item
Moodera J S and Mathon G
1999 {\it J.\ Magn.\ Magn.\ Mater.} {\bf 200} 248.
\item
Moon-Ho Jo, Mathur N D, Todd N K and Blamire M G
2000a {\it Phys.\ Rev.\ B} {\bf 61} R14905.
\item
Moon-Ho Jo, Mathur N D, Evetts J E and Blamire M G
2000b {\it Appl.\ Phys.\ Lett.} {\bf 77} 3803.
\item
Moreo A, Yunoki S and Dagotto E
1999 {\it Science} {\bf 283} 2034.
\item
Moritomo Y, Asamitsu A, Kuwahara H and Tokura Y
1996 {\it Nature (London)} {\bf 380} 141.
\item
Moritomo Y, Xu Sh, Machida A, Akimoto T, Nishibori E, Takata M and
Sakata M
2000 {\it Phys.\ Rev.\ B} {\bf 61} R7827.
\item
Morrish A H, Evans B J, Eaton J A and Leung L K
1969 {\it Can.\ J.\ Phys.} {\bf 47} 2691.
\item
Mott N F
1936 {\it Proc.\ Roy.\ Soc. (London)} {\bf 156A} 368.
\item
Mott N F
1978 {\it Rev.\ Mod.\ Phys.} {\bf 50} 203.
\item
Moutis N, Christides C, Panagiotopoulos I and Niarchos D
2001 {\it Phys.\ Rev.\ B} {\bf 64} 094429.
\item
Nadgorny B, Soulen Jr R J, Osofsky M S, Mazin I I, Laprade G, van de
Veerdonk R J M, Smits A A, Cheng S F, Skelton E F and Quadri S B
2000 {\it Phys.\ Rev.\ B} {\bf 61} R3788.
\item
Nagaev E L
1996 {\it Usp.\ Fiz.\ Nauk} {\bf 166} 833.
\item
Nagaev E L
1998 {\it Phys.\ Rev. B} {\bf 58} 12242.
\item
Nagaev E L
1999 {\it Aust.\ J.\ Phys.} {\bf 52} 305.
\item
Nagaev E L
2001 {\it Phys.\ Rep.} {\bf 346} 387.
\item
Narimanov E and Varma C M
2001 {\it cond-mat/0110047} preprint.
\item
Navarro J, Frontera C, Balcells Ll, Mart{\'\i}nez B and Fontcuberta J
2001 {\it Phys.\ Rev.\ B} {\bf 64} 092411.
\item
Niebieskikwiat D, S\'anchez R D, Caneiro A, Morales L,
V\'asquez-Mansilla M, Rivadulla F and Hueso L E
2000 {\it Phys.\ Rev.\ B} {\bf 62} 3340.
\item
Nikolaev K R, Bhattacharya A, Kraus P A, Vas'ko V A, Cooley W K and
Goldman A M
1999 {\it Appl.\ Phys.\ Lett.} {\bf 75} 188.
\item
Nikolaev K R, Krivorotov I N, Cooley W K, Bhattacharya A, Dahlberg E D
and Goldman A M
2000a {\it Appl.\ Phys.\ Lett.} {\bf 76} 478.
\item
Nikolaev K R, Dobin A Yu, Krivorotov I N, Cooley W K, Bhattacharya A,
Kobrinski A L, Glazman L I, Wentzovitch R M, Dahlberg E D and Goldman
A M
2000b {\it Phys.\ Rev.\ Lett.} {\bf 85} 3728.
\item
Nishimura K, Kohara Y, Kitamoto Y and Abe M
2000 {\it J.\ Appl.\ Phys.} {\bf 87} 7127.
\item
Noh J S, Nath T K, Eom C B, Sun J Z, Tian W and Pan X Q
2001 {\it Appl.\ Phys.\ Lett.} {\bf 79} 233.
\item
Nowak J and Rauluszkiewicz,
1992 {\it J.\ Magn.\ Magn.\ Mater.} {\bf 109} 79.
\item
Obata T, Manako T, Shimakawa Y and Kubo Y
1999 {\it Appl.\ Phys.\ Lett.} {\bf 74} 290.
\item
O'Donnell J, Onellion M, Rzchowski M S, Eckstein J N and Bozovic I
1996 {\it Phys.\ Rev.\ B} {\bf 54} R6841.
\item
O'Donnell J, Onellion M, Rzchowski M S, Eckstein J N and Bozovic I
1997a {\it Phys.\ Rev.\ B} {\bf 55} 5873.
\item
O'Donnell J, Onellion M, Rzchowski M S, Eckstein J N and Bozovic I
1997b {\it J.\ Appl.\ Phys.} {\bf 81} 4961.
\item
O'Donnell J, Rzchowski M S, Eckstein J N and Bozovic I
1998 {\it Appl.\ Phys.\ Lett.} {\bf 72} 1775.
\item
O'Donnell J, Eckstein J N and Rzchowski M S
2000 {\it Appl.\ Phys.\ Lett.} {\bf 76} 218.
\item
Ogale S B, Ghosh K, Sharma R P, Greene R L, Ramesh R 
and Venkatesan T
1998 {\it Phys.\ Rev.\ B} {\bf 57} 7823.
\item
Ogale A S, Ogale S B, Ramesh R and Venkatesan T
1999 {\it Appl.\ Phys.\ Lett.} {\bf 75} 537.
\item
Okamoto J, Mizokawa T, Fujimori A, Hase I, Nohara M, Takagi H, Takeda Y and Takano M
1999 {\it Phys.\ Rev.\ B} {\bf 60} 2281.
\item
Okimoto Y, Katsufuji T, Ishikawa T, Arima T and Tokura Y
1997 {\it Phys.\ Rev.\ B} {\bf 55} 4206.
\item
Oliver M R, Kafalas J A, Dimmock J O and Reed T B
1970 {\it Phys.\ Rev.\ Lett.} {\bf 24} 1064.
\item
Oretzki M J and Gaunt P
1970 {\it Can.\ J.\ Phys.} {\bf 48} 346.
\item
Orozco A, Ogale S B, Li Y H, Fournier P, Li E, Asano H, Smolyaninova V,
Greene R L, Sharma R P, Ramesh R and Venkatesan T
1999 {\it Phys.\ Rev.\ Lett.} {\bf 83} 1680.
\item
Osofsky M S, Nadgorny B, Soulen R J, Broussard P, Rubinstein M, Byers J,
Laprade G, Mukovskii Y M, Shulyatev D and Arsenov A
1999 {\it J.\ Appl.\ Phys.} {\bf 85} 5567.
\item
Palstra T T M, Ramirez A P, Cheong S-W, Zegarski B R, Schiffer P and Zaanen J
1997 {\it Phys.\ Rev.\ B} {\bf 56} 5104.
\item
Panagiotopoulos I, Christides C, Moutis N, Pissas M and Niarchos D
1999a {\it J.\ Appl.\ Phys.} {\bf 85} 4913.
\item
Panagiotopoulos I, Christides C, Pissas M and Niarchos D
1999b {\it Phys.\ Rev.\ B} {\bf 60} 485.
\item
Paraskevopoulos D, Meservey R and Tedrow P M
1977 {\it Phys.\ Rev.\ B} {\bf 16} 4907.
\item
Park J-H, Tjeng L H, Allen J W, Metcalf P and Chen C T
1997 {\it Phys.\ Rev.\ B} {\bf 55} 12813.
\item
Park J-H, Vescovo E, Kim H-J, Kwon C, Ramesh R and Venkatesan T
1998a {\it Phys.\ Rev.\ Lett.} {\bf 81} 1953.
\item
Park J-H, Vescovo E, Kim H-J, Kwon C, Ramesh R and Venkatesan T
1998b {\it Nature (London)} {\bf 392} 794.
\item
Park M S, Kwon S K, Youn S J and Min B I
1999 {\it Phys.\ Rev.\ B} {\bf 59} 10018.
\item
Park S K, Ishikawa T and Tokura Y
1998 {\it Phys.\ Rev.\ B} {\bf 58} 3717.
\item
Patterson F, Moeller C and Ward R
1963 {\it Inorg.\ Chem.} {\bf 2} 196.
\item
P\'enicaud M, Siberchicot B, Sommers C B and K\"ubler J
1992 {\it J.\ Magn.\ Magn.\ Mater.} {\bf 103} 212.
\item
Penney T, Shafer M W and Torrance J B
1972 {\it Phys.\ Rev.\ B} {\bf 5} 3669.
\item 
Perring T G, Aeppli G, Moritomo Y and Tokura Y
1997 {\it Phys.\ Rev.\ Lett.} {\bf 78} 3197.
\item
Perring T G, Aeppli G, Kimura T, Tokura Y and Adams M A
1998 {\it Phys.\ Rev.\ B} {\bf 58} R14693.
\item
Petrov D K, Krusin-Elbaum L, Sun J Z, Feild C and Duncombe P R
1999 {\it Appl.\ Phys.\ Lett.} {\bf 75} 995.
\item
Philipp J B, H\"ofener C, Thienhaus S, Klein J, Alff L and Gross R
2000 {\it Phys.\ Rev.\ B} {\bf 62} R9248.
\item
Pickett W E and Singh D J
1996 {\it Phys.\ Rev.\ B} {\bf 53} 1146.
\item
Pietambaram S, Kumar D, Singh R K and Lee C B
2001 {\it Appl.\ Phys.\ Lett.} {\bf 78} 243.
\item
Pinsard-Gaudart L, Suryanarayanan R, Revcolevski A, Rodriguez-Carvajal
J, Greneche J M, Smith P A I, Thomas R M, Borges R P and Coey J M D
2000 {\it J.\ Appl.\ Phys.} {\bf 87} 7118.
\item
Plaskett T S, Freitas P P, Sun J J, Sousa R C, da Silva F F, Galvao T
T P, Pinho N M, Cardoso S, da Silva M F and Soares J C
1997 {\it Magnetic Ultrathin Films, Multilayers and Surfaces}
edited by Tobin J, Chambliss D, Kubinski D, Barmak K, Dederichs P, de
Jonge W, Katayama T and Schuhl A, MRS Symposia Proceedings No. 475
(Materials Research Society, Pittsburgh, 1997) 469.
\item
Platt C L, Dieny B and Berkowitz A E
1997 {\it J.\ Appl.\ Phys.} {\bf 81} 5523.
\item
Politzer B A and Cutler P H
1972 {\it Phys.\ Rev.\ Lett.} {\bf 28} 1330.
\item
Prellier W, Smolyaninova V, Biswas A, Galley C, Greene R L, Ramesha K
and Gopalakrishnan J
2000 {\it J.\ Phys.: Condens.\ Matter} {\bf 12} 965.
\item
Prinz G A
1998 {\it Science} {\bf 282} 1660.
\item
Przyslupski P, Kole\'snik S, Dynowska E, Sko\'skiewicz T and Sawicki M
1997 {\it IEEE Trans.\ Appl.\ Superconductivity} {\bf 7} 2192.
\item
Quijada M, $\rm \check{C}$erne J, Simpson J R, Drew H D, Ahn K H, Millis A J,
Shreekala R, Ramesh R, Rajeswari M and Venkatesan T
1998 {\it Phys.\ Rev.\ B} {\bf 58} 16093.
\item
Rajeswari M, Goyal A, Raychaudhuri A K, Robson M C, Xiong G C, Kwon C,
Ramesh R, Greene R L, Venkatesan T and Lakeou S
1996 {\it Appl.\ Phys.\ Lett.} {\bf 69} 851.
\item
Raju N P, Greedan J E and Subramaniam M A
1994 {\it Phys.\ Rev.\ B} {\bf 49} 1086.
\item
Ramirez A P
1997 {\it J.\ Phys.: Condens.\ Matter} {\bf 9} 8171.
\item
Ramirez A P and Subramanian M A
1997 {\it Science} {\bf 277} 546.
\item
Ramirez A P, Cava R J and Krajewski J
1997 {\it Nature (London)} {\bf 386} 156.
\item
Rampe A, Hartmann D, Weber W, Popovic S, Reese M and G\"untherodt G
1995 {\it Phys.\ Rev. B} {\bf 51} 3230.
\item
Raquet B, Coey J M D, Wirth S and von Moln\'ar S
1999 {\it Phys.\ Rev.\ B} {\bf 59} 12435.
\item
Ray S, Kumar A, Majumdar S, Sampathkumaran E V and Sarma D D
2001a {\it J.\ Phys.: Condens.\ Matter} {\bf 13} 607.
\item
Ray S, Kumar A, Sarma D D, Cimino R, Turchini S, Zennaro S and Zema N
2001b {\it Phys.\ Rev.\ B} {\bf 87} 097204.
\item
Raychaudhuri P, Taneja P, Sarkar S, Nigam A K, Ayyub P and Pinto R
1999 {\it Physica B} {\bf 259-261} 812.
\item
Ritter C, Ibarra M R, Morellon L, Blascot J, Garc{\'\i}a J and de Teresa
J M
2000 {\it J.\ Phys.: Condens.\ Matter} {\bf 12} 8295.
\item
Rivadulla F, L\'opez-Qunitela M A, Mira J and Rivas J
2001 {\it Phys.\ Rev.\ B} {\bf 64} 052403.
\item
Rodbell D S, Lommel J M and DeVries R C
1966 {\it J.\ Phys.\ Soc.\ Japan} {\bf 21} 2430.
\item
R\"oder H, Zang J and Bishop A R
1996 {\it Phys.\ Rev.\ Lett.} {\bf 76} 1356.
\item
Rodriguez-Martinez L M and Attfield J P
1996 {\it Phys.\ Rev.\ B} {\bf 54} R15622.
\item
Roy M, Mitchell J F, Ramirez A P and Schiffer P
1999 {\it J.\ Phys.: Condens. Matter} {\bf 11} 4843.
\item
Rozenberg E, Auslender M, Felner I and Gorodetsky G
2000 {\it J.\ Appl.\ Phys.} {\bf 88} 2578.
\item
R\"udiger U, Yu J, Zhang S, Kent A D and Parkin S S P
1998a {\it Phys.\ Rev.\ Lett.} {\bf 80} 5639.
\item
R\"udiger U, Yu S, Kent A D and Parkin S S P
1998b {\it Appl.\ Phys.\ Lett.} {\bf 73} 1298.
\item
R\"udiger U, Yu J, Thomas L, Parkin S S P and Kent A D
1999a {\it Phys.\ Rev.\ B} {\bf 59} 11914.
\item
R\"udiger U, Yu J, Parkin S S P and Kent A D
1999b {\it J.\ Magn.\ Magn.\ Mater.} {\bf 198-199} 261.
\item
Santi G and Jarlborg T
1997 {\it J.\ Phys.: Condens.\ Matter} {\bf 9} 9563.
\item
Sarma D D, Shanthi N, Barman S R, Hamada N, Sawada H and Terakura K
1995 {\it Phys.\ Rev.\ Lett.} {\bf 75} 1126.
\item
Sarma D D, Mahadevan P, Saha-Dasgupta T, Ray S and Kumar A
2000 {\it Phys.\ Rev.\ Lett.} {\bf 85} 2549.
\item
Satpathy S, Popovi\'c Z S and Vukajlovi\'c F R
1996 {\it Phys.\ Rev.\ Lett.} {\bf 76} 960.
\item
Schiffer P, Ramirez A P, Bao W and Cheong S-W
1995 {\it Phys.\ Rev.\ Lett.} {\bf 75} 3336.
\item
Schelp L F, Fert A, Fettar F, Holody P, Lee S F, 
Maurice J L, Petroff F and Vaur\`es A
1997 {\it Phys.\ Rev.\ B} {\bf 56} R5747.
\item
Sheng P, Abeles B and Arie Y
1973 {\it Phys.\ Rev.\ Lett.} {\bf 31} 44.
\item
Searle C W and Wang S T
1969 {\it Can.\ J.\ Phys.} {\bf 47} 2703.
\item
Searle C W and Wang S T
1970 {\it Can.\ J.\ Phys.} {\bf 48} 2023.
\item
Seneor P, Fert A, Maurice J-L, Montaigne F, Petroff F and Vaur\`es A
1999 {\it Appl.\ Phys.\ Lett.} {\bf 74} 4017.
\item
Shang C H, Nowak J, Jansen R and Moodera J S
1998 {\it Phys.\ Rev.\ B} {\bf 58} R2917.
\item
Sharma M, Wang S X and Nickel J
1999 {\it Phys.\ Rev.\ Lett.} {\bf 82} 616.
\item
Shengelaya A, Zhao G, Keller H and M\"uller K A
1996 {\it Phys.\ Rev.\ Lett.} {\bf 77} 5296.
\item
Shimakawa Y, Kubo Y and Manako T
1996 {\it Nature (London)} {\bf 379} 53.
\item
Shimakawa Y, Kubo Y, Manako T, Sushko Y V, Argyriou D N and Jorgensen J D
1997 {\it Phys.\ Rev.\ B} {\bf 55} 6399.
\item
Shimakawa Y, Kubo Y, Hamada N, Jorgensen J D, Hu Z, Short S, Nohara M
and Takagi H
1999 {\it Phys.\ Rev.\ B} {\bf 59} 1249.
\item
Shiozaki I, Hurd C M, McAlister S P, McKinnon W R and Strobel P
1981 {\it J.\ Phys.\ C: Solid State Phys.} {\bf 14} 4641.
\item
Simmons J G
1963 {\it J.\ Appl.\ Phys.} {\bf 34} 1793.
\item
Singh D J
1997 {\it Phys.\ Rev.\ B} {\bf 55} 313.
\item
Singley E J, Weber C P, Basov D N, Barry A and Coey J M D
1999 {\it Phys.\ Rev.\ B} {\bf 60} 4126.
\item
Sinkovi{\'c} B, Shekel E and Hulbert S L
1995 {\it Phys.\ Rev. B} {\bf 52} R8696.
\item
Sleight A W, Longo J M and Ward R
1962 {\it Inorg.\ Chem.} {\bf 1} 245.
\item
Sleight A W and Weiher J F
1972 {\it J.\ Phys.\ Chem.\ Solids} {\bf 33} 679.
\item
Slonczewski J C
1989 {\it Phys.\ Rev.\ B} {\bf 39} 6995.
\item
Slonczewski J C
1996 {\it J.\ Magn.\ Magn.\ Mater.} {\bf 159} L1.
\item
Snyder G J, Hiskes R, DiCarolis S, Beasley M R and Geballe T H
1996 {\it Phys.\ Rev.\ B} {\bf 53} 14434.
\item
Soh Y-A, Aeppli G, Mathur N D and Blamire M G 2000 {\it J.\ Appl.\
Phys.} {\bf 87} 6743.
\item
Soh Y-A, Aeppli G, Mathur N D and Blamire M G 2001 {\it Phys.\ Rev.\
B} {\bf 63} 020402.
\item
Soulen Jr R J, Byers J M, Osofsky M S, Nadgorny B, Ambrose T, Cheng S
F, Broussard P R, Tanaka C T, Nowak J, Moodera J S, Barry A and Coey J
M D
1998 {\it Science} {\bf 282} 85.
\item
Srinitiwarawong C and Ziese M
1998 {\it Appl.\ Phys.\ Lett.} {\bf 73} 1140.
\item
Srinitiwarawong C and Gehring G A
2001 {\it J.\ Phys.: Condens.\ Matter} {\bf 13} 7987.
\item
Stadler S, Idzerda Y U, Chen Z, Ogale S B and Venkatesan T
1999 {\it Appl.\ Phys.\ Lett.} {\bf 75} 3384.
\item
Stadler S, Idzerda Y U, Chen Z, Ogale S B and Venkatesan T
2000 {\it J.\ Appl.\ Phys.} {\bf 87} 6767.
\item
Stearns M B
1977 {\it J.\ Magn.\ Magn.\ Mater.} {\bf 5} 1062.
\item
Steenbeck K, Eick T, Kirsch K, O'Donnell K and Steinbei{\ss} E
1997 {\it Appl.\ Phys.\ Lett.} {\bf 71} 968.
\item
Steenbeck K, Eick T, Kirsch K, Schmidt H-G and Steinbei{\ss} E
1998 {\it Appl.\ Phys.\ Lett.} {\bf 73} 2506.
\item
Steinbei{\ss} E and Steenbeck K
1998 {\it Spin News, Newsletter of the Oxide Spin Electronics Network} 
{\bf 4} (Trinity College, Dublin) p.~3.
\item
Sternlieb B J, Hill J P, Wildgruber U C, Luke G M, Nachumi B,
Moritomo Y and Tokura Y
1996 {\it Phys.\ Rev.\ Lett.} {\bf 76} 2169.
\item
Su Y-S, Kaplan T A, Mahanti S D and Harrison J F
2000 {\it Phys.\ Rev.\ B} {\bf 61} 1324.
\item
Sun J Z, Gallagher W J, Duncombe P R, Krusin-Elbaum L, Altman R A,
Gupta A, Lu Y, Gong G Q and Xiao G
1996 {\it Appl.\ Phys.\ Lett.} {\bf 69} 3266.
\item
Sun J Z, Krusin-Elbaum L, Duncombe P R, Gupta A and Laibowitz R B
1997 {\it Appl.\ Phys.\ Lett.} {\bf 70} 1769.
\item
Sun J Z
1998 {\it Phil.\ Trans.\ R.\ Soc.\ Lond.\ A} {\bf 356} 1693.
\item
Sun J Z, Abraham D W, Roche K and Parkin S S P
1998 {\it Appl.\ Phys.\ Lett.} {\bf 73} 1008.
\item
Sun J Z
1999 {\it J.\ Magn.\ Magn.\ Mater.} {\bf 202} 157.
\item
Sun J Z, Abraham D W, Rao R A and Eom C B
1999 {\it Appl.\ Phys.\ Lett.} {\bf 74} 3017.
\item
Sun J Z
2000 {\it Phys.\ Rev.\ B} {\bf 62} 570.
\item
Suzuki K and Tedrow P M
1998 {\it Phys.\ Rev.\ B} {\bf 58} 11597.
\item
Suzuki K and Tedrow P M
1999 {\it Appl.\ Phys.\ Lett.} {\bf 74} 428.
\item
Suzuki Y, Wu Y, Yu J, Ruediger U, Kent A D, Nath T K and Eom C B
2000 {\it J.\ Appl.\ Phys.} {\bf 87} 6746.
\item
Takahashi S and Maekawa S
1998 {\it Phys.\ Rev.\ Lett.} {\bf 80} 1758.
\item
Taniyama T, Nakatani I, Namikawa T and Yamazaki Y
1999 {\it Phys.\ Rev.\ Lett.} {\bf 82} 2780.
\item
Taniyama T, Kitamoto Y and Yamazaki Y
2001 {\it J.\ Appl.\ Phys.} {\bf 89} 7693.
\item
Tatara G and Fukuyama H
1997 {\it Phys.\ Rev.\ Lett.} {\bf 78} 3773.
\item
Tatara G, Zhao Y-W, Mu\~noz M and Garc{\'\i}a N
1999 {\it Phys.\ Rev.\ Lett.} {\bf 83} 2030.
\item
Taylor G R, Isin A and Coleman R V
1968 {\it Phys.\ Rev.} {\bf 165} 621.
\item
Tedrow P M and Meservey R
1971 {\it Phys.\ Rev.\ Lett.} {\bf 26} 192.
\item
Tedrow P M and Meservey R
1973 {\it Phys.\ Rev.\ B} {\bf 7} 318.
\item
Theeuwen S J C H, Caro J, Schreurs K I, van Gorkom R P, Wellock K P,
Gribov N N, Radelaar S, Jungblut R M, Oepts W, Coehoorn R and Kozub V I
2001 {\it J.\ Appl.\ Phys.} {\bf 89} 4442.
\item
Thomson W
1857 {\it Proc.\ Roy.\ Soc. (London)} {\bf 8} 546.
\item
Todo S, Siratori K and Kimura S
1995 {\it J.\ Phys.\ Soc.\ Japan} {\bf 64} 2118.
\item
Tokura Y, Urushibara A, Moritomo Y, Arima T, Asamitsu A, Kido G and
Furukawa N
1994 {\it J.\ Phys.\ Soc.\ Japan} {\bf 63} 3931.
\item
Tokura Y and Tomioka Y
1999 {\it J.\ Magn.\ Magn.\ Mater.} {\bf 200} 1.
\item
Tomioka Y, Okuda T, Okimoto Y, Kumai R, Kobayashi K-I and Tokura Y
2000 {\it Phys.\ Rev.\ B} {\bf 61} 422.
\item
Tomioka Y, Asamitsu A and Tokura Y,
2001 {\it Phys.\ Rev.\ B} {\bf 63} 024421.
\item
Trajanovic Z, Kwon C, Robson M C, Kim K-C, Rajeswari M, Lofland S E,
Bhagat S M and Fork D
1996 {\it Appl.\ Phys.\ Lett.} {\bf 69} 1005.
\item
Troyanchuk I P, Khalyavin D D, Hervieu M, Maignan A, Michel C and Petrowski K
1998 {\it phys.\ stat.\ sol.\ (a)} {\bf 169} R1.
\item
Tsujioka T, Mizokawa T, Okamato J, Fujimori A, Nohara M, Takagi H, Yamaura K and Takano M
1997 {\it Phys.\ Rev.\ B} {\bf 56} R15509.
\item
Tsymbal E Yu and Pettifor D G
1997 {\it J.\ Phys.: Condens.\ Matter} {\bf 9} L411.
\item
Tsymbal E Yu and Pettifor D G
1998 {\it Phys.\ Rev.\ B} {\bf 58} 432.
\item
Uehara M, Mori S, Chen C H and Cheong S-W
1999 {\it Nature (London)} {\bf 399} 560.
\item
Uotani M, Taniyama T and Yamazaki Y
2000 {\it J.\ Appl.\ Phys.} {\bf 87} 5585.
\item
Upadhyay S K, Palanisami A, Louie R N and Buhrman R A
1998 {\it Phys.\ Rev.\ Lett.} {\bf 81} 3247.
\item
Urushibara A, Moritomo Y, Arima T, Asamitsu A, Kido G and Tokura Y
1995 {\it Phys.\ Rev.\ B} {\bf 51}, 14103.
\item
van den Brom W E and Volger J
1968 {\it Phys.\ Lett.} {\bf 26A} 197.
\item
van der Heijden P A A, Sw\"uste C H W, de Jonge W J M, Gaines J M, van
Eemeren J T W M and Shep K M
1999 {\it Phys.\ Rev.\ Lett.} {\bf 82} 1020.
\item
van der Zaag P J, Bloemen P J H, Gaines J M, Wolf R M, van der Heijden
P A A, van de Veerdonk R J M and de Jonge W J M
2000 {\it J.\ Magn.\ Magn.\ Mater.} {\bf 211} 301.
\item
van Gorkom R P, Brataas A and Bauer G E W
1999 {\it Phys.\ Rev.\ Lett.} {\bf 83} 4401.
\item
van Gorkom R P, Caro J, Theeuwen S J C H, Wellock K P, Gribov N N and
Radelaar S
1999 {\it Appl.\ Phys.\ Lett.} {\bf 74} 422.
\item
van Hoof J B A N, Schep K M, Brataas A, Bauer G E W and Kelly P J
1999 {\it Phys.\ Rev.\ B} {\bf 59} 138.
\item
Varma C M
1996 {\it Phys.\ Rev.\ B} {\bf 54} 7328.
\item
Vas'ko V A, Larkin V A, Kraus P A, Nikolaev K R, Grupp D E, Nordman C A and Goldman A M
1997 {\it Phys.\ Rev.\ Lett.} {\bf 78} 1134.
\item
Vas'ko V A, Nikolaev K R, Larkin V A, Kraus P A and Goldman A M
1998 {\it Appl.\ Phys.\ Lett.} {\bf 73} 844.
\item
Venimadhav A, Hegde M S, Prasad V and Subramanyam S V
2000 {\it J.\ Phys.\ D: Appl.\ Phys.} {\bf 33} 2921.
\item
Ventura C I and Gusm\~{a}o M A
2001 {\it cond-mat/0106319} preprint.
\item
Versluijs J J, Ott F and Coey J M D
1999 {\it Appl.\ Phys.\ Lett.} {\bf 75} 1152.
\item
Versluijs J J, Bari M, Ott F, Coey J M D and Revcolevschi A
2000 {\it J.\ Magn.\ Magn.\ Mater.} {\bf 211} 212.
\item
Versluijs J J, Bari M A and Coey J M D
2001 {\it Phys.\ Rev.\ Lett.} {\bf 87} 026601.
\item
Verwey E J W
1939 {\it Nature (London)} {\bf 144} 327.
\item
Viret M, Vignoles D, Cole D, Coey J M D, Allen W, Daniel D S and Gregg J F,
1996 {\it Phys.\ Rev.\ B} {\bf 53} 8464.
\item
Viret M, Drouet M, Nassar J, Contour J P, Fermon C and Fert A
1997a {\it Europhys.\ Lett.} {\bf 39} 545.
\item
Viret M, Ranno L and Coey J M D
1997b {\it Phys.\ Rev.\ B} {\bf 55} 8067.
\item
Viret M, Nassar J, Drouet M, Contour J P, Fermon C and Fert A
1999 {\it J.\ Magn.\ Magn.\ Mater.} {\bf 198-199} 1.
\item
Viret M, Samson Y, Warin P, Marty A, Ott F, S{\o}nderg{\aa}rd E, 
Klein O and Fermon C
2000 {\it Phys.\ Rev.\ Lett.} {\bf 85} 3962.
\item
Volger J
1954 {\it Physica} {\bf 20} 49.
\item
von Helmolt R, Wecker J, Holzapfel B, Schultz L and Samwer K
1993 {\it Phys.\ Rev.\ Lett.} {\bf 71} 2331.
\item
Wagner P, Gordon I, Trappeniers L, Vanacken J, Herlach F, 
Moshchalkov V V and Bruynseraede Y
1998 {\it Phys.\ Rev.\ Lett.} {\bf 81} 3980.
\item
Walter T, D\"orr K, M\"uller K-H, Holzapfel B, Eckert D, Wolf M,
Schl\"afer D, Schultz L and Gr\"otzschel R
1999 {\it Appl.\ Phys.\ Lett.} {\bf 74} 2218.
\item
Wang H S and Li Q
1998 {\it Appl.\ Phys.\ Lett.} {\bf 73} 2360.
\item
Wang L, Yin J, Huang S, Hunag X, Xu J, Liu Z and Chen K
1999 {\it Phys.\ Rev.\ B} {\bf 60} R6976.
\item
Wang X and Zhang X-G
1999 {\it Phys.\ Rev.\ Lett.} {\bf 82} 4276.
\item
Wang X L, Dou S X, Liu H K, Ionescu M and Zeimetz B
1998 {\it Appl.\ Phys.\ Lett.} {\bf 73} 396.
\item
Wegrowe J-E, Kelly D, Franck A, Gilbert S E and Ansermet J-Ph
1999 {\it Phys.\ Rev.\ Lett.} {\bf 82} 3681.
\item
Wegrowe J-E, Comment A, Jaccard Y, Ansermet J-Ph, Dempsey N M and Nozi\`eres J-P
2000 {\it Phys.\ Rev.\ B} {\bf 61} 12216.
\item
Wei J Y T, Yeh N-C and Vasquez R P
1997 {\it Phys.\ Rev.\ Lett.} {\bf 79} 5150.
\item
Wei J Y T, Yeh N-C, Vasquez R P and Gupta A
1998 {\it J.\ Appl.\ Phys.} {\bf 83} 7366.
\item
Wei{\ss}e A, Loos J and Fehske H
2001 {\it Phys.\ Rev.\ B} {\bf 64} 104413.
\item
West R N
1995 {\it Proc.\ Int.\ School of Physics, `Enrico Fermi'-
Positron Spectroscopy of Solids}
edited by Dupasquier A and Mills A P (IOS Press, Amsterdam).
\item
Westerburg W, Martin F, Friedrich S, Maier M and Jakob G
1999 {\it J.\ Appl.\ Phys.} {\bf 86} 2173.
\item
Westerburg W, Reisinger D and Jakob G
2000 {\it Phys.\ Rev.\ B} {\bf 62} R767.
\item
Wiesmann H, Gurvitch M, Lutz H, Ghosh A, Schwartz B, Strongin M, Allen
P B and Halley J W
1977 {\it Phys.\ Rev.\ Lett.} {\bf 38} 782.
\item
Wolfman J, Prellier W, Simon Ch and Mercey B
1998 {\it J.\ Appl.\ Phys.} {\bf 83} 7186.
\item
Wollan E O and Koehler W C
1955 {\it Phys.\ Rev.} {\bf 100} 545.
\item
Worledge D C and Geballe T H
2000a {\it Appl.\ Phys.\ Lett.} {\bf 76} 900.
\item
Worledge D C and Geballe T H
2000b {\it Phys.\ Rev.\ Lett.} {\bf 85} 5182.
\item
Wu Y, Suzuki Y, R\"udiger U, Yu J, Kent A D, Nath T K and Eom C B
1999 {\it Appl.\ Phys.\ Lett.} {\bf 75} 2295.
\item
Xu R, Husmann A, Rosenbaum T F, Saboungi M-L, Enderby E J and
Littlewood P B
1997 {\it Nature (London)} {\bf 390} 57.
\item
Yamada Y, Hino O, Nohdo S, Kanao R, Inami T and Katano S
1995 {\it Phys.\ Rev.\ Lett.} {\bf 77} 904 (1996).
\item
Yamaguchi S, Taniguchi H, Takagi H, Arima T and Tokura Y
1995 {\it J.\ Phys.\ Soc.\ Japan} {\bf 64} 1885.
\item
Yamanaka M and Nagaosa N
1996 {\it J.\ Phys.\ Soc.\ Japan} {\bf 65} 3088.
\item
Yamamoto R, Moritomo Y and Nakamura A
2000 {\it Phys.\ Rev.\ B} {\bf 61} R5062.
\item
Yanase A and Siratori K
1984 {\it J.\ Phys.\ Soc.\ Japan} {\bf 53} 312.
\item
Yanase A and Hamada N
1999 {\it J.\ Phys.\ Soc.\ Japan} {\bf 68} 1607.
\item
Yang Z, Tan S, Chen Z and Zhang Y
2000 {\it Phys.\ Rev.\ B} {\bf 62} 13872.
\item
Yeh N-C, Vasquez R P, Fu C C, Samoilov A V, Li Y and Vakili K
1999 {\it Phys.\ Rev.\ B} {\bf 60} 10522.
\item
Yin H Q, Zhou J-S, Zhou J-P, Dass R, McDevitt J T and Goodenough J B
1999 {\it Appl.\ Phys.\ Lett.} {\bf 75} 2812.
\item
Yin H Q, Zhou J-S and Goodenough J B
2000 {\it Appl.\ Phys.\ Lett.} {\bf 77} 714.
\item
Yoon S, Liu H L, Schollerer G, Cooper S L, Han P D, Payne D A, Cheong
S-W and Fisk Z
1998 {\it Phys.\ Rev.\ B} {\bf 58} 2795.
\item
Yuan C L, Wang S G, Song W H, Yu T, Dai J M, Ye S L and Sun Y P
1999 {\it Appl.\ Phys.\ Lett.} {\bf 75} 3853.
\item
Zandbergen H W, Freisem S, Nojima T and Aarts J
1999 {\it Phys.\ Rev.\ B} {\bf 60} 10259.
\item
Zang J, Bishop A R and R\"oder H
1996 {\it Phys.\ Rev.\ B} {\bf 53} R8840.
\item
Zener C
1951 {\it Phys.\ Rev.} {\bf 82} 403.
\item
Zeng Z, Greenblatt M and Croft M
1998 {\it Phys.\ Rev.\ B} {\bf 58} R595.
\item
Zeng Z, Greenblatt M, Subramanian M A and Croft M
1999 {\it Phys.\ Rev.\ Lett.} {\bf 82} 3164.
\item
Zhang J and White R M
1998 {\it J.\ Appl.\ Phys.} {\bf 83} 6512.
\item
Zhang J, Dai P, Fernandez-Baca J A, Plummer E W, Tomioka Y 
and Tokura Y
2001 {\it Phys.\ Rev.\ Lett.} {\bf 86} 3823.
\item
Zhang S and Yang Z
1996 {\it J.\ Appl.\ Phys.} {\bf 79} 7398.
\item
Zhang S, Levy P M, Marley A C and Parkin S S P
1997a {\it Phys.\ Rev.\ Lett.} {\bf 79} 3744.
\item
Zhang N, Ding W, Zhong W, Xing D and Du Y
1997b {\it Phys.\ Rev.\ B} {\bf 56} 8138.
\item
Zhao G, Hunt M B and Keller H
1997 {\it Phys.\ Rev.\ Lett.} {\bf 78} 955.
\item
Zhao G, Smolyaninova V, Prellier W and Keller H
2000a {\it Phys.\ Rev.\ Lett.} {\bf 84} 6086.
\item
Zhao G, Wang Y S, Kang D J, Prellier W, Rajeswari M, Keller H,
Venkatesan T, Chu C W and Green R L
2000b {\it Phys.\ Rev.\ B} {\bf 62} R11949.
\item
Zhao G, Keller H, Prellier W and Kang D J
2001 {\it Phys.\ Rev.\ B} {\bf 63} 172411.
\item
Zhou J P, McDevitt J T, Zhou J-S, Yin H Q, Goodenough J B, Gim Y and
Jia Q X
1999 {\it Appl.\ Phys.\ Lett.} {\bf 75} 1146.
\item
Zhou J-S, Goodenough J B, Asamitsu A and Tokura Y
1997 {\it Phys.\ Rev.\ Lett.} {\bf 79} 3234.
\item
Zhou J-S and Goodenough J B
1998 {\it Phys.\ Rev.\ Lett.} {\bf 80} 2665.
\item
Zhou J-S and Goodenough J B
1999 {\it Phys.\ Rev.\ B} {\bf 60} R15002.
\item
Zhou J-S and Goodenough J B
2000 {\it Phys.\ Rev.\ B} {\bf 62} 3834.
\item
Zhou J-S and Goodenough J B
2001 {\it Phys.\ Rev.\ B} {\bf 64} 024421.
\item
Zhou J-S, Yin H Q and Goodenough J B
2001 {\it Phys.\ Rev.\ B} {\bf 63} 184423.
\item
Zhu J-X, Friedman B and Ting C S
1999 {\it Phys.\ Rev.\ B} {\bf 59} 9558.
\item
Zhu T, Shen B G, Sun J R, Zhao H W and Zhan W S
2001 {\it Appl.\ Phys.\ Lett.} {\bf 78} 3863.
\item
Ziese M, Sena S P, Shearwood C, Blythe H J, Gibbs M R J and Gehring G A
1998a {\it Phys.\ Rev.\ B} {\bf 57} 2963.
\item
Ziese M, Srinitiwarawong C and Shearwood C,
1998b {\it J.\ Phys.: Condens. Matter} {\bf 10} L659.
\item
Ziese M and Sena S P
1998 {\it J.\ Phys.: Condens. Matter} {\bf 10} 2727.
\item
Ziese M and Srinitiwarawong C
1998 {\it Phys.\ Rev.\ B} {\bf 58} 11519.
\item 
Ziese M
1999 {\it Phys.\ Rev.\ B} {\bf 60} R738.
\item
Ziese M and Srinitiwarawong C
1999 {\it Europhys.\ Lett.} {\bf 45} 256 .
\item
Ziese M, Heydon G, H\"ohne R, Esquinazi P and Dienelt J
1999a {\it Appl.\ Phys.\ Lett.} {\bf 74} 1481.
\item
Ziese M, Sena S P and Blythe H J
1999b {\it J.\ Magn.\ Magn.\ Mater.} {\bf 202} 292.
\item
Ziese M, Sena S P and Blythe H J
1999c {\it J.\ Appl.\ Phys.} submitted.
\item
Ziese M
2000a {\it Phil.\ Trans.\ R.\ Soc.\ Lond.\ A} {\bf 358} 137.
\item
Ziese M
2000b {\it Phys.\ Rev.\ B} {\bf 62} 1044.
\item
Ziese M and Blythe H J
2000 {\it J.\ Phys.: Condens. Matter} {\bf 12} 13.
\item
Ziese M
2001a {\it J.\ Phys.: Condens. Matter} {\bf 13} 2919.
\item
Ziese M
2001b {\it phys.\ stat.\ sol.\ (b)} {\bf 228} R1.
\item
Ziese M and Thornton M J (Eds.)
2001 {\it Spin Electronics} (Springer, Heidelberg).
\endrefs
\end{document}